\documentclass[binding=0.6cm]{sapthesis}

\usepackage{graphicx}
\usepackage{bm}
\usepackage{xcolor}
\usepackage{booktabs}
\PassOptionsToPackage{breaklinks}{hyperref}
\usepackage[linktocpage,colorlinks=true, linkcolor=bluscuro,urlcolor=bluscuro,citecolor=bluscuro]{hyperref}
\definecolor{bluscuro}{rgb}{0.15, 0.2, .85}
\usepackage{amsmath, amssymb}
\usepackage{slashed}
\usepackage{latexsym}
\usepackage{braket}
\usepackage[export]{adjustbox}
\usepackage{color}
\usepackage{multirow}
\usepackage{cite}
\usepackage{xspace}
\usepackage{comment}
\usepackage[caption=false]{subfig}
\usepackage{cancel}
\usepackage{url}
\usepackage{nicematrix}
\usepackage{needspace}
\hypersetup{pdftitle={PHD_Thesis},pdfauthor={Claudio Gambino}}

\title{From Quantum Amplitudes to Spacetime Geometry: a Multipolar Framework for Black Hole Signatures}
\author{Claudio Gambino}
\IDnumber{1793203}
\course{Dottorato di Ricerca in Fisica}
\courseorganizer{Scuola di dottorato in Scienze astronomiche, chimiche, fisiche, matematiche e della terra ``Vito Volterra''}
\cycle{XXXVIII}
\AcademicYear{2024/2025}
\advisor{Prof. Fabio Riccioni}
\authoremail{claudio.gambino@uniroma1.it}
\copyyear{2025}
\thesistype{PhD thesis}
\examdate{20/01/2026}
\examiner[]{Prof. Leonardo Gualtieri}
\examiner[]{Prof. Andrea Caputo}
\examiner[]{Prof. Luca Di Luzio}

\begin{document}

\frontmatter
\maketitle

\begin{abstract}
This thesis develops a unified framework that reconstructs the full classical content of General Relativity from the classical limit of quantum scattering amplitudes. By interpreting the analytic structure of amplitudes as the field-theoretic imprint of spacetime geometry, the work establishes a direct correspondence between quantum processes and classical gravitational observables such as metrics, deflection angles, and multipole moments.
Starting from the effective-field-theory description of gravity, the thesis shows that loop amplitudes encode not only quantum corrections but also the nonlinear classical self-interaction of the gravitational field, enabling the systematic derivation of the post-Minkowskian expansion of gravitational quantities by rewriting the Einstein equations in terms of graviton scattering processes. Building upon this foundation, the framework is applied to rotating and charged sources in arbitrary spacetime dimensions. Scattering amplitudes of massive spinning fields are used to reconstruct the metrics of Kerr, Kerr–Newman and Myers–Perry black holes, leading to the discovery of higher-dimensional stress multipoles and to an amplitude-based derivation of the universal gyromagnetic factor of charged solutions in higher dimensions. 
A momentum-space formulation of the energy–momentum tensor is then developed, introducing gravitational form factors and source multipoles that link, for the first time, the internal matter distribution to the external multipolar field in a completely relativistic framework. Furthermore, the thesis completes the transition from the microscopic amplitude picture to the macroscopic description of gravitational sources by engineering a multipole-based framework for black hole mimickers, then applied to build horizon-less compact objects mimicking the multipolar structure of Kerr black holes. 
Finally, exploiting the Kerr–Schild gauge, the Fourier transforms of rotating black hole metrics are computed in closed form, bridging perturbative and non-perturbative descriptions of gravity, and allowing to probe the multipolar structure of higher-dimensional solutions employing scattering amplitudes. 
\end{abstract}

\begin{acknowledgments}
I would like to express my deepest gratitude to Fabio Riccioni, my supervisor, for his guidance, support, and invaluable advice throughout these years. I am also sincerely thankful to Massimo Bianchi and Paolo Pani for their constant suggestions, encouragement, and for many stimulating discussions that greatly contributed to this thesis. I would like to thank Roberto Emparan for his hospitality during my stay at the Universitat de Barcelona/ICCUB, and Diego Blas for hosting me at IFAE (Institut de Física d’Altes Energies) and for many enlightening exchanges. Finally, I warmly thank the Physics Department of the University of Rome ``La Sapienza'' for providing a stimulating and inspiring environment that made this Ph.D. possible.
\end{acknowledgments}

\tableofcontents
\mainmatter

\chapter{Introduction}\label{chapter:Introduction}

General Relativity (GR) stands as one of the most profound and successful achievements of modern physics. By identifying gravity with the curvature of spacetime, it provides a unified geometrical framework in which matter and energy determine the structure of the Universe through Einstein’s field equations. Over the past century, GR has passed every experimental test with remarkable success, from the precession of Mercury to the deflection of light by the Sun and the timing of binary pulsars. 
In the last decade, however, the theory has entered a new era of empirical validation, namely the direct detection of gravitational waves by the LIGO-Virgo-KAGRA collaboration~\cite{LIGOScientific:2016aoc, LIGOScientific:2017vwq, LIGOScientific:2019fpa,KAGRA:2020tym,LIGOScientific:2025slb} and the imaging of supermassive black holes by the Event Horizon Telescope~\cite{EventHorizonTelescope:2019dse, EventHorizonTelescope:2022wkp, EventHorizonTelescope:2022xqj} have provided direct access to the strong-field regime of gravity. 
These breakthroughs not only confirmed the robustness of Einstein’s theory, but also opened a window onto a new frontier: the nonlinear, dynamical domain where gravity, electromagnetism, and rotation intertwine, and where the fundamental nature of spacetime can be probed observationally.

Yet, despite its empirical success, GR remains a classical theory that does not incorporate quantum effects. Attempts to quantize gravity as a gauge field theory, in which the interaction is mediated by a massless spin-2 boson, the graviton, lead to a non-renormalizable theory~\cite{Feynman:1963ax, DeWitt:1967yk, DeWitt:1967ub, DeWitt:1967uc, tHooft:1974toh}. Nevertheless, following the principles of effective field theories (EFTs), GR can be viewed as the leading term in a low-energy expansion of a yet-unknown quantum theory of gravity~\cite{Donoghue:1993eb, Donoghue:1994dn, Burgess:2003jk, Bjerrum-Bohr:2002fji}. In this framework, higher-derivative operators suppressed by the Planck mass encode ultraviolet corrections, while the Einstein–Hilbert term captures the infrared dynamics. This EFT perspective allows one to compute gravitational interactions perturbatively, with quantum loops generating both genuine quantum corrections (proportional to powers of $\hbar$) and purely classical terms that survive in the $\hbar\to 0$ limit~\cite{Iwasaki:1971vb, Holstein:2004dn, Donoghue:1996mt, Bjerrum-Bohr:2018xdl, Kosower:2018adc, Cheung:2018wkq, Guevara:2017csg}. The latter correspond to the post-Minkowskian (PM) expansion of the classical gravitational field in powers of Newton’s constant $G_N$, so that classical observables, such as scattering angles, time delays, or even metrics, can be reconstructed from scattering amplitudes without solving Einstein’s equations explicitly~\cite{Bjerrum-Bohr:2002fji, Bern:2019nnu, Parra-Martinez:2020dzs, Kalin:2020mvi, Cheung:2020gyp}. This feature, although conceptually striking, is a direct consequence of the long-range nature and self-interacting structure of gravity. In gauge theories like electromagnetism, the linearity of the classical field equations implies that the electromagnetic field sourced by a given conserved current is completely determined at tree level. Surely, this does not prevent loop amplitudes from giving classical contributions to composite quantities, like in the example of~\cite{Holstein:2004dn}. In gravity, however, the situation is qualitatively different. In fact the graviton couples universally to energy, including its own stress-energy, and each additional graviton exchange contributes to the nonlinear structure of the classical field. As a result, multi-loop diagrams can generate $\hbar$-independent terms corresponding to successive orders in the PM expansion of GR. This mechanism, first clarified in the seminal works of Duff~\cite{Duff:1973zz} and later by Donoghue~\cite{Donoghue:1994dn}, explains why classical observables, such as the Schwarzschild or Kerr metrics, can be systematically recovered from loop amplitudes once the $\hbar$-scaling of the external momenta and propagators is properly accounted for. The presence of classical physics in loop amplitudes thus reflects the self-coupled nature of the gravitational field and underlies the possibility of reconstructing nonlinear spacetime geometries from perturbative quantum scattering processes.

This modern amplitude-based viewpoint has revolutionized the study of classical gravity. 
Through the use of unitarity methods, on-shell recursion relations, and the double-copy correspondence between gauge and gravitational theories~\cite{Bern:2008qj, Bern:2010ue, Bern:2019prr}, scattering amplitudes have become a universal language connecting quantum field theory (QFT) and general relativity. In this language, the $S$-matrix exponentiates within the eikonal approximation, allowing the eikonal phase to be interpreted as the classical action governing the scattering process~\cite{Amati:1987uf, Bjerrum-Bohr:2018xdl, Kosower:2018adc}. Consequently, the two-body problem in GR can be reformulated in terms of quantum scattering observables, with the loop order in the amplitude corresponding to PM order in the classical expansion. Such a correspondence has been crucial for deriving PM corrections to the dynamics of compact binaries and for improving waveform models used in gravitational-wave astronomy~\cite{Buonanno:2022pgc, Bjerrum-Bohr:2022blt, Nagar:2018zoe, Varma:2018mmi, Gamba:2021ydi}. Indeed, as the sensitivity of present and future detectors~\cite{Saleem:2021iwi, Reitze:2019iox} increases, so does the need for more precise analytical predictions, making amplitude methods a powerful complement to numerical relativity. An important addressed focus in this program concerns the inclusion of spin and internal structure. Astrophysical objects are not featureless, and they possess angular momentum, magnetic fields, and higher-order multipole moments. In the amplitude framework, these features arise from the off-shell tensorial structure of the vertices describing the interaction between massive spinning particles and gravitons~\cite{Porto:2010tr, Levi:2015msa, Blanchet:2013haa, Chung:2018kqs, Guevara:2018wpp}. 
A key observation is that the spin-dependent terms in the scattering amplitude directly map to the multipole moments of the corresponding spacetime geometry, allowing one to reconstruct rotating solutions such as the Kerr metric and its generalizations from purely QFT data~\cite{Donoghue:2001qc, DOnofrio:2022cvn, Gambino:2022kvb, Gambino:2024uge}. 
From this perspective, classical spacetimes emerge as collective states of the gravitational $S$-matrix, with their multipolar structure encoded in the analytic properties of the amplitude. This idea lies at the heart of the work developed in this thesis.

The study of rotating and charged black holes (BHs) in this context is particularly compelling. 
BHs occupy a unique position in modern theoretical physics, as they are classical solutions of GR with thermodynamic properties~\cite{Bardeen:1973gs, Hawking:1974rv}, they saturate the limits of causal structure, and they provide a laboratory for testing quantum-gravitational ideas. The Schwarzschild, Kerr, Reissner–Nordström, and Kerr–Newman families describe all stationary, asymptotically flat BHs in four dimensions~\cite{Schwarzschild:1916uq, Reissner:1916cle, Kerr:1963ud, Newman:1965my}, according to the celebrated uniqueness theorems~\cite{Israel:1967wq, Carter:1971zc, Hawking:1973uf, Robinson:1975bv, Chrusciel:2012jk}. 
However, these theorems no longer hold in higher dimensions, where additional families of solutions such as black rings and black saturns appear~\cite{Emparan:2008eg, Elvang:2007rd}. 
The investigation of such spacetimes provides a valuable testing ground for the amplitude program, while also illuminating how properties like the no-hair theorem, multipolar structure, and thermodynamics generalize beyond four dimensions. At the same time, the emergence of regular or horizonless configurations that mimic BH properties has become an area of intense research~\cite{Giddings:1992hh, Lunin:2001jy, Mathur:2005zp, Hayward:2005gi, Bena:2005va, Mathur:2008nj, Frolov:2016pav, Cardoso:2019rvt, Carballo-Rubio:2018pmi, Simpson:2019cer, Bianchi:2020bxa, Mazza:2021rgq, Bianchi:2020miz, Carballo-Rubio:2025fnc}. 
These so-called black hole mimickers reproduce the far-field gravitational structure of Kerr or Schwarzschild while differing at short distances, potentially resolving curvature singularities or avoiding horizons altogether~\cite{Cardoso:2017cqb, Mark:2017dnq, Mazza:2021rgq, Casadio:2024lgw}. 
Such models can also have observable implications in gravitational-wave or electromagnetic signatures~\cite{Abedi:2016hgu, Jiang:2021ajk, Shaikh:2022ivr, Bambi:2025wjx}. A systematic understanding of their construction within the amplitude framework offers new insight into the boundary between classical and quantum gravity and into the physical content of the multipole expansion itself.

This thesis then is devoted to developing a unified, amplitude-based framework capable of connecting all these aspects. Its overarching goal is to show that the full classical phenomenology of GR, ranging from Schwarzschild and Kerr solutions to their charged, rotating, higher-dimensional, and even horizonless extensions, can be consistently reconstructed from the classical limit of quantum scattering amplitudes. The analysis evolves from perturbative to non-perturbative regimes, from minimal to non-minimal couplings, and from singular to regular sources, culminating in a comprehensive picture of how spacetime geometry emerges from quantum field–theoretic principles.

Focusing on constructing a comprehensive and systematic framework that recovers GR metrics from the classical limit of three-point scattering processes, the work extends the amplitude-based description of gravity, traditionally applied to static, spherically symmetric sources, to the far broader and physically richer domain of rotating, charged, and higher-dimensional configurations. A central focus is placed on understanding the gravitational multipolar structure of spacetime and on developing a momentum-space formalism capable of capturing these multipoles directly from the analytic properties of scattering amplitudes. 
The guiding idea is that the information encoded in amplitudes is not limited to particle interactions in the quantum regime, but it already contains, in its classical limit, the full geometrical content of gravity. By exploiting this correspondence, the thesis reformulates GR in a language that is entirely field-theoretic, where curved spacetime metrics, deflection angles, and multipole moments arise as emergent quantities from the analytic structure of the $S$-matrix. This approach provides a unified perspective on how spacetime geometry originates from quantum interactions and offers a consistent way to organize classical observables within the same formalism used to describe fundamental scattering processes. Ultimately, this line of research aims not only to reproduce known gravitational solutions within the amplitude framework, but also to gain conceptual insights into the self-interacting and multipolar nature of gravity, and to explore whether the tools of QFT can reveal new aspects of the classical gravitational dynamics across dimensions.

The first step of the thesis is to establish a precise correspondence between scattering amplitudes and classical gravitational fields. 
While most of the literature deals with on-shell processes, we adopt an off-shell formulation of the EFT of gravity and analyze the interaction between massive fields and gravitons from this broader perspective. This approach is essential when one aims to reconstruct gauge-dependent quantities, such as spacetime metrics, since it allows one to separate the physical content of the theory from pure gauge artifacts in a controlled and systematic manner. After reviewing the quantization of gravity within the EFT framework, we connect the classical limit of graviton-emission amplitudes to the conserved gravitational current of the theory, namely the energy-momentum tensor (EMT), which enters Einstein’s equations as the source of the metric. We then show that the PM expansion of a generic geometry naturally emerges from the low-energy limit of such amplitudes. This construction provides a systematic dictionary that relates the PM expansion of the EMT, and hence of the metric itself, to the classical limit of the loop amplitudes in the underlying EFT. Establishing this amplitude–geometry correspondence forms the theoretical foundation upon which the remainder of the thesis is built.

We therefore develop the general formalism and computational tools required for all subsequent analyses. We show that recovering classical gravity from quantum amplitudes is most effective when the classical limit is implemented from the outset. 
To achieve this, we define a \emph{dressed vertex}, the classical analogue of the massive Feynman vertex, which directly couples the external massive line to the graviton field. Once this ingredient is introduced, the topology of the diagrams contributing to the classical sector becomes manifest, consisting of loop configurations with a single massive line and internal tree graviton exchanges. We then outline the consistent procedure for evaluating loop amplitudes in this limit and identify a family of master integrals that govern the structure of the resulting metric. The successive terms of these integrals reconstruct the multipolar expansion of the spacetime, thereby providing an explicit, field-theoretic derivation of how the PM metric, and its spin-dependent multipoles, arise from the corresponding quantum scattering amplitudes.

After establishing the general correspondence between scattering amplitudes and classical gravitational fields, we turn our attention to the description of rotating configurations and the inclusion of spin-induced effects in arbitrary spacetime dimensions. Rotating sources represent the most natural and physically relevant generalization of spherically symmetric objects, and their study provides the first genuine test of the amplitude–geometry correspondence beyond the monopolar sector. To this end, we consider the explicit example of a massive spin-1 field interacting with gravity and construct the associated scattering amplitudes in the PM expansion~\cite{Gambino:2024uge}. The choice of a vector field is particularly meaningful, as it constitutes the simplest matter configuration whose EMT carries a non-trivial dependence on the intrinsic angular momentum of the source. Indeed, from a GR standpoint, the first physically non-trivial effects of rotation appear at the level of the quadrupole moment, and therefore this setup represents the minimal scenario in which deviations from universal results can be investigated in a fully field-theoretic framework.

The amplitude construction proceeds by dressing the massive spin-1 line with the appropriate gravitational vertices and evaluating the diagrams contributing up to quadratic order in the spin tensor. By suitably Fourier transforming the resulting amplitudes, we then extract the classical geometry generated by the rotating source. In four spacetime dimensions, the obtained expressions reproduce the weak-field limit of the Hartle–Thorne metric, which describes the exterior spacetime of a generic axisymmetric body up to second order in the angular momentum, as well as its Kerr black hole limit. When extended to higher dimensions, for some choice of the parameters involved, the same computation recovers the PM expansion of the Myers–Perry family of black hole solutions. This explicit correspondence confirms that the information about rotation and multipole structure is entirely contained in the analytic dependence of the amplitude on the spin degrees of freedom. A crucial ingredient in this analysis is the precise identification of gravitational multipoles in arbitrary dimensions. 
In higher-dimensional spacetimes, the tensorial decomposition of the metric reveals additional structures that have no counterpart in four dimensions. We clarify these definitions and show that, besides the familiar mass and current multipoles, a new hierarchy of moments naturally emerges, which we denote as \emph{stress multipoles}. These additional components characterize the spatial part of the metric and reflect the richer geometry allowed by the extra spatial dimensions. Their appearance provides the first amplitude-based derivation of higher-dimensional multipole moments and extends the Thorne formalism~\cite{Thorne:1980ru} to generic dimensions.

Within this amplitude-based picture, spin effects manifest themselves through specific tensorial structures in the matter–graviton vertices, whose coefficients directly determine the quadrupolar and higher-order multipole moments of the resulting spacetime. The framework allows one to distinguish between minimal and non-minimal couplings, where in the former case, the interaction reproduces the Kerr geometry, corresponding to the unique axisymmetric, asymptotically flat vacuum solution consistent with regularity at the horizon. In the latter, additional higher-derivative operators modify the multipole spectrum, capturing more general rotating configurations that deviate from the BH no-hair relations. The analysis carried out in this context, thus provides a complete field-theoretic characterization of the gravitational field of spinning bodies and establishes the foundations for the multipolar and momentum-space constructions developed in the rest of the thesis.

The next stage of the analysis consists of incorporating electromagnetic interactions into the amplitude framework.  This generalization is essential for describing charged and rotating configurations and for understanding how gravity and electromagnetism combine in a unified field-theoretic language. While in four spacetime dimensions the Kerr–Newman metric represents the most general stationary solution of the Einstein–Maxwell equations, in higher dimensions no exact charged rotating solutions are known in closed form. We try to address this problem by studying the scattering of massive, electrically charged spinning sources within the same EFT of gravity developed in the previous chapters~\cite{Gambino:2025iyx}. Starting from the minimal coupling of the charged field to both the graviton and the photon, we construct the corresponding tree- and loop-level amplitudes in arbitrary dimension. The computation reveals how gravitational and electromagnetic exchanges coexist in the same diagrammatic expansion and how their interference encodes the full nonlinear content of the Einstein–Maxwell system. By taking the classical limit of these amplitudes, we derive the PM expansion of the metric and of the electromagnetic potential generated by a rotating charged source, recovering in four dimensions the expected Kerr–Newman geometry.  
This analysis naturally leads to the definition of a dimension-dependent gyromagnetic factor, the dimensionless ratio between magnetic dipole moment and angular momentum, which emerges directly from the same three-point interaction vertex that couples the massive line to both graviton and photon fields.  

Remarkably, the amplitude computation shows that in four dimensions the minimal coupling reproduces the classical Kerr–Newman value $\mathfrak{g}=2$, while in spacetime dimensions $D>4$ the result becomes $\mathfrak{g}=(d-1)/(d-2)$. To reproduce the correct gyromagnetic structure then, one must extend the action by introducing a Pauli-type non-minimal coupling, involving a direct contraction between the electromagnetic field strength and the spin tensor of the source.  
In the amplitude picture, this term adds an extra tensorial component to the massive–photon vertex, whose coefficient can be fixed by matching to the four-dimensional case or by enforcing the correct multipolar fall-off of the electromagnetic field.  
Once included, this interaction restores the expected gyromagnetic behavior in any dimension and ensures consistency with the underlying multipolar expansion derived in the gravitational sector. Interestingly, the amplitude framework also provides a natural interpretation of higher-dimensional charged solutions obtained through additional gauge interactions. In particular, the inclusion of a specific Chern–Simons term, proper of the five-dimensional supergravity, leads to an exact charged rotating solution known as the Chong–Cvetič–Lü–Pope metric~\cite{Chong:2005hr}.  
From the amplitude viewpoint, such a term corresponds to adding a parity-violating cubic vertex involving one graviton and two photons, whose contribution modifies the three-point amplitude. We discuss how this structure arises naturally in the EFT description and how it connects with the amplitude-based computation of the gyromagnetic factor, thereby linking our general formalism to known higher-dimensional supergravity solutions. Beyond reproducing known results, this chapter emphasizes the conceptual unity between gravitational and electromagnetic interactions within the amplitude framework.  
Both forces emerge from the same analytic structure of the $S$-matrix, and their classical limits can be derived in parallel through the same PM expansion. 

Having established the amplitude formulation of gravitational interactions, the next step is to recast the information encoded in the scattering amplitudes into a universal description of gravitational sources. We then develop this idea by defining the most general EMT of a spinning source as a covariant expansion in the spin tensor, with each term in the series corresponding to a definite gravitational multipole~\cite{Bianchi:2024shc}.  
This construction is deeply rooted in the momentum-space formulation of GR and takes full advantage of the analytic structure of the amplitude, which naturally organizes the multipolar hierarchy in compact and manifestly covariant expressions.  
In this way, the infinite tower of spin-induced moments, mass, current, and stress, can be expressed in closed form, providing a concise field-theoretic representation of the gravitational content of an arbitrary source. A central outcome of this formalism is the identification of physical, gauge-invariant combinations associated with measurable multipoles, and their clean separation from purely local or gauge-dependent terms. This distinction is crucial, since it allows one to unambiguously extract observable quantities directly from momentum-space quantities without referring to a particular coordinate system or gauge choice. Within this framework, we introduce two key ingredients, namely the \emph{form factors} and the \emph{structure functions}. The form factors encode the asymptotic multipolar structure of the spacetime and thus determine the long-range behavior of the gravitational field, while the structure functions describe the internal distribution of energy and angular momentum that gives rise to those multipoles. Together, these functions provide a relativistic and dimension-independent generalization of the classical Newtonian multipole expansion, elevating it into a fully covariant setting. This formalism also clarifies that distinct matter configurations, differing in their local stress–energy distributions, can nonetheless generate identical external multipoles, mirroring a familiar property of electromagnetic systems.

In this regard, an important conceptual advance is the definition of \emph{source multipoles}, namely the explicit mapping between the intrinsic properties of the matter distribution and the multipolar structure of the resulting gravitational field. Indeed, in GR, multipole moments are traditionally defined as asymptotic quantities, extracted from the far-field expansion of the metric at spatial infinity~\cite{Geroch:1970cc,Geroch:1970cd,Hansen:1974zz,Thorne:1980ru}.  
As a consequence, they describe the geometry but do not directly reveal how the internal composition of the source gives rise to such geometry.  
Within the momentum-space formalism developed here, this gap is bridged for the first time, and by expanding the EMT in powers of the spin tensor, we obtain a one-to-one correspondence between the coefficients of this expansion and the physical multipole moments of the spacetime. This establishes a direct dictionary between the microscopic structure of the source, its distribution of mass, angular momentum, and stresses, and the macroscopic gravitational multipoles observed in the far field. In this sense, the source multipoles provide the relativistic analogue of the Newtonian relation between matter moments and gravitational potentials, offering a powerful and fully covariant way to characterize the internal composition of rotating compact objects from first principles. Then, once the general multipolar expansion is established, analytic resummation techniques can be applied to obtain explicit, closed-form EMTs corresponding to known BH geometries.  
By performing this resummation, we recover the linearized EMTs associated with the Kerr and Myers–Perry solutions in four and higher dimensions. Remarkably, these EMTs can be interpreted as describing rotating thin disks or higher-dimensional ellipsoids with non-trivial stress distributions, precisely reflecting the singular behavior of the corresponding classical geometries. This provides a new and elegant way to derive, from first principles, the matter sources that reproduce the external fields of spinning BHs, showing that their characteristic singularities arise directly from the multipolar structure itself.

The momentum-space framework also offers a natural setting to move beyond singular sources and to construct regular configurations that reproduce the same asymptotic multipoles as black holes. This idea is implemented by promoting the structure functions to smooth analytic profiles, thereby introducing finite-size effects while keeping the overall multipolar hierarchy unchanged~\cite{Gambino:2025xdr}.  
A particularly insightful and physically motivated choice is that of Gaussian structure functions, which smear the ring singularity of the Kerr source into a finite-width distribution.  
The resulting EMT describes an anisotropic, rotating fluid that satisfies both the energy and causality conditions at first order in the gravitational coupling.  
These configurations retain the complete Kerr multipolar structure at large distances but remain regular everywhere, thus realizing explicit examples of Kerr mimickers. Such solutions demonstrate that the multipolar content of a gravitational field does not uniquely determine the underlying matter distribution and that horizonless, smooth configurations can reproduce the external geometry of classical black holes.
From a conceptual standpoint, these results provide a consistent, amplitude-based framework for exploring the landscape of compact objects that interpolate between regular matter sources and genuine black holes.  
They offer a controlled way to test the limits of the BH uniqueness theorems within GR and to quantify the physical degeneracy between different sources that share the same external multipoles.  

The natural continuation of this program is to explore whether the amplitude-based framework can be extended beyond the perturbative domain and capture the full, non-linear structure of GR.  
In particular, one may ask whether the same analytic machinery that successfully reproduces the PM expansion of BH geometries can also describe their exact, non-perturbative form. This observation naturally motivates the use of the Kerr–Schild (KS) gauge, in which one can express a BH metric as a linear deformation of flat space along a null geodesic congruence, with the remarkable feature that this form remains exact to all orders in the gravitational coupling. As a consequence, it retains both the simplicity of a perturbative expansion and the full non-linear content of the Einstein equations. This linear–yet–exact structure allows one to reinterpret exact metrics such as Kerr, Kerr–Newman, or Myers–Perry as resummed amplitude configurations in which all gravitational self-interactions are encoded within a single effective vertex. Therefore, the KS gauge offers a unique window into how non-perturbative geometries can emerge directly from the analytic structure of scattering amplitudes.

Motivated by this, we compute the Fourier transforms of the Kerr–Newman and Myers–Perry metrics in KS gauge, obtaining compact analytic expressions for the gravitational potentials and electromagnetic fields in momentum space~\cite{Bianchi:2023lrg,Bianchi:2025xol}. These results provide the first explicit realization of how an exact rotating black hole background can be represented as a well-defined object in momentum space. From these expressions we derive the tree-level scattering amplitudes of massive scalar probes interacting with the rotating and charged backgrounds, showing that the amplitudes can be written in closed form for arbitrary orientations of the probe momentum and the BH spin. However, even though KS metrics are 1PM exact, scattering amplitudes describing physical processes are not. We can thus describe the interaction exploiting the KS gauge, which eliminates the need for multi-graviton interaction vertices and allows all gravitational dynamics to be captured by a single effective trilinear coupling. Such approach also provides an ideal setting for studying the classical scattering regime through the eikonal expansion. We identify in fact the complete set of diagrams contributing to the eikonal phase and demonstrate that the off-shell terms in the amplitude reproduce the classical contact interactions responsible for non-linear gravitational effects. This correspondence clarifies how the seemingly quantum ingredients of the amplitude reorganize into classical observables such as deflection angles, time delays, and spin–orbit couplings. Furthermore, by analyzing the momentum-space structure of the metrics, we show that the analytic kernels appearing in the amplitude, which in this case are represented by Bessel functions, encode precisely the multipolar structure of the spacetime, confirming that the same mathematical functions govern both scattering observables and the geometry of the gravitational field.

Finally, this formalism is extended to the case of higher-dimensional Myers–Perry black holes, where rotation occurs in multiple independent planes. We show that in these geometries the new hierarchy of stress multipoles naturally arises in addition to the usual mass and current moments, reflecting the richer tensorial structure of higher-dimensional spacetimes. These moments appear as specific tensor components of the KS metric in momentum space and are directly identifiable with corresponding terms in the scattering amplitude, allowing us to directly probe the multipolar structure of higher-dimensional GR solutions, employing our amplitude-based framework as a theoretical laboratory. 

The thesis is structured as follows.  
In Chapter~\ref{chapter:ClassicalGravityFromAmplitudes} we describe the general framework for extracting metrics in a PM expansion from off-shell three-point amplitude calculations, defining the conventions and theoretical tools that will be crucial for the remainder of the discussion.  
The analysis is performed within a gauge-dependent framework, which allows us to reconstruct quantities such as the metric itself and to exploit the properties of specific gauges throughout the thesis. In Chapter~\ref{chapter:RotatingMetrics} we apply this construction to rotating configurations, computing the amplitudes of spinning fundamental fields and recovering the corresponding metrics.  
We show how the gravitational multipolar structure of rotating spacetimes emerges from the tensorial form of the matter–graviton interaction and how non-minimal couplings in the EFT encode deviations from the Kerr geometry. This chapter also extends the definition of gravitational multipoles to arbitrary spacetime dimensions, introducing a new class of higher-dimensional moments referred to as stress multipoles. Chapter~\ref{chapter:GyromagneticFactor} extends the framework to electrically charged sources, incorporating the interaction between gravitons and photons within the amplitude-based approach and clarifying its correspondence with the classical Einstein–Maxwell theory. Through this formalism, we compute the gyromagnetic factor of BHs in arbitrary dimension, showing that the four-dimensional Kerr–Newman value arises only for minimal coupling, while higher-dimensional analogues require additional Pauli-type and Chern–Simons interactions. In Chapter~\ref{chapter:SourceMultipoles} we draw inspiration from the GR-amplitude correspondence to define a general EMT in momentum space at every order in the spin. This formulation captures the full tower of multipole moments of a generic gravitational source and allows us to identify physical, gauge-invariant combinations associated with observable quantities. By introducing the notions of form factors, structure functions, and source multipoles, we derive compact analytic expressions that reproduce the effective EMTs of Kerr and Myers–Perry black holes. Promoting the structure functions to smooth analytic profiles, we construct regular, horizonless configurations, such as Kerr mimickers, that share the same asymptotic multipoles as black holes, providing a controlled framework to test the limits of the uniqueness theorems. In Chapter~\ref{chapter:KerrSchildGauge} we exploit the remarkable properties of the KS gauge, using it as a non-perturbative extension of the amplitude-based framework. We compute the Fourier transforms of the Kerr–Newman and Myers–Perry metrics, derive their scattering amplitudes, and show how off-shell terms in the eikonal expansion reproduce the nonlinear gravitational interactions responsible for classical contact terms. The analysis reveals a deep correspondence between the Bessel-function kernels appearing in amplitudes and the multipolar structure of the metric, employing Myers–Perry solutions as a theoretical laboratory to probe the nature of stress multipoles. Finally, in Chapter~\ref{chapter:Conclusions} we summarize the main results and outline possible future research directions. The thesis also includes one appendix. In~\ref{chapter:App}, we discuss the harmonic reference frame and explain why it represents the standard choice in the literature for describing known solutions. We then illustrate how to move to this gauge in several relevant examples considered throughout the thesis, including the Hartle–Thorne (and Kerr) metrics, the Myers–Perry solutions in various dimensions, and black rings with a single angular momentum.

\textbf{Conventions.}  
We adopt the mostly-plus metric signature, ${\eta_{00} = -1}$, and work in natural units ${\hbar = c = 1}$ while keeping the gravitational coupling constant ${G_N}$ explicit.  
Greek indices ${\mu, \nu = 0, 1, \ldots, d}$ denote spacetime components, whereas Latin indices ${i, j = 1, \ldots, d}$ refer to spatial components.  
The total number of spacetime dimensions is ${D = d + 1}$.

\chapter{Classical gravity from quantum scattering amplitudes}\label{chapter:ClassicalGravityFromAmplitudes}

The purpose of this chapter is to review how GR can be consistently embedded within the framework of QFT and, in particular, how classical gravitational observables can be extracted from scattering amplitudes. The guiding idea is that, although a straightforward quantization of gravity leads to a non-renormalizable theory~\cite{tHooft:1974toh,Goroff:1985th}, one can nevertheless describe it as an EFT valid at energies well below a cut-off scale, usually identified as the Planck mass~\cite{Donoghue:1994dn,Burgess:2003jk}. Within this framework, gravity is mediated by a spin-2 gauge boson called graviton, and scattering amplitudes involving gravitons contain both genuine quantum corrections and purely classical contributions that survive in the ${\hbar \to 0}$ limit. Remarkably, these classical terms are precisely those that reproduce the PM expansion of GR, namely a perturbative expansion around $G_N$, so that standard spacetime metrics can be reconstructed without explicitly solving Einstein’s equations~\cite{Duff:1973zz, Bjerrum-Bohr:2002gqz,Cheung:2018wkq}.

In this context, our goal is to present the framework in which such classical solutions emerge directly from quantum scattering amplitudes and to develop a systematic approach to efficiently solve perturbative Einstein equations through QFT tools. The key observation is that when gravity is coupled to a generic massive field of spin~$s$, the interaction naturally organizes into Feynman diagrams, with the EMT of the source encoded in the corresponding \mbox{three-point} amplitudes. This perspective turns the problem of computing gravitational observables into the task of evaluating amplitudes and identifying their classical contribution. Indeed, while amplitudes generically contain loop-level quantum corrections, taking the ${\hbar \to 0}$ limit systematically isolates the classical sector, which reproduces the PM expansion of known gravitational solutions~\cite{Bjerrum-Bohr:2002gqz,Donoghue:1994dn,Cheung:2018wkq}.

Indeed, the strength of this approach lies in the possibility of enforcing the classical limit from the very beginning of the computation. Rather than carrying along the full quantum structure and projecting onto the classical sector only at the end, one can define ``classical vertices'', obtained from the ${\hbar \to 0}$ truncation of the fundamental interactions~\cite{Kosower:2018adc,Maybee:2019jus}. Similarly, the propagators and diagrammatic rules can be systematically expanded so that only the long-range, non-analytic contributions survive, while the purely quantum, short-range pieces are consistently discarded~\cite{Bjerrum-Bohr:2002fji, Holstein:2008sw}. This perspective not only streamlines the computations but also clarifies the physical origin of the various terms in the amplitude. Since the entire diagrammatic expansion is built to capture the classical PM sector, this framework allows us to understand the non-linear behavior of gravity in terms of classical loop scattering amplitudes.

The structure of the chapter is as follows. 
In section~\ref{sec:Linearized} we discuss how to quantize gravity coupled to a generic spin-$s$ massive field and how the linearized Einstein equations emerge in this framework. 
Section~\ref{sec:EMTfromAmplitudes} shows how solving Einstein equations can be reformulated as the computation of three-point amplitudes, from which the EMT can be reconstructed. 
In section~\ref{sec:ClassicalLimit} we explain how the full quantum amplitude reduces to its classical counterpart in the limit $\hbar \to 0$, valid for sources of arbitrary spin. 
Then, section~\ref{sec:DressedVertices} describes how the classical limit can be taken directly at the level of the vertices, bypassing the need to perform the limit at the very end of the computation. 
Finally, section~\ref{sec:LoopAmplitudes} discusses how loop amplitudes with massive propagators admit a well-defined classical limit for any spin, clarifying the relation between loop order and PM expansion.

\section{Quantizing linearized gravity}\label{sec:Linearized}

Let us consider a matter field coupled to gravity within GR in arbitrary spacetime dimensions. The interaction is described by the action
\begin{equation}
    S=\int d^Dx \Bigg(\sqrt{-g}\frac{2}{\kappa^2}R+\sqrt{-g}\,\mathcal{L}_m(\Phi_s, g_{\mu\nu})\Bigg)\ ,
\end{equation}
where ${\kappa^2=32\pi G_N}$, $R$ is the Ricci scalar defined as
\begin{equation}
    R=R_{\mu\nu}g^{\mu\nu}\ ,\qquad R_{\mu\nu}=R^\alpha{}_{\mu\alpha\mu}\ ,
\end{equation}
with $R_{\mu\nu}$ the Ricci tensor and $R^{\alpha}{}_{\mu\beta\nu}$ the Riemann tensor, and $\Phi_s$ denotes a generic massive field of spin~$s$. In order to quantize the theory, we expand the metric $g_{\mu\nu}$ around flat spacetime as
\begin{equation}
    g_{\mu\nu}=\eta_{\mu\nu}+\kappa\, h_{\mu\nu}\ ,
\end{equation}
so that the inverse metric reads
\begin{equation}
    g^{\mu\nu}=\eta^{\mu\nu}-\kappa\, h^{\mu\nu}+\mathcal{O}(\kappa^2)\ ,
\end{equation}
and where $h_{\mu\nu}$ defines the graviton as a massless spin-2 gauge boson. With the aim of establishing a perturbative expansion for computing scattering amplitudes, we have to define a graviton propagator that mediates the gravitational interaction, and since gravity is a gauge theory one has to add a gauge-fixing term to the action~\cite{Faddeev:1967fc,DeWitt:1967yk, DeWitt:1967ub, DeWitt:1967uc}. 

This can be done through the Faddeev–Popov procedure, following the standard treatment of gauge theories~\cite{Faddeev:1967fc}. The resulting gauge-fixed theory then admits the usual diagrammatic expansion, with propagators for the graviton and an infinite tower of self-interacting vertices. However, it is well known that GR is perturbatively non-renormalizable~\cite{tHooft:1974toh,Goroff:1985th}, but within the framework of EFTs it can be consistently treated as the leading term in a derivative expansion valid below a cut-off scale, typically associated with the Planck mass~\cite{Donoghue:1993eb,Burgess:2003jk}, where the Einstein–Hilbert term is just the first order in this higher-derivative expansion. Let us therefore impose a generic gauge-fixing condition
\begin{equation}
    F_\lambda(g_{\mu\nu})=0\ ,
\end{equation}
and introduce the corresponding gauge-fixing Lagrangian
\begin{equation}
    \mathcal{L}_{GF}=\frac{1}{\kappa^2}F_{\mu}F_{\nu}\eta^{\mu\nu}\ .
\end{equation}
The total action then becomes
\begin{equation}\label{eq:GeneralAction}
    S=\int d^Dx \Bigg(\sqrt{-g}\frac{2}{\kappa^2}R+\sqrt{-g}\,\mathcal{L}_m(\Phi_s, g_{\mu\nu})+\mathcal{L}_{GF}(g_{\mu\nu})\Bigg)\ ,
\end{equation}
which is no longer gauge invariant, but suitable for perturbation theory. Expanding around flat space in powers of the gravitational coupling, the action takes the form
\begin{equation}\label{eq:ActionExpanded}
    S=\int d^Dx\Bigg(\frac{1}{2}h_{\mu\nu}D^{\mu\nu, \rho\sigma}h_{\rho\sigma}+\frac{\kappa}{2}h_{\mu\nu}T^{\mu\nu}+\cdots\Bigg)\ ,
\end{equation}
where the first term is the kinetic operator for the graviton, the second describes the linear coupling to the matter EMT, and the ellipses denote graviton self-interactions and higher-order couplings to matter. This structure is the hallmark of gauge theories, in which the gauge field couples linearly to the Noether current of the global symmetry that is being gauged~\cite{Weinberg:1995mt}. In gravity, the gauging of global translations yields diffeomorphism invariance, and the corresponding current is the EMT. 

The linearized equations of motion can be then obtained from Eq.~\eqref{eq:ActionExpanded}, reading
\begin{equation}\label{eq:LinearizedEQsMotionRS}
    D^{\mu\nu, \rho\sigma}h_{\rho\sigma}=-\frac{\kappa}{2}T^{\mu\nu}\ ,
\end{equation}
where $D^{\mu\nu, \rho\sigma}$ is a second-order differential operator consisting of the Einstein–Hilbert kinetic term and the gauge-fixing contribution, such as
\begin{equation}
    D^{\mu\nu, \rho\sigma}=D_{EH}^{\mu\nu, \rho\sigma}+D_{GF}^{\mu\nu, \rho\sigma}\ .
\end{equation}
Moving to momentum space, in which one can write
\begin{equation}
    h_{\mu\nu}(x)=\int \frac{d^{d+1}q}{(2\pi)^{d+1}}e^{+iq^0 t-iq\cdot x}h_{\mu\nu}(q)\ ,
\end{equation}
and restricting, here and in the following of the thesis, to stationary (\textit{i.e.} time-independent) configurations such that
\begin{equation}
    h_{\mu\nu}(q)\to 2\pi\delta(q_0)h_{\mu\nu}(q)\ ,
\end{equation}
the Fourier transform simplifies to
\begin{equation}
    h_{\mu\nu}(x)=\int \frac{d^{d}q}{(2\pi)^{d}}e^{-iq\cdot x}h_{\mu\nu}(q)\ ,
\end{equation}
and Eq.~\eqref{eq:LinearizedEQsMotionRS} can be rewritten as
\begin{equation}
    \tilde{D}^{\mu\nu, \rho\sigma}h_{\rho\sigma}(q)=\frac{\kappa}{2}\frac{1}{q^2}T^{\mu\nu}(q)\ ,
\end{equation}
with $P_{\mu\nu, \rho\sigma}$ identifying the tensorial structure of the graviton propagator. Such propagator can be defined as the inverse of the differential operator $\tilde{D}_{\mu\nu, \rho\sigma}$, imposing the relation
\begin{equation}\label{eq:GenericGravPropDef}
    \tilde{D}^{\mu\nu, \rho\sigma}P_{\rho\sigma}{}^{\alpha\beta}=\frac{1}{2}\Big(\eta^{\mu\alpha}\eta^{\nu\beta}+\eta^{\mu\beta}\eta^{\nu\alpha}\Big)\ ,
\end{equation}
so that the linearized equation in momentum space reads
\begin{equation}
    \kappa\, h_{\mu\nu}(q)=\frac{\kappa^2}{2}\frac{1}{q^2}P_{\mu\nu, \rho\sigma}T^{\rho\sigma}(q)\ .
\end{equation}
Then, after a Fourier transform one obtains the relation between the metric perturbation in real space and the EMT in momentum space,
\begin{equation}\label{eq:MetricFromEMT_Generic}
    \kappa\, h_{\mu\nu}(x)=\frac{\kappa^2}{2}\int \frac{d^d q}{(2\pi)^d}\frac{e^{-iq\cdot x}}{q^2}P_{\mu\nu,\rho\sigma}T^{\rho\sigma}(q)\ ,
\end{equation}
where we are implicitly using a Cartesian reference frame ${r^2=x_1^2+\cdots +x_d^2}$ to express quantities in position space, as well as Cartesian transferred momentum ${q^2=q_1^2+\cdots+q_d^2}$, for expressing Fourier transformed quantities. 

Although Eq.~\eqref{eq:MetricFromEMT_Generic} has been derived at linear order, it can be promoted to a systematic PM expansion by accounting for the backreaction of gravity on the source itself. 
Writing both the metric perturbation and the EMT as power series in $G_N$,
\begin{equation}\label{eq:ExpansionInPM}
    h_{\mu\nu}=\sum_{n=1}^{+\infty}h_{\mu\nu}^{(n)}\ ,\qquad T_{\mu\nu}=\sum_{n=0}^{+\infty}T_{\mu\nu}^{(n)}\ ,
\end{equation}
where $n$ counts the $n$PM order, the metric at order $n+1$ is determined iteratively from
\begin{equation}\label{eq:PerturbativeEinsteinEqs}
    \kappa\, h_{\mu\nu}^{(n+1)}=\frac{\kappa^2}{2}\int \frac{d^d q}{(2\pi)^d}\frac{e^{-iq\cdot x}}{q^2}P_{\mu\nu}{}^{\rho\sigma}T_{\rho\sigma}^{(n)}\ .
\end{equation}
This recursive structure is the basis for relating PM gravity to loop amplitudes~\cite{Bjerrum-Bohr:2002gqz, Bern:2021xze, Kosower:2018adc, Cristofoli:2020uzm, Mougiakakos:2020laz, Gambino:2024uge}, as will be discussed in the following sections. Indeed, Eq.~\eqref{eq:PerturbativeEinsteinEqs} reduces the computation of the metric to a Fourier transform integral once the $nPM$ order of the EMT is computed, for which scattering amplitude techniques can be employed for such a task. 

\section{Energy-momentum tensor from amplitudes}\label{sec:EMTfromAmplitudes}

From Eq.~\eqref{eq:MetricFromEMT_Generic}, the computation of the graviton field, which corresponds to the metric perturbation in the classical limit, is reduced to the determination of the EMT in momentum space. 
The idea is to obtain this object by employing scattering amplitude techniques, so that Einstein equations are effectively replaced by the evaluation of off-shell matrix elements of the gravitational conserved current.

Let us consider a massive spin-$s$ field coupled to gravity as the source of the gravitational field. Starting from the action in Eq.~\eqref{eq:GeneralAction}, the gravitational conserved current is then defined as
\begin{equation}
    T_{\mu\nu}(q)\equiv\bra{p_2;s, \sigma'}T_{\mu\nu}(0)\ket{p_1;s, \sigma}\ ,
\end{equation}
where ${q=p_2-p_1}$ is the transferred momentum and ${\ket{p_1;s, \sigma}}$ and ${\ket{p_2;s, \sigma'}}$ are respectively the ``in'' and ``out'' quantum states of the massive particle with polarizations $\sigma$ and $\sigma'$. Then, as in any gauge theory, this current can be computed from three-point amplitudes, in which a graviton couples to the external massive states. Diagrammatically, this is represented as
\begin{equation}\label{eq:FullQuantumEMTfrom3PointAMP}
    \includegraphics[valign=c,width=0.4\textwidth]{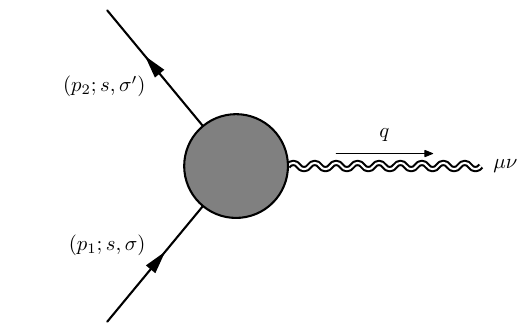}
    =\frac{i\kappa}{2}\Big(4E_1E_2\Big)^{\epsilon/2}T_{\mu\nu}(q)\delta_{\sigma,\sigma'}\ ,
\end{equation}
relating the full three-point amplitude to the quantum current, where $\epsilon=1$ is for bosonic sources and $\epsilon=0$ for fermionic ones\footnote{This difference is due to different normalization of the quantum states between fermion and boson fields.}. The relation in Eq.~\eqref{eq:FullQuantumEMTfrom3PointAMP} is fully quantum: in principle, it includes both classical and genuinely quantum contributions. 
Indeed, at the level of the EFT, this means that loop diagrams contribute not only with $\hbar$-suppressed terms but also with pieces that survive in the classical limit. 
This observation, emphasized in the modern amplitude literature~\cite{Donoghue:1993eb, Bjerrum-Bohr:2002gqz, Goldberger:2004jt, Bern:2021xze}, is at the core of the PM approach, in which the loop expansion of scattering amplitudes is systematically mapped to the expansion in powers of $G_N$. The first observation of such an unfamiliar phenomenon dates back to 1971, when Iwasaki pointed out that loop amplitudes are not strictly quantum but are indeed required to reconstruct classical gravity~\cite{Iwasaki:1971vb}. This phenomenological difference between gravity and, for example, electromagnetism stems from the non-linear nature of GR. In electromagnetism in fact, the linearity of Maxwell’s equations implies that the classical electromagnetic field sourced by a prescribed conserved current is fully determined at tree level, while classical contributions from loop amplitudes can only affect composite observables such as the EMT. In gravity, by contrast, the intrinsic non-linearity of Einstein’s equations implies that loop amplitudes are required to correctly reproduce the classical spacetime geometry beyond leading order, leading to an infinite tower of PM corrections~\cite{Duff:1973zz,Donoghue:1994dn}.

Moreover, the growing interest in rewriting gravity in terms of classical scattering amplitudes is due to the amplitudes property of being gauge invariant, allowing one to compute gravitational observables without caring about non-physical degrees of freedom. Indeed, scattering amplitudes can be defined entirely in terms of on-shell kinematics, avoiding gauge redundancies. The literature can be then broadly divided into two approaches. On the one hand, on-shell methods allow one to compute gravitational observables at very high order in perturbation theory, exploiting the absence of gauge redundancies and making use of the same vast plethora of techniques employed in Standard Model computations~\cite{Arkani-Hamed:2017jhn,Chung:2018kqs,Bern:2019crd,Cheung:2018wkq,Kosower:2018adc,Maybee:2019jus}. On the other hand, off-shell approaches, such as those employed in~\cite{Mougiakakos:2020laz,Bern:2020buy,Bianchi:2023lrg,Gambino:2024uge, Mougiakakos:2024nku}, where Feynman diagrams are used to perform calculations, are particularly useful for reconstructing the full EMT and disentangling physical from gauge contributions. Both viewpoints are of course complementary, with on-shell methods that yield compact analytic expressions, while the off-shell approach makes explicit the correspondence with the Einstein equations. In this thesis we will focus entirely on off-shell methods, which possess another important advantage, such as its validity in arbitrary spacetime dimensions. Indeed, while on-shell approaches mainly rely on spinor-helicity formalism strictly valid in $D=4$~\cite{Bjerrum-Bohr:2014lea, Guevara:2017csg, Kosower:2018adc}, employing Feynman diagrams the expression for $T_{\mu\nu}(q)$ can be computed for general spacetime dimensions, and the resulting metric can be compared with higher-dimensional solutions of the Einstein equations. This flexibility provides a direct way to study the multipolar structure of higher-dimensional compact objects from first principles, setting up a theoretical laboratory to study gravity even beyond GR.

Finally, in order to evaluate the amplitudes in Eq.~\eqref{eq:FullQuantumEMTfrom3PointAMP}, one has to extract the Feynman vertices from the action. The graviton self-interaction vertices originate from the Einstein–Hilbert action combined with the gauge-fixing term, giving rise to an infinite series of gauge-dependent vertices
\begin{equation}\label{eq:GravityVerticesExpansion}
S_{EH}+S_{GF}\rightarrow\includegraphics[valign=c,width=0.2\textwidth]{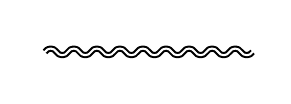}
+\includegraphics[valign=c,width=0.2\textwidth]{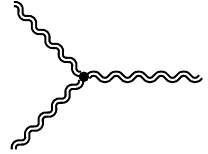}
+\includegraphics[valign=c,width=0.2\textwidth]{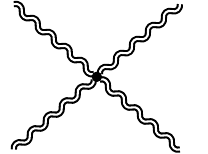}+\cdots
\end{equation}
Similarly, expanding the matter action in powers of the gravitational coupling generates the vertices describing the interaction between the massive field and the graviton
\begin{equation}\label{eq:MassiveVerticesExpansion}
S_{m}\rightarrow\includegraphics[valign=c,width=0.2\textwidth]{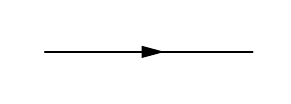}
+\includegraphics[valign=c,width=0.2\textwidth]{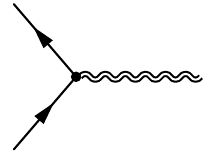}
+\includegraphics[valign=c,width=0.2\textwidth]{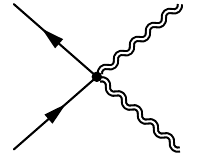}+\cdots
\end{equation}
These vertices provide the building blocks for constructing loop amplitudes, whose classical content will be analyzed in the following sections.

\section{Extracting the classical limit of amplitudes}\label{sec:ClassicalLimit}

In this section we restrict our analysis to the classical limit of Eq.~\eqref{eq:FullQuantumEMTfrom3PointAMP}. 
Our goal is to review the general framework through which classical scattering amplitudes can be constructed so as to reproduce the perturbative formulation of Einstein equations in Eq.~\eqref{eq:PerturbativeEinsteinEqs}. 
To this end, we adopt a top-down perspective, starting from the expected structure of the metric in the PM expansion. From Eq.~\eqref{eq:PerturbativeEinsteinEqs}, we expect to recover the metric perturbation order by order in $G_N$. 
Explicitly, one expects that the amplitude calculation, once the classical limit $\hbar \to 0$ is taken, will yield
\begin{equation}\label{eq:GenericExpectedMetricStructure}
    \lim_{\hbar\rightarrow 0}\kappa\, h_{\mu\nu}
    \propto\sum_{n=1}^{+\infty}(G_Nm\rho)^n
    \sum_{k=0}^{+\infty}\frac{\Lambda_{n, k}}{r^k}\ ,
\end{equation}
where $\Lambda_{n, k}$ is a scalar coefficient encoding the multipolar structure of the metric\footnote{Notice that $\Lambda_{n,k}$ contains dimensionful quantities such as the angular-momentum density of the source, thus  $[\Lambda_{n,k}]=L^{k}$.} and 
\begin{equation}\label{eq:ScalarHarmonicFunction}
    \frac{\rho(r)}{4\pi}=\frac{1}{(d-2)\Omega_{d-1}r^{d-2}}
\end{equation}
is a scalar harmonic function that generalizes the typical spatial fall-off to higher dimensions, with 
\begin{equation}\label{eq:SurfaceDsphere}
    \Omega_{d-1}=\frac{2\pi^{d/2}}{\Gamma(d/2)}
\end{equation}
being the area of the unit $(d-1)$-sphere. 

Since Eq.~\eqref{eq:GenericExpectedMetricStructure} must emerge from the Fourier transform of Eq.~\eqref{eq:PerturbativeEinsteinEqs}, one finds that, in the monopole case $k=0$ for instance,
\begin{equation}
    \kappa^2\int \frac{d^dq}{(2\pi)^d}\frac{e^{-iq\cdot x}}{q^2}T_{\mu\nu}^{(l, k=0)}
    \propto (G_Nm\rho)^{l+1}\ ,
\end{equation}
which uniquely fixes the momentum-space structure of the EMT computed from classical amplitudes to be
\begin{equation}
    T_{\mu\nu}^{(l, k=0)}\propto J_{(l)}(q)\ ,
\end{equation}
with $J_{(l)}(q)$ being the $l$-loop master integral~\cite{Mougiakakos:2020laz}
\begin{equation}\label{eq:MasterSunset}
    J_{(l)}(q)=\int\prod_{i=1}^{l}\frac{d^d\ell_i}{(2\pi)^d}
    \frac{q^2}{\left(\prod_{i=1}^{l}{\ell_i}^2\right)\left(q-\ell_1-\cdots-\ell_l\right)^2}\ .
\end{equation}
Using the Fourier transform relation~\cite{Mougiakakos:2020laz, DOnofrio:2022cvn}
\begin{equation}\label{eq:MasterFT}
    \int\frac{d^dq}{(2\pi)^d}\frac{e^{-iq\cdot x}}{q^2}J_{(l)}(q)
    =\left(\frac{\rho(r)}{4\pi}\right)^{l+1}\ ,
\end{equation}
one sees that $J_{(l)}$ serves as the master integral around which the loop expansion of the classical scattering amplitudes is organized. 
By including multipoles then, this argument generalizes to
\begin{equation}
    T_{\mu\nu}^{(l, k)}\propto q^k J_{(l)}(q)\ ,
\end{equation}
so that
\begin{equation}\label{eq:SchematicMultipoleDerivation}
    \partial^k\int \frac{d^dq}{(2\pi)^d}\frac{e^{-iq\cdot x}}{q^2}J_{(l)}(q)
    \propto (G_Nm\rho)^{l+1} \frac{\Lambda_{l+1, k}}{r^k}\ .
\end{equation}
This argument shows that the master integral in Eq.~\eqref{eq:MasterSunset} defines the unique momentum-space structure that loop amplitudes must reproduce at every order in perturbation theory. Such a statement can be verified case by case by explicit calculations of classical loop amplitudes by using integration by parts (IBPs) identities~\cite{Lee:2012cn, Lee:2013mka, Smirnov:2013dia}, in which different loop diagrams in the end reorganize into Eq.~\eqref{eq:MasterSunset}, as we will see explicitly later on.

Finally, let us examine how Eq.~\eqref{eq:FullQuantumEMTfrom3PointAMP} simplifies in the classical limit. 
As ${\hbar\rightarrow 0}$, the classical contribution is obtained by cutting the massive line, forcing all adjacent heavy propagators on-shell. 
The remaining subgraph is necessarily a tree of graviton exchanges attached to a single worldline, encoding the non-linearities of the gravitational interaction. 
By contrast, diagrams containing closed graviton loops have no compensating on-shell factor after the cut, and are therefore suppressed by explicit powers of $\hbar$, identifying them as purely quantum. 
Moreover, in the stationary sector $q^0=0$, only the potential region contributes, and radiation cuts are absent. 
Therefore, the classical EMT at each PM order is fully captured by a single massive line dressed by an internal tree~\cite{Duff:1973zz,Goldberger:2004jt,Bjerrum-Bohr:2018xdl,Gambino:2024uge}, as illustrated by
\begin{equation}\label{eq:ClassicalEMTfrom3PointAMP}
    \includegraphics[valign=c,width=0.4\textwidth]{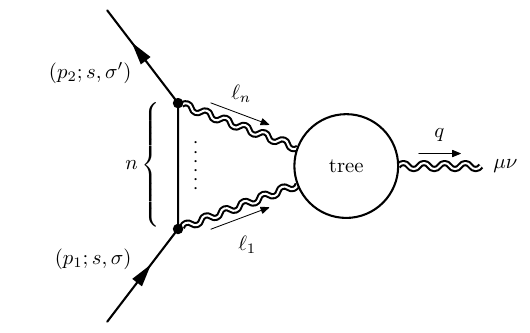}
    =\frac{i\kappa}{2}\Big(2m\Big)^{\epsilon}T_{\mu\nu}^{(n-1)}(q)\delta_{\sigma,\sigma'}\ .
\end{equation}
This analysis makes manifest that the PM expansion of the gravitational field is in one-to-one correspondence with the loop expansion of scattering amplitudes. In particular, the relation $l=n+1$ identifies the loop order $l$ with the PM order $n$. One important remark is that, for the purpose of extracting classical contributions, one requires the full tower of self-interacting graviton vertices in Eq.~\eqref{eq:GravityVerticesExpansion}, but only the three-point matter–graviton vertex for the massive sector, since higher-point matter couplings generate topologies that correspond to quantum corrections (see Ref.~\cite{Gambino:2022kvb}).

\section{Classical dressed vertices}\label{sec:DressedVertices}

Having established that Eq.~\eqref{eq:ClassicalEMTfrom3PointAMP} relates classical scattering amplitudes to the PM expansion of the gravitational field, we now develop a formalism that allows us to compute such amplitudes directly in the classical limit. 
As we already pointed out in fact, the strength of this approach lies in avoiding a two-step procedure, such as computing the full quantum amplitude and only afterwards taking its classical limit. Instead, we aim to implement the limit at the earliest stage, so that all subsequent manipulations are performed with purely classical quantities. For this purpose, we introduce the \emph{dressed vertices}, namely the classical limit of the three-point Feynman vertices entering Eq.~\eqref{eq:ClassicalEMTfrom3PointAMP}.

Let us begin by considering the three-point interaction vertex for a generic spinning particle,
\begin{equation}
    \includegraphics[valign=c,width=0.30\textwidth]{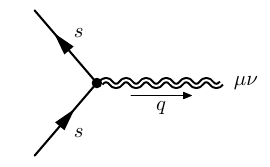}
    \;=\;(\tau_{\Phi^2 h})^{\mu\nu}_{a, b}(q)\ ,
\end{equation}
which carries graviton indices $\mu\nu$ and additional indices $a,b$ associated with the representation of the spin-$s$ field. In the amplitude of Eq.~\eqref{eq:ClassicalEMTfrom3PointAMP}, the graviton is off-shell, while the external matter states are on-shell. 
Thus, to recover the EMT, the vertex must be contracted with the polarization states of the external particles. At tree level, this gives
\begin{equation}\label{eq:treelevelamplitude}
    \frac{i \kappa}{2}(2m)^\epsilon\, T_{\mu\nu}^{(0)}(q)\,\delta_{\sigma\sigma'}
    ={}^a\bra{p_2;s,\sigma'} (\tau_{\Phi^2 h})_{\mu\nu}^{a, b}\ket{p_1;s, \sigma}^b\ ,
\end{equation}
where the polarization states may be tensorial or spinorial, depending on whether the matter field is bosonic or fermionic. Following the diagrammatic approach developed in~\cite{Bern:2020buy,Maybee:2019jus}, which captures spin effects in gravity from the classical limit of amplitudes, it is convenient to introduce the average momentum
\begin{equation}\label{eq:CapitalP}
    P^\mu = \frac{p_1^\mu+ p_2^\mu }{2}\ ,
\end{equation}
so that
\begin{equation}
    p_1^\mu=P^\mu+\frac{1}{2}q^\mu \ ,\qquad
    p_2^\mu=P^\mu-\frac{1}{2}q^\mu\ .
\end{equation}
As already mentioned, we focus only on the stationary case, where ${q^0=0}$ and energy is conserved. This implies that, since ${p_1^2=p_2^2=-m^2}$, one has ${P^\mu = m\, u^\mu + \mathcal{O}(\hbar)}$, with ${u^\mu=\delta^\mu_0}$ the velocity of the source.

To extract the purely classical terms, one then employs the scaling prescription introduced in~\cite{Kosower:2018adc}. Introducing the anti-symmetric spin tensor\footnote{Notice that in this thesis we will use spin and angular momentum as synonyms, and it will be clear from context whether we are talking about the quantum number of a particle or the actual angular momentum of an extended object.} $J_{\mu\nu}$, where in four spacetime dimensions
\begin{equation}
    J_{ij}=\varepsilon_{ijk}L_k\ ,
\end{equation}
where $\varepsilon_{ijk}$ is the Levi-Civita tensor in $d=3$ and $L_k$ is the angular momentum, the classical limit simply translates into the substitution rules
\begin{equation}\label{eq:KMOC}
    q\rightarrow\hbar\, q\ ,\qquad
    J^{\mu\nu}\rightarrow\frac{1}{\hbar} J^{\mu\nu}\ ,
\end{equation}
and it is enforced by keeping only contributions of order $\hbar^0$ as $\hbar\to 0$. Equivalently, introducing the spin-density tensor ${S^{\mu\nu}=J^{\mu\nu}/m}$, classical terms scale as ${(S\cdot q)^n}$, while terms proportional to positive powers of $q$ alone are quantum. 
In this way, spin insertions compensate factors of momentum transfer, producing non-trivial classical contributions at higher multipole order. Moreover, since $p_2=p_1+\mathcal{O}(\hbar)$, the polarization states coincide in the stationary limit~\cite{Kosower:2018adc,Chung:2018kqs,Bern:2020buy}
\begin{equation}\label{eq:ClassicalLimitPolarization}
    \ket{p_2}=\ket{p_1}+\mathcal{O}(\hbar)\ .
\end{equation}
This allows us to define the dressed vertex as
\begin{equation}\label{eq:DressedVertexGeneric}
   \bra{p_2; s, \sigma'}(\tau_{\Phi^2 h})^{\mu\nu}\ket{p_1; s, \sigma}
   =\hat{\tau}_{\Phi^2 h}^{\mu\nu}(q,S)\,\delta_{\sigma\sigma'}+\mathcal{O}(\hbar)\ ,
\end{equation}
where the explicit spin-index structure is left implicit. 

To make this construction explicit, recall that the matter states transform under Lorentz generators $M^{\mu\nu}$, satisfying~\cite{Weinberg:1995mt}
\begin{equation}
    [M^{\mu \nu}, M^{\rho \sigma}]
    =-i\big(\eta^{\mu\rho}M^{\nu\sigma}-\eta^{\nu \rho}M^{\mu \sigma}
        +\eta^{\nu \sigma}M^{\mu \rho}-\eta^{\mu\sigma}M^{\nu\rho}\big)\ .
\end{equation}
For ``in'' and ``out'' states with spin $s$ and polarizations $\sigma$ and $\sigma'$, one finds in the stationary limit
\begin{equation}\label{eq:ClassicalLimitOfGenerators}
\begin{aligned}
    \braket{p_2;s, \sigma'|p_1;s, \sigma}
        &=C(s)\,\delta_{\sigma\sigma'}+\mathcal{O}(\hbar)\ ,\\
    \braket{p_2;s, \sigma'|M^{\mu\nu}|p_1;s, \sigma}
        &=J^{\mu\nu}C(s)\,\delta_{\sigma\sigma'}+\mathcal{O}(\hbar^0)\ ,\\
    \braket{p_2;s, \sigma'|\tfrac{1}{2}\{M^{\mu\nu}, M^{\rho\sigma}\}|p_1;s, \sigma}
        &=J^{\mu\nu}J^{\rho\sigma}C(s)\,\delta_{\sigma\sigma'}+\mathcal{O}(\hbar^{-1})\ ,
\end{aligned}
\end{equation}
where $\{\cdot,\cdot\}$ denotes the anti-commutator, $C(s)$ is a normalization factor depending on the representation, and $J^{\mu\nu}$ is the classical angular momentum tensor of the source. The interpretation of Eq.~\eqref{eq:ClassicalLimitOfGenerators} is that $J^{\mu\nu}$ encodes the classical spin structure of the particle and it arises from the classical limit of the quantum spin operator identified by the generator $M^{\mu\nu}$. Moreover, in the rest frame, one has $J^{0i}=0$, so only spatial components survive. Notice that throughout the thesis we will consistently describe the angular momentum as an anti-symmetric rank-2 tensor. Indeed, in higher dimensions the concept of an angular momentum vector is lost, since is due to the duality properties between anti-symmetric rank-2 tensors and vectors, proper of $d=3$. A final important observation is that the dressed vertex of a massive spin-$s$ particle contains powers of the spin tensor up to $2s$. Consequently, the resulting metric generated by such a source features multipoles up to order $2s$, reflecting the finite multipolar content of the classical dressed vertex~\cite{Gambino:2024uge}.

\section{Loop amplitudes for any spin}\label{sec:LoopAmplitudes}

We are now in a position to extend the previous analysis to loop amplitudes. 
The key point is that, once the classical dressing procedure has been implemented, the spin dependence of a massive spin-$s$ particle is entirely encoded in the tensor structure of the dressed vertex. 
As a result, the loop calculation formally reduces to that of a scalar particle, with the only modification being the insertion of dressed vertices along the massive line. 
We illustrate this mechanism explicitly at one loop and then outline the generalization to arbitrary loop orders.

\subsection{One-loop case}

For $n=2$ (one-loop order), the contribution of the massive propagator to Eq.~\eqref{eq:ClassicalEMTfrom3PointAMP} reads
\begin{equation}\label{eq:MatterLine1Loop}
    \bra{p_2; s, \sigma'}(\tau_{\Phi^2 h})^{\mu\nu}(\ell) 
    \frac{i\, \mathcal{P}(p_1-\ell)}{(p_1-\ell)^2-m^2+i\varepsilon}
    (\tau_{\Phi^2 h})^{\rho\lambda}(q-\ell)\ket{p_1; s,\sigma}\ ,
\end{equation}
where $\mathcal{P}$ denotes the numerator of the matter propagator, whose form depends on the spin representation of the field. 
For a generic spin-$s$ particle, $\mathcal{P}$ can always be expressed as a sum over polarization states,
\begin{equation}
    \mathcal{P}(p_1-\ell)=(2m)^{1-\epsilon}
    \sum_{\sigma''}\ket{p_1-\ell;s, \sigma''}\bra{p_1-\ell;s , \sigma''}\ .
\end{equation}
Substituting this into Eq.~\eqref{eq:MatterLine1Loop} gives
\begin{equation}
\begin{aligned}
   \frac{i\, (2m)^{1-\epsilon}}{(p_1-\ell)^2-m^2+i\varepsilon}  
   \sum_{\sigma''}\bra{p_2;s,\sigma'} (\tau_{\Phi^2 h})^{\mu\nu}(\ell)\ket{p_1-\ell;s, \sigma''}\\
   \times\bra{p_1-\ell;s , \sigma''}(\tau_{\Phi^2 h})^{\rho\lambda}(q-\ell)\ket{p_1; s, \sigma}\ .
\end{aligned}
\end{equation}
Using the stationary-limit relation of Eq.~\eqref{eq:ClassicalLimitPolarization}, this expression simplifies to
\begin{equation}
\begin{aligned}
   \frac{i\, (2m)^{1-\epsilon}}{(p_1-\ell)^2-m^2+i\varepsilon}  
   \sum_{\sigma''}\bra{p_1;s,\sigma'} (\tau_{\Phi^2 h})^{\mu\nu}(\ell)\ket{p_1;s, \sigma''}\\
   \times\bra{p_1;s , \sigma''}(\tau_{\Phi^2 h})^{\rho\lambda}(q-\ell)\ket{p_1; s, \sigma}
   +\mathcal{O}(\hbar)\ ,
\end{aligned}
\end{equation}
which in terms of dressed vertices becomes
\begin{equation}
   \frac{i\, (2m)^{1-\epsilon}}{(p_1-\ell)^2-m^2+i\varepsilon} 
   \hat{\tau}_{\Phi^2 h}^{\mu\nu}(\ell, S)\,\hat{\tau}_{\Phi^2 h}^{\rho\lambda}(q-\ell,S)
   \,\delta_{\sigma \sigma'}\ .
\end{equation}
We thus conclude that, after the dressing procedure, the classical limit of a spinning one-loop amplitude reduces to a scalar-like amplitude~\cite{Mougiakakos:2020laz} in which the only spin dependence resides in the dressed vertex, modulo a normalization that depends on the particle’s statistics. 

\subsection{Generalization to all loops}

The generalization to higher loops is straightforward. 
At $l$ loops, the massive line factorizes into a product of $l+1$ dressed vertices, and the amplitude can be expressed as~\cite{Mougiakakos:2020laz,DOnofrio:2022cvn}
\begin{equation}\label{eq:EMTfromAMPfinal}
    \frac{i\, \kappa}{2}(2m)^\epsilon\,T_{\mu\nu}^{(l)}(q)
    =\frac{(i)^{l+1}}{(l+1)!}\int\prod_{i=1}^{l}\frac{d^d {\ell_i}}{(2\pi)^d}
    \frac{\prod_{i=1}^{l+1}\hat{\tau}_{\Phi^2h}^{\mu_i\nu_i}(\ell_i, S)\,
    \prod_{i=1}^{l+1}P_{\mu_i\nu_i, \alpha_i\beta_i}}{\prod_{i=1}^{l+1}{\ell_i}^2}\,
    \mathcal{M}^{\alpha_1\beta_1, \ldots, \mu\nu}\ ,
\end{equation}
where $\mathcal{M}^{\alpha_1\beta_1,\ldots,\mu\nu}$ denotes the sum of tree-level graviton diagrams appearing in Eq.~\eqref{eq:ClassicalEMTfrom3PointAMP}, $P_{\mu\nu, \alpha\beta}$ is the tensorial structure of the graviton propagator defined in Eq.~\eqref{eq:GenericGravPropDef} and where in the stationary classical limit one has $\ell_i^0=0$. From Eq.~\eqref{eq:EMTfromAMPfinal} the EMT can be systematically determined at any loop order. 
Inserting the result back into Eq.~\eqref{eq:ClassicalEMTfrom3PointAMP} then yields the metric at the corresponding PM order. 
Because the dressed vertex for a spin-$s$ field contains spin-tensor powers up to $2s$, an $l$-loop amplitude involving $n=l+1$ dressed vertices produces contributions up to order $2ns$ in $S^{\mu\nu}$. 
Thus the metric at $n$PM order expands as\footnote{For instance, $h_{\mu\nu}^{(1,1)}$ is linear in spin at 1PM order, while $h_{\mu\nu}^{(1,2)}$ is quadratic in spin at the same order, and so on.}
\begin{equation}\label{eq:MultipoleMetricExpansion}
    h_{\mu\nu}^{(n)} = \sum_{k=0}^{2ns}h_{\mu\nu}^{(n,k)}\ ,
\end{equation}
where $k$ counts the power of $S^{\mu\nu}$, coherently with Eq.~\eqref{eq:GenericExpectedMetricStructure}. 
This shows that multipole moments are reproduced up to orders $k=2s$.

In this chapter we have outlined how classical GR can be systematically reconstructed from scattering amplitudes. 
Starting from linearized gravity treated as an EFT, we showed that the metric can be obtained directly from 3-point amplitudes involving a graviton and a massive source. 
Taking the classical limit isolates the PM sector, where loop order maps to powers of $G_N$ and encodes the multipolar structure of the solution. 
The introduction of dressed vertices allows one to implement the $\hbar\to 0$ limit at the level of the Feynman rules, making the spin dependence explicit and reducing loop amplitudes to scalar-like integrals. 
Altogether, this framework provides the conceptual and technical tools to compute the energy–momentum tensor and the corresponding metric perturbations for sources of arbitrary spin, setting the stage for the applications that will be developed in the following chapters.

\chapter{Rotating metrics from scattering amplitudes in arbitrary dimensions}\label{chapter:RotatingMetrics}

In the previous chapter we discussed the general idea of recovering GR from scattering amplitudes, focusing on how perturbative Einstein equations can be obtained from off-shell graviton emission processes, and emphasizing how the EMT emerges in the classical limit with loop amplitudes encoding PM corrections. We now apply these ideas to the computation of rotating geometries sourced by spinning fields, where angular momentum introduces new physical degrees of freedom beyond the monopole mass term. A natural framework to characterize these degrees of freedom is that of gravitational multipole moments~\cite{Geroch:1970cc,Geroch:1970cd,Hansen:1974zz,Thorne:1980ru}. In four spacetime dimensions, multipoles provide the bridge between source distributions and the asymptotic structure of the metric, as codified in the classic Thorne formalism~\cite{Thorne:1980ru}. Generalizing this notion to higher dimensions reveals a richer structure, since in addition to mass and current multipoles, one encounters entirely new stress multipoles that have no counterpart in $d=3$. Their presence signals that higher-dimensional solutions cannot be fully characterized by mass and spin alone, and that new invariants are needed to distinguish inequivalent geometries. These play a central role in separating different black hole families, such as Myers–Perry solutions and black rings~\cite{Myers:1986un,Emparan:2001wn}, and in diagnosing possible horizonless mimickers~\cite{Cardoso:2019rvt}.

Moreover, it is well known that in GR, gravitational dipole moments are uniquely fixed by the asymptotic structure of spacetime~\cite{Hansen:1974zz,Thorne:1980ru}, unlike the long-range behaviour of electromagnetic fields, in which electric dipoles are not associated with any conserved charge. Therefore, in the gravitational context, the quadrupole order is the first one that allows us to discriminate between different solutions, and according to Eq.~\eqref{eq:MultipoleMetricExpansion}, this implies that the first non-trivial theory to study for probing gravitational quadrupole moments is a massive \mbox{spin-1} field coupled to the graviton~\cite{Gambino:2024uge}. Even though a comprehensive study of the full gravitational multipole tower, and how the scattering amplitude framework enhances our physical understanding of it, will be discussed in detail later in the thesis, the motivation to study geometries induced by massive \mbox{spin-1} fields is twofold. On the one hand, it enables us to explore the phenomenology of stress multipoles in the simplest setup by restricting the analysis to quadrupole order, and on the other hand, it gives us the possibility to investigate the deep relation between minimal coupling and BH physics, a subject that has attracted much attention over the last decade~\cite{Arkani-Hamed:2017jhn,Chung:2018kqs}.

The chapter is then structered as follows. In section~\ref{sec:GravitationalMultipoles} we review the multipolar description of gravitational fields. After recalling the $d=3$ case, we extend the construction to arbitrary dimensions, illustrating the emergence of stress multipoles explicitly in $d=4$. Having clarified the physical degrees of freedom, in section~\ref{sec:Spin1Calculations} we move on to show how they are realized in explicit scattering amplitude calculations. We focus on massive \mbox{spin-1} sources, which provide a controlled way to probe the quadrupole sector. Finally, in section~\ref{sec:Spin1Comparison}, we compare the rotating metric obtained from amplitude-based calculations with known higher-dimensional solutions of the Einstein equations, showing how crucial the presence of stress multipoles is for differentiating between different higher-dimensional black solutions.  

\section{Gravitational multipoles}\label{sec:GravitationalMultipoles}

The concept of multipole expansion in GR has a long history, pioneered by Geroch and Hansen~\cite{Geroch:1970cc, Geroch:1970cd, Hansen:1974zz}, who defined the gravitational multipole moments in a gauge-invariant framework, extending the Newtonian interpretation to relativistic systems for the first time. Later, Thorne introduced the so-called asymptotically cartesian mass centered (ACMC) coordinates, a particular reference frame in which one can unambiguously extract multipole moments directly from the asymptotic behavior of the metric~\cite{Thorne:1980ru}. Although not gauge invariant, Thorne's formalism is more intuitive, and it can be shown that the two definitions of multipole moments coincide~\cite{Gursel:1983nkl}. While most of the literature focuses on the $d=3$ case (see Refs.~\cite{Cardoso:2016ryw,Mayerson:2022ekj} for reviews), there have been attempts to generalize the multipole expansion to $d=4$, both in the Geroch–Hansen framework~\cite{Tanabe:2010ax} and using ACMC coordinates à la Thorne~\cite{Heynen:2023sin, Gambino:2024uge}. In this section we review the works in~\cite{Gambino:2024uge,Bianchi:2024shc}, which propose a framework to generalize Thorne's formalism to arbitrary dimensions.

Let us start from Thorne’s original construction of gravitational multipoles~\cite{Thorne:1980ru}. In this formalism, one considers the linearized metric
\begin{equation}
    g_{\mu\nu}=\eta_{\mu\nu}+\kappa\, h^{(1)}_{\mu\nu}+\cdots\ ,
\end{equation}
which corresponds to Eq.~\eqref{eq:GenericExpectedMetricStructure} truncated at first order $n=1$. It is then convenient to introduce the trace-reversed perturbation
\begin{equation}\label{eq:gammaDefinition}
    \gamma_{\mu\nu}= \kappa\, h^{(1)}_{\mu\nu}-\frac{\kappa}{2}\eta_{\mu\nu}h^{(1)}\ ,
\end{equation}
which simplifies the structure of the Einstein equations. Then, imposing the harmonic gauge condition
\begin{equation}\label{eq:HarmonicGaugeTR}
    \partial^\mu\gamma_{\mu\nu}=0\ ,
\end{equation}
the linearized field equations reduce to the remarkably simple form
\begin{equation}\label{eq:LinearizedHarmonicTReqs}
   \Box\gamma_{\mu\nu}=0\ .
\end{equation}
The task now is to find the most general solution of Eq.~\eqref{eq:LinearizedHarmonicTReqs} compatible with the harmonic gauge. A systematic way to construct such solutions is to build them from derivatives of a harmonic scalar function $\rho(r)$, defined in Eq.~\eqref{eq:ScalarHarmonicFunction}. Since this function satisfies
\begin{equation}\label{eq:BoxOfScalarHarmFunc}
    \Box \rho(r)=0\ ,
\end{equation}
where $\Box=\eta^{\mu\nu}\partial_\mu\partial_\nu$, it naturally provides the building blocks for solutions of the vacuum equations. In particular, one can generate higher multipole structures by acting with spatial derivatives on $\rho(r)$, in full analogy with the role of spherical harmonics in multipole expansions of electromagnetic fields, and coherent with the schematic picture given in Eq.~\eqref{eq:SchematicMultipoleDerivation}. The general solution of the linearized vacuum\footnote{For non-vacuum spacetimes the construction is more subtle, since the matter source modifies both the equations and the allowed multipolar structure.} equations can then be written as
\begin{equation}\label{eq:gammaMetric}
\begin{aligned}
\gamma_{00}&=\sum_{\ell=0}^{+\infty}\mathcal{M}_{A_\ell}\,\partial_{A_\ell}\rho(r)\ , \\
\gamma_{0i}&=\sum_{\ell=0}^{+\infty}\mathcal{J}_{i, A_{\ell}}\,\partial_{A_\ell}\rho(r)\ ,\\
\gamma_{ij}&=\sum_{\ell=0}^{+\infty}\mathcal{G}_{ij, A_{\ell}}\,\partial_{A_\ell}\rho(r)\ ,
\end{aligned}
\end{equation}
where $\mathcal{M}_{A_\ell}$, $\mathcal{J}_{i,A_\ell}$ and $\mathcal{G}_{ij,A_\ell}$ are constant multipole tensors. These tensors are completely symmetric and trace-free in the multi-index $A_\ell$, where we adopt the shorthand notation $A_\ell=a_1\cdots a_\ell$ with $\partial_{A_\ell}=\partial_{a_1}\cdots\partial_{a_\ell}$. Following~\cite{Thorne:1980ru}, we denote this symmetry type as $\{A_\ell\}_{\text{STF}}$ (symmetric and trace-free), noticing that the tensor $\mathcal{G}_{ij,A_\ell}$ carries additional symmetries since it is symmetric in $(i,j)$ but not necessarily traceless in those indices. This decomposition makes explicit the physical interpretation of each sector, namely, $\mathcal{M}_{A_\ell}$ encodes the mass multipoles, $\mathcal{J}_{i,A_\ell}$ the current multipoles, and $\mathcal{G}_{ij,A_\ell}$ the new stress multipoles that naturally arise in higher-dimensional spacetimes.

In $d$ spatial dimensions the multipole tensors introduced above transform under the rotation group $SO(d)$, and can be decomposed into their irreducible representations. To illustrate this in the simplest non-trivial case, let us restrict to the quadrupole ($\ell=2$) in three spatial dimensions. In this case, the trace-reversed perturbation can be decomposed as
\begin{equation} \label{eq:gammaMetricquadrupole}
\begin{aligned}
   &\gamma_{00}\big|_{\ell=2}^{d=3}=\mathcal{M}_{\{a_1a_2\}_{\text{STF}}}\,\partial_{a_1}\partial_{a_2} \left(\frac{1}{r}\right)\ ,\\
   &\gamma_{0i}\big|_{\ell=2}^{d=3}=\mathcal{J}^{(1)}_{a_1}\,\partial_{a_1}\partial_{i} \left(\frac{1}{r}\right)+\mathcal{J}^{(2)}_{\{ia_1a_2\}_{\text{STF}}}\,\partial_{a_1}\partial_{a_2} \left(\frac{1}{r}\right)+\epsilon_{ia_1a_2}\,\mathcal{J}^{(3)}_{\{a_1a_3\}_{\text{STF}}}\,\partial_{a_2}\partial_{a_3} \left(\frac{1}{r}\right)\ ,\\
   &\gamma_{ij}\big|_{\ell=2}^{d=3}=\delta_{ij}\,\mathcal{G}^{(1)}_{\{a_1a_2\}_{\text{STF}}}\,\partial_{a_1}\partial_{a_2} \left(\frac{1}{r}\right)+\mathcal{G}^{(2)}\,\partial_{i}\partial_{j} \left(\frac{1}{r}\right)+\mathcal{G}^{(3)}_{\{(i|a_1\}_{\text{STF}}}\,\partial_{|j)}\partial_{a_1} \left(\frac{1}{r}\right)\\
   &+\mathcal{G}^{(4)}_{\{ija_1a_2\}_{\text{STF}}}\,\partial_{a_1}\partial_{a_2} \left(\frac{1}{r}\right)+\epsilon_{(i|a_1a_2}\,\mathcal{G}^{(5)}_{a_1}\,\partial_{a_2}\partial_{|j)} \left(\frac{1}{r}\right)+\epsilon_{(i|a_1a_2}\,\mathcal{G}^{(6)}_{\{a_1a_3|j)\}_{\text{STF}}}\,\partial_{a_2}\partial_{a_3} \left(\frac{1}{r}\right)\ ,
\end{aligned}
\end{equation}
where all tensors are constant, the superscripts simply label the independent structures, and the symmetry properties are inherited from the STF decomposition. Then, imposing the harmonic gauge condition~\eqref{eq:HarmonicGaugeTR} results in eliminating redundant terms and fixes
\begin{equation}
    \mathcal{J}^{(2)}=0\ , \qquad \mathcal{G}^{(1)}=-\frac{1}{2}\mathcal{G}^{(3)}\ ,\qquad \mathcal{G}^{(4)}=0\ ,\qquad \mathcal{G}^{(6)}=0\ .
\end{equation}
We now investigate the effect of residual gauge freedom. Consider a generic infinitesimal coordinate transformation ${x'_\mu=x_\mu+\xi_\mu(x)}$, under which the perturbation transforms as
\begin{equation}\label{eq:MetricPrime}
    h'_{\mu\nu}=h_{\mu\nu}-(\partial_\mu\xi_\nu+\partial_\nu\xi_\mu)\ .
\end{equation}
In harmonic gauge the coordinates satisfy by definition $\Box x^\mu=0$, and in order to preserve this condition in the new frame we must require
\begin{equation}
    \Box x'^{\mu}=0\quad \Rightarrow\quad \Box \xi^\mu=0\ .
\end{equation}
Using Eq.~\eqref{eq:BoxOfScalarHarmFunc}, the most general harmonic shift can be expressed as
\begin{equation}\label{eq:InfinitesimalGaugeTransf}
    \xi^\mu=\sum_{\ell=0}^{+\infty} \mathcal{T}^{\mu,A_\ell}\,\partial_{A_\ell}\rho(r)\ ,
\end{equation}
where $\mathcal{T}$ is a constant tensor, and in this new frame the trace-reversed field transforms as
\begin{equation}\label{eq:ExampleCoordTransfMult}
    \gamma'_{\mu\nu}=\gamma_{\mu\nu}-\partial_\mu\xi_\nu-\partial_\nu\xi_\mu+\eta_{\mu\nu}\partial^\alpha\xi_\alpha\ .
\end{equation}
By exploiting this residual freedom, one can eliminate spurious structures, and choosing
\begin{equation}
    \begin{gathered}
        \xi_0=\mathcal{J}^{(1)}_{a_1}\,\partial_{a_1} \left(\frac{1}{r}\right) \ ,\\
        \xi_i=-\mathcal{G}^{(1)}_{ja_1}\,\partial_{a_1} \left(\frac{1}{r}\right)+\frac{1}{2}\epsilon_{ja_1a_2}\,\mathcal{G}^{(5)}_{a_1}\,\partial_{a_2} \left(\frac{1}{r}\right)+\frac{1}{2}\mathcal{G}^{(2)}\,\partial_{j} \left(\frac{1}{r}\right)\ ,
    \end{gathered}
\end{equation}
one arrives at the simplified form
\begin{equation}\label{eq:gammaijvanishes}
\begin{aligned}
    \gamma_{00}\big|_{\ell=2}^{d=3}&=\mathcal{M}_{\{a_1a_2\}_{\text{STF}}}\,\partial_{a_1}\partial_{a_2} \left(\frac{1}{r}\right)\ ,\\
    \gamma_{0i}\big|_{\ell=2}^{d=3}&=\epsilon_{ia_1a_2}\,\mathcal{J}^{(3)}_{\{a_1a_3\}_{\text{STF}}}\,\partial_{a_2}\partial_{a_3} \left(\frac{1}{r}\right)\ ,\\
    \gamma_{ij}\big|_{\ell=2}^{d=3}&=0\ .
\end{aligned}
\end{equation}
The lesson is that gauge freedom drastically reduces the number of independent multipolar tensors. In three spatial dimensions, the only physical degrees of freedom that remain at quadrupole order are the mass multipoles $\mathcal{M}$ and the current multipoles $\mathcal{J}$. These are precisely the invariant quantities that remain unchanged under ACMC transformations, and they form the two towers of multipoles familiar from the $d=3$ case~\cite{Thorne:1980ru}.

Let us now turn the attention to the quadrupole sector in $d=4$. In this case, the decomposition into $SO(4)$ irreducible representations introduces, beyond STF tensors, two additional structures. Following~\cite{Heynen:2023sin}, we denote by ASTF (antisymmetric and trace-free) a tensor $\mathcal{T}_{\{b_1b_2, A_\ell\}_\text{ASTF}}$ that is STF in $A_\ell$, antisymmetric in $(b_1,b_2)$, and overall trace free. Moreover, $SO(4)$ irreducible representations accommodate another class that we call RSTF (Riemann-symmetric and trace-free)~\cite{Gambino:2024uge}, denoted by $\mathcal{T}_{\{ib_1,jb_2,A_\ell\}_\text{RSTF}}$, with the symmetries of the Riemann tensor in the first four indices, STF in $A_\ell$, and fully trace free. In group-theory terms, the appearance of ASTF and RSTF terms reflects the richer tensor-product structure of $SO(4)\cong SU(2)_L\times SU(2)_R$, which allows independent (anti)symmetric combinations not available in $SO(3)$. In terms of irreducible representations, the trace-reversed perturbation then reads
\begin{equation} \label{eq:gammaMetricquadrupoleD5}
\begin{aligned}
    \gamma_{00}\big|_{\ell=2}^{d=4}&=\mathcal{M}_{\{a_1a_2\}_{\text{STF}}}\,\partial_{a_1}\partial_{a_2} \left(\frac{1}{\pi r^2}\right)\ ,\\
    \gamma_{0i}\big|_{\ell=2}^{d=4}&=\mathcal{J}^{(1)}_{a_1}\,\partial_{a_1}\partial_{i} \left(\frac{1}{\pi r^2}\right)+\mathcal{J}^{(2)}_{\{ia_1a_2\}_{\text{STF}}}\,\partial_{a_1}\partial_{a_2} \left(\frac{1}{\pi r^2}\right)+\epsilon_{ib_1b_2a_1}\,\mathcal{J}^{(3)}_{\{b_1b_2,a_2\}_{\text{ASTF}}}\,\partial_{a_1}\partial_{a_2} \left(\frac{1}{\pi r^2}\right)\ ,\\
     \gamma_{ij}\big|_{\ell=2}^{d=4}&=\delta_{ij}\,\mathcal{G}^{(1)}_{\{a_1a_2\}_{\text{STF}}}\,\partial_{a_1}\partial_{a_2} \left(\frac{1}{\pi r^2}\right)+\mathcal{G}^{(2)}\,\partial_{i}\partial_{j} \left(\frac{1}{\pi r^2}\right)+\mathcal{G}^{(3)}_{\{(i|a_1\}_{\text{STF}}}\,\partial_{|j)}\partial_{a_1} \left(\frac{1}{\pi r^2}\right)\\
     &\quad+\mathcal{G}^{(4)}_{\{ija_1a_2\}_{\text{STF}}}\,\partial_{a_1}\partial_{a_2} \left(\frac{1}{\pi r^2}\right)+\epsilon_{(i|b_1b_2a_1}\,\mathcal{G}^{(5)}_{\{b_1b_2\}_\text{ASTF}}\,\partial_{a_1}\partial_{|j)} \left(\frac{1}{\pi r^2}\right)\\
    &\quad+\epsilon_{(i|b_1b_2a_1}\,\mathcal{G}^{(6)}_{\{b_1b_2,a_2|j)\}_{\text{ASTF}}}\,\partial_{a_1}\partial_{a_2} \left(\frac{1}{\pi r^2}\right)+\mathcal{G}^{(7)}_{\{ib_1,jb_2\}_\text{RSTF}}\,\partial_{b_1}\partial_{b_2} \left(\frac{1}{\pi r^2}\right)\ .
\end{aligned}
\end{equation}
Notice that the tensor $\mathcal{G}^{(7)}$ is not manifestly symmetric in $(i,j)$ because of its Riemann-like symmetries, but after symmetrization in $(b_1,b_2)$ through contraction with derivatives, it becomes symmetric in $(i,j)$ as well. This already anticipates that, unlike in $d=3$, genuinely new spatial structures can survive the gauge fixing. Indeed, imposing the harmonic gauge and performing a suitable coordinate transformation like in Eq.~\eqref{eq:ExampleCoordTransfMult} yields
\begin{equation} \label{eq:gammaMetricquadrupoleD5final}
\begin{aligned}
    \gamma_{00}\big|_{\ell=2}^{d=4}&=\mathcal{M}_{\{a_1a_2\}_{\text{STF}}}\,\partial_{a_1}\partial_{a_2} \left(\frac{1}{\pi r^2}\right)\ ,\\
    \gamma_{0i}\big|_{\ell=2}^{d=4}&=\epsilon_{ib_1b_2a_1}\,\mathcal{J}^{(3)}_{\{b_1b_2,a_2 \}_{\text{ASTF}}}\,\partial_{a_1}\partial_{a_2} \left(\frac{1}{\pi r^2}\right)\ ,\\
     \gamma_{ij}\big|_{\ell=2}^{d=4}&=\mathcal{G}^{(7)}_{\{ib_1,jb_2\}_\text{RSTF}}\,\partial_{b_1}\partial_{b_2} \left(\frac{1}{\pi r^2}\right)\ .
\end{aligned}
\end{equation}
Since now $\gamma_{ij}\neq 0$, an extra degree of freedom appears at quadrupole order compared to $d=3$. This argument extends to arbitrary dimensions and higher multipole orders, demonstrating a new tower of independent multipoles $\mathcal{G}$, which we call \emph{stress multipoles}\footnote{The name reflects that $g_{ij}$ is sourced by $T_{ij}$, the stress part of the EMT.}. In summary, Eq.~\eqref{eq:gammaMetricquadrupoleD5final} shows that at each multipole order the spacetime is characterized by three independent tensors. In $d=3$ they reduce to two because of $SO(3)$ identities and the structure of Einstein’s equations.

Motivated by the above argument, we conjecture that an ACMC coordinate system exists in arbitrary dimensions and that a new independent multipole tensor is associated with the spatial part of the metric. Accordingly, we write the multipole expansion of a stationary metric in arbitrary dimensions as
\begin{equation}\label{eq:MultipoleExpandedMetric}
\begin{aligned}
    g_{00}&=-1+4\frac{d-2}{d-1}\sum_{\ell=0}^{+\infty}\frac{G_N m\, \rho(r)}{r^\ell}\,\mathbb{M}^{(\ell)}_{A_\ell}N_{A_\ell}+\cdots\ ,\\
    g_{0i}&=2(d-2)\sum_{\ell=1}^{+\infty}\frac{G_Nm\,\rho(r)}{r^\ell}\,\mathbb{J}^{(\ell)}_{i, A_\ell}N_{A_\ell}+\cdots\ ,\\ 
    g_{ij}&=\delta_{ij}+4\frac{d-2}{d-1}\sum_{\ell=2}^{+\infty}\frac{G_N m\, \rho(r)}{r^\ell}\,\tilde{\mathbb{G}}^{(\ell)}_{ij,A_\ell}N_{A_\ell}+\cdots\ ,
\end{aligned}
\end{equation}
with $N_{A_\ell}=x_{a_1}\cdots x_{a_\ell}/r^{\ell}$, and where the ellipses denote non–gauge-invariant contributions. Normalizing the mass $m$ and spin density $S$ to their ADM values fixes the mass monopole and spin dipole to
\begin{equation}
    \mathbb{M}^{(0)}=1\ , \qquad \mathbb{J}^{(1)}_{i a_1}=S_{ia_1}\ ,
\end{equation}
while, without loss of generality,
 \begin{equation}
    \mathbb{M}^{(1)}=0 \ , \qquad \mathbb{G}_{ij}^{(0)}=0\ , \qquad \mathbb{G}_{ij}^{(1)}=0\ .
\end{equation}
To ensure gauge invariance, quadrupole tensors are defined up to terms with explicit Kronecker deltas, such as $\delta_{a_m a_n}$, $\delta_{i a_m}$ or $\delta_{j a_m}$. Loosely speaking, in Eq.~\eqref{eq:MultipoleExpandedMetric} we neglect terms where $N_{A_\ell}$ contracts such $\delta$’s. Within a “generalized’’ ACMC transformation in arbitrary dimensions, Eq.~\eqref{eq:MultipoleExpandedMetric} is conjectured to be invariant. In this context $\mathbb{M}^{(\ell)}_{A_\ell}$ are the usual mass multipoles, $\mathbb{J}^{(\ell)}_{i,A_\ell}$ are the current multipoles, and $\tilde{\mathbb{G}}^{(\ell)}_{ij,A_\ell}$ collect the new stress tensors through the relation
\begin{equation}
    \mathbb{G}_{ij, A_\ell}^{(\ell)}=\tilde{\mathbb{G}}_{ij, A_\ell}^{(\ell)}+\frac{1}{2}\delta_{ij}\Big(\mathbb{M}^{(\ell)}_{A_\ell}-\tilde{\mathbb{G}}_{kk, A_\ell}^{(\ell)}\Big)\ ,
\end{equation}
as follows from the definition~\eqref{eq:gammaDefinition}. Finally, note that Eq.~\eqref{eq:MultipoleExpandedMetric} makes the dependence on $G_N$ explicit for dimensional reasons, and in general the multipole tensors themselves can depend on the Newton constant. Indeed, the formalism is completely general and also accommodates intrinsic (\textit{i.e.} not necessarily spin-induced) multipoles\footnote{For instance, a static body deformed away from spherical symmetry~\cite{Raposo:2018xkf,Raposo:2020yjy}.}. In that case, besides mass and angular momenta, a new length scale is present, which can be expressed in units of the fundamental scale $G_N m$, thereby introducing extra powers of $G_N$ in the expansion (an example is the black ring discussed in Sec.~\ref{sec:BlackRing}). In a bottom-up, amplitude-based approach, such intrinsic multipoles can be modeled by a non-minimal action in which, instead of universal operators quadratic in the spin tensor, one introduces independent tensors at each order, as we will discuss in the next section. 

\subsection{Multipoles in non-vacuum spacetimes}

So far we have discussed the definition of gravitational multipoles in vacuum, where the equivalence between the harmonic gauge and the ACMC frame guarantees that the multipole tensors can be unambiguously extracted from the asymptotic metric. The situation is more subtle when the spacetime is sourced by a non-localized EMT. In this case, the EMT can induce long-range contributions to the metric that spoil the ACMC condition, thus obstructing the definition of gravitational multipoles in the sense of Thorne. This issue has been emphasized in~\cite{Mayerson:2022ekj}, and it is instructive to analyze it in our framework. Consider the linearized Einstein equations in the presence of a non-vanishing EMT,
\begin{equation}
    \Box\Big(h_{\mu\nu}(x)-\tfrac{1}{2}\eta_{\mu\nu}h(x)\Big)=\frac{\kappa}{4}T_{\mu\nu}(x)\ .
\end{equation}
If the Fourier transform of the EMT contains non-analytic terms in the transferred momentum $q$, such as $1/q^2$ poles, the resulting metric at infinity inherits angular-dependent pieces that cannot be removed by a coordinate transformation. As an illustrative example, let us suppose the schematic expression
\begin{equation}
    T_{\mu\nu}(q)\sim m \, u_\mu u_\nu \,\frac{\Lambda^2}{q^2}\Big(-q_\alpha S^{\alpha}{}_\beta S^{\beta\sigma} q_\sigma\Big)^2\ ,
\end{equation}
where $\Lambda$ is a mass scale and $S_{\mu\nu}$ the spin tensor. Dimensional analysis shows that the real-space EMT decays sufficiently fast to be integrable, but the induced metric acquires contributions of the type
\begin{equation}
    h_{00}\sim G_N m \Lambda^2 \frac{\rho(r)}{r^2}\big(n_\alpha S^{\alpha}{}_\beta S^{\beta\sigma} n_\sigma\big)^2\ ,
\end{equation}
with $n^i=x^i/r$. Comparing it with Eq.~\eqref{eq:MultipoleExpandedMetric}, such a term is not compatible with the ACMC structure, which means that no consistent set of gravitational multipoles can be defined in this gauge. In other words, the equivalence between harmonic and ACMC coordinates is lost, and the multipole expansion breaks down. 

Nevertheless, not all non-localized EMTs destroy the multipole interpretation. A physically relevant case is provided by electrically charged sources in $d=3$. In this situation the EMT can be written, up to local terms, as
\begin{equation}\label{eq:ChargedEMT}
    T_{\mu\nu}(q)=m\,u_\mu u_\nu 
    + Q^2\Bigg(F^{(Q)}_{1,1} \,u_\mu u_\nu
    + F^{(Q)}_{1,2}\Big(\eta_{\mu\nu}-\frac{q_\mu q_\nu}{q^2}\Big)\Bigg)\ ,
\end{equation}
with $Q$ being the electric charge, $F^{(Q)}_{1,1}$ and $F^{(Q)}_{1,2}$ dimensionless coefficients and where charge conservation enforces $q^\mu T_{\mu\nu}(q)=0$. Computing the linearized metric in the harmonic gauge, one finds
\begin{equation}\label{eq:ChargedMetric}
    \begin{aligned}
        h_{00}(x)&=\frac{2G_Nm}{r}+\Big(F^{(Q)}_{1,1}+F^{(Q)}_{1,2}\Big)\frac{4G_NQ^2}{\pi r^2}\ , \\
        h_{ij}(x)&=\frac{2G_Nm}{r}\delta_{ij}
        +\frac{4G_NQ^2}{\pi r^4}\Bigg(\,4F^{(Q)}_{1,2}x_i x_j
        +\Big(F^{(Q)}_{1,1}-3F^{(Q)}_{1,2}\Big)r^2\delta_{ij}\Bigg)\ .
    \end{aligned}
\end{equation}
This metric is already in ACMC coordinates, so its multipoles can be directly read off from the asymptotic expansion. From Eq.~\eqref{eq:MultipoleExpandedMetric} it follows that the charge-dependent terms in~\eqref{eq:ChargedMetric} do not contribute to the mass, current, or stress multipoles. This matches the physical expectation that Reissner-Nordstr\"om and Kerr-Newman BHs share the same gravitational multipole moments as their uncharged counterparts. In summary, while generic non-localized EMTs can obstruct the definition of gravitational multipoles, special classes of sources, such as charged configurations with sufficiently fast fall-off, admit ACMC frames in which the multipoles remain well defined. The key criterion is whether the long-range structure of the EMT aligns with the asymptotic multipolar expansion of the metric. When this happens, the multipoles of the localized source are preserved, and non-local contributions become subleading, as in the Kerr-Newman case.

\section{Gravitational quadrupole moments from spin-1 fields}\label{sec:Spin1Calculations}

After having discussed how gravitational multipoles are defined in arbitrary spacetime dimensions in Eq.~\eqref{eq:MultipoleExpandedMetric}, we want to see how such structure emerges from scattering amplitude calculations. To do this, in this section we compute the metric sourced by off-shell three-point scattering amplitudes describing graviton emission from a massive spin-1 field, since from Eq.~\eqref{eq:MultipoleMetricExpansion} we know that this setup is sensitive to multipole corrections up to quadrupole order. From the amplitude perspective, this implies that the dressed matter–graviton vertex is not uniquely determined. We therefore begin by extracting the minimal dressed vertex from the minimally coupled Proca action, and subsequently introduce non-minimal interactions in order to obtain the most general conserved EMT at quadrupole order. These non-minimal couplings appear in the dressed vertex as free coefficients, which parametrize both the physical quadrupole moment and possible gauge redundancies. In order to carry out explicit calculations, as discussed in chapter~\ref{chapter:ClassicalGravityFromAmplitudes}, we must fix a gauge. To remain consistent with standard choices while preserving generality, we adopt the one-parameter family of gauges~\cite{Jakobsen:2020ksu}
\begin{equation}\label{eq:GaugeCondition}
    F^{\lambda}=(1-\alpha)\,\kappa\,\partial_{\mu} \left(h^{\mu \lambda}-\frac{1}{2}\eta^{\mu \lambda}h\right)
    +\alpha\, g^{\mu\nu}\Gamma^{\lambda}_{\mu \nu}\ ,
\end{equation}
with $h=\eta^{\mu\nu}h_{\mu\nu}$ and $\Gamma^{\lambda}_{\mu\nu}$ the Christoffel symbols. This condition interpolates smoothly between the de~Donder gauge ($\alpha=0$) and the harmonic gauge ($\alpha=1$), while keeping fixed the tensor structure of the graviton propagator in arbitrary dimensions as
\begin{equation}\label{eq:HarmonicPropagator}
    P_{\mu\nu,\rho\sigma}=\frac{1}{2}\Big(\eta_{\mu\rho}\eta_{\nu\sigma}+\eta_{\mu\sigma}\eta_{\nu\rho}-\frac{2}{d-1}\eta_{\mu\nu}\eta_{\rho\sigma}\Big)\ .
\end{equation}

\subsection{Minimal vertex}

Consider now a massive spin-1 (Proca) field minimally coupled to gravity,
\begin{equation}\label{eq:ProcaAction}
    S_{\text{min}}=\int d^{d+1}x\,\sqrt{-g}\left(-\frac{1}{4}F_{\mu\nu}F^{\mu\nu}+\frac{1}{2}m^2V_{\mu}V^{\mu}\right)\ ,
\end{equation}
with $V_\mu$ the massive vector field and $F_{\mu\nu}=\partial_{\mu}V_{\nu}-\partial_{\nu}V_\mu$ its antisymmetric field strength. Expanding this action provides the full tower of matter–graviton vertices, as depicted in Eq.~\eqref{eq:MassiveVerticesExpansion}. In particular, the three-point interaction with one graviton and two vectors takes the form~\cite{Bjerrum-Bohr:2014lea}
\begin{equation}\label{eq:spin1VertexOriginal}
\begin{aligned}
   &\varepsilon_{\beta}(p_2)\,{\tau}_{V^2h,\text{min}}^{\mu\nu, \beta\alpha}\,\varepsilon_{\alpha}(p_1)
   =-\frac{i\, \kappa}{2}\Big[
   \varepsilon(p_1) \cdot  p_2\big(p_1^\mu\varepsilon^\nu(p_2)+p_1^\nu\varepsilon^\mu(p_2)\big)\\
   &+\varepsilon(p_2) \cdot  p_1\big(p_2^\mu\varepsilon^\nu(p_1)+p_2^\nu\varepsilon^\mu(p_1)\big)-\varepsilon(p_1) \cdot \varepsilon(p_2)\big(p_1^\mu p_2^\nu+p_2^\mu p_1^\nu\big)\\
   &-\big(p_1 \cdot  p_2-m^2\big)\big(\varepsilon(p_1)^\mu\varepsilon(p_2)^\nu+\varepsilon(p_2)^\mu\varepsilon(p_1)^\nu\big)\\
   &+\eta^{\mu\nu}\Big(\big(p_1 \cdot  p_2-m^2\big)\varepsilon(p_1) \cdot \varepsilon(p_2)-p_1 \cdot \varepsilon(p_2)\,p_2 \cdot \varepsilon(p_1)\Big)\Big]\ ,
\end{aligned}
\end{equation}
where the real polarization vectors satisfy $\varepsilon(p)\cdot p=0$ on shell.  
Equation~\eqref{eq:spin1VertexOriginal} represents the minimal three-point vertex directly extracted from the Proca Lagrangian. It is convenient then to recast this result in terms of Lorentz generators, where in the vector representation these take the form
\begin{equation}
    M^{\mu\nu,\rho\sigma}=i\big(\eta^{\mu\rho}\eta^{\nu\sigma}-\eta^{\mu\sigma}\eta^{\nu\rho}\big)\ .
\end{equation}
After stripping off the external polarizations and reorganizing the kinematics in terms of $q^\mu=p_1-p_2$ and $P^\mu=(p_1+p_2)/2$, one finds (up to terms proportional to $p_1^\alpha$ or $p_2^\beta$ that vanish upon contraction with the $\varepsilon$’s)
\begin{equation}
\begin{aligned}
   &\varepsilon_{\beta}(p_2)\,\tau_{V^2h,\text{min}}^{\mu\nu,\beta\alpha}\,\varepsilon_{\alpha}(p_1)
   =\frac{i\kappa}{2}\,\varepsilon_{\beta}(p_2)\Big[
   \eta^{\alpha\beta} \left(2P^\mu P^\nu+\frac{1}{2}\eta^{\mu\nu}q^2-\frac{1}{2}q^\mu q^\nu\right)\\
   &-i q_\lambda \left(P^\mu M^{\nu\lambda,\beta\alpha}+P^\nu M^{\mu\lambda,\beta\alpha}\right)
   -\frac{1}{2}q_{\rho}q_{\sigma}\{M^{\mu\rho},M^{\nu\sigma}\}^{\beta\alpha}\Big]\varepsilon_{\alpha}(p_1)\ .
\end{aligned}
\end{equation}
This compact expression makes the Lorentz structure manifest, with monopole, dipole, and quadrupole contributions organized in powers of momentum transferred. Finally, using~\eqref{eq:ClassicalLimitOfGenerators} we can perform the dressing procedure of the vertex following Eq.~\eqref{eq:DressedVertexGeneric}, from which one obtains
\begin{equation}\label{eq:MinimalVertex}
    \includegraphics[valign=c,width=0.25\textwidth]{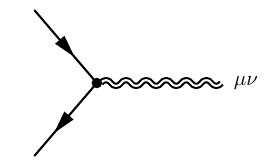}
    =\hat{\tau}_{V^2h,\text{min}}^{\mu\nu}(q)=\frac{i\, \kappa}{2}\,2m^2\Big(u^\mu u^\nu
    - \frac{i}{2}q_\lambda \big(S^{\mu\lambda}u^{\nu}+S^{\nu\lambda}u^{\mu}\big)
    - \frac{1}{2}q_\lambda q_\sigma S^{\mu\lambda}S^{\nu\sigma}\Big)\ ,
\end{equation}
with polarization normalization $C(1)=+1$. The structure of Eq.~\eqref{eq:MinimalVertex} illustrates the concept of spin universality: the monopole and dipole pieces appear automatically and coincide with results from lower spin fields~\cite{Bjerrum-Bohr:2002fji, Gambino:2024uge}, while the $O(S^2)$ terms provide the quadrupole contribution of a minimally coupled spin-1 source.

\subsection{Non-minimal vertex}

While minimal coupling already generates monopole and dipole moments, which are unique, reproducing the most general quadrupolar structure, which is needed to reproduce generic spacetime geometries,  requires extending the action to include non-minimal operators. Beyond minimal coupling, the effective action contains higher-dimensional operators whose coefficients are a priori unconstrained and therefore appear in the dressed vertex as free parameters. Deriving the full three-point vertex from each possible operator quickly becomes unwieldy. Following~\cite{Bern:2020buy}, it is more convenient to encode the effect of these additional terms in a covariant antisymmetric tensor $\mathbb{S}^{\mu\nu}$, which generalizes the notion of the spin operator as
\begin{equation}\label{eq:DefinitionSpinTensor}
\mathbb{S}^{\mu\nu}_{a,b}=S^{\mu\nu}\delta_{ab}+O(\kappa),
\end{equation}
where $a,b$ are spin indices in the chosen field representation. Since our interest lies in the long-range, non-local contributions to the metric, in the construction of the non-minimal lagrangian we can safely discard purely local terms. At tree level this corresponds to dropping all structures proportional to $q^2$, because they Fourier transform into contact terms localized at the origin. In practice, this means that terms of the form $T_{\mu\nu}(q)\propto\mathcal{O}(q^2)$ in the EMT, or equivalently $\hat{\tau}_{\mu\nu}(q)\propto\mathcal{O}(q^2)$ in the dressed vertex, do not contribute to the classical asymptotic fields and can be neglected. With these assumptions, the most general non-minimal and non-redundant action up to quadrupole order can be written as~\cite{Gambino:2024uge}
\begin{equation}\label{eq:NonMinimalAction}
\begin{aligned}
&S_{\text{non-min}}=\int d^Dx \sqrt{-g}\Big[
K_0\, R\, D^\mu V^\alpha g_{\alpha\beta} D_{\mu}V^\beta
+K_1\, R\, V^\alpha (\mathbb{S}^{\mu\nu}\mathbb{S}_{\mu\nu})_{\alpha\beta}V^\beta\\
&+K_2\, R_{\mu\nu}\, V^\alpha(\mathbb{S}^{\mu\lambda}\mathbb{S}_{\lambda}^{\ \nu})_{\alpha\beta}V^\beta
+K_3\, R_{\mu\nu\rho\sigma}\, V^\alpha(\mathbb{S}^{\mu\nu}\mathbb{S}^{\rho\sigma})_{\alpha\beta}V^\beta\\
&+K_4\, R_{\mu\nu\rho\sigma}\, D^{\nu}V^\alpha(\mathbb{S}^{\mu\lambda}\mathbb{S}_{\lambda}^{\ \sigma})_{\alpha\beta}D^{\sigma}V^\beta
+K_5\, D^{\nu}D^{\sigma}R_{\mu\nu\rho\sigma}\, V^\alpha(\mathbb{S}^{\mu\lambda}\mathbb{S}_{\lambda}^{\ \sigma})_{\alpha\beta}V^\beta\Big]\ .
\end{aligned}
\end{equation}
Other operators\footnote{Including higher-curvature and higher-derivative terms.} either contribute only locally, are suppressed as $O(\kappa^2)$, or can be related to those above by algebraic identities. Equation~\eqref{eq:NonMinimalAction} therefore captures all genuinely new structures relevant to quadrupole order. To ensure a finite classical limit, we assign appropriate $\hbar$-scalings so that only the leading non-local contributions survive. Dimensional analysis shows that $K_{1,2,3}$ are dimensionless, while $K_{0,4,5}$ carry mass dimension $M^{-2}$. Using $G_N$ and the mass $m$ as input scales, two admissible structures emerge
\begin{equation}
   \left[\frac{1}{m^2}\right]=M^{-2},
   \qquad
   \left[\frac{1}{\hbar^2}(G_N m)^{\frac{2}{d-2}}\right]=M^{-2},
\end{equation}
with the second choice yielding integer powers of $G_N$ only in $d=3,4$. Inspecting the momentum dependence of the vertices clarifies the correct scaling assignments. The $K_0$ term contributes first at $O(q^2)$ and the $K_5$ term at $O(q^4 S^2)$, so both must scale as $(G_N m)^{\frac{2}{d-2}}$. In contrast, the $K_4$ term starts at $O(q^2 S^2)$ and therefore scales as $1/m^2$. On this basis, we reparametrize the couplings as
\begin{equation}
\begin{gathered}
K_0=\frac{1}{2}\Omega_1\,(G_N m)^{\frac{2}{d-2}}\ ,\qquad
K_1=-\frac{1}{4}C_1\ ,\qquad
K_2=-\frac{1}{2}C_2\ ,\\
K_3=\frac{1-H_1}{8}\ ,\qquad
K_4=\frac{H_2}{2m^2}\ ,\qquad
K_5=\Omega_2\,(G_N m)^{\frac{2}{d-2}}\ ,
\end{gathered}
\end{equation}
introducing the dimensionless coefficients $C_{1,2}$, $H_{1,2}$ and $\Omega_{1,2}$. These parameters encode the freedom associated with non-minimal couplings and will appear explicitly in the dressed vertex.

Finally, expanding~\eqref{eq:NonMinimalAction} to $O(\kappa)$ yields the non-minimal dressed vertex
\begin{equation}\label{eq:GenericTreeQuadVertex}
\begin{aligned}
&\hat{\tau}_{V^2h,\text{non-min}}^{\mu\nu}(q)
=-\frac{i\kappa}{2}m^2\Big[
-(H_1-1)\, q_{\rho}q_{\sigma}S^{\mu\rho}S^{\nu\sigma}
+H_2\, u^{\mu}u^{\nu}\,q_{\rho}q_{\sigma}S^{\rho\lambda}S_{\lambda}^{\ \sigma}\\
&+C_1\, S^{\rho\sigma}S_{\rho \sigma}\,q^{\mu}q^{\nu}
+C_2 \left(\eta^{\mu\nu}q_{\rho}q_{\sigma}S^{\rho\lambda}S^{\sigma}{}_{\lambda}
-q^\lambda\big(q^{\mu}S_{\lambda \sigma}S^{\nu \sigma}+q^{\nu}S_{\lambda \sigma}S^{\mu \sigma}\big)\right)\Big]\ ,
\end{aligned}
\end{equation}
together with an additional ``higher-loop'' dressed vertex,
\begin{equation}\label{eq:CounterTermVertex}
    \includegraphics[valign=c,width=0.25\textwidth]{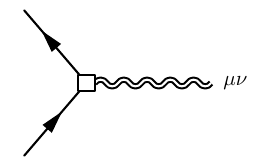}
    \;=\;
    \hat{\tau}_{V^2h,\text{HL}}^{\mu\nu}(q)
    = i\,\kappa\,(G_Nm)^{\frac{2}{d-2}}m^2
    \Big(\Omega_1\, q^{\mu}q^{\nu}
    +\Omega_2\, q^{\mu}q^{\nu}\, q_{\rho}q_{\sigma}S^{\rho\lambda}S_{\lambda}^{\ \sigma}\Big)\ .
\end{equation}
We refer to Eq.~\eqref{eq:CounterTermVertex} as a higher-loop vertex because, although it arises already at tree level in the effective action, its contribution in the PM expansion is delayed. Indeed, in $d=3$ a tree-level insertion contributes only at 3PM order, while in $d=4$ it enters at 2PM. As we shall see in the next subsection, our gauge choice in $d=3,4$ leads to divergences in intermediate expressions. In that context, $\hat{\tau}_{V^2h,\text{HL}}^{\mu\nu}$ plays the role of a counter-term that cancels these singularities~\cite{Mougiakakos:2020laz, DOnofrio:2022cvn}. The coefficients $\Omega_{1,2}$ therefore do not carry physical information, but instead act as extra gauge parameters that ensure the consistency of the formalism. Combining the minimal and non-minimal pieces, the general spin-1 dressed vertex takes the form
\begin{equation}\label{eq:FinalQuadVertex}
\begin{aligned}
\hat{\tau}_{V^2h}^{\mu\nu}(q)
&=\hat{\tau}_{V^2h,\text{min}}^{\mu\nu}(q)+\hat{\tau}_{V^2h,\text{non-min}}^{\mu\nu}(q)\\
&=-\frac{i\kappa}{2}m^2\Big[2u^\mu u^\nu
- i\, q_\lambda \big(S^{\mu\lambda}u^{\nu}+S^{\nu\lambda}u^{\mu}\big)- H_1\,q_\lambda q_\sigma S^{\mu\lambda}S^{\nu\sigma}\\
&\quad+ H_2\, u^{\mu}u^{\nu}\,q_{\rho}q_{\sigma}S^{\rho\lambda}S_{\lambda}^{\ \sigma}
+ C_1\, S^{\rho\sigma}S_{\rho \sigma}\,q^{\mu}q^{\nu}\\
&\quad+ C_2 \left(\eta^{\mu\nu}q_{\rho}q_{\sigma}S^{\rho\lambda}S^{\sigma}{}_{\lambda}
- q^\lambda\big(q^{\mu}S_{\lambda \sigma}S^{\nu \sigma}+q^{\nu}S_{\lambda \sigma}S^{\mu \sigma}\big)\right)\Big]\ .
\end{aligned}
\end{equation}
This vertex encodes the most general stationary, parity-even, rotating source at $\mathcal{O}(S^2)$ that reduces to a spherically symmetric configuration in the non-rotating limit. The normalization is such that the minimal vertex~\eqref{eq:MinimalVertex} is recovered when
\begin{equation}\label{eq:MinimalLimit}
    H_1=1,\qquad H_2=0,\qquad C_1=0,\qquad C_2=0.
\end{equation}
We will see that the coefficients $C_i$ correspond to gauge artefacts, while $H_i$ have physical significance, as they determine the multipolar content of the source. An important consistency check is conservation. Both the tree-level and higher-loop dressed vertices satisfy the Ward identity up to local terms,
\begin{equation*}
q_\mu\hat{\tau}_{V^2h}^{\mu\nu}\propto\mathcal{O}(q^2),\qquad 
q_\mu\hat{\tau}_{V^2h,\text{HL}}^{\mu\nu}\propto \mathcal{O}(q^2),
\end{equation*}
which guarantees that only long-range, non-analytic contributions survive. This condition fixes the relative coefficients multiplying $C_2$ in~\eqref{eq:GenericTreeQuadVertex} and forbids additional independent structures. In fact, even without the explicit QFT derivation, one could reconstruct~\eqref{eq:FinalQuadVertex} by demanding the most general symmetric, conserved rank-2 tensor built from $q^\mu$, $u^\mu$ and $S^{\mu\nu}$ at quadratic order in spin. The outcome is that the spin-1 analysis already captures the universal quadrupolar sector of rotating sources. 

\subsection{Loop amplitudes}

Having determined the dressed vertices of the theory, following the framework outlined in Chapter~\ref{chapter:ClassicalGravityFromAmplitudes}, we can now construct the scattering amplitudes and compute the metric order by order in the PM expansion. The non-minimal dressed vertex provides the building block for the source EMT, with the tree-level contribution~\eqref{eq:GenericTreeQuadVertex} determining $T_{\mu\nu}^{(0)}(q)$ and thus the 1PM order of the metric. Higher PM corrections originate from graviton self-interactions, entering $T_{\mu\nu}^{(n)}(q)$ through loop amplitudes according to Eq.~\eqref{eq:EMTfromAMPfinal}. For instance, at one loop the EMT takes the form
\begin{equation}
    \frac{i\kappa}{2}\,(2m)\,T_{\mu\nu}^{(1)}(q)
    =\includegraphics[valign=c,width=0.30\textwidth]{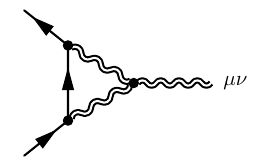}\ ,
\end{equation}
while at two loops it is given by
\begin{equation}
    \frac{i\kappa}{2}\,(2m)\,T_{\mu\nu}^{(2)}(q)
    =\;\includegraphics[valign=c,width=0.30\textwidth]{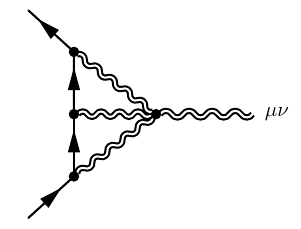}
    \;+\;3\times\includegraphics[valign=c,width=0.30\textwidth]{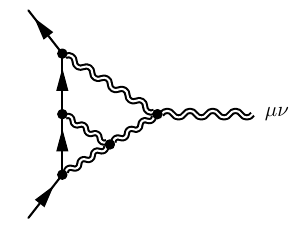}\ .
\end{equation}

Although the explicit expressions of $T_{\mu\nu}(q)$ and of the metric up to 3PM are lengthy, they carry no new physical information compared to the tree-level result, since in fact the full multipolar structure of spacetime is already encoded at 1PM order. Explicit calculations, however, reveal the presence of infrared singularities for ${d=3,4}$. These divergences are absorbed by insertions of the higher-loop vertex~\eqref{eq:CounterTermVertex}, which acts as a counterterm. At 2PM, divergences appear only in $d=4$, where a tree-level insertion of $\hat{\tau}_{V^2h,\text{HL}}^{\mu\nu}$ suffices. Parametrizing
\begin{equation}\label{eq:RenormParametrization}
\begin{aligned}
        \Omega_1\big|_{d=4}&=\frac{\Omega_1^{\mathrm{renorm.}}}{d-4}+\Omega_1^{\mathrm{free}}\ ,\\
        \Omega_2\big|_{d=4}&=\frac{\Omega_2^{\mathrm{renorm.}}}{d-4}+\Omega_2^{\mathrm{free}}\ ,
\end{aligned}
\end{equation}
and fixing
\begin{equation}\label{eq:RenormalizationChoice}
    \Omega_1^{\mathrm{renorm.}}=\frac{1}{9\pi}\ ,\qquad
    \Omega_{2}^{\mathrm{renorm.}}=\frac{H_1+2H_2-1}{60\pi}\ ,
\end{equation}
one finds that the 2PM metric in $d=4$ becomes finite, at the price of introducing logarithmic terms in $r$. The remaining $\Omega_{1,2}^{\mathrm{free}}$ are pure gauge parameters associated with the freedom in~\eqref{eq:GaugeCondition}. At 3PM, divergences arise both in $d=3$ and in $d=4$. In $d=4$ they are removed by a one-loop insertion of~\eqref{eq:CounterTermVertex}, schematically
\begin{equation}
    \frac{i\kappa}{2}\,(2m)\,\delta T_{\mu\nu}^{(1)}(q)
    =2\times\includegraphics[valign=c,width=0.30\textwidth]{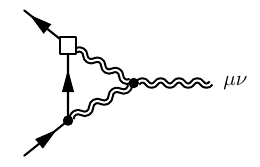}\ ,
\end{equation}
with the same renormalization prescription~\eqref{eq:RenormalizationChoice}. In contrast, in $d=3$ a tree-level insertion of~\eqref{eq:CounterTermVertex} is already sufficient to remove the singularities, and a parametrization of the form
\begin{equation}
\begin{aligned}
        \Omega_1\big|_{d=3}&=\frac{\Omega_1^{\mathrm{renorm.}}}{d-3}+\Omega_1^{\mathrm{free}}\ ,\\
        \Omega_2\big|_{d=3}&=\frac{\Omega_2^{\mathrm{renorm.}}}{d-3}+\Omega_2^{\mathrm{free}}\ ,
\end{aligned}
\end{equation}
cancels all infrared poles in the chosen gauge. Interestingly, in harmonic gauge (${\alpha=1}$) all poles cancel automatically (verified explicitly up to two loops) without fixing the free parts $\Omega_i^{\mathrm{free}}$, although no general principle is known that guarantees this cancellation at arbitrary PM orders. For instance, in $d=4$ at one loop it is possible to choose a gauge where $h_{\mu\nu}^{(2,0)}(r)$ is finite while $h_{\mu\nu}^{(2,2)}(r)$ remains divergent. From the perspective of the Einstein equations, these divergences reflect the inconsistency of enforcing a pure $\rho(r)$-series for $h_{\mu\nu}$ in certain gauges. The resolution requires including logarithmic terms, with the $\Omega_i^{\mathrm{free}}$ playing the role of gauge redundancies in~\eqref{eq:GaugeCondition}, thereby ensuring a consistent PM expansion.

\subsection{Rotating metrics in any dimensions}\label{sec:rotmetric}

We now present the explicit tree-level (1PM) metric, computed up to quadrupole order, as recovered from the amplitude-based formalism. Recalling Eq.~\eqref{eq:PerturbativeEinsteinEqs} for the perturbative construction of the metric and Eq.~\eqref{eq:MultipoleMetricExpansion} for the separation of multipole contributions, the different sectors can be organized as follows. At monopole order, corresponding to the static spherically symmetric part of the solution, we find
\begin{equation}\label{eq:MonopoleMetric}
\begin{aligned}
        h_{00}^{(1,0)}(r)&=\frac{4(d-2)}{d-1}\,G_Nm\,\rho(r)\ ,\\
        h_{0i}^{(1,0)}(r)&=0\ ,\\
        h_{ij}^{(1,0)}(r)&=\frac{4}{d-1}\,\delta_{ij}\,G_Nm\,\rho(r)\ ,
\end{aligned}
\end{equation}
which reproduces the familiar Schwarzschild-Tangherlini metric at 1PM, namely the higher-dimensional generalization of the Schwarzschild solution. At dipole order, which encodes the leading rotational effects through the spin tensor $S_{ij}$, the metric components are
\begin{equation}\label{eq:DipoleMetric}
\begin{aligned}
        h_{00}^{(1,1)}(r)&=0\ ,\\
        h_{0i}^{(1,1)}(r)&=\frac{2(d-2)}{r^2}\,x^k S^{i}{}_{k}\,G_Nm\,\rho(r)\ ,\\
        h_{ij}^{(1,1)}(r)&=0\ .
\end{aligned}
\end{equation}
This sector is completely fixed by minimal coupling and no additional parameters appear, giving a universal result for any (slow)rotating source. Finally, at quadrupole order we obtain
\begin{equation}\label{eq:QuadMetric}
\begin{aligned}
        h_{00}^{(1,2)}(r)&=-\frac{2(d-2)\big(H_2(d-2)+H_1\big)}{d-1}\,
        \frac{r^2 S_{k_1k_2}S^{k_1k_2}-d\,x^{k_1}x^{k_2}S_{k_1}{}^{k_3}S_{k_2k_3}}{r^4}\,G_Nm\,\rho(r)\ ,\\
        h_{0i}^{(1,2)}(r)&=0\ ,\\
        h_{ij}^{(1,2)}(r)&=\frac{2(d-2)}{(d-1)r^4}\Big[
        -C_1(d-1)d\,x_i x_j S_{k_1k_2}S^{k_1k_2}
        -r^2(d-1)\big(2C_2+H_1\big)S_{ik}S_{j}{}^{k}\\
        &+r^2\big(C_1(d-1)+H_1-H_2\big)S_{k_1k_2}S^{k_1k_2}\delta_{ij}
        +d\,C_2(d-1)\,x^{k_1}S_{k_1k_2}\big(x_j S_{i}{}^{k_2}+x_i S_{j}{}^{k_2}\big)\\
        &+d\,x^{k_1}x^{k_2}\Big((d-1)H_1 S_{ik_1}S_{jk_2}+(H_2-H_1)S_{k_1}{}^{k_3}S_{k_2k_3}\delta_{ij}\Big)\Big]\,G_Nm\,\rho(r)\ .
\end{aligned}
\end{equation}
Here free parameters appear for the first time. We anticipate that the coefficients $H_i$ are physical and encode the genuine gravitational quadrupole moments of the source, while $C_i$ correspond to gauge artefacts, reflecting the residual freedom in defining the EMT at $\mathcal{O}(S^2)$.

Two important features are worth stressing. First, the metric remains independent of the gauge parameter $\alpha$ introduced in~\eqref{eq:GaugeCondition}. This is due to the fact that at tree level the 1PM result is fully gauge invariant, with differences between de~Donder and harmonic gauges only arising beyond tree level, starting at 2PM from graviton self-interactions. Second, the structure of Eq.~\eqref{eq:QuadMetric} demonstrates how amplitude methods reproduce the expected multipole hierarchy. Specifically,  monopole and dipole orders are universal, while the quadrupole introduces new parameters that characterize the intrinsic structure of the source. Moreover, even though free parameters were expected in Eq.~\eqref{eq:QuadMetric}, we know from section~\ref{sec:GravitationalMultipoles} that at quadrupole order the genuine physical degrees of freedom are only two, namely the mass and stress moments. By contrast, the amplitude-based result appears to contain four independent coefficients (or six, if we also count the $\Omega$’s introduced through the renormalization procedure in $d=3,4$). We now show explicitly that only two of these parameters are physical, while the others are pure gauge redundancies.

To this end, let us work in harmonic gauge ($\alpha=1$) and consider the infinitesimal ACMC-preserving transformation~\eqref{eq:InfinitesimalGaugeTransf} with $\xi^0=0$ and
\begin{equation}\label{eq:epsilonC1C2}
  \xi^i=G_Nm\Big(A\, S^{ik}S_{k}{}^{j}+B\, S^{lm}S_{lm}\delta^{ij}\Big)\partial_j\rho(r)\ .
\end{equation}
Choosing
\begin{equation}
    A=2C_2\ ,\qquad B=C_1\ ,
\end{equation}
and using the transformation law~\eqref{eq:MetricPrime}, one finds through direct calculation that the shift~\eqref{eq:epsilonC1C2} exactly eliminates the dependence on $C_{1,2}$ in the metric. This establishes that the $C_i$ coefficients are gauge artefacts, while $H_{1,2}$ remain as the genuine physical parameters linked to the gravitational multipoles. Indeed, two further ACMC-preserving shifts affecting the quadrupole sector are relevant at higher PM order, namely
\begin{equation}\label{eq:epsilon1}
    \xi^i_1=(Gm)^{\frac{d}{d-2}}\,\tilde{\Omega}_1\,\partial^i\rho(r)\ ,
\end{equation}
and
\begin{equation}\label{eq:epsilon2}
    \xi^i_2=\frac{(Gm)^{\frac{d}{d-2}}}{m^2}\,\tilde{\Omega}_2\,S_{l}{}^{k}S_{km}\,\partial^i\partial^l\partial^m\rho(r)\ .
\end{equation}
Applying these shifts removes the explicit dependence on $\Omega_{1,2}$ in~\eqref{eq:CounterTermVertex}, confirming that these coefficients are also gauge artefacts rather than physical observables. After taking into account all ACMC-preserving gauge transformations, the quadrupole sector is therefore characterized by only two genuine physical parameters, $H_1$ and $H_2$, corresponding to the mass and stress moments. Equivalently, in~\eqref{eq:NonMinimalAction} only the $K_3$ and $K_4$ operators carry physical information, while the others can be eliminated by redefinitions.

A remarkable simplification arises in $d=3$, where the spin tensor can be dualized into a vector
\begin{equation}\label{eq:SpinDualVector}
  S^{ij}=\varepsilon^{ijk}s_k\ , 
\end{equation}
where $s_k$ is the spin vector. In this case, by choosing
\begin{equation}
    A=H_2-H_1\ ,\qquad B=-\tfrac{1}{2}H_1\ ,
\end{equation}
one finds that the metric depends on $H_1$ and $H_2$ only through the combination ${H_1+H_2}$. Hence, in four spacetime dimensions there is effectively a single independent quadrupolar parameter. This is consistent with the analysis of section~\ref{sec:GravitationalMultipoles}, where we showed that in $d=3$ the stress multipole vanishes and only one physical quadrupolar moment survives, the mass moment. Finally, matching Eq.~\eqref{eq:QuadMetric} with the general multipole expansion~\eqref{eq:MultipoleExpandedMetric}, we can extract the explicit form of the spin-induced quadrupoles in arbitrary dimensions, reading 
\begin{align}
    \mathbb{M}^{(2)}_{a_1a_2}&=-d\Big(H_1+(d-2)H_2\Big)S_{a_1k}S_{a_2}{}^{k}\ , \label{eq:MassQuadrupoleGenericD}\\[2pt]
    \mathbb{G}^{(2)}_{ij,a_1a_2}&=-d(d-1)\,H_1\,S_{(i|a_1}S_{|j)a_2}\ .\label{eq:StressQuadrupoleGenericD}
\end{align}
From these expressions it is clear that $H_1$ controls the stress quadrupole, while the linear combination $H_1+(d-2)H_2$ fixes the mass quadrupole. In $d=3$ the degeneracy between these parameters becomes manifest once the spin tensor is traded for the spin vector. In this case one obtains
\begin{align}
    \mathbb{M}^{(2)}_{a_1a_2}N_{a_1a_2}\Big|_{d=3}&=3(H_1+H_2)(S \cdot  x)^2+\cdots\ ,\label{eq:MassQuadrupoleD4}\\
    \tilde{\mathbb{G}}^{(2)}_{ij,a_1a_2}N_{a_1a_2}\Big|_{d=3}&=\delta_{ij}\,\mathbb{M}^{(2)}_{a_1a_2}N_{a_1a_2}\Big|_{d=3}+\cdots\ .\label{eq:StressQuadrupoleD4}
\end{align}
We can see explicitly that $\tilde{\mathbb{G}}^{(2)}$ is determined entirely by $\mathbb{M}^{(2)}$, and hence the stress multipole $\mathbb{G}^{(2)}$ vanishes. This property extends to all higher multipole orders in $d=3$, making manifest how the representation theory of $SO(3)$ prevents the existence of stress multipoles in four spacetime dimensions. The physical quadrupole structure of a rotating source is therefore reduced to a single mass-type moment in this case.

\section{Comparison with known solutions}\label{sec:Spin1Comparison}

We now compare the metric obtained from our amplitude-based approach with the PM expansion of known vacuum solutions in $D \geq 4$, focusing on their multipolar structure. We will see that in $D=4$, reproducing a specific solution at quadrupole order requires fixing only the combination $H_1+H_2$, while in $D=5$ both $H_1$ and $H_2$ must be fixed independently, providing direct evidence for the existence of an independent stress multipole moment. Furthermore, we define the \emph{simplest} metric as the one generated by a minimally coupled theory. We then relate this construction to explicit solutions, showing that Kerr BHs are the simplest vacuum solutions in $D=4$, while Myers–Perry BHs and black rings in $D=5$ are not.

\subsection{Hartle–Thorne metric in $D=4$}\label{sec:HartleThorneComparison}

The Hartle–Thorne metric~\cite{Hartle:1967he,Hartle:1968si} describes the spacetime generated by the most generic slowly rotating matter distribution in $D=4$, consistently expanded up to spin-induced quadrupole order, and explicitly given up to 3PM in Eq.~\eqref{app:HartleThorneMetric}.  
As shown in~\cite{Gambino:2024uge}, the Hartle–Thorne solution can be rewritten in harmonic coordinates to facilitate comparison with the amplitude-based metric. This change of coordinates introduces some free coefficients associated with gauge redundancies, which can be mapped to the free parameters of our construction. With such an identification, we verified that the two metrics match exactly up to 3PM order, while the details of such coordinate transformation are given in appendix~\ref{App:HTInHarm}. The comparison is most transparent in the frame where the angular momentum is aligned with the $z$-axis. In terms of the spin tensor this corresponds to
\begin{equation}\label{eq:d3SpinMatrix}
 S_{ij}=\begin{pNiceMatrix}[columns-width=auto]
        0 & a & 0 \\
        -a & 0 & 0 \\
        0 & 0 & 0
    \end{pNiceMatrix}\ ,
\end{equation}
where $a=J/m$ is the physical angular momentum per unit mass of the source.  
In this frame, the temporal component of the Hartle–Thorne metric in harmonic coordinates reads
\begin{equation}
    g_{00}^{\rm HT}=-1+\frac{2 G_N m}{r}-\frac{a^2 G_N m \,\zeta}{r^3}\left(3\frac{z^2}{r^2}-1\right) +\mathcal{O}(G_N^2,a^3)\ ,
\end{equation}
where $\zeta$ parametrizes the mass quadrupole\footnote{Because of axisymmetry, the quadrupole moment tensor is determined by a single parameter.}. By imposing
\begin{equation}\label{eq:HTcondition}
    H_1+H_2=\zeta\ ,
\end{equation}
the amplitude-based metric fully reproduces the Hartle–Thorne solution. In agreement with Eq.~\eqref{eq:MassQuadrupoleD4}, only the combination $H_1+H_2$ contributes to the mass quadrupole, and since the stress quadrupole vanishes in $D=4$, there is only one physical parameter at quadrupole order. This redundancy can be understood as a consequence of the special properties of $SO(3)$, which forbid stress multipoles in four spacetime dimensions.

Importantly, the Hartle–Thorne metric includes as a special case the situation in which the source is a BH. In $D=4$, the no-hair theorems state that the only asymptotically flat, stationary, vacuum solution with a regular horizon is the Kerr metric, whose multipole structure is uniquely determined by the ADM mass and angular momentum~\cite{Carter:1971zc,Robinson:1975bv}. In the Hartle–Thorne language, the Kerr solution corresponds to $\zeta=1$, so that in terms of the amplitude parameters
\begin{equation}\label{eq:Kerrcondition}
    H_1+H_2=1\ .
\end{equation}
From the EFT perspective, this condition is satisfied by infinitely many non-minimal theories, since $H_1$ and $H_2$ can be chosen arbitrarily as long as their sum equals one. This reflects the degeneracy of $D=4$, where only one physical degree of freedom survives, even though two free parameters remain in the EFT. The minimally coupled theory, however, exactly satisfies the Kerr condition, thereby reproducing the Kerr geometry. From the amplitude viewpoint, this shows that the simplest vacuum solution in $D=4$ is precisely the Kerr BH. This resonates with recent literature on the interplay between minimal coupling and BHs~\cite{Arkani-Hamed:2017jhn,Chung:2018kqs,Guevara:2018wpp,Bern:2020buy}.

\subsection{Myers–Perry black holes in $D=5$}

We now turn to higher-dimensional vacuum solutions in GR, focusing on Myers–Perry BHs in $D=5$~\cite{Myers:1986un}, which explicit metric expression is given in Eq.~\eqref{app:MPD5Metric}. In this case the rotation group is $SO(4)$, which admits two independent Casimir invariants. Physically, this means that a rotating body in five dimensions can spin independently in two orthogonal planes. Accordingly, the Myers–Perry metric in $D=5$ carries two independent angular momenta. Introducing Cartesian coordinates $(x_1,y_1,x_2,y_2)$, such that each plane $(x_i,y_i)$ is orthogonal to the angular momentum $J_i$, the spin tensor can be block-diagonalized as
\begin{equation}\label{eq:BlockDiagonalD5MEtric}
 S_{ij}=\begin{pNiceMatrix}[columns-width=auto]
        0 & a_1 & 0 & 0 \\
        -a_1 & 0 & 0 & 0 \\
        0 & 0 & 0 & a_2 \\
        0 & 0 & -a_2 & 0
    \end{pNiceMatrix}\ ,
\end{equation}
where $a_i$ are the physical angular momentum densities. Rewriting the Myers–Perry metric in harmonic coordinates (see appendix~\ref{App:MPinHarm}), and comparing it with our amplitude-based construction, we find an exact match up to the order of our expansion (namely, two loops), provided that the parameters are fixed as
\begin{equation}\label{eq:MPcondition}
   H_1=\frac{3}{8}\ ,\qquad H_2=\frac{15}{16}\ .
\end{equation}
In contrast with the $D=4$ case, here both $H_1$ and $H_2$ are independently determined, demonstrating explicitly that mass and stress quadrupoles are distinct and physical. Indeed, the corresponding multipoles are
\begin{align}
    \mathbb{M}^{(2)}_{a_1a_2}\Big|_{d=4}^{\rm MP}&=-9\,S_{a_1k}S_{a_2}{}^{k}\ , \\
    \mathbb{G}^{(2)}_{ij,a_1a_2}\Big|_{d=4}^{\rm MP}&=-\frac{9}{2}\,S_{(i|a_1}S_{|j)a_2}\ .
\end{align}
From the EFT point of view, this implies that the Myers–Perry solution cannot arise from a minimally coupled theory. Unlike $D=4$, where the Kerr solution coincides with the minimal limit and hence represents the ``simplest'' BH, in $D=5$ the Myers–Perry solution requires non-minimal couplings. This observation leads to the conclusion that, even though the Myers–Perry solution is the natural higher-dimensional generalization of Kerr~\cite{Emparan:2008eg}, from the amplitude perspective it is not the simplest vacuum solution in $D=5$.

An additional feature emerges when the two angular momenta are equal, namely $a_1=a_2$. In this case the Myers–Perry metric enjoys an enhanced \mbox{cohomogeneity-1} symmetry~\cite{Myers:1986un}, and the multipole expansion shows that the mass quadrupole vanishes~\cite{Heynen:2023sin}
\begin{equation}
   \lim_{{a}_1\rightarrow {a}_2} 
   \mathbb{M}^{(2)}_{ij}\Big|_{d=4}^{\rm MP}N_{ij}=0\ ,
\end{equation}
while, by contrast, the stress quadrupole remains finite. This phenomenon is not restricted to $D=5$. In fact, for any odd spacetime dimension $D$, when all angular momenta are equal, the mass quadrupole vanishes,
\begin{equation}
   \lim_{{a}_i\rightarrow {a}} 
   \mathbb{M}^{(2)}_{ij}\Big|_{d=\text{even}}N_{ij}=0\ ,
\end{equation}
while the stress multipoles do not~\cite{Gambino:2024uge}. We interpret this as a generic property of the gravitational field sourced by a spinning point-like mass in higher dimensions, in which equal rotation parameters suppress the mass quadrupole but not the stress quadrupole, thereby exposing the fundamentally new role of stress multipoles beyond four dimensions.

\subsection{Black rings}\label{sec:BlackRing}

As already mentioned, in $D>4$ Myers–Perry BHs are not the only asymptotically flat vacuum solutions with horizons. Another well-known family is provided by black rings~\cite{Emparan:2006mm}, whose event horizon has ring topology $S^1 \times S^2$ rather than spherical topology as in Kerr or Myers–Perry. For simplicity we restrict to the case of a single angular momentum, where the only non-vanishing components of the spin tensor are ${S_{21}=-S_{12}=a}$, with the explicit expression of the spacetime given in Eq.~\eqref{eq:BRoriginalMetric}. The solution is parametrized by three constants $(\mathcal{R},\nu,\lambda)$ with $0<\nu\leq 1$, which respectively encode the size of the ring, its shape, and its rotation velocity. As shown in~\cite{Gambino:2024uge}, one can trade $\mathcal{R}$ and $\nu$ for the physical mass $m$ and spin density $a$, leaving $(m,a,\lambda)$ as the relevant parameters. In these variables the black ring metric admits a PM expansion in harmonic coordinates, explicitly performed in appendix~\ref{app:BR}. Generally, the solution features a conical singularity unless the parameters satisfy the equilibrium condition
\begin{equation}
    \lambda=\frac{2\nu(m,a)}{1+\nu^2(m,a)}\ , \label{equilibrium}
\end{equation}
which enforces a specific relation between mass and spin. This condition ensures the absence of external struts and sets a lower bound on the angular momentum. Nevertheless, it is useful to keep $\lambda$ as a free parameter, so that the family of black ring solutions can be studied more broadly. In particular, the Myers–Perry solution is recovered in the formal limit $\lambda\to 1$ while keeping $(m,a)$ fixed, although this lies outside the equilibrium curve.

To compare the metric with Eq.~\eqref{eq:QuadMetric}, recall that in the multipole expansion~\eqref{eq:MultipoleExpandedMetric} both the mass and stress quadrupoles have dimensions of length squared. In $D=5$ they can therefore be expressed in terms of $a^2$ or $G_N m$\footnote{In higher dimensions $[G_N m]=L^{D-3}$.}. If an additional intrinsic scale $\Lambda$ characterizes the source, the corresponding quadrupole schematically takes the form
\begin{equation}
    \mathbb{M}^{(2)} \sim \Lambda = \sigma\, G_N m\ ,
\end{equation}
with $\sigma$ a dimensionless constant. Such an intrinsic quadrupole would contribute already at tree level in the amplitude-based approach, yet would enter the metric at 2PM order once expressed in terms of $G_N m$. This is precisely the case for black rings, where their ring topology induces a non-vanishing intrinsic quadrupole even in the absence of spin. Since our amplitude-based framework is designed to capture spin-induced multipoles, we cannot match intrinsic contributions at quadrupole order. For this reason we restrict the comparison to 1PM, where only spin-induced effects appear. Using the results of Eq.~\eqref{app:BRinHarm}, we find that fixing
\begin{equation}\label{eq:BRcondition}
        H_1=\frac{3}{4(1+\lambda)}\ ,\qquad 
        H_2=\frac{3(6\lambda-1)}{8(1+\lambda)}\ ,
\end{equation}
the black ring metric agrees with the amplitude-based solution at 1PM. As a check, for $\lambda=1$ these values reduce exactly to the Myers–Perry coefficients in Eq.~\eqref{eq:MPcondition}. Just as for Myers–Perry, the parameters $H_1$ and $H_2$ are independently fixed, confirming that black rings in $D=5$ also carry both mass and stress quadrupole moments. Explicitly,
\begin{align}
    \mathbb{M}^{(2)}_{a_1a_2}\Big|_{d=4}^{\text{BR}}
    &=-\frac{18 \lambda}{1+\lambda}\,S_{a_1k}S_{a_2}{}^{k}+\mathcal{O}(G_N m)\ , \\[4pt]
    \mathbb{G}^{(2)}_{ij,a_1a_2}\Big|_{d=4}^{\text{BR}}
    &=-\frac{9}{1+\lambda}\,S_{(i|a_1}S_{|j)a_2}+\mathcal{O}(G_N m)\ ,
\end{align}
where we have neglected the intrinsic (non–spin-induced) contributions.

Finally, one may ask whether there exists a value of $\lambda$ such that the black ring coincides with the simplest EFT solution, \textit{i.e.} the one generated by the minimal vertex. However, it is straightforward to verify that no value of $\lambda$ simultaneously yields $H_1=1$ and $H_2=0$. We therefore conclude that all single-spin black ring solutions necessarily correspond to non-minimally coupled theories in the scattering amplitude framework.

\subsection{The simplest metric in arbitrary spacetime dimension $D$}

For completeness, we present here the explicit form of the \emph{simplest} solution in generic spacetime dimension $D$, obtained from the minimal vertex~\eqref{eq:MinimalVertex} corresponding to $H_1=1$ and $H_2=0$. The metric at 1PM and up to quadrupole order reads
\begin{equation}\label{eq:SimplestMetric}
    \begin{aligned}
        h_{00}^{(1,2)}(r)&=\frac{2(D-3)}{D-2}
        \frac{r^2 S_{k_1k_2}S^{k_1k_2}-(D-1)\, x^{k_1}x^{k_2}S_{k_1}{}^{k_3}S_{k_2k_3}}{r^4}\,
        G_N m \rho(r)\ ,\\[4pt]
        h_{0i}^{(1,2)}(r)&=0\ ,\\[4pt]
        h_{ij}^{(1,2)}(r)&=-\frac{2(D-3)}{(D-2)r^4}\Bigg[
        -r^2(D-2)S_{ik}S_{j}{}^{k}+r^2 S_{k_1k_2}S^{k_1k_2}\delta_{ij}\\
        &\quad +(D-1)\, x^{k_1}x^{k_2}\Big((D-2)S_{ik_1}S_{jk_2}
        -S_{k_1}{}^{k_3}S_{k_2k_3}\delta_{ij}\Big)\Bigg]\,G_N m \rho(r)\ . 
    \end{aligned}
\end{equation}
As emphasized throughout this chapter, the 1PM order already captures the complete multipolar information relevant for our purposes, and the quadrupole moments of the simplest solution are given by
\begin{equation}
\begin{aligned}
    \mathbb{M}^{(2)}_{a_1a_2}\Big|^{\rm simplest}&=-(D-1)\,S_{a_1k}S_{a_2}{}^{k}\ , \\
    \mathbb{G}^{(2)}_{ij,a_1a_2}\Big|^{\rm simplest}&=-(D-1)(D-2)\,S_{(i|a_1}S_{|j)a_2}\ .
\end{aligned}
\end{equation}
In $D=4$ this metric coincides with the Kerr solution, reinforcing the idea that Kerr is the simplest BH, in the sense of being generated by minimal coupling. However, for $D=5$ the situation changes and the simplest metric does not reproduce either the Myers–Perry BH or the single-spin black ring. Our analysis therefore indicates that in higher dimensions the minimal solution is not a BH. This raises an interesting question: does there exist an exact solution of Einstein’s equations corresponding to this metric, and if so, what matter content would source it? One might imagine an analogue of the Hartle–Thorne construction in $D=4$, where a suitably deformed matter distribution reproduces the simplest geometry. Another possibility is that a BH solution exists whose spin-induced multipoles match those of the simplest metric, but whose higher-order structure differs. While at tree level the linearity of the theory would allow for the superposition of solutions, beyond this regime such a construction becomes highly nontrivial.

A further intriguing possibility is that different ``simplest'' solutions may arise in $D>4$ depending on the chosen field representation. Up to this point we have assumed symmetric, traceless representations of the Lorentz group for spin-$s$ fields. However, in higher dimensions additional representations exist. For instance, in ${D=5}$ one could consider an antisymmetric tensor field coupled minimally to gravity. The associated minimal vertex might coincide with~\eqref{eq:MinimalVertex}, reinforcing the universality of the simplest metric, or it could differ, in which case a new simplest solution would emerge, potentially closer to Myers–Perry or other higher-dimensional BHs. Similarly, one could investigate minimally coupled fields with spin $s>1$ and test whether the resulting metrics share the same universal quadrupole structure. Finally, let us emphasize that the general vertex~\eqref{eq:FinalQuadVertex}, and hence the most generic stationary metric in $D$ dimensions, could also be obtained without referring to scattering amplitudes, just by constructing directly the most general symmetric, conserved EMT in momentum space (we will discuss this in details in chapter~\ref{chapter:SourceMultipoles}). However, the virtue of the amplitude-based derivation is that it singles out a distinguished solution, namely our simplest metric, and provides a physical interpretation of BHs in terms of minimal couplings. In $D=4$, the identification of the simplest solution with Kerr resonates with the no-hair theorems, which in this language translate into the absence of extra operators in the effective action. In $D>4$, however, the lack of uniqueness theorems allows for a richer spectrum of BH solutions, and our results suggest that the correspondence between simplest metrics and BHs breaks down. Exploring this discrepancy offers a promising avenue for future work on the interplay between EFT, amplitudes, and higher-dimensional gravity.

\chapter{Gyromagnetic factor of black holes}\label{chapter:GyromagneticFactor}

In the preceding discussion, we have reconstructed the gravitational field of neutral and rotating sources from scattering amplitudes, showing how the multipolar structure of Kerr and Myers–Perry BHs arises naturally within the PM expansion. The natural next step is to investigate how this picture is modified once electric charge is included, thereby testing the amplitude framework in a richer phenomenological setting. Unlike mass and spin, which are universally dictated by minimal couplings, electromagnetic multipoles depend on additional interactions and therefore provide a sharper probe of the amplitude–geometry correspondence. In four spacetime dimensions, Einstein–Maxwell theory admits a unique family of stationary, asymptotically flat BHs with spherical horizon topology, namely the Kerr–Newman solutions~\cite{Bardeen:1973gs,Robinson:1975bv,Robinson:2004zz,Cardoso:2016ryw}. Their entire multipole tower is fixed by three conserved charges, such as the ADM mass, angular momentum, and electric charge~\cite{Geroch:1970cd,Thorne:1980ru,Hansen:1974zz}. Any amplitude-based reconstruction of charged rotating sources in four dimensions must reproduce these multipoles, with the gyromagnetic factor of Kerr–Newman serving as a key benchmark.

In higher dimensions the landscape of solutions is considerably richer. The uniqueness theorems that hold in $D=4$ no longer apply, and even though the higher-dimensional analogue of Reissner–Nordström was obtained by Tangherlini~\cite{Tangherlini:1963bw} in the static case, a closed-form for charged and rotating solution of pure Einstein–Maxwell theory remains unknown~\cite{Kunz:2017pnm}. Perturbative constructions exist in $D=5$ and beyond~\cite{Aliev:2004ec,Navarro-Lerida:2010orf,Aliev:2006yk}, but exact solutions require additional interactions, most notably the five-dimensional Chong–Cvetič–Lü–Pope (CCLP) BH~\cite{Chong:2005hr}, which is supported by a Chern–Simons (CS) term (see~\cite{Ortaggio:2023rzp,Deshpande:2024vbn} for recent reviews and developments). In this chapter we extend the amplitude framework to the case of charged, rotating sources, extending the results of the amplitude-based derivation in the non-rotating case~\cite{DOnofrio:2022cvn, Gambino:2022kvb}. The source will be described as a massive spin-$\tfrac{1}{2}$ field carrying electric charge, so that its mass and spin are encoded in the stress tensor while electromagnetic multipoles arise through photon couplings. Unlike the gravitational case, where the first non-trivial multipole appears only at quadrupole order, electromagnetic multipoles already become non-trivial at dipole order. This makes spin-$\tfrac{1}{2}$ fields the ideal setup to isolate and probe the leading electromagnetic effects beyond minimal coupling. To capture such dipole contributions, we include a Pauli-type interaction whose coefficient directly determines the gyromagnetic factor. Within this framework we recover the Kerr–Newman solution in four dimensions, as well as the CCLP BH in five dimensions, and derive the universal prediction
\begin{equation}
    \mathfrak{g} = \frac{d-1}{d-2}
\end{equation}
for the gyromagnetic factor of charged rotating BHs in $d+1$ dimensions. The key point is that while neutral multipoles are universally fixed by minimal couplings, the inclusion of electric charge qualitatively changes the picture. Only in four dimensions does minimal coupling reproduce the correct gyromagnetic factor, whereas in higher dimensions Pauli-like operators are unavoidable, reflecting the richer structure of charged rotating solutions beyond $D=4$.

The rest of this chapter is organized as follows. In Sec.~\ref{sec:ElectricCharge} we introduce the effective action for a charged spin-$\tfrac{1}{2}$ field and set up the post-Minkowskian expansion that allows us to compute both the far-field metric and the associated electromagnetic potential. This provides the foundation for analyzing how the electric charge modifies the multipolar structure relative to the neutral case discussed in the previous chapters. In Sec.~\ref{sec:KnownChargedSolution} we confront these amplitude-based results with the known charged BH solutions in four and five dimensions, such as Kerr–Newman and the CCLP BH. This comparison not only validates the framework but also highlights the role of non-minimal couplings in higher dimensions. Finally, from this analysis we extract a general formula for the gyromagnetic factor valid in arbitrary spacetime dimension, which will serve as a unifying prediction linking the amplitude perspective with the geometry of charged rotating BHs.

\section{Electromagnetic field from scattering amplitudes}\label{sec:ElectricCharge}

Let us consider the action of a massive Dirac spin-$\tfrac{1}{2}$ field $\psi$ of mass $m$ and charge $Q$, minimally coupled to both gravity and electromagnetism in arbitrary spacetime dimension as
\begin{equation}\label{eq:minimal_general_action}
    S= \int d^D x ~\mathrm{e} \left( \frac{2}{\kappa^2}R-\frac{1}{4}F_{\mu \nu} F^{\mu \nu}  +\bar{\psi}\left(i\,e^{\mu}_{~\alpha}\gamma^{\alpha} D_\mu-m \right)\psi\right)\ ,
\end{equation}
where $e^{\alpha}{}_\mu$ is the tetrad field, with $g_{\mu \nu}=\eta_{\alpha \beta}\,e^{\alpha}_{~\mu}\,e^{\beta}_{~\nu}$ and $\mathrm{e}=|\det e^{\alpha}_{~\mu}|=\sqrt{|\det g_{\mu \nu}|}$. Referring to Eq.~\eqref{eq:minimal_general_action}, $F_{\mu\nu}$ is the electromagnetic strength tensor, while the covariant derivative acting on the spinor reads
\begin{equation}\label{eq:covariant_der_def}
    D_\mu\psi=\partial_\mu \psi+ i Q A_\mu \psi - \frac{1}{2} \omega_{\mu \alpha \beta}  \Sigma^{\alpha \beta} \psi\ ,
\end{equation}
with $\Sigma^{\alpha \beta}=\tfrac{i}{4} [\gamma^\alpha,\gamma^\beta]$ the Lorentz generators in the spin-$\tfrac{1}{2}$ representation, $\omega_{\mu\alpha\beta}$ the spin connection and $A_\mu$ being the electromagnetic field. To extract long-distance observables, we expand both the metric and the gauge field in a PM expansion as in Eq.~\eqref{eq:ExpansionInPM},
obtaining
\begin{equation}\label{eq:pm_exp_a_h}
    g_{\mu \nu}(x) = \eta_{\mu \nu}+\kappa \sum_{n\geq1}h^{(n)}_{\mu \nu}(x)\ , 
    \qquad 
    A_\mu(x) = \sum_{n \geq 0} A^{(n)}_{\mu}(x)\ .
\end{equation}
To proceed with explicit calculations, in order to relate both the metric and the electromagnetic field respectively to the EMT and the conserved electromagnetic current, we must fix a gauge. For the gravitational sector we adopt the usual harmonic (de~Donder) gauge as in Eq.~\eqref{eq:GaugeCondition}, consistent with previous chapters, while for the electromagnetic field we choose the Lorenz gauge, reading 
\begin{equation}
    \mathcal{L}_{GF}^{(A)}=-\frac{1}{2}(\partial_\mu A^\mu)^2\ .
\end{equation}
With these conventions, the linearized field equations for the metric reduce to Eq.~\eqref{eq:PerturbativeEinsteinEqs}, reported here for convenience in the specific case of the harmonic gauge condition in Eq.~\eqref{eq:GaugeCondition}, reading 
\begin{equation}\label{eq:eins_eq_exp_pm_momentum}
     \kappa\,h_{\mu \nu}^{(n)}(x) = \frac{\kappa^2}{2} \int \frac{d ^d q}{(2\pi)^d}\frac{e^{-i q\cdot x}}{q^2}\left( T_{\mu \nu}^{(n-1)}(q)-\frac{1}{d-1}\eta_{\mu \nu}\, T^{(n-1)}(q) \right)\ ,
\end{equation}
and similarly, the Maxwell equations yield for the electromagnetic potential~\cite{DOnofrio:2022cvn}
\begin{equation}\label{eq:em_equations_pm_momentum}
    A_{\mu}^{(n)}(x) = \int \frac{d ^d q}{(2\pi)^d} \frac{e^{-i q\cdot x}}{q^2}\, j^{(n)}_{\mu}(q)\ ,
\end{equation}
with $j^{(n)}_{\mu}$ the electromagnetic current. In our case $j_\mu$ is simply the conserved Noether current associated with the global $U(1)$ symmetry of the Dirac action, whose lowest-order contribution reads $j_\mu = Q\,\bar{\psi}\gamma_\mu \psi$. The EMT and electromagnetic current in momentum space are then obtained by taking the classical limit of three-point graviton and photon emission amplitudes, generalizing the framework outlined in chapter~\ref{chapter:ClassicalGravityFromAmplitudes} by considering photon exchanges within the tree structure of the amplitude. Following the diagrammatic reasoning outlined in~\cite{Bjerrum-Bohr:2002fji, Mougiakakos:2020laz, DOnofrio:2022cvn, Gambino:2024uge}, diagrams at $l$-loop order with $n_g$ external gravitons and $n_p$ external photons inserted along the massive line satisfy the relation $l = n_g + n_p - 1$. This counting reflects the fact that each additional emission carries a power of $\hbar^{-1}$ in the classical limit, while loops are required to reproduce the full non-linear structure of Einstein–Maxwell theory. Generalizing Eq.~\eqref{eq:ClassicalEMTfrom3PointAMP} to include photon insertions~\cite{DOnofrio:2022cvn}, the classical contribution to the EMT is captured by diagrams of the form~\cite{Gambino:2024uge}
\begin{equation}\label{eq:metric_class_diagram_general}
\includegraphics[valign=c,width=0.45\textwidth]{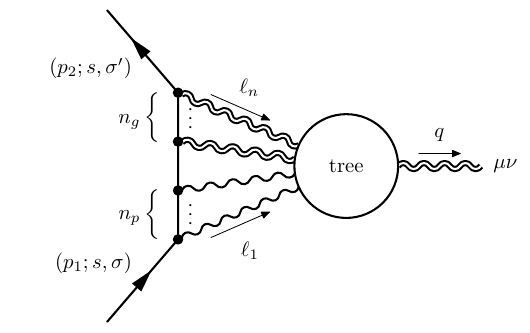}
\,=-i\frac{\kappa}{2}~ T^{(n_g+\frac{n_p}{2}-1)}_{\mu \nu}(q)~\delta_{\sigma\sigma'}\ ,
\end{equation}
while the corresponding electromagnetic current is extracted from
\begin{equation}\label{eq:potential_class_diagram_general}
\includegraphics[valign=c,width=0.45\textwidth]{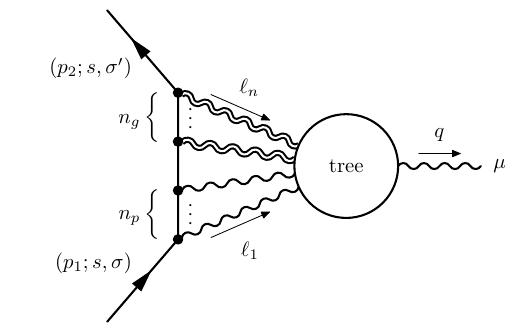}
\,=-i~ j^{(n_g+\frac{n_p+1}{2}-1)}_{\mu}(q)~\delta_{\sigma\sigma'}\ .
\end{equation}
As in the neutral case, one can show that the classical contributions reduce to terms proportional to the $l$-loop master integral of Eq.~\eqref{eq:MasterSunset}, whose Fourier transform controls the far-field metric and gauge potential.  

\begin{figure}[h]
\centering
\begin{equation*}
\includegraphics[valign=c,width=0.25\textwidth]{Diagrams_PDF/GyroDiags/Graviton_Minimal_Vertex.pdf}
+\includegraphics[valign=c,width=0.25\textwidth]{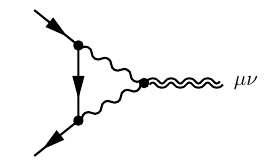}
+2\times\includegraphics[valign=c,width=0.25\textwidth]{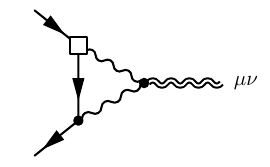}
\end{equation*}
\caption{Diagrams, with corresponding multiplicity, employed for the computation of the PM expansion of the metric up to 1PM and in dipole approximation. The fermion–fermion squared vertex denotes the Pauli coupling.}
    \label{fig:metric_one}
\end{figure}

\begin{figure}[h]
    \centering
\begin{equation*}
\includegraphics[valign=c,width=0.2\textwidth]{Diagrams_PDF/GyroDiags/Photon_Minimal_Vertex.pdf}
+\includegraphics[valign=c,width=0.2\textwidth]{Diagrams_PDF/GyroDiags/Photon_Pauli_Vertex.pdf}
+2\times\includegraphics[valign=c,width=0.2\textwidth]{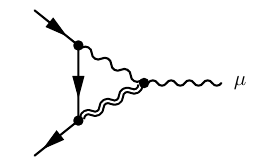}
+2\times\includegraphics[valign=c,width=0.2\textwidth]{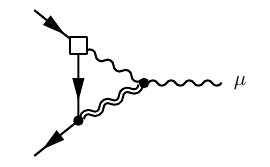}
\end{equation*}
\caption{Diagrams, with corresponding multiplicity, used in the calculation of the PM expansion of the potential up to 1PM, linear in the charge and in dipole approximation. The fermion–fermion squared vertex denotes the Pauli coupling.}
    \label{fig:potential_one}
\end{figure}

It is now worth emphasizing the structural difference between gravitational and electromagnetic multipoles. In GR, the multipole expansion is constrained so that monopole and dipole terms are uniquely fixed by mass and angular momentum, while the quadrupole moment is the first parameter sensitive to the internal structure of the source~\cite{Thorne:1980ru}. In amplitude language, this uniqueness translates into the statement that, up to dipole order, classical gravitational observables are unaffected by non-minimal couplings. By contrast, the electromagnetic sector behaves differently. Already at dipole order, non-minimal effects can appear even in gravitational systems. In fact, there exists a unique Pauli-type interaction~\cite{Pauli:1941zz}
\begin{equation}\label{eq:non_minimal_pauli_lagrangian}
{}^{\mathrm{nm}}\mathcal{L}_{\psi^2 A}
    = -i\,\zeta\frac{Q}{2 m}\,\bar{\psi}\Sigma^{\mu \nu}\psi\,F_{\mu \nu}\ ,
\end{equation}
with $\zeta$ being a dimensionless parameter, directly modifying the fermion–photon vertex and hence contributing to the electromagnetic dipole moment in the classical limit. The inclusion of such a term is essential to reproduce the correct gyromagnetic factor of charged rotating BHs and highlights how charged multipoles, unlike neutral ones, are sensitive to the details of the underlying effective action. We can now identify the classical vertices of the theory in order to compute the three-point scattering amplitudes needed for the evaluation of both the EMT and the electromagnetic current. Extracting the Feynman rules from the action in Eq.~\eqref{eq:minimal_general_action} and following the dressing procedure outlined in Eq.~\eqref{eq:DressedVertexGeneric}, the vertex associated with the minimal photon interaction reads
\begin{equation}\label{eq:01_current}
    \includegraphics[valign=c,width=0.25\textwidth]{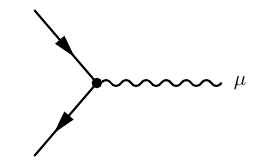}
    = ( \hat{\tau}_{\psi^2 A})_{\mu} = -i\left(Q\,\delta_\mu^0-i\,Q\,S_{\mu \nu}\,q^\nu\right)\ ,
\end{equation}
while the Pauli term in Eq.~\eqref{eq:non_minimal_pauli_lagrangian} contributes as
\begin{equation}\label{eq:01_current_nm}
    \includegraphics[valign=c,width=0.25\textwidth]{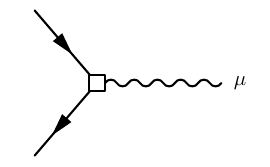}
    = \Big({}^{\mathrm{nm}}\hat{\tau}_{\psi^2 A}\Big)_{\mu} = - \zeta\,Q\,S_{\mu \nu}\,q^\nu\ .
\end{equation}
We notice that these expressions follow from the relations in Eq.~\eqref{eq:ClassicalLimitOfGenerators} as well as the Gordon identity~\cite{Itzykson:1980rh}
\begin{equation}\label{eq:gordon_identity}
    \bra{p',\sigma'} \gamma_\mu \ket{p,\sigma} = \bra{p',\sigma'} \left( \frac{p_\mu+p'_\mu}{2m} - i\frac{p^\nu-p'^\nu}{m} \Sigma_{\mu \nu} \right)\ket{p,\sigma}\ ,
\end{equation}
which is crucial for extracting the classical limit of the quantum vertices when spinors are involved. Moreover, in addition to Eqs.~\eqref{eq:01_current} and~\eqref{eq:01_current_nm}, we also have to consider the graviton interaction with the massive spinor, identified by the matter dressed vertex in Eq.~\eqref{eq:MinimalVertex} truncated at $\mathcal{O}(S)$, and reported below for completeness
\begin{equation}
    \includegraphics[valign=c,width=0.25\textwidth]{Diagrams_PDF/GyroDiags/Graviton_Minimal_Vertex.pdf}
    = (\hat{\tau}_{\psi^2h})_{\mu\nu}=\frac{i\, \kappa}{2}m\Big(u^\mu u^\nu
    - \frac{i}{2}q_\lambda \big(S^{\mu\lambda}u^{\nu}+S^{\nu\lambda}u^{\mu}\big)\Big)\ ,
\end{equation}
where we have accounted for the right spinor normalization of the vertex. We can now derive the PM expansion up to first order for both the metric and the gauge potential. Concretely, we evaluate the graviton emission diagrams in Fig.~\ref{fig:metric_one} to compute the metric, and the photon emission diagrams in Fig.~\ref{fig:potential_one} to obtain the electromagnetic potential. These diagrams include both minimal and Pauli-type vertices, as well as loop corrections, with their multiplicities indicated. 


Once $T_{\mu\nu}(q)$ and $j_\mu(q)$ are computed, from Eqs.~\eqref{eq:01_current} and~\eqref{eq:01_current_nm} one can then obtain the leading-order contributions to the metric and gauge potential. At 1PM the metric coincides with the chargeless expression in Eq.~\eqref{eq:MonopoleMetric}, while the potential reads
\begin{equation}\label{eq:potential_with_zeta}
    \begin{aligned}
        A^{(0)}_{0}(x) &= \frac{Q}{4\,\pi}\,\rho(r)\ ,\\
        A^{(0)}_{i}(x) &= (1+\zeta)(d-2)\frac{Q}{4\,\pi \,r^2}\,\rho(r)S_{ik}\,x^k\ .
    \end{aligned}
\end{equation}
The above result can be conveniently rewritten as
\begin{equation}\label{eq:magnetic_zeta_d}
    A^{(0)}_{i}(x) = (1+\zeta) \,\frac{1}{\Omega_{d-1}~r^d} Q S_{ik}\,x^k\ ,
\end{equation}
which makes contact with the generic dipole form of the electromagnetic potential such as
\begin{equation}\label{eq:dipole_potential}
    A^{\mathrm{dip.}}_i = \frac{\mathfrak{g}}{2}\frac{1}{\Omega_{d-1}\,r^d}\,Q S_{i k}\,x^k\ ,
\end{equation}
where $\mathfrak{g}$ is the gyromagnetic factor in $d+1$ dimensions. Then, by direct comparison of Eqs.~\eqref{eq:magnetic_zeta_d} and~\eqref{eq:dipole_potential}, we obtain the total gyromagnetic factor of a spin-$\tfrac{1}{2}$ fermion in a Dirac–Pauli EFT as
\begin{equation}
    \mathfrak{g}_{\mathrm{Dirac-Pauli}}=2\,(1+\zeta)\ .
\end{equation}
This result highlights the role of the Pauli coupling, namely that the minimal coupling alone gives $\mathfrak{g}=2$, but a nonzero $\zeta$ shifts the gyromagnetic factor. In the following, we will keep $\mathfrak{g}$ explicit, so that different theories (minimal or non-minimal) can be directly compared within the same formalism. The next step is to compute the loop amplitudes with multiple graviton and photon insertions on the massive line, as illustrated in Figs.~\ref{fig:metric_one} and~\ref{fig:potential_one}. The procedure follows the methods developed in chapter~\ref{chapter:ClassicalGravityFromAmplitudes}, and in particular Eq.~\eqref{eq:EMTfromAMPfinal}, where now we have to consider also the photon propagator in the Lorenz gauge when needed. Collecting all contributions, the resulting far-field metric in arbitrary spacetime dimensions, up to 1PM order and including dipole terms, within Einstein–Maxwell theory with an additional Pauli interaction, takes the form
\begin{equation}\label{eq:metric_d_dimensions}
    \begin{aligned}
        \kappa\,h_{0 0}(x) &= 4\frac{d-2}{d-1}G_N m\,\rho(r) - \frac{d-2}{d-1} \frac{G_N Q^2}{2 \,\pi} \,\rho^2(r)+\mathcal{O}(G_N^2)\ ,\\
        \kappa\,h_{0 i}(x) &= 2 (d-2) \frac{G_N m}{r^2}\rho(r)\,S_{ik}\,x^k - \frac{(d-2)^2}{d-1}\,\mathfrak{g}\,\frac{G_N Q^2}{4 \,\pi\, r^2}\rho^2(r)\,S_{ik}\,x^k+\mathcal{O}(G_N^2)\ ,\\
        \kappa\,h_{i j}(x) &= 4 \frac{1}{d-1}G_N m\,\rho(r)\,\delta_{i j} - \frac{(d-3)\,r^2\,\delta_{i j}-(d-2)^2\,x_i x_j}{(d-1) (d-4)}\,\frac{G_N Q^2}{2 \, \pi \, r^2}\rho^2(r)+\mathcal{O}(G_N^2)\ ,
    \end{aligned}
\end{equation}
while the corresponding electromagnetic potential is
\begin{equation}\label{eq:potential_d_dimensions}
    \begin{aligned}
        A_{0}(x) &= \frac{Q}{4\,\pi}\,\rho(r)-\frac{d-2}{d-1}\frac{G_N m Q}{2 \,\pi}\,\rho^2(r)+\mathcal{O}(G_N^2)\ ,\\
        A_{i}(x) &= (d-2)\,\mathfrak{g}\,\frac{Q}{8\,\pi \,r^2}\,\rho(r)\,S_{i k}\,x^k\\
        &-\frac{(d-2)^2}{(d-1)^2} \left(d\,\frac{d-1}{d-2}-\mathfrak{g} \right) \frac{G_N m Q}{4 \, \pi \, r^2} \rho^2(r)\,S_{ik}\,x^k+\mathcal{O}(G_N^2)\ .
    \end{aligned}
\end{equation}
Equations.~\eqref{eq:metric_d_dimensions} and~\eqref{eq:potential_d_dimensions} therefore describe the metric and electromagnetic potential of a charged spin-$\tfrac{1}{2}$ source with arbitrary gyromagnetic factor $\mathfrak{g}$.

\begin{figure}[h]
\centering
\begin{equation*}
    \includegraphics[valign=c,width=0.25\textwidth]{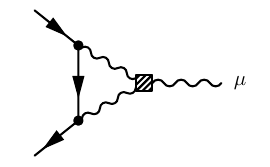}
    +2\times\includegraphics[valign=c,width=0.25\textwidth]{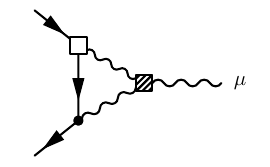}
\end{equation*}
\caption{Diagrams including the CS interaction (the three-photon vertex), inserted in five dimensions to compute the electromagnetic potential up to 1PM and dipole order.}
    \label{fig:potential_two}
\end{figure}

As already emphasized in the previous chapter, infrared divergences appear in the loop expansion. Specifically, one encounters divergences at one loop in $d=4$ and, similarly, at two loops in $d=3$~\cite{Mougiakakos:2020laz,DOnofrio:2022cvn}. These pathologies signal that the naive $\rho(r)$-series for the metric and potential is not sufficient in these dimensions, and a renormalization procedure is needed. The cure is provided by including non-minimal higher-derivative counterterms built from the Riemann tensor and spinor derivatives~\cite{Mougiakakos:2020laz,Goldberger:2004jt}, such as
\begin{equation} 
{}^{\mathrm{nm}}\mathcal{L}_{\psi^2 h}
    = K\,R~D_\mu\bar{\psi}~D^\mu\psi\ ,
\end{equation}
where $K$ is a dimensionful coefficient, in complete analogy with Eq.~\eqref{eq:NonMinimalAction} discussed in the chargless spin-1 case. These operators generate precisely the counterterms required to cancel the divergences, at the price of introducing logarithmic $\rho$-dependent terms after renormalization~\cite{Mougiakakos:2020laz,DOnofrio:2022cvn}. In addition, in odd spacetime dimensions ($d$ even), the most general effective action may include a CS term~\cite{Witten:1988hf,Birmingham:1991ty}. In particular, in $d=4$, the presence of such a term is motivated by supergravity constructions and by the existence of the CCLP BH~\cite{Chong:2005hr}. Following~\cite{Kunz:2017pnm,Deshpande:2024vbn}, the CS interaction takes the form
\begin{equation}\label{eq:chern_simons_five_D}
    \mathcal{L}_{\mathrm{CS}_{d=4}}
    = \lambda\,\frac{\kappa}{16\sqrt{6}}\,\varepsilon^{\mu \nu \alpha \beta\gamma}\,F_{\mu \nu}\,F_{\alpha \beta}\,A_\gamma\ ,
\end{equation}
where $\lambda$ is a dimensionless coefficient. In five-dimensional minimal supergravity this term is fixed with $\lambda=1$, providing the background for the charged rotating BH solution of~\cite{Cvetic:2004hs,Chong:2005hr}, while other values of $\lambda$ have been studied in the context of more general Einstein–Maxwell–CS theories~\cite{Blazquez-Salcedo:2013wka,Kunz:2017pnm}. The insertion of~\eqref{eq:chern_simons_five_D} generates a new three-photon vertex,
\begin{equation}\label{eq:chern_simons_vertex}
\includegraphics[valign=c,width=0.2\textwidth]{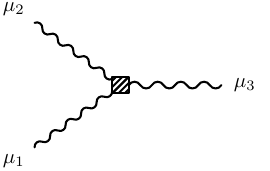} 
= \left(\tau_{A^3}\right)^{\mu_1 \mu_2 \mu_3} 
= -i\,\lambda\,\kappa\,\frac{\sqrt{6}}{4}\,\varepsilon^{\mu_1 \mu_2 \alpha \beta \mu_3}\,p_\alpha\,p'_\beta\ ,
\end{equation}
with $p_\alpha$ and $p'_\beta$ denoting incoming and outgoing momenta. At one loop this vertex affects only the electromagnetic potential, through the diagrams shown in Fig.~\ref{fig:potential_two},
whose contribution evaluates to
\begin{equation}\label{eq:02_potential_nm_and_nm2}
         A^{(\mathrm{CS})}_{\mu}(x){\Bigg |}_{d=4} 
         = - \lambda \,\mathfrak{g}\, \frac{\sqrt{G_N}\,Q^2}{16\,\sqrt3\,\pi^{7/2}\,r^6}\,
         \varepsilon_{0 j k l \mu}\,S^{j k}\,x^l\ .
\end{equation}
This term must be added to Eq.~\eqref{eq:potential_d_dimensions} to obtain the full 1PM potential. In contrast, its effect on the metric only arises at two-loop order, scaling as $Q^3 \sqrt{G_N^3}$, and therefore appears at higher PM orders.

\section{Particular cases}\label{sec:KnownChargedSolution}

We now compare our amplitude-derived metric and gauge potential with the PM expansion of known electrovacuum BH solutions in $d=3,4$. This comparison plays a dual role. On the one hand, it provides a stringent consistency check of the amplitude framework. Indeed, if our construction is correct, it must reproduce the well-established far-field expansions of classical solutions such as Kerr–Newman. On the other hand, it allows us to clarify the role of non-minimal couplings in the effective theory. In fact, while mass and spin multipoles follow universally from minimal coupling, electromagnetic multipoles are sensitive to additional operators, making them an ideal probe of the EFT structure. As we will show, the Kerr–Newman BH in four dimensions is fully captured by the minimally coupled Dirac action, corresponding to $\mathfrak{g}=2$. In contrast, in higher dimensions the situation is richer and the five-dimensional CCLP solution requires the presence of a Pauli-type interaction in the effective action, reflecting the fact that in $D>4$ the gyromagnetic factor is no longer universal and depends on non-minimal couplings. This highlights how amplitude methods not only reproduce known solutions but also shed light on the EFT principles underlying their multipolar structure.

\subsection{Kerr–Newman solution in $d=3$}

The Kerr–Newman metric describes the exterior of an asymptotically flat, stationary, charged, and rotating BH in four-dimensional Einstein–Maxwell theory~\cite{Newman:1965my,Adamo:2014baa}. In Boyer–Lindquist coordinates $(t,\tilde{r},\theta,\phi)$~\cite{Boyer:1966qh}, the line element reads~\cite{Frolov:1998wf}
\begin{equation}\label{eq:KN_metric_solution}
    d s^2 = -\frac{\Delta}{\Sigma}\left(d t - a \sin^2\theta \, d \varphi\right)^2 
    + \frac{\sin^2\theta}{\Sigma}\left(a\, d t - (\tilde{r}^2+a^2)\,d \varphi\right)^2
    +\frac{\Sigma}{\Delta} d \tilde{r}^2+\Sigma\, d\theta^2\ ,
\end{equation}
with the electromagnetic potential
\begin{equation}\label{eq:KN_potential_solution}
    A_\mu d x^\mu = \frac{Q}{4\pi}\frac{\tilde{r}}{\Sigma}\left( d t -a \sin^2 \theta\,d \varphi\right)\ ,
\end{equation}
where
\begin{equation}
    \Sigma=\tilde{r}^2+a^2 \cos^2 \theta\ , 
    \qquad 
    \Delta = \tilde{r}^2+a^2-2 m G_N \tilde{r}+\frac{1}{4\pi} G_N Q^2\ ,
\end{equation}
and $a$ is the usual spin-density parameter. Notice that in this oblate-spheroidal reference frame the radial coordinate is defined through the relation
\begin{equation}
    \frac{x^2+y^2}{\tilde{r}^2+a^2}+\frac{z^2}{\tilde{r}^2}=1\ ,
\end{equation}
and that in the chargeless limit when $Q\to 0$ the metric in Eq.~\eqref{eq:KN_metric_solution} reduces to the Kerr solution in Boyer-Lindquist coordinates. Furthermore, to directly compare such metric with the amplitude-based calculations, we have to move Cartesian harmonic coordinates up to 1PM and dipole order. Following the general procedure outlined in appendix~\ref{chapter:App}, the Kerr-Newman metric in harmonic coordinates reads
\begin{equation}\label{eq:kn_metric_pm_expansion}
    \begin{aligned}
        g^{\mathrm{KN}}_{00} &= -1 + 2 \frac{G_N m}{r} - \frac{G_N Q^2}{4\pi r^2}+ \mathcal{O}(G_N^2,Q^3,a^2)\ ,\\
	g^{\mathrm{KN}}_{0i} &= 2 \frac{G_N m}{r^3} \,S_{ik}\, x^k - \frac{G_N Q^2}{4 \pi r^4} \,S_{ik} \,x^k+ \mathcal{O}(G_N^2,Q^3,a^2)\ ,\\
	g^{\mathrm{KN}}_{ij} &= \delta_{ij}+2 \frac{G_N m}{r}\,\delta_{ij}-\frac{G_N Q^2}{4 \pi r^4}\,x_i x_j + \mathcal{O}(G_N^2,Q^3,a^2)\ ,
    \end{aligned}
\end{equation}
while the electromagnetic potential is given by
\begin{equation}\label{eq:kn_potential_pm_expansion}
    \begin{aligned}
        A^{\mathrm{KN}}_{0} &= \frac{Q}{4 \, \pi \, r} - \frac{G_N m Q}{4 \, \pi \, r^2} + \mathcal{O}(G_N^2,Q^3,a^2)\ ,\\
        A^{\mathrm{KN}}_{i} &= \frac{Q}{4\, \pi \, r^3}\,S_{ik}\, x^k-\frac{G_N m Q}{4 \, \pi \, r^4}\,S_{ik}\,x^k+\mathcal{O}(G_N^2,Q^3,a^2)\ .
    \end{aligned}
\end{equation}
On the other hand, setting $d=3$ in our general amplitude-based results, Eqs.~\eqref{eq:metric_d_dimensions} and~\eqref{eq:potential_d_dimensions} give
\begin{equation}\label{eq:metric_3_dimensions}
    \begin{aligned}
        \left.\kappa\,h_{0 0}(x)\right|_{d=3} &= 2\frac{G_N m}{r} - \frac{G_N Q^2}{4 \, \pi \, r^2}+ \mathcal{O}(G_N^2,Q^3,a^2)\ ,\\
        \left.\kappa\,h_{0 i}(x)\right|_{d=3} &= 2 \frac{G_N m}{r^3}\,S_{ik}\,x^k - \mathfrak{g}\,\frac{G_N Q^2}{8 \, \pi \, r^4}\,S_{ik}\,x^k+ \mathcal{O}(G_N^2,Q^3,a^2)\ ,\\
        \left.\kappa\,h_{i j}(x)\right|_{d=3} &= 2 \frac{G_N m}{r}\,\delta_{i j} - \frac{G_N Q^2}{4 \, \pi \, r^4}\,x_i x_j+ \mathcal{O}(G_N^2,Q^3,a^2)\ ,
    \end{aligned}
\end{equation}
and
\begin{equation}\label{eq:potential_3_dimensions}
    \begin{aligned}
        \left.A_{0}(x)\right|_{d=3} &= \frac{Q}{4 \, \pi \, r}-\frac{G_N m Q}{4 \, \pi \, r^2}+ \mathcal{O}(G_N^2,Q^3,a^2)\ ,\\
        \left.A_{i}(x)\right|_{d=3} &= \mathfrak{g}\,\frac{Q}{8 \, \pi \, r^3} \,S_{ik}\,x^k- \left(6-\mathfrak{g} \right) \frac{G_N m Q}{16 \, \pi \, r^4} \,S_{ik}\,x^k+ \mathcal{O}(G_N^2,Q^3,a^2)\ .
    \end{aligned}
\end{equation}
Finally, comparing Eqs.~\eqref{eq:kn_metric_pm_expansion}–\eqref{eq:kn_potential_pm_expansion} with Eqs.~\eqref{eq:metric_3_dimensions}–\eqref{eq:potential_3_dimensions} shows exact agreement at 1PM and dipole order if and only if
\begin{equation}
    \mathfrak{g}=2 \quad \Leftrightarrow \quad \zeta=0\ .
\end{equation}
In other words, the Kerr–Newman solution is reproduced within our framework precisely when the EFT is minimally coupled. This confirms that in four dimensions the amplitude approach recovers the expected gyromagnetic factor $\mathfrak{g}=2$, in agreement with the classical no-hair theorems.

\subsection{Chong–Cvetič–Lü–Pope solution in $d=4$}

Unlike the four dimensional case, charged rotating BHs in pure five-dimensional Einstein–Maxwell theory remain unknown in closed form~\cite{Kleihaus:2007kc}. However, in the framework of five-dimensional gauged supergravity, where the Einstein–Maxwell action is supplemented by a CS interaction with fixed coefficient $\lambda=1$, one can construct the exact CCLP solution~\cite{Cvetic:2004hs,Chong:2005hr}, representing the natural charged generalization of Myers–Perry BHs in $d=4$. In what follows, we expand the CCLP solution and compare it to our amplitude-based results, thereby testing the EFT construction in a setting where non-minimal couplings are expected to play a crucial role. In spheroidal coordinates $(t,\tilde{r},\theta,\phi,\psi)$ and vanishing cosmological constant, the CCLP metric takes the form
\begin{equation}\label{eq:cclp_metric_solution}
    \begin{aligned}
        d s^2 =& ~-d t^2-\frac{2\,\mathfrak{Q}}{\Sigma} \, d\nu\, d t+\frac{2 \mathfrak{Q}}{\Sigma}\, d\nu\, d\mu + \frac{\mathfrak{m}\, \tilde{r}^2-\mathfrak{Q}^2}{\Sigma^2}\,\left(d t-d\mu \right)^2+\frac{\Sigma}{\Delta} d \tilde{r}^2
    \\
    &+\Sigma \,d \theta^2+\left(\tilde{r}^2+\mathfrak{a}^2 \right) \sin^2\theta \,d \phi^2+\left(\tilde{r}^2+\mathfrak{b}^2 \right) \cos^2\theta\, d\psi^2\ ,
    \end{aligned}
\end{equation}
where
\begin{gather}
     d\nu=\mathfrak{b} \sin^2\theta \,d\phi+\mathfrak{a} \cos^2\theta \,d\psi\ , \qquad
    d\mu=\mathfrak{a} \sin^2\theta\,d \phi + \mathfrak{b} \cos^2\theta \,d\psi\ , \\
    \Sigma=\tilde{r}^2+\mathfrak{a}^2 \cos^2\theta + \mathfrak{b}^2 \sin^2 \theta\ , \\
    \Delta=\frac{1}{\tilde{r}^2}\left[(\tilde{r}^2+\mathfrak{a}^2)(\tilde{r}^2+\mathfrak{b}^2)+\mathfrak{Q}^2+2\, \mathfrak{a}\, \mathfrak{b}\, \mathfrak{Q} - \mathfrak{m}\, \tilde{r}^2\right]\ ,
\end{gather}
and with the electromagnetic potential given by
\begin{equation}\label{eq:cclp_potential_solution}
    A_\mu d x^\mu = \sqrt{\frac{3}{\pi \,G_N}}\frac{\mathfrak{Q}}{4\, \Sigma} \left( d t - \mathfrak{a} \sin^2\theta \, d \phi -\mathfrak{b} \cos^2\theta \,d\psi \right)\ .
\end{equation}
The parameters $\mathfrak{m}, \mathfrak{Q}, \mathfrak{a}, \mathfrak{b}$ are related to the physical mass, charge, and angular momenta through~\cite{Chong:2005hr,Aliev:2006yk}
\begin{gather}\label{eq:CCLP_parameter_physical_relations}
    \mathfrak{m}=\frac{8 \,G_N \, m}{3\,\pi}\ , \qquad \mathfrak{Q} = Q\, \sqrt{\frac{G_N}{3\, \pi^3}}\ ,\\
    \quad J_1=\frac{\pi}{4\,G_N}\left(\mathfrak{a}\, \mathfrak{m} + \mathfrak{b}\,\mathfrak{Q}\right)\ , \qquad J_2=\frac{\pi}{4 \,G_N} \left(\mathfrak{b}\, \mathfrak{m} + \mathfrak{a}\, \mathfrak{Q}\right)\ .
\end{gather}
Unlike the Myers–Perry case, these relations contain additional cross-terms proportional to $\mathfrak{Q}$, which are direct consequences of the CS interaction. They encode the mixing between charge and rotation, illustrating how the presence of the CS term modifies the link between geometric parameters and physical charges.

To make the comparison with the amplitude-derived metric feasible, we move to Cartesian harmonic coordinates following the procedure discussed in appendix~\ref{chapter:App}. Furthermore, we block-diagonalize the spin tensor as in Eq.~\eqref{eq:BlockDiagonalD5MEtric}, and expand the CCLP solution to 1PM and dipole order. The resulting far-field metric and potential are then given by
\begin{equation}\label{eq:CCLP_solution_pm_expansion_metric} \begin{aligned} g^{\mathrm{CCLP}}_{00} =&~ -1 + \frac{8\, G_N m}{3 \, \pi \, r^2} - \frac{G_N Q^2}{3 \, \pi^3 \, r^4}+ \mathcal{O}(G_N^2,Q^3,S^2)\ , \\ g^{\mathrm{CCLP}}_{0i} =& + \frac{4\,G_N m}{\pi \, r^3} \,S_{ik}\,x^k - \frac{G_N Q^2}{2 \, \pi^3 \, r^6} \,S_{ik}\,x^k+ \mathcal{O}(G_N^2,Q^3,S^2)\ , \\ g^{\mathrm{CCLP}}_{ij} =& \delta_{ij}+ \frac{4\,G_N m}{3 \, \pi \, r^2}\,\delta_{ij}-\frac{G_N Q^2}{12 \, \pi^3 \, r^4}\,\log\left( \frac{8\,G_N m}{3 \, \pi \, r^2} \right)\,\delta_{ij} \\ &-\frac{G_N Q^2}{6 \, \pi^3 \, r^6}\,x_i x_j +\frac{G_N Q^2}{3 \, \pi^3 \, r^6}\,\log\left( \frac{8\,G_N m}{3 \, \pi \, r^2} \right)\,x_i x_j + \mathcal{O}(G_N^2,Q^3,S^2)\ , \end{aligned} \end{equation} and \begin{equation}\label{eq:CCLP_solution_pm_expansion_potential} \begin{aligned} A^{\mathrm{CCLP}}_{0} =&~ \frac{Q}{4\, \pi^2 \, r^2} - \frac{G_N m Q}{3 \, \pi^3 \, r^4} + \mathcal{O}(G_N^2,Q^3,S^2) , \\ A^{\mathrm{CCLP}}_{i} =&~ \frac{3\,Q}{8 \, \pi^2 \, r^4}\,S_{ik}\,x^k-\frac{G_N m Q}{2 \, \pi^3 \, r^6}\,S_{ik}\,x^k \\ &- \frac{\sqrt{3}\,\sqrt{G_N}\,Q^2}{32\, \pi^{7/2} \,r^6}\,\varepsilon_{0 j k l i}\,S^{j k}\,x^l+ \mathcal{O}(G_N^2,Q^3,S^2)\ . \end{aligned} \end{equation}
On the amplitude side, after including the CS contribution~\eqref{eq:02_potential_nm_and_nm2} and having performed the renormalization procedure similarly to Eq.~\eqref{eq:RenormParametrization}, we obtain \begin{equation}\label{eq:metric_5_dimensions} \begin{aligned} \kappa\,h_{0 0}(x)\Big|_{d=4} &= \frac{8\, G_N m}{3 \, \pi \, r^2} - \frac{G_N Q^2}{3 \, \pi^3 \, r^4}+ \mathcal{O}(G_N^2,Q^3,S^2)\ , \\ \kappa\,h_{0 i}(x)\Big|_{d=4} &= \frac{4\,G_N m}{\pi\, r^3} \,S_{ik}\,x^k - \mathfrak{g}\,\frac{G_N Q^2}{3\, \pi^3 \,r^6}\,S_{ik}\,x^k+ \mathcal{O}(G_N^2,Q^3,S^2)\ ,\\ \kappa\,h_{i j}(x)\Big|_{d=4} &= \frac{4\,G_N m}{3\,\pi \,r^2}\,\delta_{ij}-\frac{G_N Q^2}{12\, \pi^3 \,r^4}\,\log\left( \frac{8\,G_N m}{3 \, \pi\, r^2} \right)\,\delta_{ij}\\ &-\frac{G_N Q^2}{6 \, \pi^3 \, r^6}\,x_i x_j+\frac{G_N Q^2}{3 \,\pi^3\, r^6}\,\log\left( \frac{8\,G_N m}{3 \,\pi \,r^2} \right)\,x_i x_j+ \mathcal{O}(G_N^2,Q^3,S^2)\ , \end{aligned} \end{equation} 
and
\begin{equation}\label{eq:potential_5_dimensions} \begin{aligned} A_{0}(x)\Big|_{d=4} &= \frac{Q}{4\, \pi^2 \,r^2} - \frac{G_N m Q}{3\, \pi^3 \,r^4}+ \mathcal{O}(G_N^2,Q^3,S^2)\ ,\\ A_{i}(x)\Big|_{d=4} &= \mathfrak{g}\,\frac{Q}{4\,\pi^2 \,r^4} \,S_{ik}\,x^k- \left(6-\mathfrak{g} \right) \frac{G_N m Q}{9\, \pi^3 \,r^6}\,S_{ik}\,x^k\\ &- \lambda \, \mathfrak{g} \,\frac{\sqrt{G_N}\,Q^2}{16\,\sqrt3\,\pi^{7/2}\,r^6}\,\varepsilon_{0 j k l i}\,S^{j k}\,x^l+ \mathcal{O}(G_N^2,Q^3,S^2)\ . \end{aligned} \end{equation}
A direct comparison shows perfect agreement provided the gyromagnetic factor is fixed to $\mathfrak{g}=\tfrac{3}{2}$, corresponding to a Pauli interaction with coefficient $\zeta=-\tfrac{1}{4}$. In addition, the match requires the CS coupling to take its supergravity value $\lambda=1$, consistent with the origin of the CCLP BH. This result demonstrates that, unlike Kerr–Newman in four dimensions, the five-dimensional charged rotating BH cannot be reproduced by a purely minimally coupled theory. Instead, the EFT must necessarily include both a Pauli term and a CS interaction. In this way, the amplitude framework not only reproduces the CCLP solution but also explains its structure in terms of effective operators and their physical consequences.

\subsection{Gyromagnetic factor of higher-dimensional black holes}

After having established the correspondence between known solutions and amplitude-based results, we now want to analyze how the gyromagnetic factor of BHs depends on the spacetime dimension. In particular, we follow the approach of~\cite{Aliev:2004ec,Aliev:2006yk}, where the Myers–Perry solution is taken as a neutral rotating background in higher dimensions, and a small electric charge is introduced perturbatively. Solving the Einstein–Maxwell equations order by order then yields the electromagnetic potential associated with a slowly rotating, weakly charged BH. This provides a clean setting in which to extract the leading dipolar contribution. Motivated by this, we write the magnetic part of the gauge potential in Cartesian coordinates, keeping only the dipole term, as
\begin{equation}\label{eq:potential_ansatz_q}
    A_i=\frac{\widetilde{Q}}{r^d}\,\frac{d-1}{2}\,S_{ik}\,x^k\ ,
\end{equation}
where $\widetilde{Q}$ is related to the physical electric charge $Q$. Considering then the Komar integral at infinity
\begin{equation}
    Q =  \int_{\Omega_{d-1}} \star\, F =   \widetilde{Q}\,(d-2)\,\Omega_{d-1}\ ,
\end{equation}
with $\star F$ the Hodge dual of the field strength, one solves for $\widetilde{Q}$ and rewrites~\eqref{eq:potential_ansatz_q} as
\begin{equation}\label{eq:bh_d_g}
    A_i=\frac{1}{\Omega_{d-1}\,r^d}\,\frac{d-1}{d-2}\,\frac{Q}{2}\,S_{ik}\,x^k\ .
\end{equation}
This form of the potential reproduces the Kerr–Newman expansion~\eqref{eq:kn_potential_pm_expansion} for $d=3$ and the CCLP result~\eqref{eq:CCLP_solution_pm_expansion_potential} for $d=4$. Comparing then Eq.~\eqref{eq:bh_d_g} with the amplitude-based dipole potential~\eqref{eq:dipole_potential}, immediately gives the general prediction for the gyromagnetic factor
\begin{equation}\label{eq:general_d_bh_g}
    \mathfrak{g}_{\mathrm{BH}} = \frac{d-1}{d-2}\ .
\end{equation}
This matches our explicit findings, namely $\mathfrak{g}_{\mathrm{BH}}=2$ for Kerr–Newman ($d=3$) and $\mathfrak{g}_{\mathrm{BH}}=3/2$ for CCLP ($d=4$). Importantly, only in four spacetime dimensions does the PM expansion of BH metrics emerge from a purely minimally coupled EFT, while for $d>3$ a Pauli-like interaction with coefficient 
\begin{equation}
\zeta = -\frac{d-3}{2(d-2)}
\end{equation}
is required. Summarizing our findings, for general $d+1$ spacetime dimensions, we showed that the amplitude framework predicts a universal gyromagnetic factor $\mathfrak{g}_{\mathrm{BH}}=(d-1)/(d-2)$. This result is independent of the presence of a CS interaction, since such terms do not contribute at dipole order. It therefore applies equally to perturbative constructions based on Myers–Perry backgrounds~\cite{Aliev:2004ec,Aliev:2006yk} and to exact charged rotating solutions such as CCLP~\cite{Chong:2005hr,Kunz:2017pnm}. 

\chapter{Source multipoles and black hole mimickers}\label{chapter:SourceMultipoles}

In the previous chapters, we employed off-shell scattering amplitudes to reconstruct the PM expansion of classical gravitational solutions. Within this framework, we formalized the multipolar structure of higher-dimensional spacetimes and identified a new family of stress multipoles. This construction showed how amplitude techniques offer a direct and systematic way to match perturbative gravity with the multipolar content of classical solutions, keeping full control over the gravitational degrees of freedom, as well as gauge redundancies. In this chapter we adopt a complementary perspective. Rather than starting from scattering amplitudes and reconstructing the metric from bottom-up calculations within a given theory, we now identify the most general conserved EMT in momentum space compatible with a prescribed set of physical and symmetry constraints. By working directly in momentum space, we can construct such EMT at every order in the spin expansion, organizing it in terms of constant coefficients, and computing the corresponding tree-level amplitude to obtain the complete multipole expansion of the induced gravitational field. This multipole identification naturally leads to the definition of \emph{form factors}, a new class of gauge-invariant quantities intimately connected to gravitational multipoles. In this way, the momentum space construction provides a natural and unifying language that bridges the microscopic description of the source with the macroscopic multipolar structure of spacetime.

This approach, for the first time within GR, further allows to define \emph{source multipoles} in close analogy with Newtonian gravity. Indeed, in the Newtonian framework, the multipole moments of the gravitational field can be read off directly from the matter distribution. In contrast, in GR this correspondence is far more subtle, and the non-linear structure of Einstein’s equations obscures the direct link between EMT and the gravitational multipoles of the resulting geometry. The momentum-space formalism resolves this difficulty by cleanly separating physical contributions, encoded in the form factors, from gauge artifacts, encoded in redundant coefficients that can be eliminated by coordinate transformations~\cite{Bianchi:2024shc}, acting as well as an efficient framework to disentangle local effects from the long-range contributions to the induced spacetime. A remarkable outcome of this construction is that, once the multipolar structure of the source is imposed, the resulting EMT can often be resummed to all orders in spin, making it possible to recover explicit (linearized) matter sources for rotating black holes. In particular, we show how the Kerr solution in four dimensions and the Myers-Perry family in higher dimensions naturally emerge from their corresponding linearized sources, computed from a multipole-based approach. For the Kerr spacetime, we identify its linearized source to be an equatorial thin disk originally found by Israel~\cite{Israel:1970kp,Balasin:1993kf}, rotating at superluminal speed and supported on the ring singularity. For Myers-Perry black holes in five dimensions, we instead obtain a three-ellipsoid distribution with nontrivial stresses. In both cases, the matter distribution resides on lower-dimensional submanifolds but reproduces the same singular structure as the full non-linear black hole geometry. This result indicates that the characteristic curvature singularities of rotating BHs are already encoded in their linearized multipolar content.

Beyond its conceptual significance, this analysis also bears directly on the BH uniqueness problem. In four dimensions, the no-hair theorems ensure that stationary, asymptotically flat vacuum black holes are uniquely described by the Kerr family. However, in higher dimensions no such theorem exists, and the emergence of additional families of multipoles, most notably the stress multipoles, allows for a much richer landscape of stationary solutions. The possibility of classifying gravitational configurations directly in terms of their source multipoles provides new insight into the role of these additional moments, thereby establishing a bridge between amplitude-based methods and the classification of solutions in higher-dimensional GR. Moreover, it is possible to extend our formalism to explore the physics of BH mimickers, generalizing our framework to smooth geometries. Motivated by the idea that regular, horizonless matter distributions may reproduce the same asymptotic multipolar structure as Kerr, we introduce momentum-dependent \emph{structure functions} in the EMT. These functions capture the internal profile of the source while leaving its asymptotic multipole content unchanged. By choosing suitable structure functions, one can smear the singularities of the Kerr source and obtain regular, anisotropic rotating fluids that satisfy energy and causality conditions at linearized order. As a concrete example, we analyze Gaussian profiles and show that they reproduce a Kerr-like gravitational field at large distances, while replacing the singular equatorial disk with a smooth distribution of finite extent~\cite{Gambino:2025xdr}. This approach provides a systematic method for constructing horizonless compact objects that exactly reproduce the Kerr multipolar structure, thereby offering a novel perspective on the black hole mimicker problem. It emphasizes that multipoles alone do not uniquely determine the underlying geometry, and distinct matter configurations can share the same multipolar spectrum. This ambiguity, reminiscent of Newtonian gravity where different mass distributions can yield identical multipoles, challenges the notion of geometric uniqueness in relativistic spacetimes. Moreover, since multipoles govern gravitational-wave emission, such mimickers may be observationally indistinguishable from black holes at large distances while differing substantially near the would-be horizon.

The chapter is then organized as follows. In section~\ref{sec:EMTcalc} we introduce the momentum-space construction of source multipoles and discuss the general EMT structure, emphasizing the role of form factors and their mapping to gravitational multipoles. In section~\ref{sec:GraFFsKnownSolutions} we match the full multipole tower of known solutions and determine the form factor structure of different BH solutions. In section~\ref{sec:LinearizedBHSources} we then resum the EMT expression of known BH solutions, recovering the linearized matter sources of Kerr and Myers-Perry metrics and discussing their singularity structure. Finally, in section~\ref{sec:MBFforBHM} we extend the framework to BH mimickers, focusing on Gaussian-smeared sources that reproduce the Kerr multipolar structure in four dimensions while satisfying physical requirements on the matter distribution and being horizonless.  

\section{Energy-momentum tensor at any order in the spin}\label{sec:EMTcalc}

Our goal here is to derive a general description for a rotating spin-induced source, namely the EMT in momentum space produced when the only scales (and physical objects) involved are the ADM mass $m$ and angular momentum tensor $S^{\mu\nu}$. To do so, we get inspiration from the EMT in momentum space at quadrupole order associated to a spin-1 massive particle recovered in chapter~\ref{chapter:RotatingMetrics} from a scattering amplitude approach. Such result, can be generalized to every order in the spin expansion just by taking into account every possible tensorial structure, while satisfying some requirements that we are going to discuss in the following. Recalling $u^\mu$ and $S^{\mu\nu}$ as the velocity and the anti-symmetric spin-density tensor of the source, in the stationary case it follows that $S^{\mu\nu}u_\nu=0$ and $q^\mu u_{\mu}=0$. In the simple case in which the source is sitting at the origin one has $u^0=-1$ and $u^i=0$. Then, the only combination of variables we have at our disposal to build the generic expression of the EMT in momentum space is $q S$, since we require a smooth limit in which $S\rightarrow 0$ and $m\rightarrow 0$. From an amplitude perspective, the empiric rule is that for every spin tensor $S$ there must be a transferred momentum $q$ to balance the $\hbar$-dependence. Finally, we restrict to the study of the long-range regime of the metric, and so we neglect terms in which $T_{\mu\nu}(q)\propto q^2$, that otherwise will give rise to local contributions. Indeed, such local contributions would cancel the graviton propagator in Eq.~\eqref{eq:MetricFromEMT_Generic}, leading to a term in the metric perturbation proportional to a delta function (or derivatives of delta functions). To be more precise, introducing the short-hand notation
\begin{equation}
    q\cdot S\cdot S\cdot q \equiv q^\mu S_{\mu}{}^{\nu}S_{\nu}{}^{\sigma}q_{\sigma}\ ,\quad S\cdot S=S^{\mu\nu}S_{\nu\mu}\ ,
\end{equation}
terms like
\begin{equation}
    T^{\mu\nu}(q)\propto u^{\mu}u^{\nu} q^2 S \cdot S
\end{equation}
lead to local contributions in the metric, as in 
\begin{equation}\label{eq:DeltaMetric}
    h_{\mu\nu}(x)\propto \delta(x)\ .
\end{equation}
On the other hand, terms like 
\begin{equation}
    T^{\mu\nu}(q)\propto u^{\mu}u^{\nu} q\cdot S \cdot S \cdot q\ ,
\end{equation}
are associated with non-local contributions of the gravitational field, as in 
\begin{equation}
    h_{\mu\nu}(x)\propto \frac{1}{r}\ .
\end{equation}

Within the aforementioned assumptions, the most generic expression of the EMT reads
\begin{equation}\label{eq:GenericEMT}
    \begin{aligned}
    &T^{\mu\nu}(q)=m\ u^{\mu}u^{\nu}\Bigg(1+\sum_{n=1}^{+\infty}{F}_{2n, 1}\left(-q\cdot S\cdot S\cdot q\right)^n\Bigg)+m\sum_{n=0}^{+\infty}{F}_{2n+2, 2}(S\cdot q)^\mu (S\cdot q)^\nu\left(-q\cdot S\cdot S\cdot q\right)^n\\
    &-\frac{i}{2}m\left(u^\mu(S\cdot q)^\nu+u^\nu(S\cdot q)^\mu \right)\Bigg(1+\sum_{n=1}^{+\infty}{F}_{2n+1, 3}\left(-q\cdot S\cdot S\cdot q\right)^n\Bigg)\\
    &-m\sum_{n=0}^{+\infty}{G}_{2n+2, 1}\Bigg(\eta^{\mu\nu}q\cdot S\cdot S\cdot q-(S\cdot S\cdot q)^\mu q^\nu+(S\cdot S\cdot q)^\nu q^\mu\Bigg)\left(-q\cdot S\cdot S\cdot q\right)^n\\
    &-m\sum_{n=0}^{+\infty}{G}_{2n+2, 2}q^\mu q^\nu S\cdot S\left(-q\cdot S\cdot S\cdot q\right)^n+m\sum_{n=0}^{+\infty}{G}_{2n+4, 3}q^\mu q^\nu q\cdot S\cdot S\cdot S\cdot S\cdot q\left(-q\cdot S\cdot S\cdot q\right)^n\ .
    \end{aligned}
\end{equation}
We stress that every other terms other than the ones in Eq.~\eqref{eq:GenericEMT} are local terms. Indeed, Eq.~\eqref{eq:GenericEMT} naturally organizes in a spin expansion of the source, where the $F_{n, m}$'s and $G_{n, m}$'s are constant terms, with the first index labelling the order of the spin to which each coefficient is referring to, and the second one labelling different coefficients. Their distinct nature will be discussed in detail later. Moreover, the mass and the angular momenta are normalized to the ADM values of the induced spacetime, meaning that, computing the linearized metric from Eq.~\eqref{eq:GenericEMT}, the monopole and the dipole terms are fixed in terms of $m$ and $S$ by setting $F_{0, 1}=F_{1, 3}=1$. It is possible to check that the EMT in Eq.~\eqref{eq:GenericEMT} is conserved up to local terms, namely $q_\mu T^{\mu\nu}(q)\propto  q^2$, and it reduces to a point-like mass $m$ sitting at the origin in the spinless limit. Notice that neglecting local contributions is a crucial ingredient to obtain a finite number of terms that describe the EMT at every spin-order, as well as working in momentum space. Indeed, the same argument could not be repeated in position space, in which even restricting to the long-range regime an infinite number of tensorial structures is needed in order to capture the full spin-expansion in arbitrary dimension.   

\subsection{Gauge redundant parameters}

Considering Eq.~\eqref{eq:GenericEMT}, we can ask ourselves if every term in the EMT is physical. Indeed, we can fix a gauge in which we evaluate the metric induced by the source and perform a coordinate transformation eliminating any potentially redundant gauge parameters. Generalizing to every spin order the shift in Eq.~\eqref{eq:epsilonC1C2}, in which we determined the coordinate transformation that eliminates gauge redundancies at quadrupole order, fixing the harmonic gauge and considering an infinitesimal coordinate transformation, the most general gauge shift at any spin order can be expressed as
\begin{equation}\label{eq:CoordTransf}
\begin{aligned}
    &\xi^i(x)=\frac{\kappa^2m}{8\pi}\sum_{n=0}^{+\infty}\Bigg(K_{2n+2, 1}(S\cdot S)^{iA_{2n+1}}+K_{2n+2, 2}(S\cdot S) \eta^{ia_1}(S\cdot S)^{A_{2n}}\Bigg)\partial_{A_{2n+1}}\rho\\
    &+\frac{\kappa^2m}{8\pi}\sum_{n=0}^{+\infty}\Bigg(K_{2n+4, 3}(S\cdot S\cdot S\cdot S)^{a_1a_2}\eta^{ia_3}(S\cdot S)^{A_{2n}}\Bigg)\partial_{A_{2n+3}}\rho\ ,
\end{aligned}
\end{equation}
where, here and in the following, we use the short-hand notation 
\begin{equation}
    (S\cdot S)^{A_{2n}}\partial_{A_{2n}}\rho\equiv (S\cdot S)^{a_1a_2}\cdots(S\cdot S)^{a_{n-1}a_n}\partial_{a_1}\cdots\partial_{a_n}\rho\ .
\end{equation}
Then, computing the metric in harmonic gauge and considering the shift in Eq.~\eqref{eq:CoordTransf}, by fixing 
\begin{equation}\label{eq:GaugeFixingFFF}
\begin{gathered}
    K_{2n+2, 1}=-G_{2n+2,1}\ , \qquad K_{2n+2, 2}=\frac{1}{2}G_{2n+2, 2}\ ,\\ K_{2n+4, 3}=\frac{1}{2}G_{2n+4, 3}\ ,
\end{gathered}
\end{equation}
one can show that the dependence on the $G_{n, m}$'s in the metric disappears, leaving it only dependent on the $F_{n, m}$'s. Since the latter are the only terms that cannot be canceled by a coordinate transformation, they have a physical meaning. For this reason, we dub the $G_{n, m}$ and the $F_{n, m}$ coefficients as ``residual factors'' and ``form factors'', respectively. As we shall discuss in the following, form factors are tightly related to gravitational multipoles, and they can be mapped into each other by using the higher-dimensional generalization of Thorne formalism, such as Eq.~\eqref{eq:MultipoleExpandedMetric}. Indeed, the number of independent towers of form factors is related to the fact that in higher dimensions there exist three different towers of multipoles, namely mass, current, and stress multipole moments~\cite{Gambino:2024uge}, 
and the latter can be gauged away only in $d=3$.
Thus, for spin-induced multipoles there are two physical degrees of freedom (the mass and stress) for each even-power spin term, and one degree of freedom (the current) for each odd-power contribution.

\subsection{Conserved EMT in momentum space} 

Even though Eq.~\eqref{eq:GenericEMT} is enough to reconstruct the long-range behavior of the metric induced by the source, if we are interested in computing the EMT in position space we have to make sure that it is properly conserved, including also possible local-term contributions. We can indeed consider the properly conserved EMT as
\begin{equation}\label{eq:ConservedEMT}
    \begin{aligned}
    &T^{\mu\nu}(q)=m\ u^{\mu}u^{\nu}\Bigg(1+\sum_{n=1}^{+\infty}{F}_{2n, 1}\Big(-q\cdot S\cdot S\cdot q\Big)^n\Bigg)+m\sum_{n=0}^{+\infty}{F}_{2n+2, 2}(S\cdot q)^\mu (S\cdot q)^\nu\Big(-q\cdot S\cdot S\cdot q\Big)^n\\
    &+m\sum_{n=0}^{+\infty}{G}_{2n+2, 1}\Bigg(\eta^{\mu\nu}\Big(-q\cdot S\cdot S\cdot q\Big)-(S\cdot S)^{\mu\nu}q^2+(S\cdot S\cdot q)^\mu q^\nu+(S\cdot S\cdot q)^\nu q^\mu\Bigg)\Big(-q\cdot S\cdot S\cdot q\Big)^n\\
    &+m\sum_{n=0}^{+\infty}{G}_{2n+2, 2}\Big(q^\mu q^\nu-\eta^{\mu\nu}q^2\Big)(-S\cdot S)\Big(-q\cdot S\cdot S\cdot q\Big)^n\\
    &+m\sum_{n=0}^{+\infty}{G}_{2n+4, 3}\Big(q^\mu q^\nu-\eta^{\mu\nu}q^2\Big)\Big(q\cdot S\cdot S\cdot S\cdot S\cdot q\Big)\Big(-q\cdot S\cdot S\cdot q\Big)^n\\
    &-\frac{i}{2}m\Big(u^\mu(S\cdot q)^\nu +u^\nu(S\cdot q)^\mu \Big)\Bigg(1+\sum_{n=1}^{+\infty}{F}_{2n+1, 3}\Big(-q\cdot S\cdot S\cdot q\Big)^n\Bigg)\ ,
    \end{aligned}
\end{equation}
where now $q_\mu T^{\mu\nu}(q)=0$, as requested by the fact that the EMT in coordinate space is divergence-free. 
Since the only difference with 
respect to \eqref{eq:GenericEMT} is due to terms depending on $q^2$, Eq.~\eqref{eq:ConservedEMT} gives rise to the same long-range behavior of the metric. However, by including local terms in Eq.~\eqref{eq:ConservedEMT} we have to be careful with the residual factors $G_{n, m}$, and in particular check whether or not they are still redundant parameters. Indeed, even though we previously proved that they do not physically contribute to the asymptotic behavior of the metric, they do give rise to local physical contributions in the gravitational field. This means that if one is interested in computing the coordinate-space version of the EMT, or the short-range behavior of the metric, these terms are not negligible and have to be taken into account. In fact, once the form factors are fixed, there is an infinite number of different sources that reproduce the exact same multipolar structure and differ from each other only by local contributions, both in the EMT and in the metric. Moreover, allowing for local contributions, Eq.~\eqref{eq:ConservedEMT} is no more the most generic EMT in momentum space, since there are now infinite terms (tensorial structures) contributing in such regime. Still, we can think of Eq.~\eqref{eq:ConservedEMT} as the most generic EMT modulo local terms that do not affect the multipolar structure.

\subsection{Gravitational multipoles from form factors}

We can now compare the generalized multipole expansion of the metric with the generic metric sourced by the EMT, thus establishing a relation between gravitational multipoles and the source form factors. Indeed, we can compute Eq.~\eqref{eq:MetricFromEMT_Generic} considering the EMT derived in Eq.~\eqref{eq:GenericEMT}, and imposing the harmonic gauge defined through the propagator in Eq.~\eqref{eq:HarmonicPropagator}, we obtain the metric expressed in ACMC coordinates directly comparable with the multipole expansion of Eq.~\eqref{eq:MultipoleExpandedMetric}, leading to a one-to-one identification between gravitational form factors and multipole moments. The result reads
\begin{equation}\label{eq:GravitationalMultipoles}
    \begin{aligned}
\mathbb{M}^{(2\ell)}_{A_{2\ell}}&=\frac{(d+4\ell-4)!!}{(d-2)!!}(-1)^\ell\Big(F_{2\ell, 2}+(d-2)F_{2\ell, 1}\Big)(-S\cdot S)_{A_{2\ell}}\Big|_{\rm STF}\ , \\
\mathbb{J}^{(2\ell+1)}_{i,A_{2\ell+1}}&=\frac{(d+4\ell-2)!!}{(d-2)!!}(-1)^\ell F_{2\ell+1, 3} \ S_{ia_1}(-S\cdot S)_{A_{2\ell}}|_{\rm ASTF}\ , \\
\mathbb{G}^{(2\ell)}_{ij,A_{2\ell}}&=(d-1)\frac{(d+4\ell-4)!!}{(d-2)!!}(-1)^\ell F_{2\ell, 2} \ S_{ia_1}S_{ja_2}(-S\cdot S)_{A_{2\ell-2}}|_{\rm RSTF}\ ,
    \end{aligned}
\end{equation}
with 
\begin{equation}
      \mathbb{M}^{(2\ell+1)}_{A_{2\ell+1}}=0\ ,\qquad \mathbb{J}^{(2\ell)}_{i, A_{2\ell}}=0\ ,\qquad\mathbb{G}^{(2\ell+1)}_{ij, A_{2\ell+1}}=0\ ,
\end{equation}
for every $\ell=0, 1, 2, \dots$. As expected in analogy with the $d=3$ case, mass and stress multipoles are non-vanishing only for even powers of the spin, while current multipoles are non-vanishing for odd powers. Indeed, every multipole tensor in Eq.~\eqref{eq:GravitationalMultipoles} is meant to be symmetrized in the correct way, following the symmetrization prescription of section~\ref{sec:GravitationalMultipoles}. Moreover, since the multipole moments defined in Eq.~\eqref{eq:GravitationalMultipoles} only depend on the form factors, and since through the shift in Eq.~\eqref{eq:CoordTransf} we showed that the latter are not affected by coordinate transformations, we can state that our definition of gravitational multipoles in higher dimensions in gauge invariant. Furthermore, because we are considering spin-induced multipole moments, the tensors in Eq.~\eqref{eq:GravitationalMultipoles} are axially-symmetric around each rotational plane, and they are also symmetric with respect to the exchange of every pair of angular momenta\footnote{In $d=4, 5$ with two distinguished angular momenta, this corresponds to a bi-axial symmetry, in $d=6,7$ to a tri-axial symmetry, and so on.}. These symmetries drastically reduce the number of independent components of each multipole moment, such as, for instance, the $d=3, 4$ case, in which mass moments have only a single component. However, in higher dimensions this number grows, as well as the number of independent components of the current and stress moments. Hence, using Cartesian coordinates and expressing the multipole towers in terms of tensors like in \eqref{eq:GravitationalMultipoles}, is the simplest way to deal with multipoles in higher dimensions. One of the main results that can be read off from Eq.~\eqref{eq:GravitationalMultipoles} is that, in the spin induced case, we are able to connect directly the source form factors to the gravitational multipoles. Using this recipe, one does not need to compute the metric induced by some EMT in order to find the multipole moments, but one just needs to know the form factors of the source, precisely as one would do in Newtonian gravity~\cite{Bonga:2021ouq}. This relation is the relativistic analog of the correspondence between source multipoles and Newtonian gravitational multipoles for non-relativistic systems, and in this sense form factors can be seen as an unambiguous definition of source multipoles in GR, generalizing such source-gravity multipole relation for relativistic theories for the first time. Even if Eq.~\eqref{eq:GravitationalMultipoles} is specialized for spin-induced moments, we strongly believe that such duality can be established also for generic multipole moments by extending our formalism to other fundamental multipole moments \cite{Raposo:2018xkf}. 

Later on we will match the metric induced by Eq.~\eqref{eq:ConservedEMT} with known GR solutions, in particular we will consider the Kerr metric and the Myers-Perry BH in $d=4$, in order to obtain the relative form factors, and hence the gravitational multipoles of such solutions, eventually generalizing the results to arbitrary dimensions. To this end, it is worth seeing in more details the main differences of Eq.~\eqref{eq:GravitationalMultipoles} between the $d=3$ and its higher-dimensional version. As already discussed, $d=3$ is a particularly special case, since in $SO(3)$ the dimension of the fundamental representation coincides with the adjoint one, and so an anti-symmetric rank-2 tensor is dual to a vector. This results in the possibility to define a spin vector as in Eq.~\eqref{eq:SpinDualVector}. Considering the spin vector in favor of the spin tensor into \eqref{eq:GravitationalMultipoles} one obtains
\begin{equation}\label{eq:d3Multipoles}
    \begin{aligned}
        \mathbb{M}^{(2\ell)}_{A_{2\ell}}\Big|_{d=3}&=(4\ell-1)!!\Big(F_{2\ell, 1}+F_{2\ell, 2}\Big)s_{a_1}\cdots s_{a_{2\ell}}\Big|_{TF}\ , \\         \mathbb{J}^{(2\ell+1)}_{i,A_{2\ell+1}}\Big|_{d=3}&=(4\ell)!!F_{2\ell+1, 3} \ \epsilon_{ia_1k}s_{k}s_{a_2}\cdots s_{a_{2\ell}}\Big|_{TF}\ , \\
         \mathbb{G}^{(2\ell)}_{ij,A_{2\ell}}\Big|_{d=3}&=0\ ,
    \end{aligned}
\end{equation}
where the tensors are meant to be made trace-free (TF). The first thing to be noticed is that, differently from the higher dimensional case, in $d=3$ the multipoles can be expressed in terms of just STF tensors, namely $s_{A_{\ell}}|_{\rm STF}=s_{a_1}\cdots s_{a_{\ell}}|_{TF}$. Furthermore, the stress multipoles vanish and the form factor $F_{2\ell, 2}$ is redundant. This means that only the combination $F_{2\ell, 1}+F_{2\ell, 2}$ is physical and we have more freedom to compute asymptotically equivalent EMTs. In order to explicitly see how the stress multipole tensor vanishes in $d=3$ let us discuss the quadrupole case as an example. In arbitrary dimensions, the explicit RSTF projection of the quadrupole tensor structure reads
\begin{equation}
\begin{aligned}
   &S_{ia_1}S_{ja_2}\Big|_{\rm RSTF}=S_{ia_1}S_{ja_2}-\frac{1}{3}\Big(S_{ia_1}S_{ja_2}+S_{a_1j}S_{ia_2}+S_{ji}S_{a_1a_2}\Big)\\
   &+\frac{1}{d-2}\Big(S_{a_1k}S^{k}{}_{a_2}\delta_{ij}-S_{a_1k}S^{k}{}_{j}\delta_{ia_2}-S_{ik}S^{k}{}_{a_2}\delta_{a_1j}+S_{ik}S^{k}{}_{j}\delta_{a_1a_2}\Big)\\
   &+\frac{1}{(d-2)(d-1)}\Big(S_{k_1k_2}S^{k_2k_1}\delta_{ia_2}\delta_{ja_1}-S_{k_1k_2}S^{k_2k_1}\delta_{ij}\delta_{a_1a_2}\Big)\ ,
\end{aligned}
\end{equation}
respecting manifestly all the RSTF symmetries. It is then easy  to see that for $d=3$, replacing Eq.~\eqref{eq:SpinDualVector} and expressing the product of the Levi-Civita symbols in terms of delta's, one gets exactly
\begin{equation}
    S_{ia_1}S_{ja_2}\Big|^{d=3}_{\rm RSTF}=0\ .
\end{equation}
Such an argument can be extended to higher spin orders, proving how the stress multipoles vanish in $d=3$ and how the extra asymptotic gauge redundancies arise in the computation of the four-dimensional EMT, extremely relevant for Kerr mimickers applications.

\section{Form factor structure of known solutions}\label{sec:GraFFsKnownSolutions}

While in four-dimensional spacetime Kerr is the unique stationary chargeless solution with an horizon structure~\cite{Hawking:1973uf}, ensured by the no-hair theorem, a higher-dimensional extension of the BH uniqueness theorem is lacking, even though Myers-Perry metrics~\cite{Myers:1986un} are extremely relevant due to their symmetry properties. Despite the importance of such solutions, a comprehensive study of their multipolar structure is missing in the literature due to the difficulty of dealing with the multipole definition in arbitrary dimensions. Our goal then is to employ the formalism developed so far in order to define the gravitational multipoles of the Myers-Perry solutions (which correspond to those of Kerr for $d=3$). In the following then, we will firstly review the multipole structure of the Kerr metric using our formalism, and then we will discuss the multipole expansion of the Myers-Perry solution in $d=4$. Finally we will extend such construction to arbitrary dimension. 

\subsection{The Kerr case}

Let us consider the Kerr metric in Boyer-Lindquist coordinates, obtained in the chargeless limit $Q\to 0$ of the Kerr-Newman solution in Eq.~\eqref{eq:KN_metric_solution}, and reported here for convenience as
\begin{equation}\label{eq:KerrInBL}
    d s^2 = -\frac{\Delta}{\Sigma}\left(d t - a \sin^2\theta \, d \varphi\right)^2 
    + \frac{\sin^2\theta}{\Sigma}\left(a\, d t - (\tilde{r}^2+a^2)\,d \varphi\right)^2
    +\frac{\Sigma}{\Delta} d \tilde{r}^2+\Sigma\, d\theta^2\ ,
\end{equation}
with
\begin{equation}
    \Sigma=\tilde{r}^2+a^2 \cos^2 \theta\ , 
    \qquad 
    \Delta = \tilde{r}^2+a^2-2 m G_N \tilde{r}\ .
\end{equation}
Consider now the metric obtained by the EMT in Eq.~\eqref{eq:ConservedEMT} for $d=3$. Since the $G_{n, m}$ factors do not contribute to multipoles, we can get rid of them without loss of generality and describe the long-range metric only by means of the form factors. Then, in order to fix the $F_{n, m}$ coefficients to recover the Kerr metric, we have to match Eq.~\eqref{eq:MetricFromEMT_Generic} computed from Eq.~\eqref{eq:ConservedEMT} with the linearized Kerr metric in harmonic coordinates expanded order by order in the spin, recovered by applying the general procedure outlined in appendix~\ref{chapter:App} to Eq.~\eqref{eq:KerrInBL}. Moving to a reference frame in which the $z$-axis is oriented along the rotational symmetry, as in Eq.~\eqref{eq:d3SpinMatrix}, we managed to perform this matching up to the seventh order in the spin $a$. This requires 
\begin{equation}\label{eq:KerrFixing}
\begin{gathered}
    F_{0, 2}+F_{0, 1}=1\ ,\quad F_{2, 2}+F_{2, 1}=-\frac{1}{2}\ ,\quad F_{4, 2}+F_{4, 1}=\frac{1}{24}\ ,\\ F_{6, 2}+F_{6, 1}=-\frac{1}{720}\ ,\quad
    F_{1, 3}=1\ , \quad F_{3, 3}=-\frac{1}{6}\ ,\\ \quad F_{5, 3}=\frac{1}{120}\ ,\qquad F_{7, 3}=-\frac{1}{5040}\ .
\end{gathered}
\end{equation}
As we can see, while the current moments are uniquely fixed, there is a degeneracy between $F_{2\ell, 1}$ and $F_{2\ell, 2}$, due to the fact that we are in $d=3$. Moreover, although we managed to perform this matching up to $\mathcal{O}(a^7)$, it is possible to make an ansatz for the series at every order in spin as 
\begin{equation}\label{eq:d3Series}
    F_{2\ell, 2}+F_{2\ell, 1}=\frac{(-1)^\ell}{(2\ell)!}\ ,\qquad F_{2\ell+1, 3}=\frac{(-1)^\ell}{(2\ell+1)!}\ .
\end{equation}
Such series can be resummed introducing a dummy variable $\zeta$, and it leads to
\begin{equation}\label{eq:d3FormFactor}
\begin{aligned}
    F_{2}^{(d=3)}(\zeta)+F_{1}^{(d=3)}(\zeta)&=\sum_{\ell=0}^{+\infty}\Big(F_{2\ell, 2}+F_{2\ell, 1}\Big)\zeta^{2\ell}=\cos\zeta\ , \\
    F_{3}^{(d=3)}(\zeta)&=\sum_{\ell=0}^{+\infty}F_{2\ell+1, 3}\zeta^{2\ell}=\frac{\sin\zeta}{\zeta}\ .
\end{aligned}
\end{equation}
The form factor coefficients then can be extracted from the series expansion around $\zeta=0$ of such expressions, that generates only even power of the dummy variable. 
Furthermore, for later purposes, it is important to notice that the resummed form factors can indeed be expressed in terms of spherical Bessel functions $j_n(\zeta)$ as 
\begin{equation}\label{eq:d3SphericalB}
\begin{gathered}
    F_{3}^{(d=3)}(\zeta)=j_0(\zeta)\ , \\ F_{2}^{(d=3)}(\zeta)+F_{1}^{(d=3)}(\zeta)=j_0(\zeta)-\zeta\, j_1(\zeta)\ .
\end{gathered}
\end{equation}
Interestingly the resummed form factors resemble the expressions of the Fourier Transform of the metric in the Kerr-Schild gauge that appear in the scattering amplitudes of~\cite{Bianchi:2023lrg, Bianchi:2025xol}, extensively discussed in chapter~\ref{chapter:KerrSchildGauge}. Finally, Eq.~\eqref{eq:d3Series} can be used in Eq.~\eqref{eq:d3Multipoles}, providing some closed form relations for the well-known infinite multipole towers of the Kerr metric as~\cite{Hansen:1974zz}
\begin{equation}\label{eq:KerrMultipoles}
    \begin{aligned}
        \mathbb{M}^{(2\ell)}_{A_{2\ell}}\Big|_{d=3}&=\frac{(4\ell-1)!!}{(2\ell)!}s_{a_1}\cdots s_{a_{2\ell}}\Big|_{TF}\ , \\         \mathbb{J}^{(2\ell+1)}_{i,A_{2\ell+1}}\Big|_{d=3}&=\frac{(4\ell)!!}{(2\ell+1)!}\ \epsilon_{ia_1k}s_{k}s_{a_2}\cdots s_{a_{2\ell}}\Big|_{TF}\ , \\
         \mathbb{G}^{(2\ell)}_{ij,A_{2\ell}}\Big|_{d=3}&=0\ .
    \end{aligned}
\end{equation}

\subsection{Myers-Perry in $d=4$}

We can repeat the argument performed in the Kerr case for the Myers-Perry solution in $d=4$. Considering the explicit metric expression in Eq.~\eqref{app:MPD5Metric} and following the procedure outlined in section~\ref{App:MPinHarm}, it is possible to express such solution in harmonic gauge as in Eq.~\eqref{eq:MPmetricHarmonic}, in order to match it with the gravitational perturbation computed from Eq.~\eqref{eq:ConservedEMT} by fixing the form factors. Moving to Myers-Perry coordinates and expressing the spin tensor as Eq.~\eqref{eq:BlockDiagonalD5MEtric}, we were able to perform such matching up to the seventh order in the spin. The form factors then read
\begin{equation}\label{eq:ExplicitMPFF}
    \begin{gathered}
      F_{0, 1}=1\ ,   \quad F_{2, 1}=-\frac{15}{32}\ ,\quad F_{4, 1}=\frac{63}{1024}\ ,\\
      F_{6, 1}=-\frac{243}{65536}\ ,\quad
      F_{0, 2}=0\ , \quad  F_{2, 2}=-\frac{3}{16}\ ,\\
      F_{4, 2}=\frac{9}{256}\ ,\quad F_{6, 2}=-\frac{81}{32768}\ ,\quad
        F_{1, 3}=1\ ,\\
        F_{3, 3}=-\frac{9}{32}\ ,\quad F_{5, 3}=\frac{27}{1024}\ ,\quad F_{7, 1}=-\frac{81}{65536}\ .
    \end{gathered}
\end{equation}
As we did previously, we can make an ansatz for the series of form factors for every spin order. Indeed, one can verify that the above sequence can be reproduced by 
\begin{equation}\label{eq:d4Series}
\begin{gathered}
        F_{2\ell+2, 2}=-\frac{2}{3}\frac{(-1)^\ell}{(\ell)!(\ell+2)!}\left(\frac{3}{4}\right)^{2\ell+2}\ ,\\F_{2\ell+1, 3}=\frac{4}{3}\frac{(-1)^\ell}{(\ell)!(\ell+1)!}\left(\frac{3}{4}\right)^{2\ell+1}\ ,\\
        F_{2\ell,1}= F_{2\ell,2}+F_{2\ell+1,3}\ ,
\end{gathered}
\end{equation}
from which, using again a dummy variable $\zeta$, the form factors can be resummed as
\begin{equation}\label{eq:d4FormFactors}
\begin{aligned}
    F_{2}^{(d=4)}(\zeta)&=-\frac{2}{3}J_{2}\left(\frac{3}{2}\zeta\right)\ ,\\
    F_{3}^{(d=4)}(\zeta)&=\frac{4}{3\zeta}J_{1}\left(\frac{3}{2}\zeta\right)\ ,\\
    F_{1}^{(d=4)}(\zeta)&= F_{2}^{(d=4)}(\zeta)+ F_{3}^{(d=4)}(\zeta)\ ,
\end{aligned}
\end{equation}
where $J_\alpha(\zeta)$ is the Bessel function of the first kind. 
Once again these expressions resemble the ones that appear in the higher dimensional scattering amplitudes in the Kerr-Schild gauge (see chapter~\ref{chapter:KerrSchildGauge}). Finally, replacing Eq.~\eqref{eq:d4Series} into \eqref{eq:GravitationalMultipoles}, we can give the complete series of the Myers-Perry gravitational multipoles in $d=4$ as
\begin{equation}
    \begin{aligned}
        \mathbb{M}^{(2\ell+2)}_{A_{2\ell+2}}\Big|_{d=4}&=\frac{(4+4\ell)!!}{(\ell+1)!^2}\left(\frac{3}{4}\right)^{2\ell+2}(-S\cdot S)_{A_{2\ell+2}}\Big|_{\rm STF}\ , \\
         \mathbb{J}^{(2\ell+1)}_{i,A_{2\ell+1}}\Big|_{d=4}&=\frac{1}{2}\frac{(2+4\ell)!!}{\ell!(\ell+1)!} \left(\frac{3}{4}\right)^{2\ell} \ S_{ia_1}(-S\cdot S)_{A_{2\ell}}|_{\rm ASTF}\ , \\
         \mathbb{G}^{(2\ell+2)}_{ij,A_{2\ell+2}}\Big|_{d=4}&=\frac{(4+4\ell)!!}{\ell!(\ell+2)!} \left(\frac{3}{4}\right)^{2\ell+2} \ S_{ia_1}S_{ja_2}(-S\cdot S)_{A_{2\ell}}|_{\rm RSTF}\ .
    \end{aligned}
\end{equation}
We notice that while mass and current multipoles were already found in~\cite{Heynen:2023sin}, we derived for the first time the complete tower of stress multipoles associated to the Myers-Perry solution in $d=4$.

\subsection{Myers-Perry in arbitrary dimensions}

From Eq~\eqref{eq:d4FormFactors}, a recurrent structure with respect to the $d=3, 4$ cases can be noticed. Indeed, defining
\begin{equation}
    \mathcal{Z}_n^{(d)}(\zeta)=\frac{1}{2}2^{d/2}\Gamma(d/2) \frac{J_{n+\frac{d-2}{2}}\left(\frac{d-1}{2}\zeta\right)}{(\frac{d-1}{2}\zeta)^{\frac{d-2}{2}}}\ ,
\end{equation}
it is possible to express the form factors as\footnote{Notice that the functions $\mathcal{Z}_n^{(d)}$ can be related to the spherical Bessel functions in arbitrary dimension.}
\begin{equation}\label{eq:d4SphericalB}
     F_{2}^{(d=4)}(\zeta)=-\frac{1}{2}\zeta\, \mathcal{Z}_1^{(d=4)}(\zeta)\ , \qquad
    F_{3}^{(d=4)}(\zeta)=\mathcal{Z}_0^{(d=4)}(\zeta)\ .
\end{equation}
Then, from Eqs.~\eqref{eq:d3SphericalB} and \eqref{eq:d4SphericalB}, noticing the dependence on the spatial dimension in the resummed form factors, we conjecture that the form factors of the Myers-Perry solutions in arbitrary dimensions can be extracted as the coefficients of the series expansion of the following expressions
\begin{equation}\label{eq:dGenericSphericalB}
\begin{gathered}
     F_{2}^{(d)}(\zeta)=-\frac{1}{2}\zeta\, \mathcal{Z}_1^{(d)}(\zeta)\ , \qquad
    F_{3}^{(d)}(\zeta)=\mathcal{Z}_0^{(d)}(\zeta)\ ,\\
    F_{1}^{(d)}(\zeta)=F_{2}^{(d)}(\zeta)+F_{3}^{(d)}(\zeta)\ .
\end{gathered}
\end{equation}
As a sanity check, we can see that for $d=3$ the above definition reproduces Eq.~\eqref{eq:d3SphericalB}, 
\begin{equation}
\begin{aligned}
    F_{1}^{(d=3)}(\zeta)+F_{2}^{(d=3)}(\zeta)&=2F_{2}^{(d=3)}(\zeta)+F_{3}^{(d=3)}(\zeta)\\
    &=\mathcal{Z}_0^{(d=3)}(\zeta)-\zeta\, \mathcal{Z}_{1}^{(d=3)}(\zeta)\ .
    \end{aligned}
\end{equation}
Moreover, from \eqref{eq:dGenericSphericalB} we can extract the infinite series of Myers-Perry form factors in arbitrary dimensions as
\begin{equation}\label{eq:FFMParbitraryD}
\begin{aligned}
    F_{2\ell+2, 2}&=-\frac{1}{2}\Gamma(d/2)\frac{(-1)^\ell}{\ell!\ \Gamma\left(\ell+2+\frac{d-2}{2}\right)}\left(\frac{d-1}{4}\right)^{2\ell+1}\ , \\
    F_{2\ell+1, 3}&=\Gamma(d/2) \frac{(-1)^\ell}{\ell!\ \Gamma\left(\ell+1+\frac{d-2}{2}\right)}\left(\frac{d-1}{4}\right)^{2\ell}\ ,\\
    F_{2\ell+2, 1}&=F_{2\ell+2, 2}+F_{2\ell+1, 3}\ ,
\end{aligned}
\end{equation}
from which, replacing into Eq.~\eqref{eq:GravitationalMultipoles}, we finally get the gravitational multipoles of Myers-Perry BHs in arbitrary dimensions
\begin{equation}\label{eq:MPGraFFsDgeneric}
    \begin{aligned}
        \mathbb{M}^{(2\ell+2)}_{A_{2\ell+2}}&=\frac{d-1}{4}\frac{(d+4\ell)!!\ (d+2\ell)}{(d-2)!!\ (\ell+1)!\ \Gamma\left(\ell+2+\frac{d-2}{2}\right)}\left(\frac{d-1}{4}\right)^{2\ell+1+\frac{d-2}{2}}(-S\cdot S)_{A_{2\ell+2}}\Big|_{\rm STF}\ , \\
         \mathbb{J}^{(2\ell+1)}_{i,A_{2\ell+1}}&=\frac{(d+4\ell-2)!!}{(d-2)!!\ \ell!\ \Gamma\left(\ell+1+\frac{d-2}{2}\right)}\left(\frac{d-1}{4}\right)^{2\ell+\frac{d-2}{2}} \ S_{ia_1}(-S\cdot S)_{A_{2\ell}}|_{\rm ASTF}\ , \\
         \mathbb{G}^{(2\ell+2)}_{ij,A_{2\ell+2}}&=\frac{d-1}{2}\frac{(d+4\ell)!!}{(d-2)!!\ \ell!\ \Gamma\left(\ell+2+\frac{d-2}{2}\right)}\left(\frac{d-1}{4}\right)^{2\ell+1+\frac{d-2}{2}} \ S_{ia_1}S_{ja_2}(-S\cdot S)_{A_{2\ell}}|_{\rm RSTF}\ .
    \end{aligned}
\end{equation}
Although the validity of Eq.~\eqref{eq:d3SphericalB} is conjectured, we managed to prove it also for $d=5$ up to $\mathcal{O}(S^5)$. Indeed, even though writing the Myers-Perry metric in harmonic coordinates is challenging in this case, we can still move to a generic ACMC reference frame as in Eq.~\eqref{eq:MPACMCmetricExpr}, then verifying Eq.~\eqref{eq:MPGraFFsDgeneric} in $d=5$ up to $\mathcal{O}(S^5)$.

\section{Linearized matter sources}\label{sec:LinearizedBHSources}

Once the explicit expression for the form factors that generates the Myers-Perry solution are found in Eq.~\eqref{eq:dGenericSphericalB}, an infinite number of EMTs can be defined such that they all share the same multipolar structure, differing between each other by local contributions. Among all the infinite choices of sources, in this section we want to compute the Fourier transform of the EMT in a specific configuration. Indeed, in analogy with the description of conserved charges in QFT, form factors characterize the internal structure of the source, and considering them as constants coefficients corresponds to impose that the object we are describing is a point-like particle. 
In this spirit, in the following we show that the Israel EMT that sources the Kerr metric corresponds to a point-like distribution with all vanishing residual factors. By analogy we compute the EMT for the Myers-Perry solution in $d=4$ in a similar setup. 

Moving to Myers-Perry coordinates in which the radial distance can be expressed in terms of Cartesian coordinates as
\begin{equation}
\begin{aligned}
r^2&=\sum_{k=1}^{\frac{d}{2}}(x_k^2+y_k^2)\quad \text{for} \quad d=\text{even}\ ,\\
r^2&=\sum_{k=1}^{\frac{d-1}{2}}(x_k^2+y_k^2)+z^2\quad \text{for} \quad d=\text{odd}\ ,
\end{aligned}
\end{equation}
each angular momenta is perpendicular to the plane $(y_k, x_k)$. Then, we can set
\begin{equation}\label{eq:PhysSpectralDef}
    \zeta^\mu=i(S\cdot q)^\mu \qquad \mathrm{and} \qquad\zeta^\mu\zeta_\mu=\zeta^2=-q\cdot S\cdot S\cdot q=\sum_{k}q_{\perp, k}^{2}a_k^2\ ,
\end{equation}
where the sum is performed over every angular momenta $a_k$ of the BH, and with ${q_{\perp, k}^2=q_{y_k}^2+q_{x_k}^2}$. We can now resum the EMT in Eq.~\eqref{eq:ConservedEMT}, from which we obtain\footnote{Notice that in the stress part of the EMT we have to consider an extra $\zeta^{-2}$ factor since the expansion starts at quadrupole order.}
\begin{equation}\label{eq:ConservedEMTresummed}
    T^{\mu\nu}(q)=m\ u^{\mu}u^{\nu}F_1^{(d)}(\zeta)-m \frac{{F}^{(d)}_{2}(\zeta)}{\zeta^2}\zeta^\mu \zeta^\nu-\frac{1}{2}m\Big(u^\mu\zeta^\nu +u^\nu\zeta^\mu \Big){F}^{(d)}_{3}(\zeta)\ ,
\end{equation}
where the functions $F_n$ are defined in Eq.~\eqref{eq:dGenericSphericalB}. 
Then, considering the explicit form factor function that we found for the Myers-Perry solution, for $d>3$ one ends up with
\begin{equation}\label{eq:mpEMTresumDgeneric}
    \begin{aligned}
    &T^{00}(q)=m \left(\mathcal{Z}_{0}^{(d)}(\zeta)-\frac{1}{2}\zeta\ \mathcal{Z}_{1}^{(d)}(\zeta)\right)\ , \\
     &T^{0i}(q)=\frac{1}{2}m\, \zeta^i \mathcal{Z}_{0}^{(d)}(\zeta)\ ,\\
    &T^{ij}(q)=\frac{1}{2}m\, \zeta^i \zeta^j \frac{\mathcal{Z}_{1}^{(d)}(\zeta)}{\zeta}\ ,
    \end{aligned}
\end{equation}
while for $d=3$ we should consider the expression
\begin{equation}
    \begin{aligned}
    &T^{00}(q)\big|_{d=3}=m \left(\mathcal{Z}_{0}^{(d=3)}(\zeta)-F_{2}^{(d=3)}(\zeta)\right)\ , \\
    &T^{0i}(q)\big|_{d=3}=\frac{1}{2}m\, \zeta^i\mathcal{Z}_{0}^{(d=3)}(\zeta)\ ,\\
    &T^{ij}(q)\big|_{d=3}=-m\, \zeta^i \zeta^j \frac{F_2^{(d=3)}(\zeta)}{\zeta^2}\ ,
    \end{aligned}
\end{equation}
where $F_2^{(d=3)}(\zeta)$ is a free parameter that we can arbitrarily fix without changing the asymptotic multipolar structure of the induced spacetime. Surely, both cases possess a cylindrical symmetry in momentum space, and we can set up the calculation of the EMT in a generic way. In fact, from Eq.~\eqref{eq:ConservedEMTresummed}, in the case of $d=\text{even}$ one gets
\begin{equation}\label{eq:EMTmsGenericD}
    \begin{aligned}
    T^{00}(x)&=m \prod_k\int_0^{+\infty}\frac{dq_{\perp, k}}{2\pi}q_{\perp, k} J_0(q_{\perp, k}\rho_k)\Big(F_2^{(d)}(\zeta)+F_3^{(d)}(\zeta)\Big)\ , \\
    T^{0i}(x)&=-\frac{1}{2}m(S\cdot \partial)^i\prod_k\int_0^{+\infty}\frac{dq_{\perp, k}}{2\pi}q_{\perp, k}J_0(q_{\perp, k}\rho_k)F_3^{(d)}(\zeta)\ , \\
    T^{ij}(x)&=m(S\cdot \partial)^i (S\cdot \partial)^j \prod_k\int_0^{+\infty}\frac{dq_{\perp, k}}{2\pi}q_{\perp, k} J_0(q_{\perp, k}\rho_k)\frac{F_2^{(d)}(\zeta)}{\zeta^2}\ ,
    \end{aligned}
\end{equation}
where here and in the following we identify with $\rho_k^2=y_k^2+x_k^2$ the radial coordinate in each rotational plane, and where in the case of $d=\text{odd}$ we just need to add an extra $\delta(z)$ in front of each contribution.

\subsection{The Kerr case}

Let us now focus on the $d=3$ case. First of all, since there is only one angular momenta we have $\zeta=q_\perp a$, where we consider $a>0$ without loss of generality. Moreover, we can exploit the extra freedom reminiscent of the vanishing stress multipoles, and move to a ``pressure-less gauge'' in which we set $F_2^{(d=3)}(\zeta)=0$. In this case, the EMT simply reads
\begin{equation}\label{eq:EMTkerrMP}
    \begin{aligned}
    &T^{00}(x)=m\  \delta(z)\int_0^{+\infty}\frac{dq_{\perp}}{2\pi}q_{\perp}J_0(q_{\perp}\rho)\cos(q_\perp a)\ , \\
    &T^{0i}(x)=-\frac{1}{2}m(S\cdot \partial)^i\delta(z)\int_0^{+\infty}\frac{dq_{\perp}}{2\pi}q_{\perp}J_0(q_{\perp}\rho)\frac{\sin(q_\perp a)}{q_\perp a}\ , \\
    &T^{ij}(x)=0\ .
    \end{aligned}
\end{equation}
We start with the explicit computation of the mass density $T^{00}(x)$. Considering the following tabulated integral involving Bessel functions~\cite{Gradshteyn:1943cpj} 
\begin{equation}
    \int_{0}^{+\infty}dz\ z \cos(c_1 z)J_0(c_2 z)=-\frac{c_1}{(c_1^2-c_2^2)^{3/2}}\Theta(c_1-c_2)\ ,
\end{equation}
where $\Theta(x)$ represents the Heaviside step function, we can compute the mass density energy as
\begin{equation}\label{eq:T00Kerr}
    T^{00}(x)=-\frac{m}{2\pi} \delta(z)\frac{a}{(a^2-\rho^2)^{3/2}}\Theta(a-\rho)\ .
\end{equation}
The energy density distribution turns out to be a thin equatorial disk with radius $a$. Moreover for $\rho=a$ there is a singularity that coincides with the curvature (ring-)singularity of Kerr.
Indeed, Eq.~\eqref{eq:T00Kerr} is in perfect agreement with the original result of Israel~\cite{Israel:1970kp}  and with the derivation in~\cite{Balasin:1993kf}. However, while in the original work of Israel the shape of the energy distribution was assumed to be a disk of a finite radius, in~\cite{Balasin:1993kf} no assumption on the EMT was made, but still, in both derivations the full non-linear Kerr metric was employed. Remarkably, in the present derivation, we did not make use of any assumptions on the EMT, except for requiring a vacuum solution of the Einstein equations and a source endowed with a mass and angular momentum.
Then, the only ingredient to derive Eq.~\eqref{eq:T00Kerr} is the particular multipolar structure of Kerr, which uniquely leads to its characteristic curvature (ring-)singularity.

It is worth noticing that the energy density in Eq.~\eqref{eq:T00Kerr} is negative and hence violates the weak energy condition. We can in fact consider a static observer with four-velocity ${U^\mu=(-1, 0, 0, 0)}$, from which 
\begin{equation}\label{eq:WEC}
    T_{\mu\nu}U^\mu U^\nu<0\ .
\end{equation}
This means that Eq.~\eqref{eq:T00Kerr} is not physical, even besides its singular nature. Moreover, due to the singularity at $\rho=a$, Eq.~\eqref{eq:T00Kerr} is defined up to a delta function. To get the total mass $m$ by integrating the mass density function we need to consider a regularized mass density function as
\begin{equation}
    \tilde{T}^{00}(x)=-\frac{m}{2\pi} \delta(z)\frac{a}{(a^2-\rho^2)^{3/2}}\Theta\Big(a(1-\epsilon)-\rho\Big)+\frac{m}{\sqrt{2\epsilon}}\delta(z)\frac{\delta(\rho-a)}{2\pi\rho}\ ,
\end{equation}
such that one gets
\begin{equation}
   \lim_{\epsilon\rightarrow 0} \int d^3 x \tilde{T}^{00}(x)=m\ .
\end{equation}
Finally, in~\cite{Balasin:1993kf} it has been shown that \eqref{eq:T00Kerr} correctly reproduces the Schwarzschild limit for $a\rightarrow 0$. Indeed, while it is easy to reproduce such limit in momentum space by looking at Eq.~\eqref{eq:EMTkerrMP}, it is more subtle to see it in coordinate space due to the distributional nature of the expression. 

We can now move to the computation of the current part of the EMT. Since the $T^{0z}(x)$ component is vanishing, we need to compute only
\begin{equation}
\begin{aligned}
    T^{0x}(x)&=\frac{m}{4\pi\rho}y\, \delta(z)\int_0^{+\infty}dq_{\perp}q_{\perp}J_1(q_{\perp}\rho)\sin(q_\perp a)\ ,\\
    T^{0y}(x)&=-\frac{m}{4\pi\rho}x\, \delta(z)\int_0^{+\infty}dq_{\perp}q_{\perp}J_1(q_{\perp}\rho)\sin(q_\perp a)\ ,
\end{aligned}
\end{equation}
and using the known integral~\cite{Gradshteyn:1943cpj}
\begin{equation}
    \int_0^{+\infty}dz\ z \sin(c_1 z)J_1(c_2 z)=-\frac{c_2}{(c_1^2-c_2^2)^{3/2}} \Theta(c_1-c_2)\ ,
\end{equation}
one gets
\begin{equation}\label{eq:T0iKerr}
\begin{aligned}
    T^{0x}(x)&=-\frac{m}{4\pi} \delta(z)\frac{y}{(a^2-\rho^2)^{3/2}} \Theta(a-\rho)\ ,\\
    T^{0y}(x)&=+\frac{m}{4\pi} \delta(z)\frac{x}{(a^2-\rho^2)^{3/2}} \Theta(a-\rho)\ .
\end{aligned}
\end{equation}
Out of Eq.~\eqref{eq:T0iKerr} we can extract the angular momentum of the source by considering 
\begin{equation}
    \frac{d\vec{L}}{d\rho}=\int dz (2\pi \rho)\  \vec p\times \vec x\ ,
\end{equation}
where $p^i=T^{0i}(x)$ is the momentum density, leading to 
\begin{equation}\label{eq:AngMomKerr}
    dL_z=-\frac{m}{2}\delta(z)\frac{\rho^3}{(a^2-\rho^2)^{3/2}}d\rho \Theta(a-\rho)\ ,
\end{equation}
where $dL_x=dL_y=0$. This result matches exactly with what derived in~\cite{Israel:1970kp}. Similar to the energy density term, even the current part of the EMT is non-physical, since for a stationary observer at infinity the angular velocity near the boundary of the disk can reach superluminal speed. Moreover, Eq.~\eqref{eq:AngMomKerr} needs to be regulated exactly as the energy density function. Indeed, by defining
\begin{equation}
    d\tilde{L}_z=-\frac{m}{2}\delta(z)\frac{\rho^3}{(a^2-\rho^2)^{3/2}}d\rho \Theta\Big(a(1-\epsilon)-\rho\Big)+\frac{m a}{2\sqrt{2\epsilon}}\delta(z)\frac{\delta(\rho-a)}{2\pi \rho}\ ,
\end{equation}
one gets 
\begin{equation}
    \lim_{\epsilon\rightarrow 0}\int  d\tilde{L}_z= ma\ .
\end{equation}
In summary, the derived mass and angular-momentum densities exactly match the classical results of~\cite{Israel:1970kp,Balasin:1993kf}, confirming that the entire structure follows solely from the Kerr multipolar content. Despite its non-physical nature and distributional singularities, this construction establishes how the multipolar framework encodes both the ring singularity and the correct global charges of Kerr, leaving the door open for interesting applications in the case of non-singular sources.

\subsection{Myers-Perry in $d=4$}

Let us now focus on the EMT of Myers-Perry solutions in $d=4$. In this case there are two independent angular momenta\footnote{Notice that here we are using the physical ADM normalization for the angular momenta, which are related to the spin parameters entering the Myers-Perry metric expression in Eq.~\eqref{app:MPD5Metric} through the relations~\eqref{eq:MPd4SpinRelations}.} $a_1$ and $a_2$, and in the case in which $a_1\neq a_2$ the EMT reads
\begin{equation}\label{eq:EMTmpd4}
    \begin{aligned}
    &T^{00}(x)=m \int_0^{+\infty}\frac{dq_{\perp, 1}}{2\pi}q_{\perp, 1}J_0(q_{\perp, 1}\rho_1)\int_0^{+\infty}\frac{dq_{\perp, 2}}{2\pi}q_{\perp, 2}J_0(q_{\perp, 2}\rho_2)\left(\frac{4}{3}\frac{J_1\left(\frac{3}{2}\zeta\right)}{\zeta}-\frac{2}{3}J_2\left(\frac{3}{2}\zeta\right)\right)\ , \\
    &T^{0i}(x)=-\frac{1}{2}m(S\cdot \partial)^i\int_0^{+\infty}\frac{dq_{\perp, 1}}{2\pi}q_{\perp, 1}J_0(q_{\perp, 1}\rho_1)\int_0^{+\infty}\frac{dq_{\perp, 2}}{2\pi}q_{\perp, 2}J_0(q_{\perp, 2}\rho_2)\left(\frac{4}{3}\frac{J_1\left(\frac{3}{2}\zeta\right)}{\zeta}\right)\ , \\
    &T^{ij}(x)=m(S\cdot \partial)^i (S\cdot \partial)^j \int_0^{+\infty}\frac{dq_{\perp, 1}}{2\pi}q_{\perp, 1}J_0(q_{\perp, 1}\rho_1)\int_0^{+\infty}\frac{dq_{\perp, 2}}{2\pi}q_{\perp, 2}J_0(q_{\perp, 2}\rho_2)\left(-\frac{2}{3}\frac{J_2\left(\frac{3}{2}\zeta\right)}{\zeta^2}\right)\ .
    \end{aligned}
\end{equation}
We start the computation by considering the $T^{00}(x)$ component of the EMT. Using the Bessel identity 
\begin{equation}
    J_2(\tfrac{3}{2}\zeta)=\frac{4}{3}\frac{J_1(\tfrac{3}{2}\zeta)}{\zeta}-J_0(\tfrac{3}{2}\zeta)\ ,
\end{equation}
we can rewrite
\begin{equation}\label{eq:MPT00}
    T^{00}(x)=\frac{m}{(2\pi)^2}\Big(\frac{4}{9}A_1(x)+\frac{2}{3}A_0\Big)\ ,
\end{equation}
where 
\begin{equation}\label{eq:Idefinition}
    \begin{aligned}
A_1&=\int_0^{+\infty}dq_{\perp, 1}q_{\perp, 1}J_0(q_{\perp, 1}\rho_1)\int_0^{+\infty}dq_{\perp, 2}q_{\perp, 2}J_0(q_{\perp, 2}\rho_2)\frac{J_1\left(\frac{3}{2}\zeta\right)}{\zeta}\ ,\\
A_0&=\int_0^{+\infty}dq_{\perp, 1}q_{\perp, 1}J_0(q_{\perp, 1}\rho_1)\int_0^{+\infty}dq_{\perp, 2}q_{\perp, 2}J_0(q_{\perp, 2}\rho_2)J_0\left(\frac{3}{2}\zeta\right)\ .
    \end{aligned}
\end{equation}
Considering then the Bessel integral~\cite{Gradshteyn:1943cpj}
\begin{equation}\label{eq:BesselSQRT}
    \int_0^{+\infty}dt\ J_{c_2}(\beta t)\frac{J_{c_1}\left(\alpha \sqrt{t^2+u^2}\right)}{\sqrt{(t^2+u^2)^{c_1}}}t^{c_2+1}=\frac{\beta^{c_2}}{\alpha^{c_1}}\Bigg(\frac{\sqrt{\alpha^2-\beta^2}}{u}\Bigg)^{c_1-c_2-1}J_{c_1-c_2-1}\left(u\sqrt{\alpha^2-\beta^2}\right)\Theta(\alpha-\beta)\ ,
\end{equation}
we get
\begin{equation}\label{Eq:I1}
    A_1=\frac{4}{3}\delta\Big(a_1^2\rho_2^2+a_2^2\rho_1^2-(\tfrac{3}{2}a_1a_2)^2\Big)\Theta(\tfrac{3}{2}a_1-\rho_1)\Theta(\tfrac{3}{2}a_2-\rho_2)\ ,
\end{equation}
where we used the relation 
\begin{equation}\label{eq:DeltaIdentity}
    \frac{\delta\left(\rho_2-\frac{a_2}{a_1}\sqrt{(\textstyle \frac{3}{2}a_1)^2-\rho_1^2}\right)}{2a_1^2\rho_2}=\delta\Big(a_1^2\rho_2^2+a_2^2\rho_1^2-(\textstyle \frac{3}{2}a_1a_2)^2\Big)\ ,
\end{equation}
and the Bessel ortogonality relation
from which 
\begin{equation}
    \int dx\ x\ J_p(c_1x)J_p(c_2x)=\frac{\delta(c_1-c_2)}{c_1}\ ,
\end{equation}
with $c_1, c_2, p\in \mathbb{Z}$. Then, in order to compute $A_0$ and ensure ourselves to get a symmetric relation, we notice that Eq.~\eqref{eq:Idefinition} exhibits a bi-axial symmetry, and the result that one obtains integrating first in $q_{\perp, 1}$ (or $q_{\perp, 2}$) can be symmetrized in 
\begin{equation}\label{Eq:I0}
    A_0=\frac{1}{2}\Bigg(\frac{4}{3}\frac{\pi}{a_2}\delta(y_1)\delta(x_1)\delta(\tfrac{3}{2}a_2-\rho_2)+\frac{4}{3}\frac{\pi}{a_1}\delta(y_2)\delta(x_2)\delta(\tfrac{3}{2}a_1-\rho_1)\Bigg)\ .
\end{equation}
The resulting mass-energy distribution in Eq.~\eqref{eq:MPT00} is then noticed to be singular for ${a_1^2\rho_2^2+a_2^2\rho_1^2=(\textstyle \frac{3}{2}a_1a_2)^2}$, and vanishing everywhere else, describing a 3-ellipsoid embedded in $\mathbb{R}^4$ of semi-axis $\rho_1=\frac{3}{2}a_1$ and $\rho_2=\frac{3}{2}a_2$.
As a sanity check, it is easy to show that 
\begin{equation}\label{eq:MassNormalization}
    \int d\rho_1\, 2\pi\rho_1\int d\rho_2\, 2\pi\rho_2\ T^{00}=m\ .
\end{equation}

We can now compute the current part of the EMT in Eq.~\eqref{eq:EMTmpd4}. Since the integral is equal to the one already computed for $A_1$, we can already write
\begin{equation}
    T^{0i}(x)=-(S\cdot \partial)^i\frac{m}{(2\pi)^2}\frac{8}{9}\delta\Big(a_1^2\rho_2^2+a_2^2\rho_1^2-(\textstyle \frac{3}{2}a_1a_2)^2\Big)\Theta(\textstyle \frac{3}{2}a_1-\rho_1)\Theta(\textstyle \frac{3}{2}a_2-\rho_2)\ ,
\end{equation}
and for specific components one gets
\begin{equation}\label{eq:EMTcurrentComponentsMPd4}
\begin{aligned}
    T^{0y_1}(x)&=\frac{x_1}{\rho_1}\frac{m}{(2\pi)^2}\frac{8}{9}\frac{\pi}{a_1}\delta(y_2)\delta(x_2)\delta(\textstyle \frac{3}{2}a_1-\rho_1)\ ,\\
     T^{0x_1}(x)&=-\frac{y_1}{\rho_1}\frac{m}{(2\pi)^2}\frac{8}{9}\frac{\pi}{a_1}\delta(y_2)\delta(x_2)\delta(\textstyle \frac{3}{2}a_1-\rho_1)\ ,
\end{aligned}
\end{equation}
and similar for $T^{0y_2}(x)$ and $T^{0x_2}(x)$. We can recover the angular momentum density tensor defined as
\begin{equation}
    l^{ij}=T^{0i}x^j-T^{0j}x^i\ ,
\end{equation}
where the angular momenta densities associated to the respective rotational planes can be identified as ${l_k=l^{y_kx_k}}$.
The explicit expressions then read 
\begin{equation}\label{eq:InfinitesimalAngMomSimplified}
\begin{aligned}
    l_1&=\frac{\rho_1}{a_1} \frac{m}{2\pi}\frac{4}{9}\delta(\textstyle\frac{3}{2}a_1-\rho_1)\delta(y_2)\delta(x_2)\ ,\\
    l_2&=\frac{\rho_2}{a_2} \frac{m}{2\pi}\frac{4}{9}\delta(\textstyle\frac{3}{2}a_2-\rho_2)\delta(y_1)\delta(x_1)\ .
\end{aligned}
\end{equation}
It is then easy to show that integrating such expressions one recovers the ADM angular momenta of the source
\begin{equation}
    L_k=\int d^4x\ l_k=a_k\, m\ .
\end{equation}
Finally we can focus on the stress part of the EMT. Limiting for simplicity to the study of
\begin{equation}
    P_k=T^{x_k x_k}+T^{y_k y_k}\ ,
\end{equation}
since the above expression only depends on the Laplacian, we can use the fact that the integrand does not depend on any angle, so that in 2 dimensions the radial part of the operator reads $\partial_{x_k}^2+\partial_{y_k}^2=\frac{1}{\rho_k}\partial_{\rho_k}\rho_k\partial_{\rho_k}$.
So then let us consider
\begin{equation}
    P_1=-a_1^2\frac{2}{3}\frac{m}{(2\pi)^2}\frac{1}{\rho_1}\partial_{\rho_1}\rho_1\partial_{\rho_1}\int_0^{+\infty}dq_2 \ q_2 J_0(\rho_2 q_2)\int_0^{+\infty}dq_1 \ q_1 J_0(\rho_1 q_1)\frac{J_2(\zeta^2)}{\zeta^2}\ .
\end{equation}
Following the same steps as the previous calculation of the energy and current density, we can write
\begin{equation}
\begin{aligned}
    &P_1=\frac{m}{(2\pi)^2}\Bigg(\frac{1}{a_1}\frac{8}{9}\pi\delta(y_2)\delta(x_2)\delta(\tfrac{3}{2}a_1-\rho_1)\\
    &-\frac{32}{27}\delta\Big(a_1^2\rho_2^2+a_2^2\rho_1^2-(\tfrac{3}{2}a_1 a_2)^2\Big)\Theta(\tfrac{3}{2}a_1-\rho_1)\Theta(\tfrac{3}{2}a_2-\rho_2)\Bigg)\ ,
\end{aligned}
\end{equation}
and similar for $P_2$. We can also compute the integrated value, leading to 
\begin{equation}
    \int d^4x\ P_k=0\ .
\end{equation}
Moreover we can notice that $\delta_{ij}T^{ij}(x)=P_1(x)+P_2(x)$, from which considering Eq.~\eqref{eq:MassNormalization} we can make the following gauge invariant statement
\begin{equation}
    \int d^4x\, \eta_{\mu\nu}T^{\mu\nu}(x)=m\ .
\end{equation}

Let us summarize some results. Equation~\eqref{eq:MPT00} shows that the particle-like mass-energy density that sources a Myers-Perry BH in $d=4$ exactly resembles the curvature singularity of the full nonperturbative solution. Indeed, the 3-ellipsoid in Eq.~\eqref{eq:MPT00} whose explicit expression can be read from Eq.~\eqref{Eq:I1},  is exactly the hyper-surface in which the Myers-Perry solution is singular~\cite{Myers:1986un}. The same singularity structure is then present even in the current and stress parts of the EMT. This is quite an interesting result since it shows that BH singularities are not strictly a nonperturbative gravitational effect, but they arise already at linearized level. Moreover they are intrinsically connected to the gravitational multipole expansion, establishing a surprising relationship between IR and UV BH physics. 

\subsection{Singularity structure in arbitrary dimensions}

Looking at Eq.~\eqref{eq:EMTmsGenericD} then, the generalization to an arbitrary number of dimensions should be straightforward. However, in the chosen setup of residual factors, higher-order Bessel functions make the possibility of deriving some analytical expression in terms of distribution less obvious. Restricting ourselves to the structure of the singularity though, we can still generalize it to an arbitrary number of dimensions and state that from Eq.~\eqref{Eq:I1} one gets
\begin{equation}\label{eq:SingularityD}
\begin{aligned}
T^{00}\Big|_{d=\text{even}}&\propto \frac{m}{(2\pi)^\frac{d}{2}}\frac{1}{\prod_k a_k^2}\delta\Big(\tfrac{\rho_k^2}{a_k^2}-(\tfrac{d-1}{2})^2\Big)\prod_k\Theta(\tfrac{d-1}{2}a_k-\rho_k)+\cdots\ ,\\
T^{00}\Big|_{d=\text{odd}}&\propto \frac{m}{(2\pi)^\frac{d-1}{2}}\frac{1}{\prod_k a_k^2}\delta(z)\delta\Big(\tfrac{\rho_k^2}{a_k^2}-(\tfrac{d-1}{2})^2\Big)\prod_k\Theta(\tfrac{d-1}{2}a_k-\rho_k)+\cdots\ ,
\end{aligned}
\end{equation}
where the ellipses stand for other contributions to the EMT, similar to Eq.~\eqref{Eq:I0} for instance. We conclude by noting that Eq.~\eqref{eq:SingularityD} coincides exactly with the curvature singularity structure of Myers-Perry BHs in arbitrary spacetime dimensions~\cite{Myers:1986un}, proving that the tight relationship between singularity and gravitational multipoles is a general feature independent of the number of dimensions. 

\section{Multipole-based framework for black hole mimikers}\label{sec:MBFforBHM}

The reasons why black holes stand out among other compact objects are due to several distinctive features, including their thermodynamic behavior~\cite{Bardeen:1973gs,Hawking:1974rv}, the no-hair theorems~\cite{Israel:1967wq,Carter:1971zc}, and the presence of event horizons and curvature singularities. Classical solutions such as Schwarzschild and Kerr possess horizons where light undergoes infinite redshift for distant observers, and central singularities where curvature diverges. These features hint that our present understanding of gravity may be incomplete in extreme curvature regimes, suggesting that a quantum theory of gravity could predict objects that mimic black holes at macroscopic scales but differ at short distances~\cite{Giddings:1992hh,Lunin:2001jy,Mathur:2005zp,Hayward:2005gi,Bena:2005va,Bena:2006kb,Mathur:2008nj,Frolov:2016pav,Cano:2018aod,Carballo-Rubio:2018pmi,Simpson:2018tsi,Simpson:2019cer,Bianchi:2020bxa,Bianchi:2020miz} (see also~\cite{Buoninfante:2024oxl,Carballo-Rubio:2025fnc} for recent reviews). This line of thought has motivated the study of BH mimickers, namely horizonless compact objects that reproduce the macroscopic properties of black holes while differing in their internal structure~\cite{Cardoso:2017cqb,Mark:2017dnq,Carballo-Rubio:2018jzw,Mazza:2021rgq,Cardoso:2022fbq,Casadio:2024lgw}. Beyond their theoretical appeal, such objects are of growing observational interest, as they may leave subtle but measurable imprints in gravitational-wave signals or black hole images~\cite{Cardoso:2019rvt,Abedi:2016hgu,Jiang:2021ajk,Shaikh:2022ivr,Bambi:2025wjx}.

A natural strategy for constructing black-hole mimickers is to demand that they share the same multipolar structure as classical black holes~\cite{Ryan:1995wh,Pappas:2012ns}. While the Schwarzschild geometry is fully determined by its mass monopole, the Kerr spacetime possesses an infinite tower of mass and current multipoles uniquely fixed by the mass and angular momentum~\cite{Geroch:1970cc,Geroch:1970cd,Hansen:1974zz,Thorne:1980ru,Gursel1983}. However, as in Newtonian gravity, a given set of multipole moments does not uniquely determine the matter source. Therefore, observing a gravitational field with the Kerr multipolar structure does not necessarily imply that the central object is a Kerr black hole. Moreover, it remains an open question whether stable, physically consistent matter configurations can reproduce the Kerr geometry asymptotically~\cite{Friedman:1978ygc,Cardoso:2007az,Moschidis:2016zjy}. While spherical objects generically approach the Schwarzschild solution at large distances, stationary and axisymmetric configurations admit greater freedom. The uniqueness theorems ensure that, in four dimensions, the Kerr family exhausts all stationary, asymptotically flat vacuum black holes~\cite{Israel:1967wq,Hawking:1971vc,Carter:1971zc,Hawking:1973uf,Robinson:1975bv}, yet they can be evaded by relaxing either the vacuum or horizon assumptions. Identifying horizonless sources that precisely reproduce the Kerr asymptotics is therefore a nontrivial problem, whose study can illuminate both the origin of these theorems and the link between Kerr uniqueness and multipole structure. In higher dimensions, where no strict uniqueness theorem exists~\cite{Emparan:2025wsh}, the problem becomes even richer. Our aim then is to explore whether a rotating matter distribution can (i) generate a gravitational field that reproduces the full Kerr multipolar structure at large distances, and (ii) satisfy energy and causality conditions for physical viability.

To address these questions, we build upon the momentum-space formalism developed so far. We showed that the Fourier transform of the linearized EMT can be expanded in terms of spin-dependent form factors that encode the gravitational multipoles of the induced spacetime. However, fixing them only determines the asymptotic multipolar structure, and does not uniquely determine the internal profile of the source. Consequently, infinitely many distinct EMTs can correspond to the same multipole spectrum. These differences are captured by \emph{structure functions}, analytic functions of the transferred momentum that describe the internal composition of the source. Setting all structure functions to unity corresponds to a point-like singular object, and this prescription reproduces a ring-like source generating the linearized Kerr metric. In the following, we extend that construction by introducing non-trivial structure functions that smear the singularity, yielding a regular EMT whose gravitational field still matches the Kerr asymptotics.

As a concrete example, we consider Gaussian structure functions, leading to an anisotropic, rotating fluid that sources a Kerr-like spacetime while remaining regular everywhere. The source is characterized by its mass, angular momentum, and a hierarchy of length scales $R_n$, whose values must be constrained to ensure physical consistency. We identify a region of parameter space where the EMT remains real-valued and satisfies the energy and causality conditions at linearized order. Within this framework, we analyze the EMT and the corresponding linearized metric in harmonic gauge, explicitly showing that the curvature singularity is replaced by a smooth distribution of finite extent, while the asymptotic gravitational field preserves the full Kerr multipolar structure. Although our construction is exact in the angular momentum expansion, it remains perturbative in the gravitational coupling ${G_N}$. Extending this formalism beyond the linear regime, so that the resulting geometry satisfies Einstein’s equations non-perturbatively, remains a central open challenge, and achieving this would be essential to determine whether regular, horizonless mimickers can exist as exact solutions of GR supported by physically realistic matter.

\subsection{General construction}\label{sec:GeneralConstruction}

Our goal now is to relate local properties of the EMT in momentum space to the long-range behavior of the linearized metric. To this end, consider a stationary source described by the EMT in momentum space in Eq.~\eqref{eq:ConservedEMTresummed}. As already discussed, the form factors are in one-to-one correspondence with the gravitational multipoles of the induced spacetime, however, fixing them does not uniquely determine the source. In the linearized regime, this ambiguity arises in two ways: either by modifying the tensorial structure of the EMT through the addition of terms that do not contribute asymptotically, or by altering the local structure of the source through momentum-dependent functions. Focusing on the latter, one observes that constant form factors describe ``point-like'' sources, and it is possible to promote them to analytic functions of ${q^2}$, thereby giving the source a non-trivial structure without modifying its multipolar content. To construct such an equivalence class explicitly, here and in the following discussion we restrict to ${d = 3}$ spatial dimensions, although the method generalizes to higher-dimensional spacetimes. Let us consider Cartesian coordinates ${\vec{x} = (x, y, z)}$ and move to a frame in which the spin is aligned along the ${z}$-axis. Considering the set of form factors $F_{n}(aq_\perp)$, for $n=1, 2, 3$,  we can now formally add a local structure that corresponds to promoting these functions to analytic functions of ${q^2}$,
\begin{equation}\label{eq:promotion}
F_n(a q_\perp) \rightarrow F_n(a q_\perp)\, K_n(q^2)\ ,
\end{equation}
where ${K_n(q^2)}$ are the structure functions, such as
\begin{equation}
    K_n(q^2)=1+\sum_{i=1}^{+\infty}a_{i}^{(n)}\, q^{2i}\ .
\end{equation}
Equation~\eqref{eq:promotion} is only a specific example of such promotion, indeed more involved constructions are certainly possible and may lead to richer physical scenarios. Then, introducing the spin vector ${s^i = (0, 0, a)}$, we can rewrite the EMT as
\begin{equation}\label{eq:EMTinMomentum}
\begin{aligned}
T^{00}(q) &= m\, F_1(a q_\perp)\, K_1(q^2)\ , \\
T^{ij}(q) &= m\, (s \times q)^i (s \times q)^j\, F_2(a q_\perp)\, K_2(q^2)\ , \\
T^{0i}(q) &= -\frac{i}{2} m\, (s \times q)^i\, F_3(a q_\perp)\, K_3(q^2)\ .
\end{aligned}
\end{equation}

Our goal now is to perform the Fourier transform of Eq.~\eqref{eq:EMTinMomentum} and obtain a compact expression for the EMT in coordinate space, valid for arbitrary form factors and structure functions. Let us first focus on the ${T^{00}}$ component. Using
\begin{equation}
K_1(q^2) = \int d^3 x'\, e^{i q \cdot x'} K_1(r'^2)\ ,
\end{equation}
and substituting it into the definition of ${T^{00}}$, we find
\begin{equation}
T^{00}(\rho, z) = \int \frac{d^3 q}{(2\pi)^3} e^{-i q \cdot x} F_1(a q_\perp) \int d^3 x'\, e^{i q \cdot x'} K_1(\rho'^2 + z'^2)\ .
\end{equation}
Since the integrand does not depend on ${q_z}$, the integration over ${q_z}$ yields a delta function ${\delta(z - z')}$, and the expression becomes
\begin{equation}
T^{00}(\rho, z) = \int_0^{+\infty} dq_\perp\, q_\perp \int_0^{+\infty} d\rho'\, \rho'\, J_0(q_\perp \rho)\, J_0(q_\perp \rho')\, F_1(a q_\perp)\, K_1(\rho'^2 + z^2)\ ,
\end{equation}
where we switched to cylindrical coordinates ${\vec{x} = (\rho, \phi, z)}$.
Expanding the form factor function order by order in angular momentum, we then obtain
\begin{equation}
T^{00}(\rho, z) = \sum_{\ell=0}^{+\infty} F_{2\ell, 1} a^{2\ell} \int_0^{+\infty} dq_\perp\, q_\perp^{2\ell+1} \int_0^{+\infty} d\rho'\, \rho'\, J_0(q_\perp \rho)\, J_0(q_\perp \rho')\, K_1(\rho'^2 + z^2)\ .
\end{equation}
Using the identity
\begin{equation}
\frac{1}{\rho} \partial_\rho \Big( \rho\, \partial_\rho J_0(q_\perp \rho) \Big) = \nabla_\rho^2 J_0(q_\perp \rho) = -q_\perp^2 J_0(q_\perp \rho)\ ,
\end{equation}
where ${\nabla_\rho^2}$ is the radial part of the Laplacian in cylindrical coordinates,
\begin{equation}
\nabla^2 = \nabla_\rho^2 + \frac{1}{\rho^2} \frac{\partial^2}{\partial \phi^2} + \frac{\partial^2}{\partial z^2}\ ,
\end{equation}
we can rewrite the expression as
    \begin{equation}
T^{00}(\rho, z) = \sum_{\ell=0}^{+\infty} (-1)^\ell F_{2\ell, 1} a^{2\ell} (\nabla_\rho^2)^\ell \int_0^{+\infty} dq_\perp\, q_\perp \int_0^{+\infty} d\rho'\, \rho'\, J_0(q_\perp \rho)\, J_0(q_\perp \rho')\, K_1(\rho'^2 + z^2)\ .
\end{equation}
Then, using the orthogonality of Bessel functions one finally obtains
\begin{equation}
T^{00}(\rho, z) = \sum_{\ell=0}^{+\infty} (-1)^\ell F_{2\ell, 1} a^{2\ell} (\nabla_\rho^2)^\ell K_1(\rho^2 + z^2)\ .
\end{equation}
Repeating the same argument for the other components, we obtain the full EMT in coordinate space, valid at every order in the spin expansion
\begin{equation}\label{eq:EMTwithRotationFromNR}
\begin{aligned}
T^{00}(\rho, z) &= m \sum_{\ell=0}^{+\infty} (-1)^\ell F_{2\ell, 1} a^{2\ell} (\nabla_\rho^2)^\ell K_1(\rho^2 + z^2)\ , \\
T^{ij}(\rho, z) &= -m\, (s \times \partial)^i (s \times \partial)^j \sum_{\ell=0}^{+\infty} (-1)^\ell F_{2\ell+2, 2} a^{2\ell} (\nabla_\rho^2)^\ell K_2(\rho^2 + z^2)\ , \\
T^{0i}(\rho, z) &= \frac{1}{2} m\, (s \times \partial)^i \sum_{\ell=0}^{+\infty} (-1)^\ell F_{2\ell+1, 3} a^{2\ell} (\nabla_\rho^2)^\ell K_3(\rho^2 + z^2)\ .
\end{aligned}
\end{equation}
To interpret the real-space structure functions, consider the non-rotating case in which ${a = 0}$, where the EMT reduces to ${T_{\mu\nu}(r) = u_\mu u_\nu \epsilon(r)}$, with ${\epsilon(r)=m K_1(r^2)}$ corresponding to the energy density. More generally, the functions ${K_n(r^2)}$ encode physical properties of the energy distribution, such as rotational velocity and pressure. This means that, when restricting to spin-induced multipoles, starting from a spherically symmetric source, Eq.~\eqref{eq:EMTwithRotationFromNR} provides the recipe to make it rotate exactly in the way required to reproduce a specific multipolar structure determined by the form factors. Notice that, up to this point, the construction has been entirely general and does not assume any particular form for the multipole coefficients or the structure functions. 

We now focus on a specific case of study, namely a source with a Gaussian energy-density profile. Consider a Gaussian structure function of the form
\begin{equation}
    K_n(q^2) = e^{-q^2 R_n^2}\ ,
\end{equation}
which leads to the energy density
\begin{equation}\label{eq:EnergyDensityDef}
    \epsilon(r) = m \int \frac{d^3 q}{(2\pi)^3} e^{-i q \cdot x} K_1(q^2) = \frac{m}{8\pi^{3/2} R_1^3} e^{-\frac{r^2}{4 R_1^2}}\ ,
\end{equation}
where the ${R_n}$'s are new characteristic length scales of the system. The Gaussian energy profile not only smooths out potential singularities, as any well-defined analytic structure function does, but also coincides with the type of matter distribution used in noncommutative-inspired models~\cite{Nicolini:2005vd}, where it emerges as an effective way to regularize curvature divergences in specific quantum gravity approaches. Moreover, this choice proves particularly convenient for analytic calculations, as it diagonalizes the power series of the Laplacian operator in Eq.~\eqref{eq:EMTwithRotationFromNR}. Indeed, one can show that
\begin{equation}\label{eq:EigenValueRelation}
    (\nabla_\rho^2)^\ell \epsilon(r) = \left( \frac{m\, e^{-\frac{z^2}{4 R_1^2}}}{8\pi^{3/2} R_1^3} \right) \frac{(-1)^\ell \ell!}{R_1^{2\ell}}\, {}_1F_1\left(\ell+1, 1, -\frac{\rho^2}{4 R_1^2} \right)\ ,
\end{equation}
where ${ {}_1F_1(a, b, z) }$ is the confluent hypergeometric function (Kummer function). To prove Eq.~\eqref{eq:EigenValueRelation}, note that the Gaussian can be written in terms of a Kummer function as in
\begin{equation}
    \epsilon(r) = \left( \frac{m}{8\pi^{3/2} R_1^3} e^{-\frac{z^2}{4 R_1^2}} \right)\, {}_1F_1\left(1, 1, -\frac{\rho^2}{4 R_1^2} \right)\ .
\end{equation}
Now consider the action of the radial Laplacian on a function ${f\Big(-\frac{\rho^2}{4 R_1^2}\Big)}$, such as
\begin{equation}
    \nabla_\rho^2 f(\chi) = -\frac{\chi f'' + f'}{R_1^2}\ ,
\end{equation}
where ${\chi = -\frac{\rho^2}{4 R_1^2}}$. If ${f(\chi) = {}_1F_1(a, b, \chi)}$, then it satisfies the differential equation
\begin{equation}
    \chi f'' + (b - \chi) f' = a f\ .
\end{equation}
For the case ${a = b = 1}$, this simplifies to
\begin{equation}
    -\frac{\chi f'' + f'}{R_1^2} = -\frac{\chi f' + f}{R_1^2}\ ,
\end{equation}
which, combined with the identity
\begin{equation}
    \chi \frac{\partial}{\partial \chi} \Big({}_1F_1(a, b, \chi)\Big) + a\Big({}_1F_1(a, b, \chi)\Big) = a\Big( {}_1F_1(a + 1, b, \chi)\Big)\ ,
\end{equation}
yields
\begin{equation}
    \nabla_\rho^2 \epsilon(r) = -\left( \frac{m}{8\pi^{3/2} R_1^3} e^{-\frac{z^2}{4 R_1^2}} \right)\, \frac{1}{R_1^2} {}_1F_1\left(2, 1, -\frac{\rho^2}{4 R_1^2} \right)\ .
\end{equation}
Then, higher-order derivatives can be obtained recursively, thus proving Eq.~\eqref{eq:EigenValueRelation} by induction. Substituting this result into Eq.~\eqref{eq:EMTwithRotationFromNR}, the EMT sourced by a Gaussian structure function and arbitrary form factors takes the form
\begin{equation}\label{eq:GeneralGaussianEMT}
\begin{aligned}
    T^{00}(\rho, z) &= \left( \frac{m}{8\pi^{3/2} R_1^3} e^{-\frac{z^2}{4 R_1^2}} \right) \sum_{\ell=0}^{+\infty} \ell!\, F_{2\ell, 1} \left( \frac{a^2}{R_1^2} \right)^\ell {}_1F_1\left(\ell+1, 1, -\frac{\rho^2}{4 R_1^2} \right)\ , \\
    T^{ij}(\rho, z) &= -\left( \frac{m}{8\pi^{3/2} R_2^3} e^{-\frac{z^2}{4 R_2^2}} \right) (s \times \partial)^i (s \times \partial)^j \sum_{\ell=0}^{+\infty} \ell!\, F_{2\ell+2, 2} \left( \frac{a^2}{R_2^2} \right)^\ell {}_1F_1\left(\ell+1, 1, -\frac{\rho^2}{4 R_2^2} \right)\ , \\
    T^{0i}(\rho, z) &= \frac{1}{2} \left( \frac{m}{8\pi^{3/2} R_3^3} e^{-\frac{z^2}{4 R_3^2}} \right) (s \times \partial)^i \sum_{\ell=0}^{+\infty} \ell!\, F_{2\ell+1, 3} \left( \frac{a^2}{R_3^2} \right)^\ell {}_1F_1\left(\ell+1, 1, -\frac{\rho^2}{4 R_3^2} \right)\ .
\end{aligned}
\end{equation}
The above expression provides a closed-form representation of a linearized EMT, valid to all orders in the angular momentum expansion, for a rotating source with a Gaussian-like profile. By construction, it induces a gravitational field with multipole moments matching exactly those given in Eq.~\eqref{eq:d3Multipoles}. While the construction presented here holds for arbitrary multipole moments encoded in the form factors ${F_{\ell,n}}$, we now specialize to the case in which these are chosen to reproduce exactly the multipolar structure of a Kerr BHs. This will allow us to study in detail the properties of a specific mimicker sourced by a rotating anisotropic fluid with a Gaussian profile.

\subsection{Kerr case}\label{sec:KerrCase}

As already pointed out, our goal is to apply the general construction of Eq.~\eqref{eq:EMTwithRotationFromNR} to build a physically reasonable linearized source that mimics the Kerr geometry. Starting from the Gaussian-like source introduced in the previous section, and given that we are free to impose any asymptotic structure, constructing a Kerr mimicker simply requires us to impose the Kerr multipolar structure on Eq.~\eqref{eq:GeneralGaussianEMT}. Recalling Eq.~\eqref{eq:d3FormFactor}, the resummed expression of the form factors corresponding to the Kerr multipoles reads
\begin{gather}
    F_{1}(a q_\perp) + (a q_\perp)^2 F_{2}(a q_\perp) = \cos(a q_\perp)\ , \label{eq:KerrFF}\\
    F_{3}(a q_\perp) = \frac{\sin(a q_\perp)}{a q_\perp}\ ,\label{eq:KerrFFcurrent}
\end{gather}
noticing once again that that in ${d = 3}$ the mass and stress form factors are redundant. Therefore, respecting the constraint of Eq.~\eqref{eq:KerrFF}, we are free to choose them arbitrarily without altering the gravitational multipole content, effectively identifying an equivalence class of EMTs all inducing the same asymptotic geometry. Due to this redundancy, we can fix $F_2$ conveniently to simplify the EMT in Eq.~\eqref{eq:GeneralGaussianEMT}. A particularly interesting case is obtained by setting ${F_2(a q_\perp) = 0}$, which corresponds to ${T^{ij}(x) = 0}$. We refer to the resulting configuration as the Gaussian-smeared Israel source, since in the limit ${R_n \to 0}$, the EMT reduces to the singular distribution identified by Israel in Eqs.~\eqref{eq:T00Kerr} and~\eqref{eq:T0iKerr}. We thus expect our Gaussian-smeared generalization, controlled by the scales ${R_n}$, to regularize the ring singularity and recover the Israel/Kerr limit as ${R_n \to 0}$. To this end, we define two master functions
\begin{equation}\label{eq:MasterIsreaelIntegrals}
\begin{aligned}
    \mathcal{M}(\rho, z; R) &= \left( \frac{m}{8\pi^{3/2} R^3} e^{-\frac{z^2}{4 R^2}} \right) \sum_{\ell=0}^{+\infty} (-1)^\ell \frac{\ell!}{(2\ell)!} \left( \frac{a^2}{R^2} \right)^\ell {}_1F_1\left(\ell+1, 1, -\frac{\rho^2}{4 R^2} \right)\ , \\
    \mathcal{J}(\rho, z; R) &= \left( \frac{m}{8\pi^{3/2} R^3} e^{-\frac{z^2}{4 R^2}} \right) \sum_{\ell=0}^{+\infty} (-1)^\ell \frac{\ell!}{(2\ell+1)!} \left( \frac{a^2}{R^2} \right)^\ell {}_1F_1\left(\ell+1, 1, -\frac{\rho^2}{4 R^2} \right)\ ,
\end{aligned}
\end{equation}
so that the EMT compactly reads
\begin{equation}\label{eq:EMTphenoImplicit}
\begin{aligned}
    T^{00}(\rho, z) &= \mathcal{M}(\rho, z; R_1)\ , \\
    T^{0i}(\rho, z) &= \frac{1}{2} (s \times \partial)^i\, \mathcal{J}(\rho, z; R_3)\ , \\
    T^{ij}(\rho, z) &= 0\ .
\end{aligned}
\end{equation}

Although we will focus primarily on the Gaussian-smeared Israel source, it is worth briefly mentioning another natural and well-motivated choice for the stress form factors. In higher dimensions, where the stress multipoles do contribute to the gravitational field, the Myers–Perry BH provides a generalization to arbitrary dimensions of the Kerr solution. In this case, a fully consistent set of form factors reproducing the multipolar structure of Myers–Perry can be written as in Eq.~\eqref{eq:dGenericSphericalB}. In particular, when applied to ${d = 3}$, these form factors yield
\begin{equation}
\begin{aligned}
    F_1(a q_\perp) &= \frac{1}{2} \left( \cos(a q_\perp) + \frac{\sin(a q_\perp)}{a q_\perp} \right)\ , \\
    F_2(a q_\perp) &= \frac{1}{2} \left( \cos(a q_\perp) - \frac{\sin(a q_\perp)}{a q_\perp} \right)\ , \\
    F_3(a q_\perp) &= \frac{\sin(a q_\perp)}{a q_\perp}\ .
\end{aligned}
\end{equation}
This configuration provides then an alternative EMT, effectively representing the ${d=3}$ limit of a Myers–Perry mimicker source in higher dimensions, reproducing once again the asymptotic structure of Kerr BHs. In summary, by fixing the form factors to match the Kerr multipolar structure and choosing a Gaussian structure function, we have constructed a smooth, analytic EMT that mimics the linearized Kerr spacetime. In the following, we analyze the physical properties of this source, such as energy conditions, rotation velocity, and causality, across different parameter configurations.

\subsection{Source phenomenology}\label{sec:EMT_KerrPheno}

Now we explicitly express Eq.~\eqref{eq:EMTphenoImplicit} in cylindrical coordinates and match the resulting EMT to that of an anisotropic rotating fluid, laying the groundwork for a possible non-perturbative generalization. We then explore different parameter configurations, showing that, at linear order, the Gaussian-smeared Israel source can satisfy both energy and causality conditions. Let us work in cylindrical coordinates with
\begin{equation}
    x = \rho \cos\phi\ , \qquad y = \rho \sin\phi\ ,
\end{equation}
where the flat background metric reads ${\eta_{\mu\nu} = \text{diag}(-1, 1, \rho^2, 1)}$. Since the spin is aligned along the ${z}$-axis, the tensorial structure of the EMT simplifies to
\begin{equation}\label{eq:EMTansatzPL}
T^{\mu\nu} =
\begin{pNiceMatrix}[columns-width=10pt]
    \mathcal{M}(R_1) & 0 & \frac{a}{2\rho}\partial_\rho \mathcal{J}(R_3) & 0 \\
    0 & 0 & 0 & 0 \\
    \frac{a}{2\rho}\partial_\rho \mathcal{J}(R_3) & 0 & 0 & 0 \\
    0 & 0 & 0 & 0
\end{pNiceMatrix}\ .
\end{equation}
By construction, the source shares the same symmetries as the Kerr geometry, namely axial and equatorial symmetry, so the EMT is independent of ${\phi}$. Moreover, from here on, we will suppress the ${\rho}$ and ${z}$ dependence in favor of a simplified notation, keeping only the explicit dependence on ${R_n}$. We further interpret the source as a covariant rotating anisotropic fluid, modeled by the ansatz
\begin{equation}\label{eq:EMTAnsatz}
    T^{\mu\nu} = \epsilon\, u^\mu u^\nu + p_\rho\, l_\rho^{\mu} l_\rho^{\nu} + p_\phi\, l_\phi^\mu l_\phi^\nu\ ,
\end{equation}
where ${u^\mu = \gamma(1, 0, \Omega, 0)}$ is the fluid four-velocity, normalized as ${u^\mu u_\mu = -1}$, with ${\gamma = (1 - \rho^2 \Omega^2)^{-1/2}}$ the Lorentz factor. Here, ${\Omega = \Omega(\rho, z)}$ is the angular velocity, and ${l_\rho^\mu = (0, 1, 0, 0)}$, ${l_\phi^\mu = \gamma(\rho \Omega, 0, 1/\rho, 0)}$ are unit space-like vectors orthogonal to ${u^\mu}$ and to each other, aligned with the ${\rho}$ and ${\phi}$ directions respectively. The eigenvalues of $T^\mu{}_\nu$ are then identified as the energy density ${\epsilon}$ and the anisotropic pressures ${p_\rho}$ and ${p_\phi}$. We now specify what are the positive energy and causality conditions we are going to consider. The weak energy condition is defined by requiring 
\begin{equation}\label{eq:GenericEnergCond}
T^{\mu\nu} U_\mu U_\nu \geq 0    
\end{equation}
for any timelike vector ${U^\mu = \alpha_1 u^\mu + \alpha_2 l_\rho^\mu + \alpha_3 l_\phi^\mu}$ satisfying ${U^\mu U_\mu = -1}$, which implies ${\alpha_1^2 = 1 + \alpha_2^2 + \alpha_3^2}$. Substituting then Eq.~\eqref{eq:EMTAnsatz} into Eq.~\eqref{eq:GenericEnergCond}, we obtain the conditions
\begin{equation}\label{eq:EnergyCondition}
    \epsilon \geq 0\ , \qquad \xi_\rho = \epsilon + p_\rho \geq 0\ , \qquad \xi_\phi = \epsilon + p_\phi \geq 0\ .
\end{equation}
Causality on the other hand, requires all characteristic speeds to be subluminal. Since the tangential rotational speed is ${v = \rho \Omega}$, and causality imposes ${|v| < 1}$. Additionally, the sound speeds, defined as ${c_k^2 = \partial p_k / \partial \epsilon}$ for ${k = \rho, \phi}$, must also satisfy ${c_k^2 < 1}$. In certain configurations however, we will encounter regions where ${c_k^2 < 0}$, signaling a linear instability of the fluid~\cite{Rezzolla:2013dea, Romatschke:2017ejr}. We will not treat this as a definitive rule-out condition, since stability might be restored in the full non-perturbative regime. Finally, matching Eq.~\eqref{eq:EMTansatzPL} with the fluid ansatz in Eq.~\eqref{eq:EMTAnsatz}, we obtain the following relations
\begin{equation}\label{eq:FuncDef}
\begin{gathered}
    \epsilon = \frac{\mathcal{M}(R_1) + \sqrt{ \mathcal{M}(R_1)^2 - \left( a \partial_\rho \mathcal{J}(R_3) \right)^2 }}{2}\ , \\
    \Omega = \frac{\mathcal{M}(R_1) - \sqrt{ \mathcal{M}(R_1)^2 - \left( a \partial_\rho \mathcal{J}(R_3) \right)^2 }}{a \rho\, \partial_\rho \mathcal{J}(R_3)}\ , \\
    p_\phi = \frac{ -\mathcal{M}(R_1) + \sqrt{ \mathcal{M}(R_1)^2 - \left( a \partial_\rho \mathcal{J}(R_3) \right)^2 }}{2}\ , \\
    p_\rho = 0\ ,
\end{gathered}
\end{equation}
and for these quantities to be real, the following condition must hold
\begin{equation}\label{eq:RealityCond}
    \mathcal{M}(R_1)^2 \geq \left( a \partial_\rho \mathcal{J}(R_3) \right)^2\ .
\end{equation}
To satisfy Eq.~\eqref{eq:RealityCond}, we parametrize ${R_1 = R}$ and ${R_3 = \alpha R}$, and we find that the inequality is satisfied for all ${\rho}$ when ${\alpha < 1}$ and ${R > R^*_\alpha}$ for some threshold ${R^*_\alpha}$. 

\begin{figure}[h]
\centering
\includegraphics[width=0.48\textwidth, valign=c]{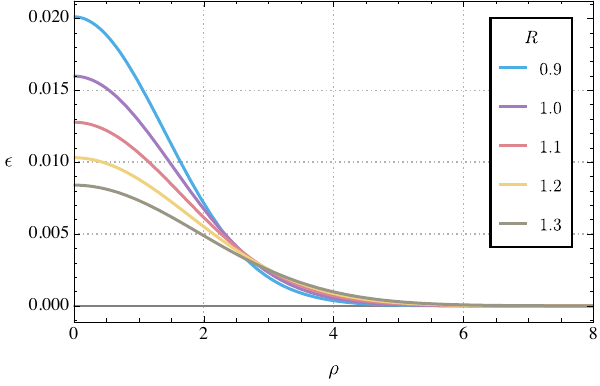}
\includegraphics[width=0.48\textwidth, valign=c]{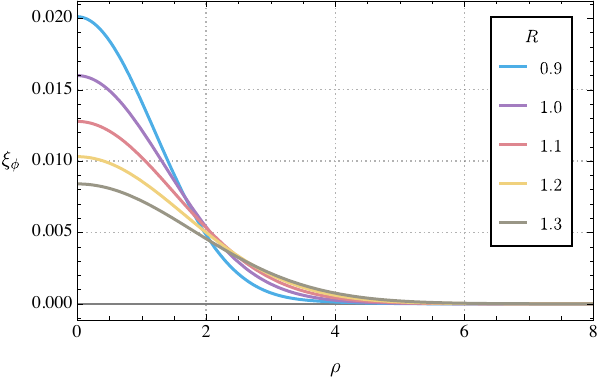}
\caption{On the left the energy density and on the right the $\xi_\phi$ variable for the study of the energy condition of the Gaussian-smeared Israel source for ${a=0.8}$, ${z=0}$ and ${\alpha=0.99}$ in units of ${G_N=m=1}$.}
\label{fig:EnergyCondA099}
\end{figure}

\begin{figure}[h]
\centering
\includegraphics[width=0.48\textwidth, valign=c]{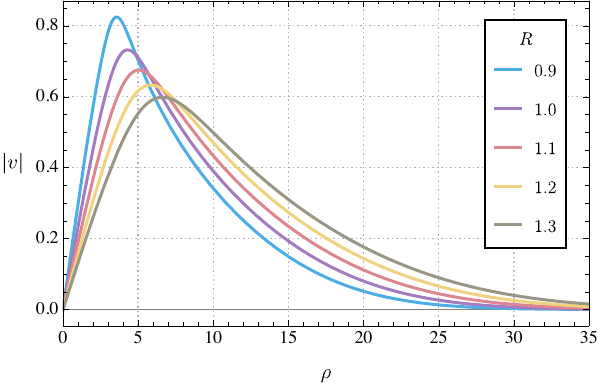}
\includegraphics[width=0.48\textwidth, valign=c]{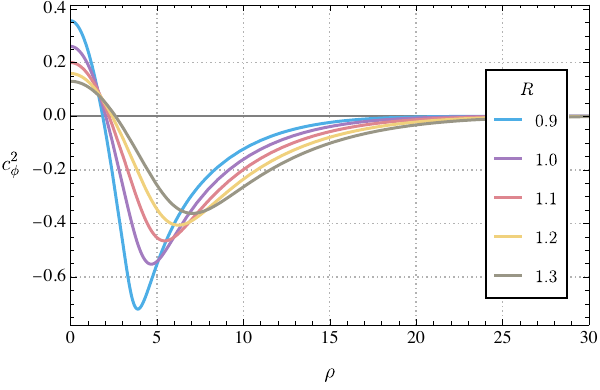}
\caption{On the left the tangential rotational speed and on the right the sound speed in the $\phi$-direction of the Gaussian-smeared Israel source for ${a=0.8}$, ${z=0}$ and ${\alpha=0.99}$ in units of ${G_N=m=1}$.}
\label{fig:CausalityCondA099}
\end{figure}

\begin{figure}[h]
\centering
\includegraphics[width=0.48\textwidth, valign=c]{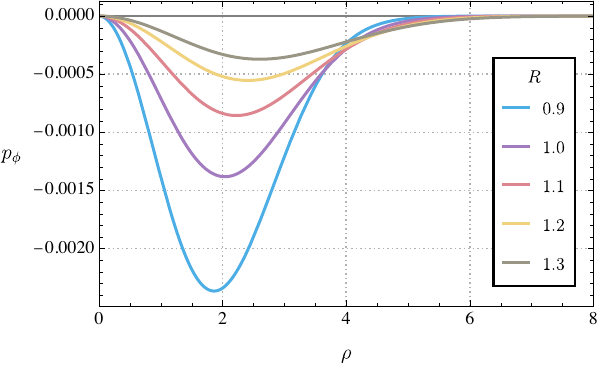}
\caption{Pressure in the $\phi$-direction of the Gaussian-smeared Israel source with ${a = 0.8}$, ${z = 0}$ and ${\alpha = 0.99}$ in units of ${G_N=m = 1}$.}
\label{fig:PressureA099}
\end{figure}

The first case of study is for ${\alpha=0.99}$, just below the unity threshold value. Fixing the angular momentum density to a reference value of ${a=0.8}$, the weak energy condition, and hence the energy-density distribution, are studied in Fig.~\ref{fig:EnergyCondA099} for different values of the $R$ parameter. For this configuration the threshold value for which Eq.~\eqref{eq:RealityCond} is satisfied is ${R_{\alpha=0.99}^*\approx 0.84}$, so then we will consider only values ${R>R_{\alpha=0.99}^*}$. Moreover notice that from here on we will study the source phenomenology keeping the angular momentum fixed for different values of $R$ and $\alpha$, since in Eq.~\eqref{eq:MasterIsreaelIntegrals} we can see that the angular momentum always enters in a ratio with the $R$ parameter. Likewise, the dependency on $z$ in Eq.~\eqref{eq:MasterIsreaelIntegrals} enters just as a Gaussian damping factor, hence, for simplicity, we will always consider ${z=0}$ in the study of the EMT. From Fig.~\ref{fig:EnergyCondA099} one can see that the energy condition is satisfied since both $\epsilon$ and $\xi_\phi$ are always positive. Moreover, the energy-density distribution preserves a Gaussian-like profile even in the presence of rotation and increases its central value with $R$ becoming smaller.
Causality is tested in Fig.~\ref{fig:CausalityCondA099}, where we plot the rotational tangential speed and the sound speed.
The rotational speed satisfies ${|v| < 1}$ throughout and decays to zero at large ${\rho}$. Moreover, as ${R \rightarrow R^*_{\alpha=0.99}}$, the maximum rotational speed approaches unity for some $\rho$ value. On the other hand, the sound speed develops an imaginary part. Indeed, its real part remains subluminal, while the imaginary part develops a peak around $\rho\approx 4R$ and decays at large distances. While this suggests an instability, we do not exclude such configurations, anticipating that non-linear effects could restore stability. It is crucial to emphasize that, within the presented model, such instability arises for every parameter choice, due to the intrinsic behavior of the energy density and tangential pressure. Specifically, since ${\epsilon}$ monotonically decreases while ${p_\phi}$ vanishes at the origin exhibiting a stationary point at ${r > 0}$, then inevitably exists a critical radius beyond which ${c_\phi^2 < 0}$ holds. Thus, a completely ``safe'' parameter region without such instabilities is absent within the current linearized description. Finally the pressure profile is shown in Fig.~\ref{fig:PressureA099}.
We can see that the pressure vanishes at the origin and reaches a peak around ${\rho \approx 2 R}$, similarly to the rotational velocity phenomenology as expected. Indeed, the behavior of these quantities suggest the source to have a shell-like shape (ring-like in the equatorial plane), typical of regularized Kerr-cores.

\begin{figure*}[h]
\centering
\includegraphics[width=0.48\textwidth, valign=c]{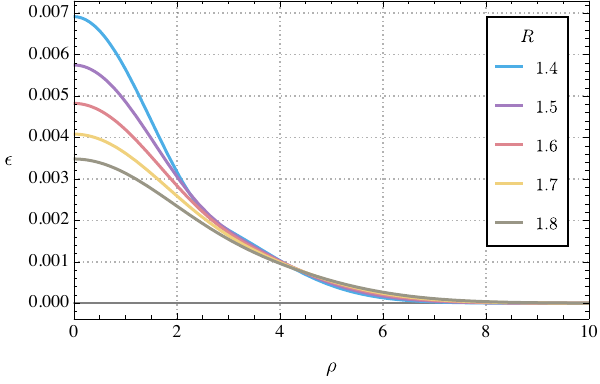}
\includegraphics[width=0.48\textwidth, valign=c]{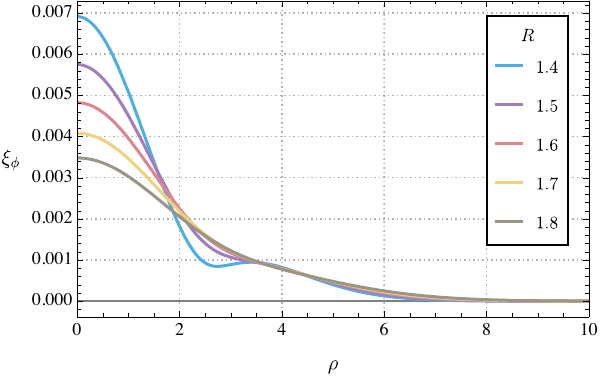}
\includegraphics[width=0.48\textwidth, valign=c]{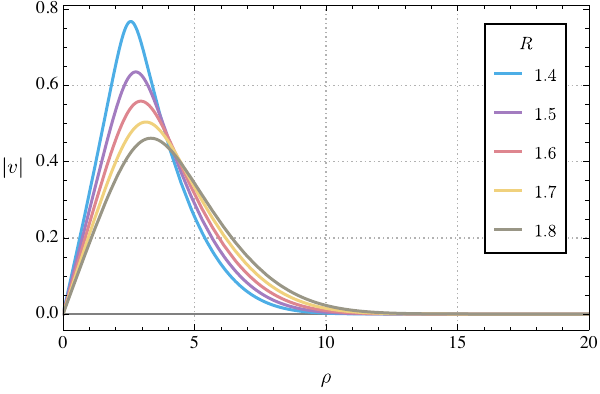}
\includegraphics[width=0.48\textwidth, valign=c]{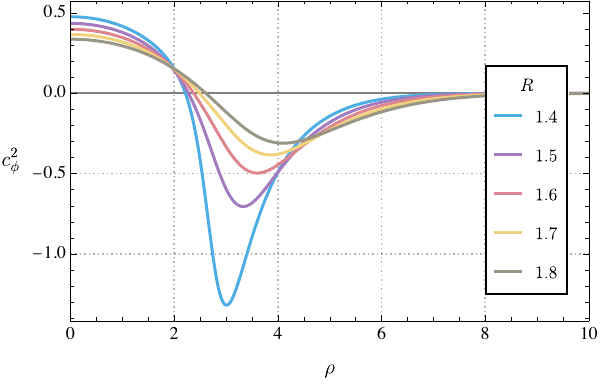}
\includegraphics[width=0.48\textwidth, valign=c]{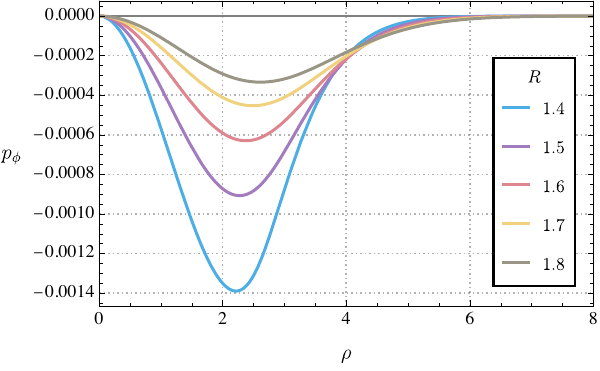}
\caption{On the top-left the energy density, on the top-right the $\xi_\phi$ variable for the study of the energy condition, on the mid-left the tangential rotational speed, on the mid-right the sound speed in the $\phi$-direction and on the bottom the pressure in the $\phi$-direction of the Gaussian-smeared Israel source for ${a=0.8}$, ${z=0}$ and ${\alpha=0.8}$ in units of ${G_N=m=1}$.}
\label{fig:PlotA08}
\end{figure*}

We now consider the case for ${\alpha = 0.8}$, with the corresponding results shown in Fig.~\ref{fig:PlotA08}. In this case, the threshold is ${R^*_{\alpha=0.8} \approx 1.35}$, and since we are only interested in physically viable configurations we consider only the parameter space for which ${R>R_{\alpha=0.8}^*}$.
As in the previous case, all energy and causality conditions are satisfied. Moreover, the larger threshold ${R^*_{\alpha=0.8} > R^*_{\alpha=0.99}}$ reflects the fact that when ${R_1 > R_3}$, their difference must be balanced by larger absolute values to maintain real-valued quantities. Ultimately, we can conclude that the phenomenology remains qualitatively similar, with differences only in peak amplitudes and spatial profiles. Condensing our results, the Gaussian-smeared Israel source defined in Eqs.~\eqref{eq:EMTAnsatz} and~\eqref{eq:FuncDef} provides, within a certain parameter range, a physically viable EMT that induces a gravitational field with the exact Kerr multipolar structure. Although we have focused on a particular example, the general procedure is broadly applicable, and other satisfactory mimickers could be constructed by varying the structure functions or the form factor choices. Ultimately, this framework lays the foundation for developing a full non-perturbative solution of Einstein's equations. Indeed, in the static limit (${a = 0}$), an exact solution is already known~\cite{Nicolini:2005vd}, demonstrating that the Gaussian-smeared Israel source admits a global completion at least in the non-rotating case.

\subsection{Metric phenomenology}\label{sec:MetricPheno}

Having established that the Gaussian-smeared Israel source defines a physically reasonable configuration at least at linear order, it is instructive to explore the phenomenology of the resulting metric, considering also that this approximation dominates the long-range expansion of the full spacetime. Consider the EMT in Eq.~\eqref{eq:EMTphenoImplicit} defined onto a flat background spacetime, where
\begin{equation}
    T = \eta^{\mu\nu} T_{\mu\nu} = -T^{00}\ .
\end{equation}
Working in cylindrical coordinates and imposing the harmonic gauge, the linearized metric takes the form
\begin{equation}\label{eq:MetricImplicitComponents}
\begin{aligned}
    h_{00} &= 8\pi G_N \int \frac{d^3 q}{(2\pi)^3} e^{-i q \cdot x} \frac{1}{q^2} T_{00}(q)\ , \\
    h_{0\phi} &= 8\pi G_N \int \frac{d^3 q}{(2\pi)^3} e^{-i q \cdot x} \frac{1}{q^2} T_{0\phi}(q)\ , \\
    h_{ij} &= \eta_{ij} h_{00}\ .
\end{aligned}
\end{equation}
As expected, the linearized metric has only two independent components, namely temporal and angular sectors. We now examine them separately.

\begin{figure*}[htbp!]
\centering
\includegraphics[width=0.48\textwidth, valign=c]{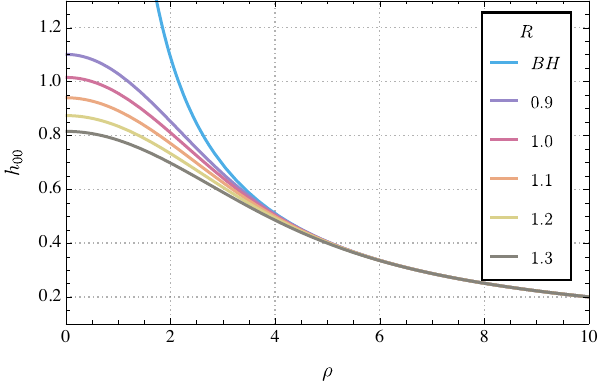}
\includegraphics[width=0.48\textwidth, valign=c]{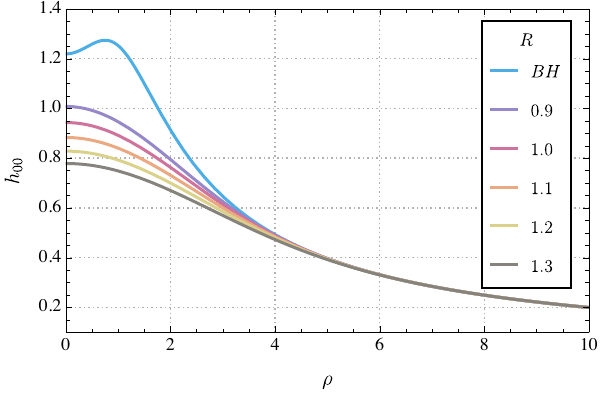}
\caption{Temporal component of the linearized metric sourced by the Gaussian-smeared Israel EMT for different values of ${R}$ and for ${z = 0}$ on the left and ${z = 1}$ on the right, with ${a = 0.8}$ and in units of ${G_N=m=1}$.}
\label{fig:Metric00}
\end{figure*}

Replacing Eq.~\eqref{eq:KerrFF} into Eq.~\eqref{eq:MetricImplicitComponents}, the temporal component becomes
\begin{equation}
    h_{00} = 8\pi G m \int \frac{d^3 q}{(2\pi)^3} e^{-i q \cdot x} e^{-q^2 R^2} \cos(a q_\perp)\ .
\end{equation}
Switching to cylindrical coordinates and integrating over the angular part, we obtain
\begin{equation}
    h_{00} = 8\pi G m \int_{-\infty}^{+\infty} \frac{d q_z}{2\pi} e^{-i q_z z} \int_0^{+\infty} \frac{d q_\perp\, q_\perp}{2\pi} J_0(q_\perp \rho) \frac{e^{-q_z^2 R^2}}{q_z^2 + q_\perp^2} e^{-q_\perp^2 R^2} \cos(a q_\perp)\ .
\end{equation}
The ${q_z}$ integral can be performed analytically, resulting in a single-integral expression for the temporal component as in
\begin{equation}\label{eq:TemporalMetricMimicker}
    h_{00} = G m \int d q_\perp\, J_0(q_\perp \rho)\, \cos(a q_\perp)\, \left[ e^{-q_\perp z} \mathrm{Erfc}\left(q_\perp R - \tfrac{z}{2R}\right) + e^{q_\perp z} \mathrm{Erfc}\left(q_\perp R + \tfrac{z}{2R} \right) \right]\ ,
\end{equation}
where ${\mathrm{Erfc}(x)}$ is the complementary error function
\begin{equation}
    \mathrm{Erfc}(x) = 1 - \frac{2}{\pi} \int_0^x dt\, e^{-t^2}\ .
\end{equation}
As far as we know, the integral in Eq.~\eqref{eq:TemporalMetricMimicker} does not admit a closed-form expression, and must be evaluated numerically. However, in the ${R \to 0}$ limit, corresponding to the Kerr singular case, the integral simplifies to
\begin{equation}\label{eq:KerrTempImp}
    h_{00}^{Kerr} = 2G m \int d q_\perp\, J_0(q_\perp \rho)\, \cos(a q_\perp)\, e^{-q_\perp |z|}\ ,
\end{equation}
which can be analytically integrated by means of tabulated integrals listed in~\cite{Gradshteyn:1943cpj}, leading to  
\begin{equation}\label{eq:MetricMasterInt}
   h_{00}^{Kerr}=2Gm\frac{\sqrt{l_+^2-a^2}}{l_+^2-l_-^2}\ ,
\end{equation}
where
\begin{equation}\label{eq:lsDef}
\begin{aligned}
    l_+&=\frac{\sqrt{(a+\rho)^2+z^2}+\sqrt{(a-\rho)^2+z^2}}{2}\ ,\\ 
    l_-&=\frac{\sqrt{(a+\rho)^2+z^2}-\sqrt{(a-\rho)^2+z^2}}{2}\ .
\end{aligned}
\end{equation}
Equation~\eqref{eq:MetricMasterInt} is singular on the ring ${z = 0}$ and ${\rho = a}$, corresponding to the Kerr curvature singularity. For finite ${R}$ however, such singularity is smeared in Eq.~\eqref{eq:TemporalMetricMimicker} and a comparison between the exact Kerr and the mimicking metric is shown in Fig.~\ref{fig:Metric00}.
As expected, the mimicking metric approaches the Kerr profile asymptotically, and for finite ${R}$, the curvature singularity is smeared into a smooth core. We can see that the central value of the metric depends on the $R$ parameter, and that differently from the EMT phenomenology the dependence on $z$ is no longer a simple dumping factor, hence it is worth depicting the gravitational field for different slices of $z$.

\begin{figure*}[htbp!]
\centering
\includegraphics[width=0.48\textwidth, valign=c]{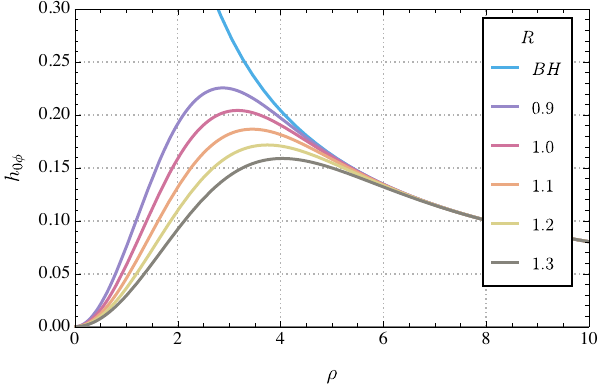}
\includegraphics[width=0.48\textwidth, valign=c]{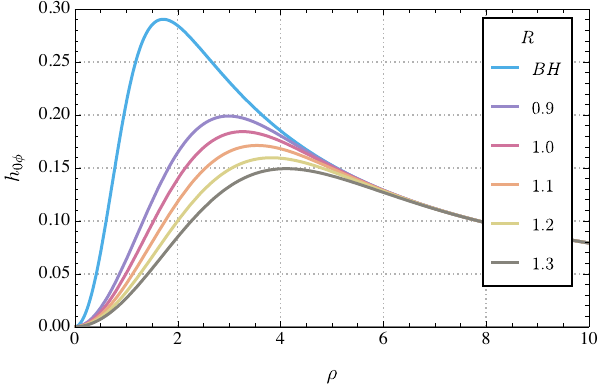}
\caption{Angular component of the linearized metric sourced by the Gaussian-smeared Israel EMT for different values of ${\alpha R}$ and for ${z = 0}$ on the left and ${z = 1}$ on the right, with ${a = 0.8}$ and in units of ${G_N=m=1}$.}
\label{fig:Metric0phi}
\end{figure*}

We now consider the angular component of the metric, which in integral form is given by
\begin{equation}
    h_{0\phi} = -4\pi G m a \rho\, \partial_\rho \int \frac{d q_z}{2\pi} e^{-i q_z z} \int \frac{d q_\perp\, q_\perp}{2\pi} J_0(q_\perp \rho)\, \frac{e^{-q_z^2 R^2}}{q_z^2 + q_\perp^2}\, e^{-q_\perp^2 R^2} \frac{\sin(a q_\perp)}{a q_\perp}\ .
\end{equation}
Again, the ${q_z}$ integration can be carried out analytically, yielding
\begin{equation}\label{eq:AngularPart}
    h_{0\phi} = \frac{G m \rho}{2} \int d q_\perp\, J_1(q_\perp \rho)\, \sin(a q_\perp)\, \left[ e^{-q_\perp z} \mathrm{Erfc}\left(q_\perp R - \tfrac{z}{2R} \right) + e^{q_\perp z} \mathrm{Erfc}\left(q_\perp R + \tfrac{z}{2R} \right) \right]\ .
\end{equation}
As with the temporal component, no closed-form solution is known for Eq.~\eqref{eq:AngularPart}, however, in the ${R \to 0}$ limit the angular component simplifies to
\begin{equation}
    h_{0\phi}^{Kerr} = G m \rho \int d q_\perp\, J_1(q_\perp \rho)\, \sin(a q_\perp)\, e^{-q_\perp |z|}\ ,
\end{equation}
which can be integrated analytically by using known tabulated integrals~\cite{Gradshteyn:1943cpj} as in 
    \begin{equation}
        h_{0\phi}^{Kerr} =  \frac{G m \rho^2 a}{l_+^2}\frac{\sqrt{l_+^2-a^2}}{l_+^2-l_-^2}\ ,
    \end{equation}
where $l_+$ and $l_-$ are defined in Eq.~\eqref{eq:lsDef}.
The behavior of the angular component for finite ${R}$ is then shown in Fig.~\ref{fig:Metric0phi}.
As in the temporal sector, the mimicking metric converges to the Kerr solution asymptotically, and for finite ${R}$, the ring singularity is regularized into a smooth, extended geometry. We can notice that the central value of the angular metric component is always vanishing, while the peak depends parametrically on $R$. In summary, the Gaussian-smeared source generates a linearized metric that smoothly reproduces the Kerr geometry at large distances while regularizing its curvature singularity. Both the temporal and angular components converge asymptotically to their Kerr counterparts, but remain finite at the origin, giving rise to a smooth core whose size is controlled by the smearing scale $R$. The metric thus interpolates between a regular interior and the asymptotic Kerr regime, preserving the correct multipolar structure. This demonstrates that the smeared source provides a physically consistent, singularity-free approximation of the Kerr spacetime at linear order.

\chapter{Scattering amplitudes in Kerr-Schild gauge}\label{chapter:KerrSchildGauge}

The analysis so far has shown that, within the momentum-space formulation of GR, the linearized regime already captures several key physical features of gravitational systems, ranging from source characterization to multipole identification. However, while the linearized multipolar expansion faithfully reproduces the essential structure of the gravitational field, its connection to exact spacetime solutions is often obscured by higher PM corrections. Indeed, in a generic gauge, the proliferation of interaction terms at successive orders makes any resummation of the perturbative expansion practically unfeasible. From the amplitude perspective, this difficulty stems from the appearance of new graviton self-interaction vertices at each loop order (see Eq.~\eqref{eq:GravityVerticesExpansion}), preventing the construction of recursive relations for an all-order resummation. Nonetheless, there have been attempts to reformulate the non-linearities of gravity by adopting suitable gauge choices that simplify the interaction structure. A notable example is the work of~\cite{Mougiakakos:2024nku}, where a cubic formulation of GR allows one to resum all loop orders for a static, non-rotating source, successfully recovering the exact Schwarzschild metric. Although extending this approach to spinning sources is considerably more challenging, one can exploit another special gauge, applicable to specific physical configurations, that enables a closed-form characterization of fully non-perturbative geometries. This is the Kerr–Schild (KS) gauge, a coordinate system in which the exact metric remains linear in the gravitational coupling constant. The KS representation is particularly advantageous: it allows exact metrics to be described already at first PM order and, for scalar probes interacting with the KS background, rewrites the multitude of gravitational self-interactions into a single trilinear vertex, thereby making the definition of recursive amplitude relations possible. This remarkable property, already well known in the Kerr and Kerr–Newman cases~\cite{Bianchi:2023lrg}, extends naturally to the Myers–Perry family~\cite{Bianchi:2025xol}. In this setting, the metric can be written in terms of a harmonic function that fully encodes the gravitational potential, while its Fourier transform can be computed analytically, preserving the key features of the gauge even in momentum space, where scattering amplitudes are naturally formulated. 

In this chapter then, we exploit the simplifying features of the KS gauge to study scattering processes for specific geometries, with particular emphasis on the Myers–Perry family. Working directly in momentum space, we perform the Fourier transform of the metric in hyperspherical KS coordinates and express the result in terms of Bessel functions. This formulation not only enables the computation of the complete tree-level amplitude but also makes explicit the correspondence between its analytic structure and the multipolar expansion of the gravitational field. The KS gauge thus provides an ideal framework for evaluating gravitational observables even in higher dimensions, allowing one to construct physical scenarios that explicitly probe gravitational stress multipoles, trace their imprint in observables, and infer information about the source through scattering processes. Moreover, as will become clear later, the same Bessel functions that appear in the amplitude encode the multipole hierarchy of the spacetime, thereby linking the amplitude description to the momentum-space multipole formalism developed in the previous chapters, and culminating in the identification of the Myers–Perry form factors in Eq.~\eqref{eq:dGenericSphericalB}.

Within this gauge, we first compute the Fourier transform of the Kerr–Newman and Myers–Perry metrics and then evaluate the tree-level scattering amplitude for a scalar probe in such backgrounds. We subsequently move to impact-parameter space, showing that the leading contribution reproduces the classical deflection angle, while the subleading terms correspond to higher-order PM corrections encoded in the eikonal phase, a gauge-invariant function generating gravitational observables such as deflection angles, time delays, and energy–momentum transfer. Because higher-order interaction vertices between the probe and the background are absent in this gauge, all loop diagrams contributing in the classical limit reduce to simple comb-like topologies. Beyond tree level, loop corrections can be organized systematically through the eikonal phase, and radiative or self-force effects, relevant for extreme-mass-ratio inspirals, can be incorporated using modern on-shell and EFT techniques~\cite{Travaglini:2022uwo,Buonanno:2022pgc,Adamo:2022dcm}. Finally, this analysis carries direct implications even for the physics of exotic compact objects and black hole mimickers~\cite{Cardoso:2019rvt,Gambino:2025xdr}. Even in $D=4$, horizonless configurations sharing the same multipole moments as Kerr have been proposed~\cite{Bonga:2021ouq,Gambino:2025xdr}. Moreover, when electromagnetic fields are included, the Kerr–Newman metric~\cite{Newman:1965tw} retains the same gravitational multipoles as Kerr~\cite{Sotiriou:2004ud}, differing only in the spin–charge bound. In higher dimensions, the KS framework provides then a systematic way to test whether the multipolar signatures of black holes can be distinguished from those of fuzzballs or other exotic compact objects through scattering observables such as deflection angles or gravitational-wave phase shifts~\cite{Raposo:2018xkf,Raposo:2020yjy,Bena:2020see,Bianchi:2020bxa,Bena:2020uup,Bianchi:2020miz,Fransen:2022jtw}, making it an ideal theoretical laboratory to probe gravitational effects. 

The rest of the chapter is structured as follows. In section~\ref{Sec:KerrNewmanKSFT} we introduce the KS gauge and summarize its main geometric properties. We then compute the Fourier transform of the Kerr-Newman solution in such a gauge. In section~\ref{sec:MPKS} we compute the Fourier transform of the Myers-Perry metric in the same setup and express it in a form suitable for scattering amplitude calculations. Section~\ref{Sec:Eik_Expansion} is devoted to the computation of the tree-level amplitudes and the definition of the eikonal phase as a tool to link amplitude calculations with gravitational observables. Then in section~\ref{sec:SubEik} we study how to formally account for sub-leading contributions and explicitly show how to compute loop amplitudes in such a formalism, giving an explicit example for a simple set-up. We conclude in section~\ref{sec:MPprobe} by discussing how to employ Myers-Perry black holes as a probe of stress multipoles in higher dimensions, giving explicit results for suitable physical configurations. 

\section{Kerr–Newman solution in Kerr–Schild gauge and its Fourier Transform}\label{Sec:KerrNewmanKSFT}

A simple and elegant way to introduce the KS gauge is to regard it as a reference frame in which the metric is written as a linear deformation of flat spacetime. In this form, all non-linear gravitational effects are encoded in a single scalar function multiplying a null vector field. Specifically, the metric in such gauge takes the form
\begin{equation}\label{eq:GeneralKSdef}
    g_{\mu\nu} = \eta_{\mu\nu} + \Phi\, K_\mu K_\nu\ ,
\end{equation}
where $\Phi$ plays the role of an effective gravitational potential, while $K_\mu$ is a null vector field defined both with respect to the background and the full metric. This implies the condition
\begin{equation}\label{eq:KSnullCondition}
    g^{\mu\nu} K_\mu K_\nu = \eta^{\mu\nu} K_\mu K_\nu = 0\ .
\end{equation}
The double null condition in Eq.~\eqref{eq:KSnullCondition} ensures that $K_\mu$ defines a congruence of null geodesics that is preserved under the deformation of the metric. As a consequence, the KS ansatz~\eqref{eq:GeneralKSdef} linearizes Einstein’s equations in the gravitational coupling constant, even though the resulting metric can describe fully non-linear geometries. In this gauge, the gravitational field can thus be treated as an exact linear perturbation over Minkowski spacetime, making the KS form particularly suited for amplitude-based or momentum-space approaches to gravity. From the condition in Eq.~\eqref{eq:KSnullCondition}, one can immediately derive the inverse metric
\begin{equation}\label{eq:InverseKSdef}
    g^{\mu\nu} = \eta^{\mu\nu} - \Phi\, K^\mu K^\nu\ ,
\end{equation}
which, together with the property $\det(g_{\mu\nu}) = \det(\eta_{\mu\nu}) = -1$, ensures a highly tractable structure. As a consequence, perturbative computations around a KS background can be performed in a way that remains effectively linear in $\Phi$. This feature plays a central role in the amplitude formalism developed in the previous chapters. In particular, within the probe limit, it implies that only the three-point interaction vertex contributes to the expansion of Eq.~\eqref{eq:MassiveVerticesExpansion}. As a result, gravitational interactions can be described to all orders in a formally exact way within an EFT framework, with the entire non-linear dynamics encoded in repeated insertions of a single trilinear vertex. 

Historically, the class of physically relevant geometries known to satisfy the KS ansatz is rather limited. In particular, only spacetimes of Petrov type~D are compatible with this construction, meaning that the KS gauge is not a universal coordinate choice such as the harmonic or de~Donder gauges. Instead, it applies only to specific families of highly symmetric solutions in which the gravitational field can be expressed as an exact linear perturbation of flat spacetime. Remarkably, all standard BH metrics of physical interest, including the Schwarzschild, Kerr, and Kerr-Newman solutions in four dimensions, and their higher-dimensional generalizations of the Myers-Perry type, belong to this class~\cite{Debney:1969zz,Adamo:2014baa,Monteiro:2014cda}. These spacetimes are characterized by the presence of horizons and regions of non-smooth curvature, yet they all admit a KS representation, making the ansatz an exceptionally powerful tool for studying rotating and charged black holes in both classical and amplitude-based formalisms.

For Kerr-Newman spacetime, the effective gravitational potential reads
\begin{equation}\label{eq:KNKSpotential}
\Phi = G_N \frac{2 m r - Q^2}{r^2 + a^2 \cos^2\vartheta}\ ,
\end{equation}
where $m$ is the mass of the BH, $a = J/m \leq m$ its angular momentum per unit mass (aligned with the $z$–axis), and $Q$ its electric charge, where a set of oblate spheroidal coordinates
\begin{equation}
x \pm i y = \sqrt{r^2 + a^2}\,\sin\vartheta\, e^{\pm i\varphi} \quad \text{and} \quad z = r\cos\vartheta\ ,
\end{equation}
satifying the relation
\begin{equation}
\frac{x^2 + y^2}{r^2 + a^2} + \frac{z^2}{r^2} = 1\ ,
\end{equation}
are employed. These coordinates are particularly convenient for describing axisymmetric and rotating spacetimes, since they encode the deformation of spherical symmetry produced by angular momentum through the parameter $a$, with the corresponding null vector taking the form
\begin{equation}
K_\mu = \left(1,\, \frac{r x + a y}{r^2 + a^2},\, \frac{r y - a x}{r^2 + a^2},\, \frac{z}{r}\right)\ .
\end{equation}
Since the Kerr-Newman BH carries electric charge, the spacetime must also sustain an electromagnetic field that satisfies the Maxwell equations in curved space. Within the KS framework, the electromagnetic potential can be expressed in an especially compact and elegant form as
\begin{equation}\label{eq:KSpotentialDef}
A_\mu = \phi\, K_\mu\ ,
\end{equation}
where the same null vector $K_\mu$ appears in both the metric and the gauge potential. The scalar function $\phi$ acts as an effective electric potential and is given by
\begin{equation}
\phi = \frac{Q r}{r^2 + a^2 \cos^2\vartheta}\ .
\end{equation}
Equations~\eqref{eq:GeneralKSdef} and~\eqref{eq:KSpotentialDef} together define the full Kerr-Newman solution of the Einstein-Maxwell theory. Moreover, the structural parallel between the metric $g_{\mu\nu}$ and the gauge potential $A_\mu$ is far from accidental, as it embodies the so-called classical double-copy correspondence, according to which the gravitational field can be interpreted, schematically, as the ``square'' of a gauge field~\cite{Monteiro:2014cda,Luna:2015paa}. This duality lies at the heart of modern amplitude-based approaches to gravity and will be revisited later in its classical formulation.

We now compute the Fourier transform of the Kerr-Newman solution in the KS gauge, considering both the metric and the electromagnetic potential. Since the metric is stationary, the temporal component of the wave vector vanishes, and the Fourier transform can be written as
\begin{equation}
\tilde{h}^{\text{KN}}_{\mu\nu}(q) = \int d^3x\, e^{i q \cdot x}\, \Phi(x)\, K_{\mu}(x)\, K_{\nu}(x)\ .
\end{equation}
Switching to oblate spheroidal coordinates, the Jacobian for the transformation from Cartesian coordinates reads
\begin{equation}
d^3x = dx\,dy\,dz = (r^2 + a^2 \cos^2\vartheta)\, \sin\vartheta\, dr\, d\vartheta\, d\varphi\ .
\end{equation}
Remarkably, the factor $(r^2 + a^2 \cos^2\vartheta)$ exactly cancels the denominator of the potential $\Phi(x)$ in Eq.~\eqref{eq:KNKSpotential}, and since the latter is independent of the azimuthal angle $\varphi$ (owing to axial symmetry), the expression simplifies to
\begin{equation}
\tilde{h}^{\text{KN}}_{\mu\nu}(q) = G_N \int dr\, (2mr - Q^2)\, \sin\vartheta\, d\vartheta\, d\varphi\, e^{i q \cdot x}\, K_{\mu}(r,\vartheta)\, K_{\nu}(r,\vartheta)\ .
\end{equation}
To compute the Fourier transform more efficiently, it is convenient to replace the spatial coordinate $\vec{x}$ with a derivative operator acting in momentum space, $\vec{x} \to i \partial_{\vec{q}}$, while keeping explicit dependence on the radial coordinate $r$. This yields
\begin{equation}
\tilde{h}^{\text{KN}}_{\mu\nu}(q) = G_N \int dr\, (2mr - Q^2)\, \sin\vartheta\, d\vartheta\, d\varphi\,
\tilde{K}_\mu(r, \vec{x} = i \partial_{\vec{q}})\, \tilde{K}_\nu(r, \vec{x} = i \partial_{\vec{q}})\, e^{i q \cdot x}\ ,
\end{equation}
where the KS null vector becomes a differential operator in $\vec{q}$–space acting on the exponential, as in  
\begin{equation}
\tilde{K}_\mu \left(r, \vec{x} = i \partial_{\vec{q}}\right) =
\left(1,\, i\frac{r\partial_{q_x} + a\partial_{q_y}}{r^2 + a^2},\,
i\frac{r\partial_{q_y} - a\partial_{q_x}}{r^2 + a^2},\, i\frac{\partial_{q_z}}{r}\right)\ .
\end{equation}
Consider now the phase in the exponential, given by
\begin{equation}
q \cdot x = (q_x \cos\varphi + q_y \sin\varphi)\, \sin\vartheta\, \sqrt{r^2 + a^2} + q_z r \cos\vartheta\ .
\end{equation}
The angular part of the integral can now be isolated. The factor ${\sin\vartheta\, d\vartheta\, d\varphi = d\Omega}$ corresponds to the rotation-invariant measure on the unit 2-sphere, and the argument of the exponential can be conveniently rewritten as
\begin{equation}
(q_x \cos\varphi + q_y \sin\varphi)\, \sin\vartheta\, \sqrt{r^2 + a^2} + q_z r \cos\vartheta
= \vec{\mathfrak{u}} \cdot \vec{n}\ ,
\end{equation}
where we defined
\begin{equation}
\vec{n} = (\sin\vartheta \cos\varphi,\, \sin\vartheta \sin\varphi,\, \cos\vartheta)
\end{equation}
as the standard unit vector on the sphere, and
\begin{equation}
\vec{\mathfrak{u}} = \big(q_x \sqrt{r^2 + a^2},\, q_y \sqrt{r^2 + a^2},\, q_z r\big)
\end{equation}
as a momentum-dependent auxiliary vector whose magnitude
\begin{equation}
\mathfrak{u} = |\vec{\mathfrak{u}}| = \sqrt{r^2 q^2 + a^2 q_\perp^2}\ ,
\end{equation}
encodes both the rotation parameter $a$ and the transferred momentum components. The integration over the solid angle is elementary and gives
\begin{equation}
\int d\Omega\, e^{i \vec{\mathfrak{u}} \cdot \vec{n}} = 4\pi\, \frac{\sin\mathfrak{u}}{\mathfrak{u}} = 4\pi\, j_0(\mathfrak{u})\ ,
\end{equation}
where $j_0$ is the spherical Bessel function of the first kind. Substituting this result inside the Fourier transform integral, we obtain the compact expression
\begin{equation}\label{FTmetric}
\tilde{h}^{\text{KN}}_{\mu\nu}(q) =
4\pi G_N \int_0^{\infty} dr\, (2mr - Q^2)\,
\tilde{K}_\mu(r, \vec{x} = i \partial_{\vec{q}})\,
\tilde{K}_\nu(r, \vec{x} = i \partial_{\vec{q}})\,
j_0(\mathfrak{u})\ .
\end{equation}
To perform the radial integration, it is convenient to change variable from $r$ to $\mathfrak{u}$ using
\begin{equation}
q^2 r^2 = \mathfrak{u}^2 - a^2 q_\perp^2\ , \qquad
q^2 (r^2 + a^2) = \mathfrak{u}^2 + a^2 q_z^2\ , \qquad
r\,dr = \frac{\mathfrak{u}\, d\mathfrak{u}}{q^2}\ .
\end{equation}
Because $\mathfrak{u}$ itself depends on $q$, one must carefully evaluate how the operator $\tilde{K}_\mu$ acts on a generic function $F(\mathfrak{u})$, with its explicit action reading
\begin{equation}
\tilde{K}_\mu F(\mathfrak{u}) =
\Big(1,\, i(r(\mathfrak{u}) q_x + a q_y),\, i(r(\mathfrak{u}) q_y - a q_x),\, i r(\mathfrak{u}) q_z\Big)\,
\frac{1}{\mathfrak{u}}\, \frac{d F(\mathfrak{u})}{d\mathfrak{u}}\ .
\end{equation}
The remaining radial integrals can be expressed in terms of a family of master integrals~\cite{Gradshteyn:1943cpj}
\begin{equation}\label{masterintegral}
C_n = \int_{q_\perp a}^{+\infty} d\mathfrak{u}\,
\frac{\mathfrak{u}^{1-n}\, j_{n}(\mathfrak{u})}
{\sqrt{\mathfrak{u}^2 - q_\perp^2 a^2}}
= \frac{\pi}{2}\,
\frac{J_n(q_\perp a)}{(q_\perp a)^n}\ ,
\qquad n = 0, 1, 2\ ,
\end{equation}
where $J_n$ are the ordinary Bessel functions of the first kind. These integrals will be used later to express the Fourier-transformed metric components and to draw parallels with the structure of scattering amplitudes in the eikonal regime.

Let us first focus on the Kerr case by setting $Q=0$, so that
\begin{equation}
    \lim_{Q\to 0}\tilde{h}^{KN}_{\mu\nu}(q)=8\pi G_N\tilde{h}_{\mu\nu}(q)\ .
\end{equation}
Dropping the overall factor common to all $\tilde{h}_{\mu\nu}(q)$ (to be reinstated when computing the amplitude), the Fourier transform conveniently takes the form
\begin{equation}
\tilde{h}_{\mu\nu}(q) = \int_{q_\perp a}^\infty \frac{\mathfrak{u}\, d\mathfrak{u}}{q^2}\, \tilde{K}_\mu\, \tilde{K}_\nu\, j_0(\mathfrak{u})\ .
\end{equation}
Using the master integrals in Eq.~\eqref{masterintegral}, and introducing a regulator like $e^{-\epsilon\,  \mathfrak{u}}$ to control the large–$\mathfrak{u}$ behavior (removed at the end), one obtains
\begin{equation}\label{htilde}
\begin{aligned}
 \tilde{h}_{00}(q) & = \frac{1}{q^2}\, \cos|\vec{a}{\times}\vec{q}\,|\ , \\  
 \tilde{h}_{0i}(q) & = -i\,\frac{q_i}{q^3}\,\frac{\pi}{2}\,J_0(|\vec{a}{\times}\vec{q}\,|)\;+\;i\,\frac{(\vec{a}\times \vec{q}\,)_i}{q^2}\,j_0(|\vec{a}{\times}\vec{q}\,|)\ ,\\
\tilde{h}_{ij}(q)&= -\frac{1}{q^2}\,\frac{j_1(|\vec{a}{\times}\vec{q}\,|)}{|\vec{a}{\times}\vec{q}\,|}\,(\vec{a}\times\vec{q}\,)_i(\vec{a}\times\vec{q}\,)_j
+ \frac{j_0(|\vec{a}{\times}\vec{q}\,|)}{q^2} \left(\delta_{ij}-2\frac{q_i q_j}{q^2}\right)\\
&\qquad + \frac{1}{q^3}\,\frac{\pi}{2}\,\frac{J_1(|\vec{a}{\times}\vec{q}\,|)}{|\vec{a}{\times}\vec{q}\,|}\Bigl(q_i (\vec{a}\times \vec{q}\,)_j+q_j (\vec{a}{\times}\vec{q}\,)_i\Bigr)\ ,
\end{aligned}
\end{equation}
where in order to reach the above compact form we exploited the identities 
\begin{equation}
(-a q_y,\, a q_x,\, 0)=\vec a \times \vec q\qquad \text{and}\qquad 
q_\perp a = |\vec{a}{\times}\vec{q}\,|\ .
\end{equation}
Moreover, by symmetry, the $y$–components follow from the $x$–components under $q_x\to q_y$ and $q_y\to -q_x$. As a consistency check, one verifies that $\eta^{\mu\nu}\tilde{h}_{\mu\nu}=0$, a direct consequence of $K_\mu K^\mu=0$, showing that the defining KS property is preserved in momentum space. For charged Kerr-Newman BHs, the gravitational potential $\Phi$ in Eq.~\eqref{eq:KNKSpotential} contains an additional contribution proportional to $Q^2$. Inserting it into Eq.~\eqref{FTmetric} changes the radial dependence relative to the uncharged case. Technically, this alters the parity of the integrand and leads to an exchange of roles between spherical and ordinary Bessel functions in the final expressions. Up to an overall factor, the relevant shift due to the electric charge is
\begin{equation}
\Delta\tilde{h}_{\mu\nu}(q) = \int_{q_\perp a}^\infty \frac{\mathfrak{u}\, d\mathfrak{u}}{q^2}\, \frac{1}{r}\, \tilde{K}_\mu\, \tilde{K}_\nu\, j_0(\mathfrak{u})\ .
\end{equation}
Evaluating the integrals as before with the master integrals in Eq.~\eqref{masterintegral}, we find
\begin{equation}\label{Deltahtilde}
\begin{aligned}
\Delta\tilde{h}_{00}(q) &= \frac{1}{q}\,\frac{\pi}{2}\,J_0(|\vec a \times \vec q\,|)\ , \\ 
\Delta \tilde{h}_{0i}(q) & = -i\,\frac{q_i}{q^2}\,j_0(|\vec a \times \vec q\, |)
\;+\; i\,\frac{(\vec a \times \vec q\,)_i}{q}\,\frac{\pi}{2}\,\frac{J_1(|\vec a \times \vec q\,|)}{|\vec a \times \vec q\,|}\ , \\
\Delta \tilde{h}_{ij}(q)&= -\frac{1}{q}\,\frac{\pi}{2}\,\frac{J_2(|\vec a \times \vec q\,|)}{|\vec a \times \vec q\,|^2}\,(\vec{a}\times \vec q\,)_i(\vec{a}\times \vec q\,)_j
+ \frac{1}{q}\,\frac{\pi}{2}\,\frac{J_1(|\vec a \times \vec q\,|)}{|\vec a \times \vec q\,|} \left(\delta_{ij}-\frac{q_i q_j}{q^2}\right)\\
&\qquad + \frac{1}{q^2}\,\frac{j_1(|\vec a \times \vec q\,|)}{|\vec a \times \vec q\,|}\Bigl(q_i(\vec a\times \vec q\,)_j+q_j(\vec a\times \vec q\,)_i\Bigr)\ ,
\end{aligned}
\end{equation}
and the condition $\eta^{\mu\nu}\Delta \tilde h_{\mu\nu}=0$ follows from the Bessel identity
\begin{equation}\label{RecurrenceRelation_Bessel}
J_0(x)+J_2(x)-2\frac{J_1(x)}{x}=0\ .
\end{equation}
Restoring the overall coefficients, Eqs.~\eqref{htilde} and~\eqref{Deltahtilde} give the momentum-space version of the Kerr-Newman metric in KS gauge,
\begin{equation}
\tilde{h}^{KN}_{\mu\nu}(\vec{q}\,) = 8\pi G_N m \tilde{h}_{\mu\nu}(q) - {4\pi G_N Q^2} 
\Delta\tilde{h}_{\mu\nu}(q)\ ,
\end{equation}
which we shall employ in the following to derive scattering amplitudes off a KN background. Finally, for completeness, we write the Fourier transform of the gauge potential. In the KS ansatz one may compute it efficiently by exploiting the structural relation $A_\mu \propto h_{\mu0}$. Starting from
\begin{equation}
\tilde{A}_\mu(q) = \int d^3x\, e^{iq\cdot x}\, A_\mu(x)\ ,
\end{equation}
and repeating the same steps as for the metric, one finds
\begin{equation}
\tilde{A}_\mu(q) = 4\pi Q \int_{q_\perp a}^\infty \frac{\mathfrak{u}\, d\mathfrak{u}}{q^2}\, \tilde{K}_\mu\, j_0(\mathfrak{u})\ .
\end{equation}
The integrals are then identical to those entering $\tilde{h}_{0\mu}$ up to overall factors, yielding the compact result
\begin{equation}
\begin{aligned}
 & \tilde A_0(q)=\frac{4 \pi Q}{q^2}\,\cos(|\vec a \times \vec q\,|)\ ,\\  
 & \tilde{A}_i(q)=-i\,\frac{4\pi Q}{q^2}\left(\frac{q_i}{q}\,\frac{\pi}{2}\,J_0(|\vec a \times \vec{q}\,|)-j_0(|\vec a \times \vec{q}\,|)\,(\vec{a}\times \vec q\,)_i\right)\ . \label{Atilde}
\end{aligned}
\end{equation}

\section{Myers–Perry solution in Kerr–Schild gauge}\label{sec:MPKS}

We now turn to rotating solutions in higher dimensions. As discussed in Chapter~\ref{chapter:GyromagneticFactor}, within pure higher-dimensional Einstein–Maxwell gravity no exact, closed-form solutions are currently known that describe a regular, rotating, and charged configuration analogous to the Kerr–Newman spacetime in $D>4$ dimensions. Therefore, by extending the logic developed in the previous section, we restrict our attention to neutral geometries. In particular, we focus on the Myers-Perry BHs~\cite{Myers:1986un}, which generalize the Kerr metric to higher spacetime dimensions and represent the most general asymptotically flat, stationary, vacuum solutions with spherical horizon topology.

\subsection{Various coordinate systems}

Before delving into the computation of the Fourier transform of the Myers-Perry metric, it is useful to give the explicit expression of the metric in arbitrary spacetime dimension and review the various coordinate systems in which this solution can be represented. In particular, we introduce a new family of coordinates, hereafter referred to as \emph{spectral coordinates}, which will prove especially convenient for evaluating the Fourier transform in a manifestly symmetric and computationally transparent framework. This preliminary discussion will not only streamline the forthcoming derivation but also shed light on how the higher-dimensional structure of the MP geometry naturally extends the familiar four-dimensional Kerr spacetime.

\subsubsection{Myers–Perry coordinates}

Let us begin by recalling the standard form of the Myers-Perry metric written in its natural coordinates. This reference frame can be regarded as the higher-dimensional analogue of Boyer–Lindquist coordinates, since in the limit $d=3$ the Myers-Perry metric exactly reduces to the Kerr metric in Boyer–Lindquist form. Due to the distinct algebraic properties of the rotation group in even and odd spatial dimensions, one must treat the two cases separately from the outset. In oblate spheroidal coordinates $(t, r, \mu_i, \phi_i, \mu_0)$ with $i=1,\dots,n$, for odd spatial dimensions $d = 2n + 1 \geq 3$, the metric reads
\begin{equation}\label{eq:MPoblatesphODD}
    ds^2 = -dt^2 + \frac{\mu r}{\Pi F}\left(dt + \sum_{i=1}^{n}\mathfrak{a}_i \mu_i^2 d\phi_i\right)^2 + \frac{\Pi F}{\Pi - \mu r}\,dr^2 + \sum_{i=1}^{n}(r^2 + \mathfrak{a}_i^2)(d\mu_i^2 + \mu_i^2 d\phi_i^2) + r^2 d\mu_0^2\ ,
\end{equation}
where the auxiliary functions are defined as
\begin{equation}\label{eq:MPfunctionDefinition}
    F = 1 - \sum_{i=1}^{n}\frac{\mathfrak{a}_i^2 \mu_i^2}{r^2 + \mathfrak{a}_i^2}\ , 
    \qquad
    \Pi = \prod_{i=1}^{n}(r^2 + \mathfrak{a}_i^2)\ ,
\end{equation}
and $\mu$ and $\mathfrak{a}_i$ denote, respectively, the mass and angular-momentum parameters of the solution. It is important to emphasize that these parameters do not directly correspond to the physical ADM quantities. In fact the physical angular momenta are instead related to the parameters appearing in the metric by
\begin{equation}\label{eq:MPvsPhys}
    S_{\text{MP}} = \frac{d-1}{2}\, S_{\text{phys}}\ ,
\end{equation}
where $S_{\text{MP}}$ represents the antisymmetric angular-momentum tensor, which can be block-diagonalized as
\begin{equation}
    S_{\text{MP}} =
\begin{pmatrix}
0 & \mathfrak{a}_1 & 0 & 0 & \cdots \\
-\mathfrak{a}_1 & 0 & 0 & 0 & \cdots \\ 
0 & 0 & 0 & \mathfrak{a}_2 & \cdots \\
0 & 0 & -\mathfrak{a}_2 & 0 & \cdots \\ 
\vdots & \vdots & \vdots & \vdots & \ddots
\end{pmatrix}\ .
\end{equation}
Notice that, for odd $d$, the last row and column vanish, reflecting the fact that there is always one unpaired spatial direction that remains unaffected by rotations. Similarly, the mass parameter $\mu$ can be rewritten in terms of the physical ADM mass $m$ using
\begin{equation}
    \mu = \frac{\kappa^2 m}{2}\,\frac{1}{(d-1)\Omega_{d-1}}\ ,
\end{equation}
where $\kappa^2 = 32\pi G_N$ and $\Omega_{d}$ denotes the surface area of the unit $d$–sphere, defined in Eq.~\eqref{eq:SurfaceDsphere}. It is often convenient to introduce Cartesian-like coordinates associated with the Myers-Perry metric,
\begin{equation}
    x_i = \sqrt{r^2 + \mathfrak{a}_i^2}\,\mu_i \cos\phi_i\ , \qquad 
    y_i = \sqrt{r^2 + \mathfrak{a}_i^2}\,\mu_i \sin\phi_i\ , \qquad 
    z = r \mu_0\ ,
\end{equation}
which satisfy
\begin{equation}
    \sum_{i=1}^{n}\frac{x_i^2 + y_i^2}{r^2 + \mathfrak{a}_i^2} + \frac{z^2}{r^2} = 1\ .
\end{equation}
This embedding generalizes the four-dimensional oblate spheroidal relation and proves useful when mapping the metric into the KS form. For an even number of spatial dimensions, $d = 2n + 2 \geq 4$, and in oblate spheroidal coordinates $(t, r, \mu_i, \phi_i)$ with $i = 1, \dots, n$, the Myers-Perry metric takes the slightly modified form
\begin{equation}\label{eq:MPoblatesphEVEN}
    ds^2 = -dt^2 + \frac{\mu r^2}{\Pi F}\left(dt + \sum_{i=1}^{n}\mathfrak{a}_i \mu_i^2 d\phi_i\right)^2 + \frac{\Pi F}{\Pi - \mu r^2}\,dr^2 + \sum_{i=1}^{n}(r^2 + \mathfrak{a}_i^2)(d\mu_i^2 + \mu_i^2 d\phi_i^2)\ ,
\end{equation}
where the definitions of $F$ and $\Pi$ remain the same as in Eq.~\eqref{eq:MPfunctionDefinition}. The associated Cartesian coordinates are
\begin{equation}
    x_i = \sqrt{r^2 + \mathfrak{a}_i^2}\,\mu_i \cos\phi_i\ , \qquad 
    y_i = \sqrt{r^2 + \mathfrak{a}_i^2}\,\mu_i \sin\phi_i\ , \qquad 
    z = r \mu_0\ ,
\end{equation}
and they satisfy the constraint
\begin{equation}
    \sum_{i=1}^{n}\frac{x_i^2 + y_i^2}{r^2 + \mathfrak{a}_i^2} = 1\ .
\end{equation}
This coordinate structure clearly exhibits the topological and algebraic differences between the even– and odd–dimensional cases, which play a crucial role in the behavior of the rotation group and in the number of independent spin parameters $\mathfrak{a}_i$. These subtleties will later translate into distinct analytic features of the Fourier-transformed metric components.

\subsubsection{Kerr–Schild coordinates and the double copy}

We now express the Myers-Perry metric in KS gauge, giving the explicit expression of the gravitational potential and the null vector field in coordinate space. Indeed, it can be shown that a suitable coordinate transformation exists in arbitrary spacetime dimensions that brings the Myers-Perry solution into the KS form of Eq.~\eqref{eq:GeneralKSdef}, preserving all its defining properties. Hence, to describe the geometry in this gauge, one only needs to identify the two fundamental ingredients, namely the gravitational potential $\Phi$ and the null vector $K_\mu$. For odd spatial dimensions, the gravitational potential takes the form
\begin{equation}
    \Phi(r) = \frac{\kappa^2 m}{2}\,\frac{1}{(d-1)\Omega_{d-1}}\,\frac{r}{\Pi F}\ ,
\end{equation}
while the associated null vector reads
\begin{equation}
    K_\mu = \Bigg(1,\,
    \frac{r x_1+\mathfrak{a}_1 y_1}{r^2+\mathfrak{a}_1^2},\,
    \frac{r y_1-\mathfrak{a}_1 x_1}{r^2+\mathfrak{a}_1^2},\,\ldots,\,
    \frac{z}{r}\Bigg)\ .
\end{equation}
In the even-dimensional case, the potential becomes
\begin{equation}
    \Phi(r) = \frac{\kappa^2 m}{2}\,\frac{1}{(d-1)\Omega_{d-1}}\,\frac{r^2}{\Pi F}\ ,
\end{equation}
with a null vector of the form
\begin{equation}
    K_\mu = \Bigg(1,\,
    \frac{r x_1+\mathfrak{a}_1 y_1}{r^2+\mathfrak{a}_1^2},\,
    \frac{r y_1-\mathfrak{a}_1 x_1}{r^2+\mathfrak{a}_1^2},\,\ldots\Bigg)\ .
\end{equation}
Remarkably, although a consistent electrically charged MP black hole does not exist within pure Einstein–Maxwell theory, the validity of Eq.~\eqref{eq:GeneralKSdef} in arbitrary dimensions implies that one can nevertheless define a formal gauge field satisfying the Maxwell equations in the KS gauge. This observation underlies the so-called \emph{double copy} construction, in which the metric tensor can be viewed as the ``square'' of a gauge field. Concretely, given $\Phi$ and $K_\mu$, one can define a vector potential
\begin{equation}\label{eq:FormalGaugeKSphoton}
    A_\mu = h_{\mu 0} = \Phi\, K_\mu\ ,
\end{equation}
which satisfies the Maxwell equations and is therefore interpreted as the \emph{single copy} of the gravitational field. It is important to emphasize that this gauge field does not correspond to a physical electromagnetic source, since the black hole under consideration is neutral. Rather, Eq.~\eqref{eq:FormalGaugeKSphoton} highlights that, in the KS gauge, the tensorial structure of gravity effectively reduces to a vector description. Consequently, solving Einstein’s equations in this context becomes formally equivalent to solving Maxwell’s equations~\cite{Monteiro:2014cda,Luna:2015paa}. This structural analogy can be pushed one step further. By discarding the vector structure altogether, one can introduce a \emph{zeroth copy} in which the gravitational potential itself is interpreted as a scalar field obeying the Laplace equation,
\begin{equation}
    \Box\,\Phi = 0\ .
\end{equation}
Hence, within the KS framework, one observes a progressive reduction in the number of physical degrees of freedom required to construct the spacetime geometry. Starting from a scalar harmonic potential $\Phi$, one can systematically build the single copy and the double copy through simple algebraic products with the null vector $K_\mu$, as in 
\begin{equation}
\begin{gathered}
    \text{zeroth copy} \;\rightarrow\; \text{single copy} \;\rightarrow\; \text{double copy}\ ,\\[2mm]
    \Phi \;\rightarrow\; A_\mu = \Phi\, K_\mu \;\rightarrow\; h_{\mu\nu} = \Phi\, K_\mu K_\nu\ .
\end{gathered}
\end{equation}
While the quantum realization of the double copy connects non-Abelian gauge theories to gravity through intricate color–kinematics dualities and algebraic relations~\cite{Bern:2010ue,Bern:2019prr,Bern:2022wqg}, its classical limit exhibits a much simpler ``abelianized'' structure. In this limit, the mapping between the scalar, vector, and tensor levels of the theory acquires the elegant and transparent formulation above. Although the complete theoretical foundation of the classical double copy remains under active investigation, this framework already provides profound insight into the geometric and algebraic unity underlying gauge and gravitational interactions. In particular, the existence of a KS representation for the Myers-Perry family of metrics strongly supports the idea that many exact GR solutions admit a gauge-theoretic interpretation, namely that the gravitational solution can be re-expressed or understood as emerging from the structure of a gauge theory, an observation that continues to inspire efforts to uncover the fundamental building blocks of BH spacetimes and the nature of gravity itself.

\subsubsection{Spectral coordinates}

Since our ultimate goal is to compute the Fourier transform of the Myers–Perry metric in the KS gauge, it is convenient to introduce a new family of coordinates that we shall refer to as spectral coordinates. The name stems from their direct connection with the spectral (momentum–space) representation of the metric. As we will see, this construction allows one to express the spacetime geometry in arbitrary dimensions through a single compact formula, thereby avoiding the need to distinguish explicitly between even and odd $d$. Moreover, these coordinates naturally encode the rotational structure of the spacetime and greatly simplify the algebra involved in the Fourier transform computation. Let us start from the Cartesian–like Myers-Perry coordinates defined by
\begin{equation}
    \vec{x} = (x_1, y_1, \ldots)\ ,
\end{equation}
where, as already discussed, each pair $(x_i, y_i)$ identifies an independent two–dimensional plane of rotation. In order to connect position space to its spectral counterpart, we introduce a set of rescaled position variables that capture the relative scaling of each plane with respect to its characteristic rotation parameter $\mathfrak{a}_i$
\begin{equation}
    \vec{\chi} = \Bigg(\frac{x_1}{r^2 + \mathfrak{a}_1^2},\ \frac{y_1}{r^2 + \mathfrak{a}_1^2},\ \ldots\Bigg)\ .
\end{equation}
In components, this can be written more explicitly as
\begin{equation}
    \chi_{2i-1} = \frac{x_i}{r^2 + \mathfrak{a}_i^2}\ , \qquad 
    \chi_{2i} = \frac{y_i}{r^2 + \mathfrak{a}_i^2}\ .
\end{equation}
The variables $\chi_i$ thus define a coordinate basis that effectively separates the radial dependence from the rotational degrees of freedom in each plane. We anticipate that this scaling will prove instrumental in writing the KS null vector and the metric in a form that closely mirrors their Fourier–transformed expressions. We can now introduce a covariant generalization of these coordinates by extending the definition to spacetime indices and enforcing the condition $\chi_0 = 0$, so that
\begin{equation}
    \chi_\mu = \Bigg(0,\ \frac{x_1}{r^2 + \mathfrak{a}_1^2},\ \frac{y_1}{r^2 + \mathfrak{a}_1^2},\ \ldots\Bigg)\ .
\end{equation}
Let us also recall the definition of the time–like unit vector ${u^\mu = \delta_0^\mu}$, which in the KS framework represents the asymptotic velocity of the source. Physically, this corresponds to the rest frame of the massive rotating object generating the field, and in momentum space $u^\mu$ identifies the reference direction along which the energy flow of the source is defined. With these definitions, the KS null vector can be compactly written in terms of the spectral coordinates as
\begin{equation}\label{eq:NullVecInSpectralPosition}
    K_\mu = u_\mu + r\, \chi_\mu + \big(S_{\text{MP}} \cdot  \chi\big)_\mu\ ,
\end{equation}
where $S_{\text{MP}}$ is the Myers-Perry spin tensor introduced in Eq.~\eqref{eq:MPvsPhys}. Moreover, because ${S_{\text{MP}}^{0\mu} = 0}$, it follows that ${(S_{\text{MP}} \cdot  \chi)_0 = 0}$, ensuring that $K_\mu$ remains null in any spacetime dimension. Substituting Eq.~\eqref{eq:NullVecInSpectralPosition} into the KS ansatz~\eqref{eq:GeneralKSdef}, the metric can be elegantly rewritten in its spectral form as
\begin{equation}\label{eq:SpectralMetricOblate}
\begin{aligned}
    g_{\mu\nu} &= \eta_{\mu\nu} + \Phi(r)\Bigg\{
    u_\mu u_\nu
    + \Big[u_\mu (S_{\text{MP}} \cdot  \chi)_\nu + u_\nu (S_{\text{MP}} \cdot  \chi)_\mu\Big]
    + (S_{\text{MP}} \cdot  \chi)_\mu (S_{\text{MP}} \cdot  \chi)_\nu \\[2mm]
    &\quad + r\Big[\chi_\mu (S_{\text{MP}} \cdot  \chi)_\nu + \chi_\nu (S_{\text{MP}} \cdot  \chi)_\mu\Big]
    + r\big(u_\mu \chi_\nu + u_\nu \chi_\mu\big)
    + r^2\, \chi_\mu \chi_\nu
    \Bigg\}\ .
\end{aligned}
\end{equation}
This compact expression encodes the multipolar structure of the Myers-Perry spacetime in terms of the rescaled coordinates $\chi_\mu$ and the spin tensor $S_{\text{MP}}$. In particular, the term $\propto u_\mu u_\nu$ is associated with the mass multipoles, the terms $\propto u_\mu \chi_\nu$ describe the current multipole tower while terms $\propto \chi_\mu \chi_\nu$ encode stress moments. It is important to remark, however, that the metric in Eq.~\eqref{eq:SpectralMetricOblate}, although manifestly adapted to the spectral representation, is not expressed in ACMC coordinates, and hence one cannot extract gravitational multipoles directly from the metric expression. Indeed, to read off the multipole moments explicitly, one must either move to an ACMC system, such as the harmonic gauge, or compute the Fourier transform and determine the corresponding gravitational form factors. The relation between these form factors and the multipole tensors has been established in Eq.~\eqref{eq:GravitationalMultipoles}, and will be employed again later. The important advantage of the spectral representation is that it simplifies the computation of the Fourier transform while preserving the physical transparency of the construction. In this framework, the structure of Eq.~\eqref{eq:SpectralMetricOblate} closely resembles its momentum–space counterpart, making the subsequent analysis of the spectral metric both technically straightforward and conceptually intuitive. As will become clear in the following, this correspondence allows one to seamlessly interpret the momentum–space expressions of the Myers-Perry metric in terms of their geometric and multipolar content.

\subsection{Fourier transform in arbitrary dimensions}

Our goal now is to compute the Fourier transform of the Myers–Perry metric written in KS coordinates
\begin{equation}
    \tilde{h}_{\mu\nu} = \int d^d x\, e^{i q \cdot x}\, h_{\mu\nu}\ ,
\end{equation}
by exploiting the spectral representation developed previously. As anticipated, we will demonstrate that in momentum space the metric admits a unified and compact expression valid in any number of spacetime dimensions, without the need to distinguish between even and odd $d$. To proceed, let us introduce a momentum–space auxiliary variable $\vec{\mathfrak{u}}$ defined analogously to the $d=3$ case through
\begin{equation}
    \vec{q}\cdot\vec{x} = \vec{\mathfrak{u}}\cdot\vec{n} = \mathfrak{u}\cos\theta\ ,
\end{equation}
where $\vec{n}$ is a unit vector such as ${|\vec{n}| = 1}$. The variable $\vec{\mathfrak{u}}$ is then implicitly defined as
\begin{equation}
    \vec{\mathfrak{u}} = \big(q_{x_1}\sqrt{r^2 + \mathfrak{a}_1^2},\, q_{y_1}\sqrt{r^2 + \mathfrak{a}_1^2},\, \ldots\big)
\end{equation}
in the even–dimensional case, and as
\begin{equation}
    \vec{\mathfrak{u}} = \big(q_{x_1}\sqrt{r^2 + \mathfrak{a}_1^2},\, q_{y_1}\sqrt{r^2 + \mathfrak{a}_1^2},\, \ldots,\, q_z r\big)
\end{equation}
in the odd–dimensional case. These definitions ensure that the inner product $\vec{q}\cdot\vec{x}$ is consistently rewritten in terms of $\mathfrak{u}$ and $\theta$ in any dimension. We can then express the squared modulus of $\vec{\mathfrak{u}}$ in arbitrary dimensions as
\begin{equation}
    |\vec{\mathfrak{u}}|^2 = \mathfrak{u}^2 = \sum_{i=1}^{n}(q_{x_i}^2 + q_{y_i}^2)\mathfrak{a}_i^2 + q^2 r^2\ ,
\end{equation}
with
\begin{align}
    q^2 &= \sum_{i=1}^{n}(q_{x_i}^2 + q_{y_i}^2) \qquad &&\text{for } d \text{ even}\ ,\\
    q^2 &= \sum_{i=1}^{n}(q_{x_i}^2 + q_{y_i}^2) + q_z^2 \qquad &&\text{for } d \text{ odd}\ .
\end{align}
For later convenience, we define a further quantity that will play a central role in the computation, namely
\begin{equation}\label{eq:MPSpectralDef}
    \xi_\mu = i\,(S_{\text{MP}}\cdot q)_\mu\ .
\end{equation}
By construction, this vector satisfies
\begin{equation}
    \xi_\mu \xi^{\mu} = \xi^2 = -\,q \cdot S_{\text{MP}}\cdot S_{\text{MP}}\cdot q
    = \sum_{i=1}^{n}(q_{x_i}^2 + q_{y_i}^2)\mathfrak{a}_i^2\ ,
\end{equation}
allowing us to express $\mathfrak{u}$ compactly as
\begin{equation}
    \mathfrak{u}^2 = \xi^2 + q^2 r^2\ .
\end{equation}
Moreover, the variable $\xi_\mu$ defined in Eq.~\eqref{eq:MPSpectralDef} is related to the physical spin–momentum variable introduced earlier in Eq.~\eqref{eq:PhysSpectralDef} through
\begin{equation}
    \xi_\mu = \frac{d-1}{2}\,\zeta_\mu\ ,
\end{equation}
which follows directly from the normalization given in Eq.~\eqref{eq:MPvsPhys}. With these definitions, the Fourier transform of the metric can be compactly written as
\begin{equation}
    \tilde{h}_{\mu\nu} = \int d^d x\, e^{i \mathfrak{u} \cos\theta}\, K_\mu K_\nu\, \Phi(r)\ ,
\end{equation}
where $\theta$ denotes the angle between $\vec{n}$ and $\vec{\mathfrak{u}}$. This formulation mirrors the structure encountered in the Kerr–Newman analysis and paves the way for a dimension–independent treatment of the integral.

A key reason why the explicit Fourier transform computation was feasible in the $d=3$ case lies in a remarkable cancellation. Indeed, the Jacobian of the transformation from Cartesian to oblate spheroidal coordinates exactly cancels the denominator of the gravitational potential $\Phi(r)$ in the KS form of the metric. We now show that this property extends to Myers-Perry BHs in arbitrary dimensions, and in fact, it can be demonstrated\footnote{For a detailed derivation, see Appendix~B of Ref.~\cite{Bianchi:2025xol}.} that the Jacobian of the coordinate transformation reads
\begin{equation}
    \mathcal{J} = \left|\det \left(\frac{\partial(x_i, y_i)}{\partial(r, \mu_i, \phi_i)}\right) \right| =
    \begin{cases}
        \displaystyle \frac{\Pi F}{r}\,\prod_{k=1}^{n-1}\mu_k & \text{for } d = 2n\ ,\\[2mm]
        \displaystyle \Pi F\,\prod_{k=1}^{n-1}\mu_k & \text{for } d = 2n + 1\ .
    \end{cases}
\end{equation}
Considering, for definiteness, the even–dimensional case, the integration measure transforms as
\begin{equation}
    d^d x = \mathcal{J}\, \Pi F\, dr \prod_{k=1}^{n-1}\mu_k\, d\mu_k\, \prod_{k=1}^{n-1} d\phi_k\ .
\end{equation}
Upon substituting the explicit form of the gravitational potential, the Fourier transform of the metric becomes
\begin{equation}
    \tilde{h}_{\mu\nu} =
    \frac{\kappa^2 m}{2}\,\frac{1}{(d-1)\Omega_{d-1}}
    \int r\, dr \prod_{k=1}^{n-1}\mu_k\, d\mu_k \prod_{k=1}^{n-1} d\phi_k\;
    e^{i \mathfrak{u}\cos\theta}\, K_\mu K_\nu\ .
\end{equation}
Next, exploiting the symmetries of the angular integration, one can factor out the surface integral over the unit $(d-2)$–sphere as
\begin{equation}
    \prod_{k=1}^{n-1}\mu_k\, d\mu_k \prod_{k=1}^{n-1} d\phi_k
    = \sin^{d-2}\theta\, d\theta\, d\Omega_{d-2}\ .
\end{equation}
Computing then the basic integral over $\theta$
\begin{equation}
    \mathcal{I}_0 = \int_0^\pi d\theta\, e^{i \mathfrak{u}\cos\theta} = \pi\, J_0(\mathfrak{u})\ ,
\end{equation}
higher–order angular integrals can be evaluated recursively by integration by parts, yielding
\begin{equation}
    \mathcal{I}_{d-2} = \int_0^\pi \sin^{d-2}\theta\, d\theta\, e^{i \mathfrak{u}\cos\theta}
    = (-1)^n (d-3)!! \left(\frac{1}{\mathfrak{u}}\frac{d}{d\mathfrak{u}}\right)^n \pi\, J_0(\mathfrak{u})\ ,
\end{equation}
where $n!!$ denotes the double factorial of the integer $n$ defined as 
\begin{equation}
    n!!=n(n-2)(n-4)\cdots\ .
\end{equation}
Using the Bessel recurrence identity
\begin{equation}\label{eq:RecurrenceRelationBessel}
    \left(\frac{1}{\mathfrak{u}}\frac{d}{d\mathfrak{u}}\right)^{\beta}
    \left(\frac{J_{\alpha}(\mathfrak{u})}{\mathfrak{u}^{\alpha}}\right)
    = (-1)^{\beta}\,\frac{J_{\alpha+\beta}(\mathfrak{u})}{\mathfrak{u}^{\alpha+\beta}}\ ,
\end{equation}
one arrives at a compact expression for the Fourier integral of the metric for even $d$ as
\begin{equation}\label{eq:NonGenericMPeven}
     \tilde{h}_{\mu\nu} =
     \frac{\kappa^2 m}{2}\,
     \frac{\Omega_{d-2}}{(d-1)\Omega_{d-1}}\,
     (d-3)!!\, \pi
     \int_0^{+\infty} r\, dr\, K_\mu K_\nu\,
     \frac{J_{\frac{d-2}{2}}(\mathfrak{u})}{\mathfrak{u}^{\frac{d-2}{2}}}\ ,
\end{equation}
leaving implicit only the radial integral. This expression is the higher–dimensional analogue of Eq.~\eqref{FTmetric} for the Kerr case and will serve as the starting point for the analysis of the Myers-Perry metric in momentum space. As we shall see in the following, this compact form clearly exposes the spin dependence through $\xi_\mu$ and provides a natural bridge between the multipolar expansion of the metric and its spectral (Fourier) representation. Even though we performed the explicit calculation for the $d=\text{even}$ case, one can verify that the odd–dimensional sector proceeds identically and leads to the same functional form as Eq.~\eqref{eq:NonGenericMPeven}, differing only by an overall parity–dependent coefficient. Collecting these prefactors, we write
\begin{equation}\label{eq:ExplicitParityFactor}
\mathcal{C}_{d=\text{even}}=\frac{\Omega_{d-2}}{\Omega_{d-1}}(d-3)!!\ \pi\ , \qquad
    \mathcal{C}_{d=\text{odd}}=\frac{\Omega_{d-2}}{\Omega_{d-1}}(d-3)!! \sqrt{2\pi}\ ,
\end{equation}
and observe that, in arbitrary dimensions, by defining
\begin{equation}
    \mathcal{C}=\frac{1}{2}\,2^{d/2}\,\Gamma(d/2)\ ,
\end{equation}
one reproduces Eq.~\eqref{eq:ExplicitParityFactor} upon fixing the parity of $d$. Considering this general normalization in Eq.~\eqref{eq:NonGenericMPeven}, generalizing to arbitrary spacetime dimensions, and changing variables from $r$ to $\mathfrak{u}$, one gets
\begin{equation}\label{eq:TildeMetricBeforeRadialInt}
    \tilde{h}_{\mu\nu}=\frac{\kappa^2m}{2}\Bigg(\frac{1}{2}\frac{2^{d/2}\Gamma(d/2)}{d-1}\Bigg)\frac{1}{q^2}\int_{\xi}^{+\infty}d\mathfrak{u}\  \mathfrak{u}\  K_\mu K_\nu \frac{J_{\frac{d-2}{2}}(\mathfrak{u})}{\mathfrak{u}^{\frac{d-2}{2}}}\ .
\end{equation}

We are thus reduced the problem to evaluating the radial integral
\begin{equation}
    \mathcal{I}_{\mu\nu}=\int_{\xi}^{+\infty}d\mathfrak{u}\  \mathfrak{u}\  K_\mu K_\nu \frac{J_{\frac{d-2}{2}}(\mathfrak{u})}{\mathfrak{u}^{\frac{d-2}{2}}}\ .
\end{equation}
To compute it efficiently, we promote the null vector $K_\mu$ to a differential operator in momentum space,
\begin{equation}
    K_\mu\rightarrow \tilde{K}_\mu=\Bigg(1, -i\frac{r \partial_{q_{x_1}}-\mathfrak{a}_1 \partial_{q_{y_1}}}{r^2+\mathfrak{a}_1^2}, -i\frac{r \partial_{q_{y_1}}+\mathfrak{a}_1 \partial_{q_{x_1}}}{r^2+\mathfrak{a}_1^2}, ...\Bigg)\ .
\end{equation}
Likewise, the rescaled variable $\chi_\mu$ becomes
\begin{equation}
    \chi_\mu\rightarrow\tilde{\chi}_\mu=-i\Bigg(0, \frac{\partial_{q_{x_1}}}{r^2+\mathfrak{a}_1^2}, \frac{\partial_{q_{y_1}}}{r^2+\mathfrak{a}_1^2}, ....\Bigg)\ ,
\end{equation}
so that the KS decomposition retains its form,
\begin{equation}
    \tilde{K}_\mu=u_\mu+r \tilde{\chi}_\mu+\Big(S_{\text{MP}}\cdot \tilde{\chi}\Big)_\mu\ .
\end{equation}
Because $\tilde{K}_\mu$ now acts as a differential operator and $\mathfrak{u}=\mathfrak{u}(q)$, it is convenient to rewrite the needed integral as
\begin{equation}
\begin{aligned}
    \mathcal{I}_{\mu\nu}&=\int_{\xi}^{+\infty}d\mathfrak{u}\  \mathfrak{u}\Bigg\{ u_\mu u_\nu+\Bigg[u_\mu \Big(S_{\text{MP}}\cdot \tilde{\chi}\Big)_\nu+u_\nu \Big(S_{\text{MP}}\cdot \tilde{\chi}\Big)_\mu\Bigg]+\Big(S_{\text{MP}}\cdot \tilde{\chi}\Big)_\mu\Big(S_{\text{MP}}\cdot \tilde{\chi}\Big)_\nu\\
    &\quad +r\Bigg[ \tilde{\chi}_\mu \Big(S_{\text{MP}}\cdot \tilde{\chi}\Big)_\nu+\tilde{\chi}_\nu \Big(S_{\text{MP}}\cdot \tilde{\chi}\Big)_\mu\Bigg]+r\Big(u_\mu \tilde{\chi}_\nu+u_\nu\tilde{\chi}_\mu\Big)+r^2\tilde{\chi}_\mu \tilde{\chi}_\nu\Bigg\}\frac{J_{\frac{d-2}{2}}(\mathfrak{u})}{\mathfrak{u}^{\frac{d-2}{2}}}\ ,
\end{aligned}
\end{equation}
where ${\tilde{\chi}_\mu}$ acts on functions of the type $f(\mathfrak{u})$. In particular,
\begin{equation}
    \tilde{\chi}_i f(\mathfrak{u})=-iq_i \frac{1}{\mathfrak{u}}\frac{d}{d\mathfrak{u}}f(\mathfrak{u})\ ,
\end{equation}
and
\begin{equation}\label{eq:DoubleChiTilde}
    \tilde{\chi}_i\tilde{\chi}_j f(\mathfrak{u})=-\Bigg(\frac{\delta_{ij}}{r^2+a_i^2}\frac{1}{\mathfrak{u}}\frac{d}{d\mathfrak{u}}+q_iq_j\Big(\frac{1}{\mathfrak{u}}\frac{d}{d\mathfrak{u}}\Big)^2\Bigg)f(\mathfrak{u})\ .
\end{equation}
We now evaluate separately the three sectors of $\mathcal{I}_{\mu\nu}$, starting with the scalar piece that is given by
\begin{equation}
    \mathcal{I}_{00}=\int_{\xi}^{+\infty}d\mathfrak{u}\  \mathfrak{u}\frac{J_{\frac{d-2}{2}}(\mathfrak{u})}{\mathfrak{u}^{\frac{d-2}{2}}}\ .
\end{equation}
Using the master integral~\cite{Gradshteyn:1943cpj}
\begin{equation}
    \mathcal{I}(\alpha, n)=\int_\xi^{+\infty}d\mathfrak{u}\  \mathfrak{u}(\mathfrak{u}^2-\xi^2)^{\,n+1/2} \frac{J_{\alpha}(\mathfrak{u})}{\mathfrak{u}^{\alpha}}=\Gamma \big(3/2+n\big)\,2^{\,n+1/2}\,\frac{J_{\alpha-n-3/2}(\xi)}{\xi^{\alpha-n-3/2}}\ ,
\end{equation}
we immediately obtain
\begin{equation}
    \mathcal{I}_{00}=\mathcal{I}\Big(\alpha=\tfrac{d-2}{2},\, n=-\tfrac{1}{2}\Big)\ .
\end{equation}
Similarly, the vector piece reads
\begin{equation}
    \mathcal{I}_{0i}=\int_{\xi}^{+\infty}d\mathfrak{u}\  \mathfrak{u}\Bigg\{r \tilde{\chi}_i+\Big(S_{\text{MP}}\cdot \tilde{\chi}\Big)_i\Bigg\}\frac{J_{\frac{d-2}{2}}(\mathfrak{u})}{\mathfrak{u}^{\frac{d-2}{2}}}\ ,
\end{equation}
and applying the action of $\tilde{\chi}_i$ on $f(\mathfrak{u})$ and using
\begin{equation}
    r=\frac{\sqrt{\mathfrak{u}^2-\xi^2}}{q}\ ,
\end{equation}
we find
\begin{equation}
    \mathcal{I}_{0i}=-\int_{\xi}^{+\infty}d\mathfrak{u}\  \mathfrak{u}\Bigg\{i\frac{\sqrt{\mathfrak{u}^2-\xi^2}}{q}\,q_i+\xi_i\Bigg\}\frac{1}{\mathfrak{u}}\frac{d}{d\mathfrak{u}}\frac{J_{\frac{d-2}{2}}(\mathfrak{u})}{\mathfrak{u}^{\frac{d-2}{2}}}\ .
\end{equation}
Using the recurrence relation in Eq.~\eqref{eq:RecurrenceRelationBessel} yields
\begin{equation}
    \mathcal{I}_{0i}=\int_{\xi}^{+\infty}d\mathfrak{u}\  \mathfrak{u}\Bigg\{i\frac{\sqrt{\mathfrak{u}^2-\xi^2}}{q}\,q_i+\xi_i\Bigg\}\frac{J_{\frac{d-2}{2}+1}(\mathfrak{u})}{\mathfrak{u}^{\frac{d-2}{2}+1}}\ ,
\end{equation}
from which
\begin{equation}
    \mathcal{I}_{0i}=i\frac{q_i}{q}\sqrt{\frac{\pi}{2}}\,
    \mathcal{I} \Big(\alpha=\tfrac{d-2}{2}+1,\, n=0\Big)+\xi_i\,
    \mathcal{I} \Big(\alpha=\tfrac{d-2}{2}+1,\, n=-\tfrac{1}{2}\Big)\ .
\end{equation}
Finally, the stress sector reads
\begin{equation}
\begin{aligned}
    \mathcal{I}_{ij}&=\int_{\xi}^{+\infty}d\mathfrak{u}\  \mathfrak{u}\Bigg\{\Big(S_{\text{MP}}\cdot \tilde{\chi}\Big)_i\Big(S_{\text{MP}}\cdot \tilde{\chi}\Big)_j\\
    &\quad +r\Bigg[ \tilde{\chi}_i \Big(S_{\text{MP}}\cdot \tilde{\chi}\Big)_j+\tilde{\chi}_j \Big(S_{\text{MP}}\cdot \tilde{\chi}\Big)_i\Bigg]+r^2\tilde{\chi}_i \tilde{\chi}_j\Bigg\}\frac{J_{\frac{d-2}{2}}(\mathfrak{u})}{\mathfrak{u}^{\frac{d-2}{2}}}\ ,
\end{aligned}
\end{equation}
which can be recasted as
\begin{equation}
    \mathcal{I}_{ij}=\int_{\xi}^{+\infty}d\mathfrak{u}\  \mathfrak{u}\Bigg\{S_{\text{MP}}^{i a}S_{\text{MP}}^{j b}+r\Big[ \delta^{i a}S_{\text{MP}}^{j b}+\delta^{j a}S_{\text{MP}}^{i b}\Big]+r^2\delta^{i a}\delta^{j b}\Bigg\}\tilde{\chi}_a\tilde{\chi}_b\frac{J_{\frac{d-2}{2}}(\mathfrak{u})}{\mathfrak{u}^{\frac{d-2}{2}}}\ .
\end{equation}
Using Eq.~\eqref{eq:DoubleChiTilde} and the identities
\begin{equation}
    \Big(-S_{\text{MP}}\cdot S_{\text{MP}}\Big)_{ij}=a_i^2\delta_{ij}\ , \qquad S^{ij}+S^{ji}=0\ ,
\end{equation}
one obtains
\begin{equation}
\begin{aligned}
\mathcal{I}_{ij}&=\delta_{ij}\,\mathcal{I} \Big(\alpha=\tfrac{d-2}{2}+1,\, n=-\tfrac{1}{2}\Big)+\xi_i\xi_j\,\mathcal{I} \Big(\alpha=\tfrac{d-2}{2}+2,\, n=-\tfrac{1}{2}\Big)\\
&-\frac{q_i q_j}{q^2}\,\mathcal{I} \Big(\alpha=\tfrac{d-2}{2}+2,\, n=+\tfrac{1}{2}\Big)+i\frac{q_i\xi_j+q_j\xi_i}{q}\sqrt{\frac{\pi}{2}}\,\mathcal{I} \Big(\alpha=\tfrac{d-2}{2}+2,\, n=0\Big)\ .
\end{aligned}
\end{equation}
Putting all pieces together, the momentum–space metric in covariant form ultimately reads
\begin{equation}\label{eq:MPKSMetricFinal}
\begin{aligned}
    \tilde{h}_{\mu\nu}&=\frac{\kappa^2m}{2}\Bigg(\frac{1}{2}\frac{2^{d/2}\Gamma(d/2)}{d-1}\Bigg)\frac{1}{q^2}\Bigg\{u_\mu u_\nu \Bigg(\frac{J_{\frac{d-2}{2}-1}(\xi)}{\xi^{\frac{d-2}{2}-1}}+\frac{J_{\frac{d-2}{2}}(\xi)}{\xi^{\frac{d-2}{2}}}\Bigg)\\
    &+\Big(u_\mu \xi_\nu+u_\nu \xi_\mu\Big)\frac{J_{\frac{d-2}{2}}(\xi)}{\xi^{\frac{d-2}{2}}}+i\sqrt{\frac{\pi}{2}}\frac{u_\mu q_\nu+u_\nu q_\mu}{q}\frac{J_{\frac{d-2}{2}-\frac{1}{2}}(\xi)}{\xi^{\frac{d-2}{2}-\frac{1}{2}}}\\
    &+\xi_\mu\xi_\nu\frac{J_{\frac{d-2}{2}+1}(\xi)}{\xi^{\frac{d-2}{2}+1}}+\Bigg(\eta_{\mu\nu}-2\frac{q_\mu q_\nu}{q^2}\Bigg)\frac{J_{\frac{d-2}{2}}(\xi)}{\xi^{\frac{d-2}{2}}}+i\sqrt{\frac{\pi}{2}}\frac{\xi_\mu q_\nu+\xi_\nu q_\mu}{q}\frac{J_{\frac{d-2}{2}+\frac{1}{2}}(\xi)}{\xi^{\frac{d-2}{2}+\frac{1}{2}}}\Bigg\}\ .
\end{aligned}
\end{equation}

\section{The eikonal expansion}\label{Sec:Eik_Expansion}

\begin{figure}[h]
\centering
\includegraphics[width=0.4\textwidth, valign=c]{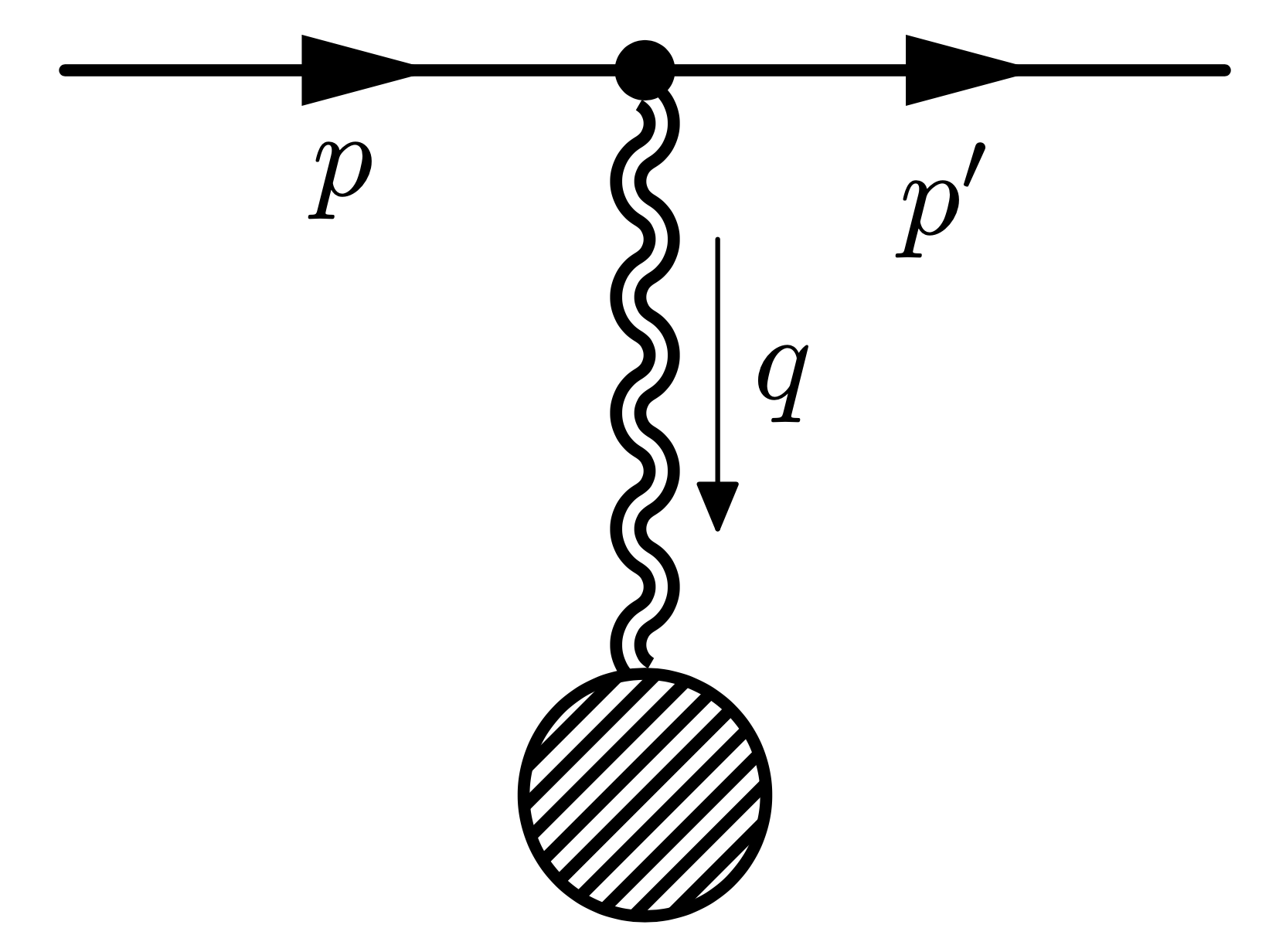}
\caption{Trilinear scattering amplitude between a scalar probe and a fixed Myers-Perry background.}
\label{TrilinearAmpKS}
\end{figure}

Let us now consider the interaction of a probe with the Myers–Perry background in the KS gauge. From Eq.~\eqref{eq:ActionExpanded}, the trilinear interaction term always takes the form
\begin{equation}\label{eq:InteractionLagrangian}
    \mathcal{L}_{\text{int}}=\frac{1}{2}\,h_{\mu\nu}T^{\mu\nu}\ ,
\end{equation}
where, in general, $h_{\mu\nu}$ denotes the graviton field and $T_{\mu\nu}$ is the EMT of the source that induces the spacetime geometry. The same expression, however, can be used to describe the coupling between a probe and a fixed background. Consider for instance a massive extended object scattering off a stationary BH geometry. In the probe limit in which $m_p \ll m$, with $m$ the BH mass and $m_p$ the probe mass, the EMT of the probe can be modeled as that of a scalar field
\begin{equation}\label{eq:EMTscalarKS}
    T_{\mu\nu}=p_\mu p'_\nu+p_\nu p'_\mu-\eta_{\mu\nu}\,(p \cdot  p'-m_p^2)\ ,
\end{equation}
where $p$ and $p'$ are respectively the ingoing and outgoing probe momenta and where in this context we have 
\begin{equation}
    q=p-p'\ ,\qquad \ell=p+p'\ ,
\end{equation}
such that $q\cdot\ell=0$. Then the trilinear scattering amplitude represented in Fig.~\ref{TrilinearAmpKS} becomes
\begin{equation}\label{eq:TriLinearAMPKS}
    i\mathcal{M}=i\,\frac{1}{2}\,\tilde{h}^{\mu\nu}T_{\mu\nu}=i\,\tilde{h}^{\mu\nu}p_\mu p'_\nu\ ,
\end{equation}
where the background metric in momentum space is given by Eq.~\eqref{eq:MPKSMetricFinal} and where Eq.~\eqref{eq:EMTscalarKS} has been used in the last step. Crucially, to derive Eq.~\eqref{eq:TriLinearAMPKS} we exploit the distinctive KS property 
\begin{equation}
    \tilde{h}\equiv \eta^{\mu\nu}\tilde{h}_{\mu\nu}=0\ ,
\end{equation}
which removes the trace term from the contraction.

\begin{figure}[h]
\centering
\includegraphics[width=0.5\textwidth, valign=c]{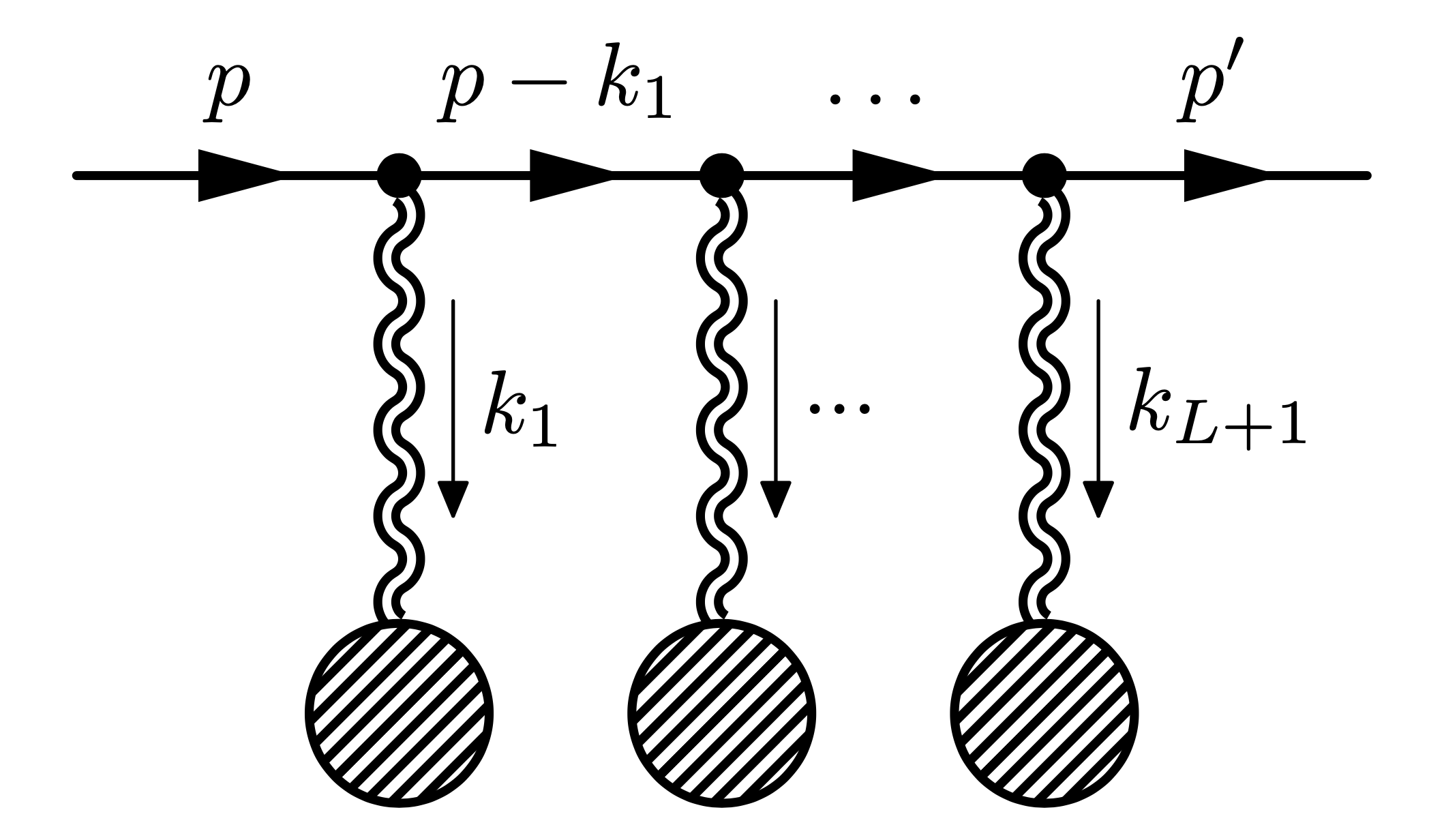}
\caption{Full scattering amplitude describing $(L + 1)$-graviton exchange between the probe and the background source.}
\label{L_loop_Amp}
\end{figure}

We can now write the explicit form of Eq.~\eqref{eq:TriLinearAMPKS} for a Myers–Perry background. Considering
\begin{equation}
    i\mathcal{M}(p,p',q)=i\,\tilde{h}^{\mu\nu}p_\mu p'_\nu
    =-i\Big(\tilde{h}_{00}E^2+\tilde{h}_{0i}\,E\,(p'_i+p_i)+\tilde{h}_{ij}\,p_i p'_j\Big)\ ,
\end{equation}
where $E=E'$ is the energy of the probe (in the elastic limit), the amplitude evaluates to
\begin{equation}\label{eq:OffShellMPKSAMP}
\begin{aligned}
    i\mathcal{M}(p,p',q)&=\frac{\kappa^2 m}{2}\Bigg(\frac{1}{2}\frac{2^{d/2}\Gamma(d/2)}{d-1}\Bigg)\frac{1}{q^2}\Bigg\{E^2 \Bigg(\frac{J_{\frac{d-2}{2}-1}(\xi)}{\xi^{\frac{d-2}{2}-1}}+\frac{J_{\frac{d-2}{2}}(\xi)}{\xi^{\frac{d-2}{2}}}\Bigg)\\
    &\quad+\xi \cdot (p+p')\,\frac{J_{\frac{d-2}{2}}(\xi)}{\xi^{\frac{d-2}{2}}}
    +i\sqrt{\frac{\pi}{2}}\frac{q \cdot  (p+p')}{q}\,\frac{J_{\frac{d-2}{2}-\frac{1}{2}}(\xi)}{\xi^{\frac{d-2}{2}-\frac{1}{2}}}\\
    &\quad+(\xi \cdot  p)(\xi \cdot  p')\,\frac{J_{\frac{d-2}{2}+1}(\xi)}{\xi^{\frac{d-2}{2}+1}}
    +\Bigg(p \cdot  p'-2\frac{(q \cdot  p)(q \cdot  p')}{q^2}\Bigg)\frac{J_{\frac{d-2}{2}}(\xi)}{\xi^{\frac{d-2}{2}}}\\
    &\quad+i\sqrt{\frac{\pi}{2}}\frac{(\xi \cdot  p)(q \cdot  p')+(\xi \cdot  p')(q \cdot  p)}{q}\,
    \frac{J_{\frac{d-2}{2}+\frac{1}{2}}(\xi)}{\xi^{\frac{d-2}{2}+\frac{1}{2}}}\Bigg\}\ .
\end{aligned}
\end{equation}
Notice that in the expression above the momenta $p$ and $p'$ are kept off shell, since for later purposes we will employ this amplitude within loop calculations. However, in the on-shell limit where $p_i=p'_i$ (hence $q \cdot p=q \cdot p'=0$), the amplitude simplifies to
\begin{equation}
\begin{aligned}
    &i\mathcal{M}_{\text{on-shell}}(p,p',q)=\frac{\kappa^2 m}{2}\Bigg(\frac{1}{2}\frac{2^{d/2}\Gamma(d/2)}{d-1}\Bigg)\frac{1}{q^2}\Bigg\{E^2 \Bigg(\frac{J_{\frac{d-2}{2}-1}(\xi)}{\xi^{\frac{d-2}{2}-1}}+\frac{J_{\frac{d-2}{2}}(\xi)}{\xi^{\frac{d-2}{2}}}\Bigg)\\
    &+\xi \cdot (p+p')\,\frac{J_{\frac{d-2}{2}}(\xi)}{\xi^{\frac{d-2}{2}}}
    +(\xi \cdot  p)(\xi \cdot  p')\,\frac{J_{\frac{d-2}{2}+1}(\xi)}{\xi^{\frac{d-2}{2}+1}}
    +m_p^2\,\frac{J_{\frac{d-2}{2}}(\xi)}{\xi^{\frac{d-2}{2}}}\Bigg\}\ ,
\end{aligned}
\end{equation}
since all terms proportional to $q\cdot p$ or $q \cdot p'$ vanish. Indeed, the contributions multiplying the half–integer–order Bessel functions are precisely those that carry these kinematic prefactors, and hence they drop out on shell. The key aspect in such a construction, is that in the case of a massive scalar probe, the defining properties of the KS gauge guarantee that Eq.~\eqref{eq:TriLinearAMPKS} does not merely represent the tree-level interaction, but it actually corresponds to the fundamental (and only) building block needed to construct loop amplitudes. Indeed, since the KS gauge describes the exact BH geometry already at 1PM accuracy, this means that in amplitude language the trilinear interaction vertex in Fig.~\ref{TrilinearAmpKS} is the only vertex appearing in the perturbative expansion of Eq.~\eqref{eq:MassiveVerticesExpansion}, and all higher-order contributions are obtained by iterating such trilinear building block. Diagrammatically, every higher-order process is thus represented by ``comb-like'' loop diagrams made exclusively of these single vertices connected along the probe’s world-line, as illustrated in Fig.~\ref{L_loop_Amp}. 
This property implies that the $L$-loop scattering amplitude can be entirely reconstructed from the tree-level building block in Eq.~\eqref{eq:OffShellMPKSAMP}, namely
\begin{equation}\label{Amp_Lloop_generic}
i\mathcal{M}^{(L+1)}=\int \prod_{i=1}^{L}\frac{d^d k_i}{(2\pi)^d}\prod_{i=1}^{L+1}i\mathcal{M}(p_{i-1}, p_i,\vec{k}_i\, ) 
\prod_{i=1}^{L}\frac{i}{p_i^2-m_p^2+i\varepsilon}\ ,
\end{equation}
where $p_{i-1}=p-\sum_{j=1}^{i-1}k_j$, $p_i=p_{i-1}-k_i$, and $k_{L+1}=q-\sum_{i=1}^{L}k_i$.  
This recursive structure encapsulates the entire tower of loop corrections in terms of trilinear vertices only, a remarkable simplification specific to the KS representation of the background geometry. 

To connect these scattering amplitudes to physical observables, we now turn the attention to the standard eikonal expansion~\cite{Amati:1987wq,Amati:1987uf,Amati:1990xe,Verlinde:1991iu,Kabat:1992tb,Levy:1969cr,AccettulliHuber:2020oou}.  
In the eikonal approximation, the $S$-matrix is expressed as
\begin{equation}
\widetilde{S}(p,\vec{b}\,)=1+i\,\widetilde{\mathcal{T}}(p,\vec{b}\,)=\Big(1+2i\Delta(p, \vec{b}\,)\Big)e^{2i\,\delta(p,\vec{b}\,)}\ ,
\end{equation}
where $\delta(p,\vec{b}\,)$ is the eikonal phase depending on the classical probe momentum $p$ and the impact parameter $\vec{b}$, and $\Delta(p, \vec{b}\,)$ encode non-exponentiating pieces. Notice that the remainder term $\Delta$ is only relevant starting from 3PM~\cite{DiVecchia:2023frv}, hence, in this thesis, we can set $\Delta=0$ since we will consider explicit calculations of the eikonal phase only up to one-loop.
Expanding perturbatively in $G_N$, one obtains
\begin{equation}\label{seriesexpeikonal}
i\,\widetilde{\mathcal{T}}(p,\vec{b}\,)=i\sum_{n=1}^{+\infty}\widetilde{\mathcal{M}}^{(n)}(p,\vec{b}\,)
=\sum_{k=1}^{+\infty}\frac{1}{k!}\left(2i\sum_{n=1}^{+\infty}\delta^{(n)}(p,\vec{b}\,)\right)^k\ ,
\end{equation}
with the index $n$ labeling the PM order, and with $i\widetilde{\mathcal{M}}^{(n)}$ denoting the amplitude in impact-parameter space, defined as
\begin{equation}\label{Amp_in_IPS}
\widetilde{\mathcal{M}}^{(n)}(p,\vec{b}\,)=\frac{1}{2|\vec{p}\,|}\int \frac{d^{d-1}q}{(2\pi)^2}\,e^{i\vec{q}\cdot\vec{b}}\,\mathcal{M}^{(n)}\ ,
\end{equation}
where the integration runs over the momentum transfer $\vec{q}$ on the plane orthogonal to the longitudinal direction $\vec{\ell}$. From the eikonal phase $\delta(p,\vec{b}\,)$ one can then extract various classical observables. One example is the deflection angle between the incoming and outgoing trajectories of the scattered particle, given by~\cite{Amati:1990xe, DiVecchia:2023frv}
\begin{equation}\label{Deflection_Ang}
2\sin\frac{\vartheta(p,\vec{b}\,)}{2}=-\frac{2}{|\vec{p}\,|}\frac{\partial\delta(p,\vec{b}\,)}{\partial b}\ ,
\end{equation}
where $b=|\vec{b}\,|$. So then, at each loop order $L$, by computing the corresponding amplitude and using Eq.~\eqref{seriesexpeikonal}, one obtains the eikonal phase at each PM order, thus establishing a direct link between scattering amplitudes and classical gravitational observables. Specifically, at 1PM the expansion of Eq.~\eqref{seriesexpeikonal} yields
\begin{equation}\label{1PM_Eikonal_Phase}
\widetilde{\mathcal{M}}^{(1)}(p,\vec{b}\,)=2\,\delta^{(1)}(p,\vec{b}\,)\ ,
\end{equation}
while at 2PM order one finds
\begin{equation}\label{2PM_Eikonal_Formal}
\widetilde{\mathcal{M}}^{(2)}(p,\vec{b}\,)=2\,\delta^{(2)}(p,\vec{b}\,)
-\frac{i}{2}\big(2i\,\delta^{(1)}(p,\vec{b}\,)\big)^2\ ,
\end{equation}
and analogous relations hold for higher PM orders. Starting from 2PM, the amplitude in impact-parameter space naturally decomposes into two classes of contributions. The first type, the so-called \emph{hyper-classical} terms, arises from combinations of lower-order phases and reflects the exponentiation of the eikonal phase itself. The second type, the genuinely \emph{classical} terms, stems from the new PM corrections to $\delta^{(n)}$ and represents the nontrivial physical information of the scattering process. In standard on-shell approaches, classical terms originate from amplitudes that are massive-particle irreducible, while hyper-classical ones arise from reducible diagrams.  
By contrast, in our KS formulation, no vertices with multiple gravitons exist, thus the entire eikonal expansion is built from the repeated insertion of the single trilinear vertex shown in Fig.~\ref{L_loop_Amp}, using the amplitude~\eqref{eq:OffShellMPKSAMP}.  
Our next goal then is to clarify how such contributions are systematically organized and computed. Before diving into that, we notice that the eikonal phase $\delta$ contains in principle both quantum and classical pieces, and we wish to isolate the latter following the KMOC formalism of Ref.~\cite{Kosower:2018adc}. Since the eikonal phase is related to the classical action through
\begin{equation}
\delta \sim \frac{1}{\hbar}\,S_{\text{cl.}}\ ,
\end{equation}
it scales as $\mathcal{O}(\hbar^{-1})$. Therefore, to extract the classical contribution from the full amplitude, one must keep track of the appropriate $\hbar$ dependence of each term. Analogously to the discussion in Chapter~\ref{chapter:ClassicalGravityFromAmplitudes}, we treat every internal momentum as a ``quantum’’ quantity by performing the substitution
\begin{equation}
q,\;k_i\ \rightarrow\ \hbar\, q,\;\hbar\,k_i\ .
\end{equation}
More explicitly, the momentum of the intermediate scalar line scales as
\begin{equation}
p_i \rightarrow \frac{1}{2}\,\ell+\frac{1}{2}\,\hbar q - \hbar\sum_{j=1}^{i}k_j\ ,
\end{equation}
where $\vec{\ell}$ denotes the fixed longitudinal momentum characterizing the physical process.  
At the same time, we rescale the spin tensor according to
\begin{equation}
S_{\mu\nu}\ \rightarrow\ \frac{1}{\hbar}\,S_{\mu\nu}\ ,
\end{equation}
so that the quantity $\xi_\mu$ introduced earlier remains purely classical.  
Furthermore, each vertex contributes an additional inverse power of $\hbar$. The resulting eikonal phase, obtained after consistently applying these $\hbar$-counting rules, is then a function of the longitudinal momentum $\ell$, and at every PM order it can be replaced in favor of either the incoming or outgoing momentum, since their difference is of quantum order and thus vanishes in the classical limit.

Let us now sketch how to organize the calculations in this framework. At tree level the ``in'' and ``out'' scalar momenta are on shell, and Eq.~\eqref{Amp_Lloop_generic} reduces trivially to $i\mathcal{M}^{(1)}=i\mathcal{M}(p, p-q, \vec{q}\,)$. Using Eq.~\eqref{1PM_Eikonal_Phase}, we see that this tree amplitude in impact-parameter space is directly proportional to $\delta^{(1)}$, which scales as $\mathcal{O}(\hbar^{-1})$. From the $\hbar$-counting rules introduced above, obtaining an $\mathcal{O}(\hbar^{-1})$ eikonal phase requires the momentum–space amplitude to scale as $\mathcal{O}(\hbar^{-d})$. This is precisely achieved by taking
\begin{equation}
    i\mathcal{M}^{(1)}=i\mathcal{M}(p, p, \vec{q}\,)
\end{equation}
and replacing $\vec{\ell}$ with $2\vec{p}$ (or $2\vec{p}\, '$), since the difference between these choices is of quantum order and therefore irrelevant in the classical limit. Then, at one loop the amplitude reads
\begin{equation}\label{Amp_2loops}
i\mathcal{M}^{(2)}=\int \frac{d^d k}{(2\pi)^d}\,i\mathcal{M}(p, p-k,\vec{k}\, ) \,\frac{i}{(p-k)^2-m_p^2+i\varepsilon}\, i\mathcal{M}(p-k, p-q,\vec{q}-\vec k\, ) \ ,
\end{equation}
and from Eq.~\eqref{2PM_Eikonal_Formal}, it yields two distinct contributions once mapped to impact-parameter space. The hyper-classical term in Eq.~\eqref{2PM_Eikonal_Formal} must scales as $\mathcal{O}(\hbar^{-2})$ and is then reconstructed by extracting
\begin{equation}
i\mathcal{M}^{(2)}\Big|_{\text{hyp.\,cl.}}=\mathcal{O}(\hbar^{-(d+1)})\ \longrightarrow\ i\widetilde{\mathcal{M}}^{(2)}\Big|_{\text{hyp.\,cl.}}=2 i\big(\delta^{(1)}\big)^2=\mathcal{O}(\hbar^{-2})\ .
\end{equation}
More generally, at any PM order the entire hyper-classical sector is recovered by suitable convolutions of lower-order momentum–space amplitudes. Explicitly, at 2PM the relevant contribution is
\begin{equation}
i \mathcal{M}^{(2)}\Big|_{\text{hyp.\,cl.}}=\frac{1}{2}\int\frac{d^dk}{(2\pi)^d}\,i\mathcal{M}(p, p, \vec k\, )\,i \mathcal{M}(p, p, \vec q-\vec k\, )\;2\pi\, \delta(\vec \ell \cdot  \vec k\, )\ ,
\end{equation}
where the delta function arises from expanding the propagator at order $\mathcal{O}(\hbar^{-1})$ (see Eq.~(5.4) of~\cite{Brandhuber:2021eyq}). Integrating over the transverse loop momenta and transforming to impact-parameter space, one obtains
\begin{equation}
\begin{aligned}
i\widetilde{\mathcal{M}}^{(2)}\Big|_{\text{hyp.\,cl.}}
&=\frac{1}{2}\frac{1}{4 p^2}\int\frac{d^{d-1} q}{(2\pi)^{d-1}}\int \frac{d^{d-1}k}{(2 \pi)^{d-1}}\,e^{i\vec q\cdot \vec b}\,i\mathcal{M}_{KN}(p, p, \vec k\, )\,i\mathcal{M}_{KN}(p,p,  \vec q-\vec k\,)\\
&=\frac{1}{2}\Big(i \widetilde{\mathcal{M}}^{(1)}\Big)^2=\frac{1}{2}\Big(2 i \delta^{(1)}\Big)^2\ ,
\end{aligned}
\end{equation}
in exact agreement with Eq.~\eqref{2PM_Eikonal_Formal}. The same strategy extends straightforwardly to higher orders, including mixed terms built from different lower-PM ingredients. On the contrary, the genuinely classical contribution at a given PM order is the nontrivial part of the eikonal phase, namely the piece that cannot be reconstructed by exponentiation. From the $\hbar$ power counting, at 2PM one finds the same scaling of the leading order contribution, leading to the conclusion that at every loop order one has
\begin{equation}
i \mathcal{M}^{(L)}\Big|_{\text{cl.}}=\mathcal{O}(\hbar^{-d})\ .
\end{equation}
Now for the physical interpretation of the KS gauge, and how it relates to known results in literature, it is useful to split the building-block amplitude as
\begin{equation}\label{M0_Mextra}
i\mathcal{M}=i\mathcal{M}_0+i\mathcal{M}_{\text{extra}}\ ,
\end{equation}
where $\mathcal{M}_0$ collects all terms that survive on shell, while $\mathcal{M}_{\text{extra}}$ contains those contributions that vanish for on-shell external scalars but nonetheless contribute to Eq.~\eqref{Amp_2loops}. With this decomposition, the one-loop classical piece naturally splits into three parts
\begin{equation}\label{MKN_0extra}
    i\mathcal{M}^{(2)}\Big|_{\text{cl.}}=i\mathcal{M}^{(2)}\Big|_{\text{cl.}}^{(0,0)}+i\mathcal{M}^{(2)}\Big|_{\text{cl.}}^{(\text{extra},\text{extra})}+2\,i\mathcal{M}^{(2)}\Big|_{\text{cl.}}^{(\text{extra},0)}\ .
\end{equation}
For the $(\text{extra},\text{extra})$ term, one can show that the massive propagator cancels out, and the resulting structure behaves as if arising from an effective contact vertex, schematically represented as
\begin{equation}\label{Eq:ContactTerm}
i\mathcal{M}^{(2)}\Big|_{\text{cl.}}^{(\text{extra},\text{extra})} \sim \includegraphics[width=0.2\textwidth, valign=c]{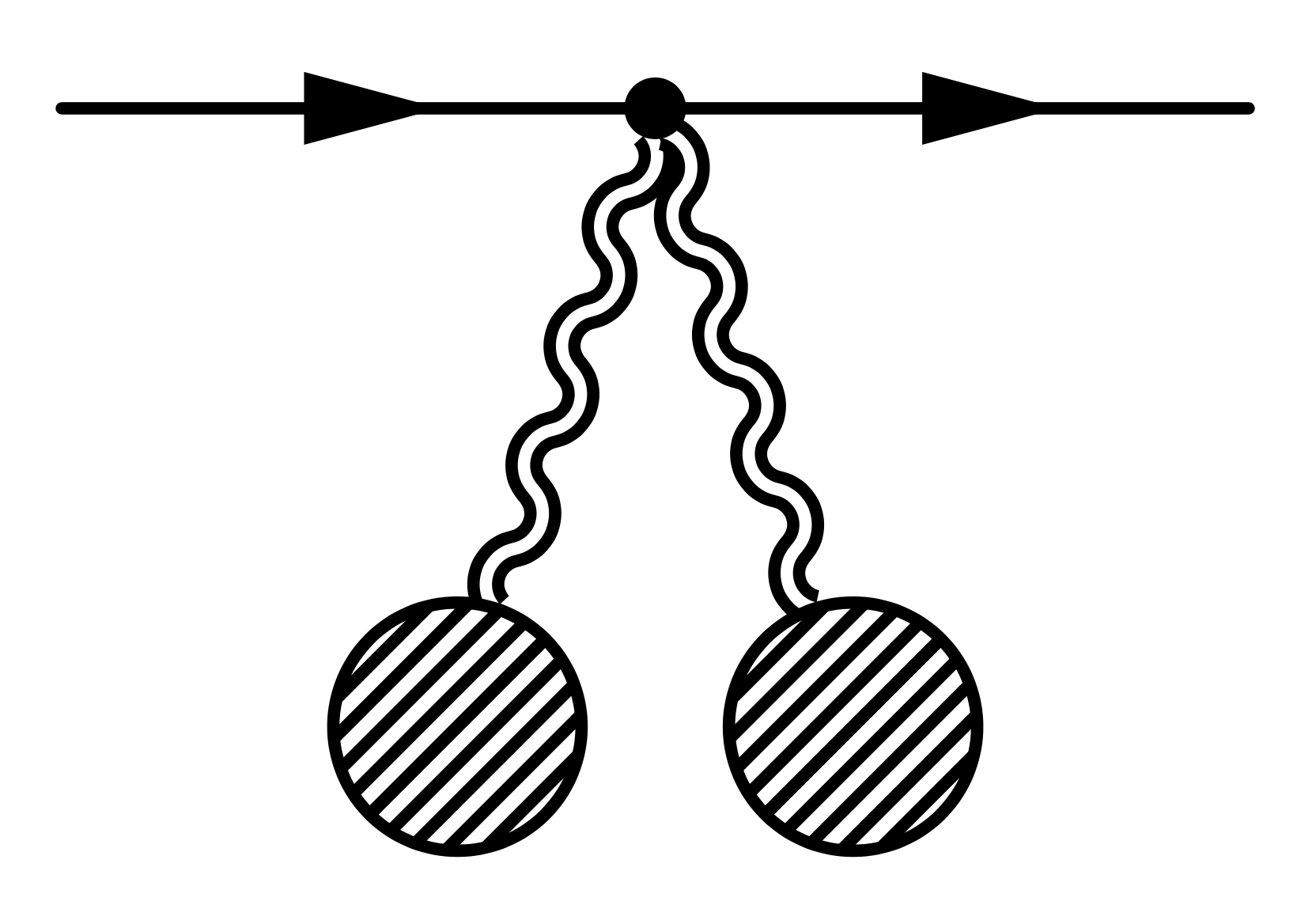}\ .
\end{equation}
By contrast, the mixed term $(\text{extra},0)$ produces an imaginary contribution which vanishes upon integration. This is consistent with the fact that genuine radiation effects, and hence nonvanishing imaginary parts, start at 3PM~\cite{Damour:2020tta}. Consequently, the classical 2PM eikonal phase is determined by the sum of the $(0,0)$ term (where the propagator remains explicit) and the $(\text{extra},\text{extra})$ term (which mimics the needed contact interaction present in standard literature). Specializing to Kerr BHs for simplicity, the two pieces $\mathcal{M}_{0}$ and $\mathcal{M}_{\text{extra}}$ take the form
\begin{equation}\label{Mamplitude0}
\begin{aligned}
&i \mathcal{M}_0 (p_{i-1}, p_i, \vec k_i  )= i\frac{8\pi G_N m}{|\vec k_i|^2}\Bigg\{ E^2\cos|\vec{a}{\times}\vec{k}_i|+iE\,
 j_0(|\vec{a}{\times}\vec{k}_i|)\,(\vec{a}\times \vec{k}_i) \cdot ( \vec p_i + \vec p_{i-1})  \\
&+\,{j_0(|\vec{a}{\times}\vec{k}_i|)}\left(\vec p_i  \cdot  \vec p_{i-1}-2\frac{\vec k_i \cdot  \vec p_i\ \vec k_i \cdot  \vec p_{i-1}}{|\vec k_i|^2}\right)
-\frac{j_1(|\vec{a}{\times}\vec{k}_i|)}{|\vec{a}{\times}\vec{k}_i|}\,(\vec{a}\times\vec{k}_i) \cdot  \vec p_i \,(\vec{a}\times\vec{k}_i) \cdot  \vec p_{i-1}\Bigg\} \ ,
\end{aligned}
\end{equation}
and
\begin{equation}
\begin{aligned}
&i \mathcal{M}_{\text{extra}} (p_{i-1}, p_i, \vec k_i  )= i\frac{4\pi^2 G_N m}{|\vec k_i|^2}\Bigg\{ -iE\,\frac{\vec k_i  \cdot  (\vec p_i + \vec p_{i-1})}{|\vec k_i|}\,J_0(|\vec{a}{\times}\vec{k}_i|) \label{Mamplitudeextra} \\
&+\,\frac{1}{|\vec k_i|}\frac{J_1(|\vec{a}{\times}\vec{k}_i|)}{|\vec{a}{\times}\vec{k}_i|}\Bigl(\vec k_i \cdot  \vec p_{i-1}\, (\vec{a}\times \vec{k}_i) \cdot  \vec p_i+\vec k_i \cdot  \vec p_i \, (\vec{a}{\times}\vec{k}_i) \cdot  \vec p_{i-1}\Big)\Bigg\}\ ,
\end{aligned}
\end{equation}
which makes manifest the fact that, in $d=3$, the ordinary Bessel terms populate precisely the $\mathcal{M}_{\text{extra}}$ sector, while the spherical Bessel functions characterize the $\mathcal{M}_0$ piece of the amplitude. In what follows, we will illustrate explicitly how the loop integrals are performed for recoverying the classical term of the amplitude. Specifically, we will extract the leading eikonal contribution for Kerr BHs, and then analyze the first subleading correction, the one-loop (2PM) term, in detail for the Schwarzschild case. We stress that the method is completely general and, in principle, allows one to determine the deflection angle and other observables at arbitrary PM order thanks to a closed-form analytic expression.

\section{Subleading eikonal contributions}\label{sec:SubEik}

Let us now discuss the central feature of our KS approach, namely that higher orders in the PM expansion, including the classical pieces, arise solely from the comb–like diagram of Fig.~\ref{L_loop_Amp}. Since the hyper–classical terms have already been addressed in Sec.~\ref{Sec:Eik_Expansion}, we focus here on the classical contributions of Eq.~\eqref{MKN_0extra}. We first treat the Schwarzschild case and show how our method reproduces known results, and then briefly outline the extension to Kerr. We also comment on why graviton self–interaction can be neglected in the probe–limit approximation.

\subsection{2PM Schwarzschild classical term }

As discussed in Sec.~\ref{Sec:Eik_Expansion}, the 2PM classical eikonal phase stems from those terms in the one–loop amplitude \eqref{Amp_2loops} that scale as $\mathcal{O}(\hbar^{-d})$. Using the decomposition \eqref{M0_Mextra}, the 1–loop amplitude is organized as in Eq.~\eqref{MKN_0extra}, and the computation reduces to evaluating these three contributions. For simplicity, we now take the background to be Schwarzschild, namely considering Eqs.~\eqref{Mamplitude0} and~\eqref{Mamplitudeextra} in the limit $a\to 0$, further restricting to the $d=3$ case. We begin with the $(0,0)$ contribution in Eq.~\eqref{MKN_0extra}. Adopting the $\hbar$–expansion rules for the scalar propagator, one finds
\begin{equation}
\frac{1}{(p-\hbar k)^2-m_p^2+i\varepsilon}=\frac{1}{\hbar \,\vec \ell \cdot  \vec k +i\varepsilon}+\frac{|\vec k\, |^2- \vec k \cdot \vec q}{(\vec k \cdot  \vec \ell\, )^2}+\mathcal{O}(\hbar)\ ,
\end{equation}
so that the classical piece is selected as
\begin{equation}\label{00_Explicit_Amp}
\mathcal{M}^{(2)}\Big|^{(0, 0)}_{cl.}=8G_N^2m^2\pi^2\,(3|\vec p\, |^2+4 E^2)\int\frac{d^3k}{(2\pi)^3}\frac{1}{|\vec k\, |^2\,|\vec k -\vec q\, |^2}\ ,
\end{equation}
where integration by part (IBP) reduction techniques have been performed with \texttt{LiteRed}~\cite{Lee:2012cn,Smirnov:2013dia,Lee:2013mka}. Using the master integral~\cite{Mougiakakos:2020laz,DOnofrio:2022cvn}
\begin{equation}
\int\frac{d^3k}{(2\pi)^3}\frac{1}{|\vec k\, |^2\,|\vec k -\vec q\, |^2} =\frac{1}{8}\frac{1}{|\vec q\, |}\ ,
\end{equation}
we then obtain
\begin{equation}\label{M_00_explicit}
\mathcal{M}^{(2)}\Big|^{(0, 0)}_{cl.}=G_N^2m^2\pi^2\,(3|\vec p\, |^2+4 E^2)\,\frac{1}{|\vec q\, |}\ .
\end{equation}
This contribution represents the classical part of the pure comb–like diagram in our KS setup in the $d=3$ Schwarzschild case. We next address the $(\text{extra},\text{extra})$ piece of Eq.~\eqref{MKN_0extra}. This is the characteristic KS–gauge contribution, built from the terms $i\mathcal{M}_{extra}$ that vanish on shell but become visible off shell. These ``extra’’ pieces mimic contact interactions within comb–like diagrams, as in Eq.~\eqref{Eq:ContactTerm}. Indeed, one finds the exact cancellation of the propagator\footnote{The same cancellation arises if one uses $i\mathcal{M}_{extra}(p-k, p-q, \vec q-\vec k)$ instead.}
\begin{equation}\label{scalar_contact}
i\mathcal{M}_{extra}(p, p-k, \vec k)\,\frac{i}{(p-k)^2-m_p^2+i\varepsilon}=i\,\frac{4 G_N m \pi^2 E}{|\vec k\, |^3}\ ,
\end{equation}
leading to
\begin{equation}
\mathcal{M}^{(2)}\Big|^{(extra, extra)}_{cl.}=-16\, G_N^2m^2\pi^4 E^2 \int \frac{d^3 k}{(2\pi)^3}\frac{\vec k \cdot  (\vec q-\vec k\, )}{|\vec k\, |^3\,|\vec q- \vec k|^3}\ .
\end{equation}
Considering then
\begin{equation}
\int\frac{d^3 k}{(2\pi)^3}\frac{\vec k \cdot  (\vec q-\vec k\, )}{|\vec k\, |^3\,|\vec q- \vec k|^3}=-\frac{1}{2\pi^2}\,\frac{1}{|\vec q\, |} \ , 
\end{equation}
we arrive at
\begin{equation}\label{M_extraextra_Explicit}
\mathcal{M}^{(2)}\Big|^{(extra, extra)}_{cl.}=8\, G_N^2m^2\pi^2 E^2\,\frac{1}{|\vec q\, |}\ .
\end{equation} 
Finally, consider the mixed $(\text{extra},0)$ term. A direct evaluation gives
\begin{equation}\label{M_0extra_explicit}
2\,\mathcal{M}^{(2)}\Big|^{(extra, 0)}_{cl.}=i\,  G_N^2m^2\pi^3E \int \frac{d^3k}{(2\pi)^3}\ \vec k \cdot  \vec p\ I(\vec k, \vec q\, )\ ,
\end{equation}
with $I$ parametrizing a scalar quantity independent of $\vec p$. By tensor decomposition of the integrand it follows that
\begin{equation}
\int\frac{d^3k}{(2\pi)^3} \vec k \cdot  \vec p\ I(\vec k,\vec q\, )= \vec q \cdot  \vec p\ \tilde{I}(\vec q\, )=0\ ,
\end{equation}
and hence the mixed contribution vanishes identically. As anticipated in Sec.~\ref{Sec:Eik_Expansion}, this is consistent with the fact that \eqref{M_0extra_explicit} is purely imaginary, while genuine radiative (imaginary) effects enter only at 3PM~\cite{Damour:2020tta}.

Collecting the nonvanishing pieces \eqref{M_00_explicit} and \eqref{M_extraextra_Explicit}, the 2PM eikonal phase for the Schwarzschild $d=3$ case ultimately reads
\begin{equation}
\begin{aligned}
\delta^{(2)}(p, \vec b\,)\Big|_{a=0}&=\frac{1}{4|\vec p\, |}\int\frac{d^2q}{(2\pi)^2}e^{i\vec q \cdot \vec b}\Big(\mathcal{M}^{(2)}\Big|^{(0,0)}_{cl.}+\mathcal{M}^{(2)}\Big|^{(extra, extra)}_{cl.}\Big)\\
&=\frac{3\,G_N^2m^2\pi\, E}{8\,v\,b}\,(4+v^2)\ ,
\end{aligned}
\end{equation}
in agreement with~\cite{Damour:2017zjx,KoemansCollado:2018hss}. The same methodology can be, in principle, extended to spinning backgrounds, where the $(0,0)$ channel continues to encode the pure comb–like classical piece, while $(\text{extra},\text{extra})$ reproduces the effective contact–term structure induced by the KS representation. Indeed, in standard on-shell approaches, such as those based on generalized unitarity or EFT matching, the same physical information that we have extracted here from the off-shell amplitude in KS gauge is encoded within the on-shell Compton amplitude involving two massive scalars and two gravitons. In this picture, the contact-term content of the interaction, that is the part of the amplitude that cannot be factorized into lower-point exchanges, is not described by explicit off-shell vertices, but rather emerges from the consistent on-shell completion of the Compton process through minimal and non-minimal couplings of the external states. More precisely, while in the KS formalism contact terms appear automatically as off-shell contributions to the trilinear building block, in the on-shell framework they are reconstructed from the double-copy structure of the Compton amplitude, whose low-energy (classical) limit reproduces the same PM physics. The spin-dependent contributions that in our approach arise from the $\mathcal{M}_{\text{extra}}$ sector of Eq.~\eqref{M0_Mextra} correspond, in the language of on-shell methods, to the non-minimal higher-multipole operators appearing in the gravitational Compton amplitude~\cite{Arkani-Hamed:2017jhn, Guevara:2018wpp, Chung:2018kqs, Aoude:2020onz, Kosmopoulos:2021zoq}. These operators encode the same physical effects, such as tidal and stress multipole interactions, that here are generated directly by the off-shell structure of the KS background. In other words, both formalisms describe the same classical physics, and in the on-shell approach the relevant information is distributed among factorizable Compton subamplitudes and effective higher-derivative operators, while in the off-shell KS gauge all such contributions are geometrically organized into the background metric itself. This correspondence highlights the deep relation between the off-shell KS picture and the on-shell amplitude construction of the gravitational interaction, offering two complementary yet equivalent descriptions of the same underlying dynamics. Moreover, in the probe limit, graviton self–interactions do not contribute to the classical eikonal through 2PM because all required multi–graviton vertices are absent in the KS expansion, and any residual diagrams are suppressed by powers of the small mass ratio. We will discuss this later on.

\subsection{2PM Kerr classical term}

With additional effort, the previous analysis can be extended to the case of rotating black holes. In general, in the KS gauge, it is always convenient to split the tree–level building block of Eq.~\eqref{M0_Mextra} into two components, $i\mathcal{M}_0$ and $i\mathcal{M}_{extra}$, defined in Eqs.~\eqref{Mamplitude0} and~\eqref{Mamplitudeextra}, respectively.\footnote{Here the discussion is restricted to Kerr black holes, but it can be straightforwardly generalized to KN black holes.} Once these ingredients are substituted into Eq.~\eqref{Amp_2loops}, the actual evaluation of the resulting integrals becomes rather intricate. Therefore, although the conceptual structure of the KS approach remains simple and transparent, obtaining an explicit analytic expression for $\delta^{(2)}$, even in the Kerr case, is beyond the scope of this section. Nevertheless, the structure of the integrals allows us to make several general and physically relevant observations. Referring to Eq.~\eqref{Amp_2loops}, the hyper–classical contribution is reproduced by those terms in $i\mathcal{M}^{(2)}$ of order $\mathcal{O}(\hbar^{-4})$, which can be reconstructed through a convolution of the 1PM amplitudes. Hence, the only genuinely new information at this order lies once again in the purely classical term. Its structure follows the pattern of Eq.~\eqref{MKN_0extra}, and, as in the Schwarzschild case, for Kerr three distinct pieces must be computed, each associated with integrals of a different type.

The first $(0,0)$ term involves integrals of the general form
\begin{equation}
i\mathcal{M}^{(2)}\Big|_{(0,0)}^{cl.}=G_N^2m^2 \int \frac{d^3 k}{(2\pi)^3}\,
\frac{j_{a}(|\vec a\times\vec k|)\,j_b(|\vec a\times(\vec q-\vec k)|)}{|\vec k|^{m}|\vec q-\vec k|^{n}}\,
I^{(0,0)}_{a,b,m,n}(\vec k,\vec q,\vec\ell,\vec a\,)\ ,
\end{equation}
where, in the following, $a,b=0,1$, $m,n=0,\dots,4$, and $I^{(0,0)}_{a,b,m,n}(\vec k,\vec q,\vec\ell,\vec a\,)$ are rational functions of the indicated vector and scalar products.  
The $(\text{extra},0)$ contribution has the structure
\begin{equation}
2i\mathcal{M}^{(2)}\Big|_{(extra,0)}^{cl.}=G_N^2m^2 \int \frac{d^3 k}{(2\pi)^3}\,
\frac{J_{a}(|\vec a\times\vec k|)\,j_b(|\vec a\times(\vec q-\vec k)|)}{|\vec k|^{m}|\vec q-\vec k|^{n}}\,
I^{(ex,0)}_{a,b,m,n}(\vec k,\vec q,\vec\ell,\vec a\,)\ ,
\end{equation}
while the $(\text{extra},\text{extra})$ term reads
\begin{equation}
i\mathcal{M}^{(2)}\Big|_{(extra,extra)}^{cl.}=G_N^2m^2 \int \frac{d^3 k}{(2\pi)^3}\,
\frac{J_{a}(|\vec a\times\vec k|)\,J_b(|\vec a\times(\vec q-\vec k)|)}{|\vec k|^{m}|\vec q-\vec k|^{n}}\,
I^{(ex,ex)}_{a,b,m,n}(\vec k,\vec q,\vec\ell,\vec a\,)\ .
\end{equation}
Although the explicit evaluation of these integrals is more involved than in the non–rotating case, their structure still conveys useful general features. First, in direct analogy with the Schwarzschild case, both amplitudes 
${i\mathcal{M}_{extra}(p,p-k)}$ and ${i\mathcal{M}_{extra}(p-k,p-q)}$ exactly cancel the massive propagator, effectively reproducing the contact–term behavior typical of the KS gauge. This cancellation reads explicitly
\begin{equation}
\begin{gathered}
i\mathcal{M}_{extra}(p,p-k,\vec k\,)\,\frac{i}{(p-k)^2-m_p^2+i\varepsilon}\\[3pt]
=\frac{2G_N m \pi^2}{|\vec a\times \vec k\,|\,|\vec k\,|^3}
\Big(2 i E\,|\vec a\times \vec k\,|\,J_0(|\vec a\times\vec k\,|)
-J_1(|\vec a\times\vec k\,|)\,\vec a\times \vec k \cdot (\vec\ell+\vec q)\Big)\ ,
\end{gathered}
\end{equation}
which smoothly reproduces Eq.~\eqref{scalar_contact} in the limit $a\to0$, noticing that an analogous expression holds for $i\mathcal{M}_{extra}(p-k,p-q)$.  

Furthermore, as already observed for Schwarzschild, the $(\text{extra},0)$ mixed term vanishes at leading order, and the same property persists at least up to quadratic order in the BH spin. Indeed, one can prove that
\begin{equation}
2i\mathcal{M}^{(2)}\Big|_{(extra,0)}^{cl.}=O(a^2)\ .
\end{equation}
A full demonstration that this cancellation extends to all orders in the spin parameter $a/b$ is more tedious, but the evidence up to $O(a^2)$ strongly supports this expectation. In summary, although a closed analytic expression for the 2PM eikonal phase $\delta^{(2)}$ in the Kerr case is not yet available, the KS–gauge formalism offers a conceptually clear and computationally structured pathway to compute PM corrections iteratively. The organization of the amplitude into $(0,0)$, $(\text{extra},0)$, and $(\text{extra},\text{extra})$ sectors provides a natural bookkeeping system, and the pattern observed for Schwarzschild continues to hold for rotating black holes. The KS approach therefore appears particularly promising for systematically extending the eikonal expansion to higher orders and to spinning backgrounds.

\subsection{Graviton self-interaction contribution}

Finally, one might wonder whether diagrams involving the exchange of virtual gravitons could already contribute at 2PM. To address this, let us expand the metric around a background written in the KS gauge as
\begin{equation}
\bar{g}_{\mu\nu}=g_{\mu\nu}+\kappa\,\delta h_{\mu\nu}
=\eta_{\mu\nu}+h_{\mu\nu}+\kappa\,\delta h_{\mu\nu}\ ,
\end{equation}
where $\delta h_{\mu\nu}$ denotes the quantum fluctuation (the graviton field), while $h_{\mu\nu}$ represents the classical background.  
When this expansion is substituted into the Einstein–Hilbert action, one obtains interaction vertices both among the gravitons themselves and between the gravitons and the source.  
Because the background metric satisfies the Einstein equations, the trilinear vertex involving two background fields and one graviton fluctuation vanishes
\begin{equation}
\tau_{\delta h,h,h}=0\ ,
\end{equation}
as do all the higher mixed vertices of the type $\tau_{\delta h,h,\ldots,h}=0$ containing an arbitrary number of background insertions.  
This observation is sufficient for Schwarzschild and Kerr spacetimes. For Kerr–Newman, however, one must also include the gauge potential, which is expanded as
\begin{equation}
\bar{A}_\mu=A_\mu+\delta A_\mu\ ,
\end{equation}
where $\delta A_\mu$ is the photon field and $A_\mu$ is the classical background potential in KS gauge, given in Eq.~\eqref{eq:KSpotentialDef}.  
Since $A_\mu$ satisfies the Maxwell equations, the correct combined relation becomes
\begin{equation}
\tau_{\delta h,h,h}+\tau_{\delta h,A,A}=0\ ,
\end{equation}
together with
\begin{equation}
\tau_{\delta A,A,h}=0
\end{equation}
for charged sources.

\begin{figure}[h]
\centering
\includegraphics[width=0.4\textwidth, valign=c]{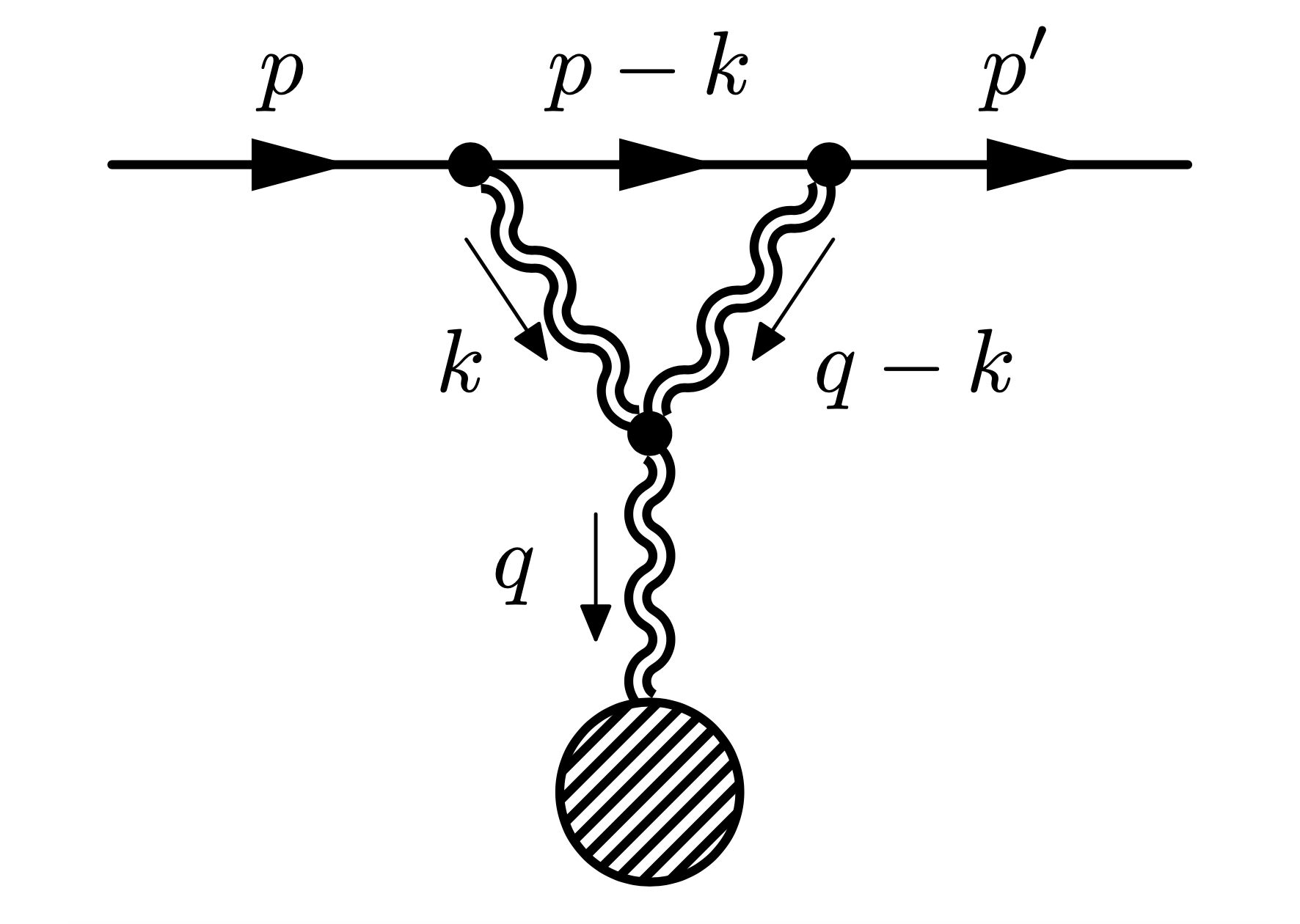}
\caption{Diagram in which the probe interacts with the background through a loop of virtual gravitons.}
\label{3Grav_Diagram}
\end{figure}

At 2PM there is, however, another vertex to consider, namely $\tau_{\delta h,\delta h,h}$, which leads to the one–loop diagram shown in Fig.~\ref{3Grav_Diagram}.
To estimate the contribution of this diagram, we use the trilinear vertex $\tau_{\delta h,\delta h,h}$ as given in~\cite{Donoghue:1995cz}, and the scalar–graviton vertex
\begin{equation}
\tau_\phi^{\mu\nu}(p,p')=-\frac{i}{2}\,\kappa\,\widetilde{T}^\phi_{\mu\nu}(p,p')\ ,
\end{equation}
so that the amplitude reads
\begin{equation}
\begin{aligned}
i\mathcal{M}^{(2)}\Big|_{\delta h,\delta h,h}
=\int\frac{d^4k}{(2\pi)^4}&\,
\frac{i\tau_\phi(p,p-k)\tau_\phi(p-k,p-q)}{(p-k)^2-m_p^2+i\varepsilon}\,
\tau_{\delta h,\delta h,h}(k,q-k)\\
&\times\frac{iP}{k^2+i\varepsilon}\,
\frac{iP}{(q-k)^2+i\varepsilon}\,
i\kappa^{-1}\tilde{h}^{KN}(\vec q\,)\ ,
\end{aligned}
\end{equation}
where $P$ denotes the tensor structure of the graviton propagator, and contractions are left implicit. To extract the classical piece of this amplitude, we recall that it corresponds to
\begin{equation}
i\mathcal{M}^{(2)}\Big|^{cl.}_{\delta h,\delta h,h}=\mathcal{O}(\hbar^{-3})\ .
\end{equation}
This condition requires putting the massive external legs on shell and reducing the scalar propagator to its on–shell form, $\delta(\ell \cdot  k)$, so that the temporal component of the internal momentum is fixed as
\begin{equation}
k_0=\frac{\vec p \cdot \vec k}{E}\ .
\end{equation}
After this reduction, the amplitude becomes
\begin{equation}
i\mathcal{M}^{(2)}\Big|^{cl.}_{\delta h,\delta h,h}
=-i\,\frac{1}{4E\kappa} \int \frac{d^3k}{(2\pi)^3}
\frac{\tau_\phi(p,p)\,\tau_\phi(p,p)\,\tau_{\delta h,\delta h,h}(k,q-k)\,
\tilde{h}^{KN}(\vec q\,)\,PP}
{\big((\frac{\vec p \cdot \vec k}{E})^2-|\vec k\,|^2\big)
\big((\frac{\vec p \cdot \vec k}{E})^2-|\vec q-\vec k\,|^2\big)}\ .
\end{equation}
The explicit evaluation of this integral is unnecessary. Indeed, from its structure, one immediately observes that it scales as $\mathcal{O}(G_N^2m\ m_p)$, implying that in the probe limit $m\gg m_p$ this contribution is negligible compared to the dominant classical terms discussed earlier in this section. This conclusion could have been anticipated directly from the topology of the diagram in Fig.~\ref{3Grav_Diagram}, since it represents a radiative self–interaction of the background field rather than a classical exchange process. In principle, two analogous diagrams must also be considered, namely those involving $\tau_{\delta A,\delta A,h}$ and $\tau_{\delta h,\delta A,A}$, where photons circulate in the loop.  
However, the same reasoning applies and one can see that these diagrams are suppressed in the probe limit. We therefore conclude that, at 2PM order, the only contributions to the eikonal phase $\delta^{(2)}$ are the classical ones arising from the comb–like diagram, while all graviton (and photon) self–interaction effects remain subleading and can be safely neglected in this regime.

\section{Myers–Perry black holes as probes of stress multipoles}\label{sec:MPprobe}

We now turn the focus on the application of the formalism developed so far on Myers–Perry black holes, treating them as probes of the multipole structure of gravity in higher dimensions. We specialize the analysis to the $d=4$ case, which represents the simplest non–trivial setup in which stress multipoles contribute in a physically meaningful way. The $d=4$ case, corresponding to a five-dimensional spacetime, is of particular importance because it represents the lowest dimension in which rotation can occur simultaneously in more than one independent plane. Unlike the four-dimensional Kerr geometry, where the angular momentum is characterized by a single parameter $a$, the five-dimensional Myers–Perry solution admits two independent rotation parameters, $\mathfrak{a}_1$ and $\mathfrak{a}_2$, associated with the two orthogonal planes $(x_1, y_1)$ and $(x_2, y_2)$. This richer rotational structure gives rise to new physical effects that have no analogue in lower dimensions, namely the two spins can combine constructively or destructively, generating distinctive current and stress multipole moments whose interplay determines the geometry’s multipolar hierarchy. In particular, the presence of two angular momenta allows one to explore how the stress quadrupole and higher-order moments arise from the relative orientation and magnitude of the rotation planes, providing an ideal framework to test the connection between the source’s internal structure and the gravitational field it generates. 

For this reason, we focus on two specific and complementary configurations of the rotation parameters, $\mathfrak{a}_1=\mathfrak{a}_2$ and $\mathfrak{a}_2=0$, which allow us to perform explicit calculations. The first corresponds to a maximally symmetric configuration in which the two spins are equal in magnitude, resulting in a cohomogeneity-one geometry that retains full rotational invariance. This setting is especially useful because it isolates the effects of the current dipole and the stress quadrupole, while all higher multipoles vanish, offering a clean probe of the first non-trivial stress contributions in higher-dimensional gravity. The second configuration, $\mathfrak{a}_2=0$, breaks the rotational symmetry and describes a singly rotating Myers-Perry BH, the closest analogue in five dimensions of the four-dimensional Kerr solution. Examining this case enables a direct comparison between higher-dimensional and four-dimensional dynamics and clarifies how the additional rotational degree of freedom modifies the eikonal phase and the associated multipolar structure. Together, these two scenarios provide the simplest yet most informative arena for investigating how multiple angular momenta shape the long-range gravitational interaction in higher-dimensional spacetimes.

\subsection{$\mathfrak{a}_1=\mathfrak{a}_2$ case}

Let us first consider the case in which $\mathfrak{a}_1=\pm \mathfrak{a}_2=\mathfrak{a}$. As already mentioned, this configuration is remarkably simple and physically appealing because the metric becomes cohomogeneity–one, namely it depends only on the radial coordinate, hence exhibiting full rotational symmetry despite the presence of two non–zero angular momenta. This symmetry implies that all higher gravitational multipoles vanish except for the current dipole and the stress quadrupole, thereby isolating the minimal set of non–trivial multipole contributions. To evaluate the eikonal phase, we start from
\begin{equation}
    \delta_1=\frac{1}{2\ell}\int \frac{d^{d-1}q}{(2\pi)^{d-1}}\,e^{iq\cdot b}\,\mathcal{M}_{\text{on-shell}}\ ,
\end{equation}
where, as usual, $\vec b \cdot \vec \ell=\vec q \cdot \vec \ell=0$. In the present setup, one has $\xi=\mathfrak{a} q$, and using the master integral
\begin{equation}
    \mathcal{F}(d-1,\nu)=\int \frac{d^{d-1}q}{(2\pi)^{d-1}}\,e^{iq\cdot b}\,q^{2\nu}
    =\frac{2^{2\nu}}{\pi^{\frac{d-1}{2}}}\,
    \frac{\Gamma(\nu+\frac{d-1}{2})}{\Gamma(-\nu)}\,
    \frac{1}{b^{2\nu+d-1}}\ ,
\end{equation}
one finds that only the leading (zeroth–order) term in the Bessel–function expansion contributes to the non–local part of the eikonal phase, that is the long–range component surviving in the $r \to +\infty$ limit.  
Indeed, since
\begin{equation}
    \frac{J_n(aq)}{(aq)^n}=\frac{1}{2^n n!}+\mathcal{O}(\mathfrak{a}^2 q^2)\ ,
\end{equation}
focusing on the long–distance regime yields
\begin{equation}
    \delta_1=\frac{1}{2\ell}\int \frac{d^3q}{(2\pi)^3}\,
    e^{iq\cdot b}\,
    \frac{32\pi G_Nm }{3}\,\frac{1}{q^2}
    \Bigg(E^2+\frac{1}{8}\ell^2+\frac{i}{2}E\,\xi \cdot \ell
    -\frac{1}{32}(\xi \cdot \ell)^2\Bigg)\ .
\end{equation}
Replacing now
\begin{equation}
    \xi \cdot \ell\ \rightarrow\ -i\,S_{MP}^{ij}\ell_j\,(\partial_b)_i\ ,
\end{equation}
and applying this differential operator to the master integral ${\mathcal{F}(3,-1)=\tfrac{1}{4\pi b}}$, one obtains
\begin{equation}
    \vec{\mathfrak{u}}(\partial_b)\cdot\vec \ell\,\frac{1}{b}
    =\frac{i}{b^3}\,\vec \ell\cdot S_{MP} \cdot \vec b\ ,\qquad 
    \Big(\vec{\mathfrak{u}}(\partial_b) \cdot \vec \ell\Big)^2\frac{1}{b}
    =-3\,\frac{(\vec \ell \cdot S_{MP} \cdot \vec b)^2}{b^5}
    -\frac{\vec \ell \cdot S_{MP} \cdot S_{MP} \cdot \vec \ell}{b^3}\ ,
\end{equation}
where in this case the spin tensor simply reads
\begin{equation}
    S_{\text{MP}} =
\begin{pmatrix}
0 & \mathfrak{a} & 0 & 0 \\
-\mathfrak{a} & 0 & 0 & 0 \\ 
0 & 0 & 0 & \mathfrak{a} \\
0 & 0 & -\mathfrak{a} & 0 \\ 
\end{pmatrix}\ .
\end{equation}
Hence the eikonal phase takes the compact and physically transparent form
\begin{equation}
    \delta_1=\frac{4G_Nm}{3\ell b}\Bigg(
    E^2+\frac{1}{8}\ell^2
    -\frac{1}{2}E\,\frac{\ell \cdot S_{MP} \cdot b}{b^2}
    +\frac{1}{32}\Bigg[
    3\,\frac{(\ell \cdot S_{MP} \cdot b)^2}{b^4}
    +\frac{\ell \cdot S_{MP} \cdot S_{MP} \cdot \ell}{b^2}\Bigg]\Bigg)\ .
\end{equation}
This expression reveals several key physical insights. The leading dependence on the rotation parameter $\mathfrak{a}$ appears through the coupling between the spin of the BH and the orbital angular momentum of the probe, ${{\ell}_{ij} \sim b_i \ell_j-b_j \ell_i}$, which corresponds to the standard spin–orbit interaction. The subleading terms, scaling with $\mathfrak{a}^2$, are instead associated with the stress quadrupole moment of the rotating geometry. Because the mass quadrupole vanishes for the symmetric configuration $\mathfrak{a}_1=\pm \mathfrak{a}_2$, these higher–order terms provide a clean diagnostic of the gravitational stress contribution encoded in the MP background. From a broader perspective, this calculation demonstrates how the eikonal phase acts as a sensitive probe of multipolar corrections beyond the monopole and dipole structure of the gravitational field.

\subsection{$\mathfrak{a}_2=0$ case}

We now consider a slightly different configuration, namely $\mathfrak{a}_1=\mathfrak{a}$ and $\mathfrak{a}_2=0$, which is of particular interest because it represents the closest higher–dimensional analogue of the four–dimensional Kerr geometry. This asymmetric setup breaks the perfect balance between the two rotation planes, producing a genuinely axisymmetric background and allowing us to study how individual spin parameters influence long–range observables. To simplify the analysis we furthermore restrict to an equatorial scattering process, choosing
\begin{equation}
    \vec \ell=(\ell,0,0,0)\ ,\qquad 
    \vec b=(0,b,0,0)\ ,
\end{equation}
so that the momentum and the impact parameter are orthogonal, as in standard deflection configurations. Following~\cite{Bianchi:2023lrg}, the eikonal integral can be written as
\begin{equation}
    \delta_1=-\frac{1}{2}\int_{-\infty}^{+\infty}d\xi\,
    h_{\mu\nu}(\vec x=\vec b+\xi \vec \ell)\,p^\mu p^\nu\ .
\end{equation}
Performing the integration for the above kinematics then yields
\begin{equation}
    \delta_1=\frac{G_Nm}{3\mathfrak{a}^2}
    \Bigg(-b\ell+4\mathfrak{a}E+\frac{(b\ell-2\mathfrak{a}E)^2}{\ell\sqrt{b^2-\mathfrak{a}^2}}\Bigg)\ ,
\end{equation}
valid for $b>\mathfrak{a}$. Although this expression appears complicated, its long–distance expansion reveals the same physical structure encountered in the symmetric case, namely
\begin{equation}
    \delta_1=\frac{4G_Nm}{3\ell b}
    \Bigg(E^2+\frac{1}{8}\ell^2-\frac{\mathfrak{a}\ell E}{2b}
    +\frac{\mathfrak{a}^2(3\ell^2+16E^2)}{32b^2}
    +\mathcal{O} \left(\frac{1}{b^3}\right) \Bigg)\ .
\end{equation}
Up to overall normalization factors, this coincides with the leading eikonal result for $\mathfrak{a}_1=\pm \mathfrak{a}_2$. Indeed, both configurations reproduce, in the large–impact–parameter limit, the universal eikonal phase for a non–rotating Schwarzschild(–Tangherlini) black hole, confirming that the long–range scattering behavior is insensitive to rotation at leading order. Only subleading terms encode the characteristic spin–orbit and stress–quadrupole corrections, which thus provide a clear theoretical handle to isolate the genuine multipolar content of higher–dimensional rotating black holes. From a conceptual standpoint, these results illustrate that the KS–gauge framework not only reproduces known deflection phenomena but also provides a systematic and gauge–controlled avenue to connect scattering observables with the multipolar decomposition of gravitational sources.  
The approach naturally separates the physical contributions associated with the spin–dipole and stress–quadrupole moments, offering a unified picture of how rotational degrees of freedom deform the eikonal structure of the gravitational interaction.

Another interesting kinematical configuration for $\mathfrak{a}_2=0$ is the case of a probe incident with $\vec{\ell}$ along the $(x_2,y_2)$ plane orthogonal to the plane $(x_1,y_1)$ with non-trivial angular momentum. Without loss of generality one can take $\vec{\ell}=(0,0,0,\ell)$ and integrate over $\vec{q}=(q_1,q_2,q_\perp,0)$ orthogonal to $\vec{\ell}$ and with $q_1^2+q_2^2=q_\parallel^2$, where parallel and orthogonal are intended with respect to the rotational place $(x_1, y_1)$. Switching to polar coordinates $(q_\parallel,\varphi_\parallel)$ in the $(q_1,q_2)$ plane and taking into account that $\vec \xi=(\mathfrak{a} q_2,-\mathfrak{a} q_1,0,0)$ so that ${\xi^2 =\mathfrak{a}^2q_\parallel^2}$, one has 
\begin{equation}
\delta_1 =  \frac{2 \pi G_N m}{3\ell\pi^3} \int \frac{q_\parallel dq_\parallel d\varphi_\parallel dq_\perp}{q_\parallel^2+q_\perp^2}e^{iq_\parallel b_\parallel\cos\varphi_\parallel + i q_\perp b_\perp} \Bigg(E^2 J_0(\xi)+\tfrac{1}{4}{\ell}^2\frac{J_1(\xi)}{\xi}\Bigg)  \ ,
\end{equation}
since in this configuration $\vec{\xi}\cdot \vec \ell=0$. The angular integral produces a Bessel function 
\begin{equation}
\int_0^{2\pi} d\varphi_\parallel e^{iq_\parallel b_\parallel\cos\varphi_\parallel} = 2\pi J_0(q_\parallel b_\parallel) \ ,
\end{equation}
while the integral over $q_\perp$ can be computed with the method of residues 
\begin{equation}
\int_{-\infty}^{+\infty} {dq_\perp \over q_\parallel^2+q_\perp^2}e^{i q_\perp b_\perp} =
{\pi \over q_\parallel} e^{-q_\parallel|b_\perp|} \ ,
\end{equation}
leading to the expression 
\begin{equation}
    \delta_1=\frac{4GM}{3\ell}\int_0^{+\infty}dq_\parallel J_0(b_\parallel q_\parallel)e^{-q_\parallel|b_\perp|}\Bigg(E^2 J_0(\mathfrak{a} q_\parallel)+\tfrac{1}{4}\ell^2\frac{J_1(\mathfrak{a} q_\parallel)}{\mathfrak{a} q_\parallel}\Bigg)\ .
\end{equation}
One is then left with integrals of the form
\begin{equation}
\int_0^\infty  dq e^{-q_\parallel|b_\perp|} J_0(q_\parallel b_\parallel) {J_n(q_\parallel \mathfrak{a}) \over (q_\parallel \mathfrak{a})^n}
\end{equation}
with $n=0,1$ that can be mapped into the known integral~\cite{Gradshteyn:1943cpj}
\begin{equation}
\label{eq:masterintegral626}
\begin{aligned}
    \mathcal{I}(\lambda; \mu,\nu; \alpha, \beta, \gamma)=&\int_0^{+\infty} dx\,  x^{\lambda-1}e^{-\alpha x} J_\mu(\beta x)J_\nu(\gamma x)=\frac{\beta^\mu\gamma^\nu}{\Gamma(\nu+1)}2^{-\nu-\mu}\alpha^{-\lambda-\mu-\nu}\\
    &\times\sum_{m=0}^{+\infty}\frac{\Gamma(\lambda+\mu+\nu+2m)}{m!\Gamma(\mu+m+1)}{}_2F_1(-m, -\mu-m;\nu+1;\tfrac{\gamma^2}{\beta^2})\Big(-\frac{\beta^2}{4\alpha^2}\Big)^m\ ,
\end{aligned}
\end{equation}
so that the eikonal phase reads
\begin{equation}\label{eq:eikonalorthogonal}
    \delta_1=\frac{4G_Nm}{3\ell}\Bigg(E^2 \mathcal{I}(1, 0, 0, |b_\perp|, b_\parallel, \mathfrak{a})+\frac{\ell^2}{4\mathfrak{a}}\mathcal{I}(0, 0, 1, |b_\perp|, b_\parallel, \mathfrak{a})\Bigg)\ .
\end{equation}
Notice that ${}_2F_1(-m, -\mu-m;\nu+1;z)$ above are (Jacobi) polynomials of degree $m$. Moreover in the special case in which $\lambda=1$ and $\mu=\nu=\sigma$ the integral in \eqref{eq:masterintegral626} simplifies into
\begin{equation}
\mathcal{I}(1; \sigma,\sigma; |b_\perp|, b_\parallel, \mathfrak{a})=\int_0^\infty dq_\parallel e^{-|b_\perp| q_\parallel} J_\sigma(q_\parallel b_\parallel) J_\sigma(q_\parallel \mathfrak{a}) = {1\over \pi \sqrt{\mathfrak{a}b_\parallel} } Q_{\sigma-{1\over2}} \left( {\mathfrak{a}^2 + b^2 \over 2 \mathfrak{a}b_\parallel}\right)
\end{equation}
where $b^2=b_\parallel^2+b_\perp^2$ and  ${Q_{\mu}(x)}$ are the associated Legendre functions of the second kind, that
can be expressed in terms of hypergeometric functions as
\begin{equation}
 Q^{\mu}_{\nu} (x) = e^{i\pi \mu} {\pi^{1/2} \Gamma(\mu+\nu+1) (x^2-1)^{\mu/2} \over x^{\mu+\nu+1} \Gamma\left(\nu + {3\over 2}\right)}    {}_2F_1\left({\mu+\nu \over 2} +1, {\mu+\nu +1\over 2} ; \nu + {3\over 2}; {1\over x^2}\right)             
\end{equation}
In particular for $\mu=0$ one has $Q_\nu(x) = Q^0_\nu(x)$, so that for $\sigma= 0$ (\textit{i.e.} $\nu=-1/2$) one finds
\begin{equation}
{1\over \pi \sqrt{\mathfrak{a}b_\parallel}} Q_{-{1\over2}} \left( {\mathfrak{a}^2 + b^2 \over 2 \mathfrak{a}b_\parallel}\right) = 
{\sqrt{2} \over \sqrt{\mathfrak{a}^2 + b^2  }} {}_2F_1\left({3\over 4}, {1\over 4} ; 1; { 4 \mathfrak{a}^2b_\parallel^2 \over (\mathfrak{a}^2 + b^2 )^2 }\right) \ .
\end{equation} 
Expanding Eq. \eqref{eq:eikonalorthogonal} for $\mathfrak{a}\ll b_\parallel,|b_\perp|$ one finally has 
\begin{equation}
 \delta_1=\frac{4G_Nm}{3 \ell b } \left(E^2 + {1\over 8} \ell^2\right)\left(1  + \mathfrak{a}^2{(b^2-3b_\perp^2) \over 8b^4}\frac{\ell^2+16E^2}{\ell^2+8E^2}  + \mathcal{O}(\mathfrak{a}^4)\right) 
\end{equation}
where the leading term reproduces the spherically symmetric case.
It is interesting to compare this result with the four-dimensional scattering off a Kerr BH with incident momentum parallel to the axis of rotation in which one finds
\begin{equation}
\delta^{\rm Kerr}_1 = -{G_N m \over \ell} \left(E^2+{\ell^2 \over 4}\right)\log { \mu^2|\vec{b}|^2} \ ,
\end{equation}
with $\mu$ an IR regulator and no dependence on $a$ for this very special kinematic configuration, so that the eikonal phase becomes effectively equal to the non-rotating case.
The important result then is the fact that in dimensions higher than four, in the limit of axial incidence of the probe there is still  an angular momentum dependence left in the expression of the eikonal phase, differently with respect to the Kerr case. Therefore,  even though in this configuration the stress multipoles vanish, this difference in phenomenology proves once again how going to higher dimensions enriches gravitational physics due to geometrical arguments.

Summarizing our results, the detailed analysis of the cases $\mathfrak{a}_1=\pm \mathfrak{a}_2$ and $\mathfrak{a}_2=0$ clarified how the multipolar structures contribute to the scattering process. In the symmetric configuration the metric becomes cohomogeneity-one, and only the current dipole and the stress quadrupole survive, giving rise to a clean two-parameter hierarchy of physical effects. In the asymmetric case the rotational symmetry is broken, but the large impact parameter limit correctly reproduces the Schwarzschild(-Tangherlini) behaviour, confirming the consistency of the approach. The study of the orthogonal configuration, where the probe travels along the axis orthogonal to the rotation plane, provided an explicit expression for the eikonal phase in terms of Bessel and Legendre functions. Interestingly, even in this limit the eikonal phase retains a residual dependence on the rotation parameter $\mathfrak{a}$, contrary to the four-dimensional Kerr case where the corresponding configuration leads to a purely spherically symmetric result. This persistence of spin dependence highlights how extra dimensions enrich the gravitational phenomenology and give rise to new angular couplings that have no analogue in $D=4$. Altogether, the results obtained in this chapter demonstrate that the KS formalism is not only computationally efficient but also conceptually illuminating, allowing one to disentangle classical and hyper-classical contributions to the amplitude, to interpret the eikonal phase as a direct measure of the spacetime’s multipolar content, and to connect scattering observables with the geometric properties of the background in a fully covariant manner. The Myers-Perry solution thus emerges as an ideal probe of the deep relation between rotation, multipolar structure, and higher-dimensional gravitational dynamics.

\chapter{Conclusions}\label{chapter:Conclusions}

The research line presented in this thesis has been guided by a single unifying goal: to recover the full classical content of GR from the classical limit of quantum scattering amplitudes, unveiling new insights from the momentum-space structure of the \mbox{$S$-matrix} when applied to gravity. Through this perspective, the familiar language of QFT and the geometric formulation of gravity are brought together within a single conceptual framework. The central message emerging from this work is that the analytic structure of scattering amplitudes already contains, in its classical limit, all the information needed to reconstruct spacetime geometries, their multipolar structure, and their physical observables. 
This idea rests on a profound yet subtle observation, namely that loop amplitudes in gravity can give rise to contributions that survive in the classical limit, reflecting the breakdown of naive $\hbar$-counting in the presence of massive sources. Even though this phenomenon occurs for any QFT coupled to massive sources, what makes gravity qualitatively different is the non-linear nature of the classical field itself. Since the graviton couples to energy and momentum, the spacetime metric generated by a conserved current acts as its own source. Consequently, reproducing the classical gravitational field beyond leading order requires an infinite tower of graviton exchanges, appearing as loop amplitudes that encode the iterative back-reaction of the geometry on itself.
This feature blurs the boundary between the quantum and classical regimes and makes the diagrammatic language of amplitudes a natural way to organize the dynamics of curved spacetime. The guiding philosophy of this thesis is therefore that scattering amplitudes, when properly interpreted in their off-shell and low-energy limits, encode both the microscopic interaction rules of gravitons and the macroscopic structure of spacetime. By exploiting this dual interpretation, we have developed a systematic, amplitude-based framework capable of reconstructing metrics, EMTs, and gravitational multipoles from first principles. Within this unified picture, the Einstein equations emerge as an effective description of the classical sector of the quantum theory, while quantities such as deflection angles, tidal effects, and multipole moments appear as natural classical quantities derived from amplitude data. In doing so, this thesis shows how classical geometries, ranging from static to spinning, charged, and higher-dimensional configurations, can be obtained directly from quantum scattering processes, revealing the deep continuity between the field-theoretic and geometric formulations of gravity. This correspondence not only provides a consistent bridge between quantum and classical regimes but also offers a new, conceptually transparent language in which the fundamental structure of GR can be re-expressed and extended.

After establishing the conceptual foundations of the amplitude–geometry correspondence, this thesis developed a concrete algorithm to extract the PM expansion of the metric from off-shell three-point scattering amplitudes. In this framework the classical limit is implemented at the earliest possible stage through the definition of a dressed vertex, a classical analogue of the quantum matter–graviton interaction. This prescription isolates the gauge-dependent components needed for metric reconstruction and identifies a restricted set of diagram topologies that generate the full PM hierarchy. As a result, gravitational observables such as potentials and curvature tensors can be systematically obtained from the classical parts of multi-loop amplitudes, providing a direct solution of Einstein’s equations in momentum space. 

The formalism was then applied to spinning sources in arbitrary dimensions by computing graviton-emission amplitudes of a massive spin-1 field. This analysis reproduces, in four dimensions, the Hartle–Thorne metric and, in higher dimensions, its generalizations within the Myers–Perry family. It also led to the discovery of a new class of stress multipoles, additional to the usual mass and current moments, which appear naturally for $D>4$ and characterize the anisotropic stresses of higher-dimensional rotating geometries. From the amplitude viewpoint these arise from higher-rank tensor structures in the dressed vertex, while in the EFT description they correspond to specific higher-derivative non-minimal couplings. This provided a clear and compact dictionary between EFT operators and gravitational multipole moments, clarifying how deviations from minimal coupling translate into measurable deformations of the spacetime geometry. Moreover, including electromagnetic interactions allowed us to describe charged rotating configurations and to determine their gyromagnetic factors in generic dimension. While in four dimensions the minimal coupling reproduces the Kerr–Newman solution, in higher dimensions a non-minimal Pauli-type term is needed for recovering known solutions, and in particular, in five dimensions, a further Chern–Simons interaction is required to recover the CCLP geometry. From the amplitude perspective this analysis yielded, for the first time, an analytic derivation of the gyromagnetic BH factor and explained its origin in terms of EFT couplings.

Building on these results, a momentum-space formalism for the EMT of spinning sources at every order in the spin was constructed. Within this framework the EMT is expressed as a covariant expansion in the spin tensor, whose coefficients define gravitational form factors, encoding the asymptotic multipoles, and structure functions, describing the internal distribution of energy and spin. This provided, for the first time, a relativistic and gauge-consistent definition of source multipoles, establishing a direct link between matter distributions and the multipolar structure of the induced spacetime in a complete relativistic setup. Analytic resummation of these quantities reproduces the effective EMTs of Kerr and Myers–Perry black holes, interpreted respectively as thin rotating disks and higher-dimensional ellipsoids. We then considered non-trivial structure functions in order to extend such a description to smooth horizonless geometries, with the aim of building upon the developed amplitude-based framework to recover physically reasonable Kerr BH mimickers. Indeed, considering a source with a Gaussian-smeared energy-density profile, we were able to build a physically viable mimicker source at linearized level, capable of regularizing the original Kerr singularity into a regular finite-size core while reproducing the full multipolar structure of Kerr BHs. 

Finally, taking advantage of the special properties of the Kerr–Schild gauge, we computed the exact Fourier transform of the Kerr–Newman and Myers–Perry metrics, obtaining analytic tree-level amplitudes in arbitrary dimension. Because the metric is linear in the gravitational potential, all higher-order diagrams reduce to ladder graphs, allowing a transparent identification of the eikonal phase and its classical contact terms, culminating in a closed-form expression for the scattering amplitude at every loop order. We then used scalar test particles scattering off Myers-Perry backgrounds in order to probe gravitational multipoles in a higher-dimensional setup. In particular, we probe the fundamental nature of stress multipoles, showing their distinctive imprint in scattering processes and proving how amplitude data can constrain gravitational quantities, formally elevating our multipole-based framework for distinguishing BHs geometries from exotic compact objects through amplitude-derived data.  

The results obtained in this thesis open several directions for future research, both conceptual and phenomenological. At the most fundamental level, it would be natural to push the amplitude–geometry correspondence beyond the perturbative regime and investigate whether generic classical solutions (even beyond the BH paradigm) can be fully resummed from scattering amplitudes. In this respect, the Kerr–Schild framework provides a promising starting point. Indeed, its linear structure in the gravitational potential suggests that a non-perturbative formulation of amplitude methods in GR may be within reach. Understanding how higher-loop contributions reorganize into exact geometric quantities could also clarify the interplay between infrared and ultraviolet regimes in gravity, and perhaps shed light on how spacetime emerges from quantum dynamics.

From a technical perspective, the framework developed here can be extended in several complementary ways. One of them is the possibility of generalizing the notion of stress multipoles to theories beyond GR, such as massive gravity, higher-derivative models, or theories with extra fields. This could reveal how additional degrees of freedom leave imprints in the multipolar hierarchy, and elucidate to us possible special conditions that would make stress multipoles non-vanishing even in four-dimensional spacetimes, such as environmental effects for instance. Another important step will be to compute amplitude-based observables for composite or deformed objects, going beyond spin-induced multipoles to include intrinsic or tidal deformations. Such generalizations could connect directly with the physics of binary systems and the emission of gravitational waves, providing analytic templates complementary to numerical relativity. Indeed, the amplitude program has already shown its capacity to compute PM corrections to the conservative and radiative dynamics of two-body systems, and the methods developed in this thesis may substantially contribute to this effort.

The formalism of source multipoles also opens the way to a systematic classification of black hole mimickers. By tuning the structure functions that define the internal distribution of the source, one can build regular, horizonless objects that reproduce the exact multipolar structure of Kerr or Myers–Perry spacetimes. Extending these models to the non-linear regime could improve our knowledge on the process regarding the transition from regular compact configurations to black hole geometries, giving a new interpretation of the no-hair theorems in four dimensions, and attempting a higher-dimensional generalization of the BH uniqueness theorem. A natural continuation of this program is to explore whether such objects can form dynamically and whether they admit horizons under specific limits of the parameters, thus bridging the gap between horizonless and regular black holes. Finally, future research could apply the techniques developed here to probe exotic compact objects and fuzzball-like configurations in higher dimensions, as well as to study scattering processes in curved or asymptotically (A)dS backgrounds. The combination of the Kerr–Schild gauge and the amplitude-based expansion appears especially well-suited for these problems. In all these directions, the guiding principle remains the same, namely to extract the geometric and phenomenological content of gravity directly from its most fundamental quantum description. In this sense, the work presented in this thesis provides not only concrete results but also a coherent strategy for future explorations, deepening our understanding of how classical spacetime and its multipolar structure emerge from the microscopic language of scattering amplitudes.

\appendix

\chapter{Harmonic coordinate system}\label{chapter:App}

In this appendix we outline the general strategy to express multi-axially symmetric metrics, such as the Myers–Perry family, in harmonic coordinates. The relevance of this coordinate system is twofold. On the one hand, the harmonic gauge is universal, since any spacetime metric can in principle be written in a coordinate frame satisfying the harmonic condition
\begin{equation}
    \partial_\nu (\sqrt{-g}\, g^{\mu\nu}) = 0\ ,
\end{equation}  
this makes this gauge a standard choice for comparing different solutions directly at the level of the metric.
On the other hand, for vacuum and asymptotically flat solutions, harmonic coordinates represent a concrete realization of ACMC coordinates, as already mentioned in section~\ref{sec:GravitationalMultipoles}. This property is particularly important when one aims to extract the gravitational multipole moments directly from the metric following Thorne’s formalism, because working in harmonic coordinates automatically ensures that the frame is ACMC. This correspondence between harmonic and ACMC frames is not merely of conceptual value, it also provides a practical computational advantage. Indeed, the high degree of symmetry exhibited by the class of metrics under consideration makes this program feasible. In particular, the presence of multiple axial symmetries, each associated with an independent rotation plane, naturally constrains the metric structure and allows one to construct the coordinate transformation explicitly. As a result, the metric components can be written in harmonic form through direct algebraic manipulations, without the need for solving differential gauge conditions case by case.

Even though the explicit implementation of the procedure depends on the specific solution and on the number of spacetime dimensions, the overall strategy follows a common pattern. Let us consider, for definiteness, a Myers–Perry–like coordinate system in the case of $d$-even spatial dimension, where the coordinates are denoted as $x^\mu = (t, x_i, y_i)$. In the $d$-odd case, an additional coordinate $z$ is present, but the underlying logic of the construction remains the same. The goal is to perform a coordinate transformation that acts only on the spatial components,
\begin{equation}
    (t, x_i, y_i) \ \longrightarrow\ (t, X_i, Y_i)\ ,
\end{equation}
such that the resulting coordinates satisfy the harmonic condition
\begin{equation}\label{app:GenericHarmCondition}
    g^{\mu\nu} D_\mu \partial_\nu (t, X_i, Y_i) = 0 \ ,
\end{equation}
where $D_\mu$ denotes the spacetime covariant derivative compatible with the background metric $g_{\mu\nu}$. To exploit the symmetry structure of the solution, it is convenient to introduce a spherical-like parametrization of the harmonic coordinates,
\begin{equation}
    X_i = R\, \mu_i(\vec{\Theta}) \cos\Phi_i \ , 
    \qquad 
    Y_i = R\, \mu_i(\vec{\Theta}) \sin\Phi_i \ ,
\end{equation}
where the $\Phi_i$ represent the azimuthal angles associated with each rotation plane, while 
$\vec{\Theta} = (\Theta_1, \ldots, \Theta_{\frac{d-2}{2}})$ are polar angles. Within this parametrization, we can formulate a general ansatz for the coordinate transformation involving only the radial and polar variables, such as
\begin{equation}\label{app:GenricHarmTransf}
    r = f(R, \vec{\Theta}) \ , 
    \qquad 
    \vec{\theta} = \vec{g}(R, \vec{\Theta}) \ .
\end{equation}
Substituting this ansatz into Eq.~\eqref{app:GenericHarmCondition} yields a coupled system of $1 + \frac{d-2}{2}$ independent partial differential equations, which determine the unknown functions $f$ and $\vec{g}$.

In practice, solving these equations in closed form is generally not feasible for rotating spacetimes. However, the high degree of symmetry and the analytic structure of the Myers–Perry class make it possible to determine the transformation perturbatively. The standard approach is to expand the functions $f$ and $\vec{g}$ in powers of the spin parameters and, when necessary, in a PM expansion around flat space. At each order, the differential equations reduce to algebraic relations among the expansion coefficients, allowing one to construct the harmonic coordinate system recursively to arbitrarily high order in both spin and gravitational coupling. This general strategy is applied throughout the appendix to a number of relevant examples. In four dimensions, we first express the Hartle–Thorne metric (and its Kerr limit) in harmonic coordinates, illustrating how the slow–rotation expansion simplifies the transformation. Next, we analyze the five–dimensional Myers–Perry black hole, for which the presence of two independent angular momenta requires solving two coupled transformation equations (as in the lower-dimensional case), one per rotation plane. We briefly comment the Myers-Perry solution in $d=5$, which due to its spacetime dimensionality is less straightforward to put in harmonic gauge. We then restrict ourselves to transforming it in generic ACMC coordinates in order to be simply able to read gravitational multipoles directly from the metric. Finally, we extend the same procedure to the case of the singly–spinning black ring, showing how the harmonic representation naturally highlights the ring’s multipolar structure and its relation to the amplitude–based expansion.

\section{Hartle-Thorne metric in harmonic gauge}\label{App:HTInHarm}

We start by analyzing the four–dimensional case ($d = 3$), where the goal is to express the Hartle–Thorne metric in harmonic coordinates. This metric represents the most general stationary and axisymmetric vacuum solution describing the exterior spacetime of a slowly rotating body, accurate up to spin–induced quadrupole order and exact in the gravitational constant $G_N$~\cite{Hartle:1967he,Hartle:1968si}. Physically, it captures the geometry generated by any isolated rotating object, such as a neutron star or a compact black hole mimicker, up to second order in the rotation parameter. For $\zeta = 1$ the metric reproduces the Kerr limit, while deviations $\zeta \neq 1$ encode arbitrary quadrupolar deformations of the source. Given a spherical coordinate system $(t, r, \theta, \phi)$, the explicit expression of the metric up to 3PM order is given by
\begin{equation}\label{app:HartleThorneMetric}
    \begin{aligned}
    g_{tt}&=-1+\frac{2 G_N m}{r}+\frac{a^2 G_N m \zeta (3
   \cos (2 \theta )+1)}{2 r^3}\\
   &-\frac{a^2G_N^2 m^2 ((3
   \zeta -2) \cos (2 \theta )+\zeta -2)}{2 r^4} -\frac{a^2G_N^3m^3
   (4 \zeta -11)  (3 \cos (2 \theta )+1)}{7 r^5}+\mathcal{O}(G_N^4, a^3)\ ,\\
    g_{t\phi}&=- \frac{2 a G_N m \sin^2(\theta)}{r}+\mathcal{O}(G_N^4, a^3)\ ,\\
    g_{rr}&=1+ \frac{2 G_N m}{r}-\frac{(a^2G_N m \zeta  (3 \cos (2 \theta )+1)) }{2
   r^3}+ \frac{4 G_N^2m^2}{r^2}+\frac{a^2G_N^2m^2
   (-5 \zeta -3 (5 \zeta -8) \cos (2 \theta )+4) }{2
   r^4}\\
   &+ \frac{8 G_N^3m^3}{r^3}+\frac{a^2G_N^3m^3
   (-60 \zeta -9 (20 \zeta -27) \cos (2 \theta )+25)}{7
   r^5}+\mathcal{O}(G_N^4, a^3)\ ,\\
   g_{\theta\theta}&=r^2+\frac{a^2G_N m\zeta   (3 \cos (2 \theta
   )+1)}{2 r}-\frac{a^2G_N^2m^2 (5 \zeta -1) (3
   \cos (2 \theta )+1)}{4 r^2}\\
   &-\frac{18 a^2G_N^3m^3 (\zeta -1)
    (3 \cos (2 \theta )+1)}{7 r^3}+\mathcal{O}(G_N^4, a^3)\ ,\\
    g_{\phi\phi}&=r^2 \sin ^2(\theta ) -\frac{a^2G_N m \zeta  (3 \cos (2 \theta
   )+1) \sin ^2(\theta )}{2 r}-\frac{\left(a^2G_N^2m^2 (5 \zeta -1) (3 \cos (2 \theta )+1) \sin
   ^2(\theta )\right) a^2}{4 r^2}\\
   &-\frac{18 \left(a^2G_N^3m^3 (\zeta -1) (3 \cos (2 \theta )+1) \sin
   ^2(\theta )\right)}{7r^3}+\mathcal{O}(G_N^4, a^3)\ .
    \end{aligned}
\end{equation}
Here $a$ denotes the angular momentum per unit mass, and $\zeta$ is the dimensionless parameter controlling the spin–induced quadrupole moment. Notice that, despite being derived under the assumption of slow rotation, the structure of each term already reflects the hierarchy of multipole moments, monopole, dipole, and quadrupole, consistent with the amplitude–based expansion discussed in the main text.

To move to harmonic coordinates, we introduce a new set of spherical harmonic variables $(t, R, \Theta, \Phi)$, together with their Cartesian counterparts,
\begin{equation}
    \begin{cases}
    X=R\sin\Theta\cos\Phi\ ,\\
    Y=R\sin\Theta\sin\Phi\ ,\\
    Z=R\cos\Theta\ ,
    \end{cases}
\end{equation}
which define a coordinate system that asymptotically coincides with Minkowski space and is regular on the symmetry axis. In these coordinates, the harmonic gauge condition takes the explicit form
\begin{equation}\label{app:HarmonicBoxd3}
    g^{\mu\nu}D_\mu \partial_\nu (t, X, Y, Z)=0\ ,
\end{equation}
where each coordinate is treated as a scalar function on the curved background. Equation~\eqref{app:HarmonicBoxd3} thus provides four equations, but only two of them are independent. Indeed, the equation corresponding to the time component is automatically satisfied since the temporal coordinate remains unaltered, while the two equations associated with $X$ and $Y$ are equivalent due to the underlying axial symmetry of the spacetime. Consequently, the harmonic condition effectively reduces to two coupled partial differential equations determining the radial and polar components of the transformation, in full agreement with the general analysis outlined at the beginning of the appendix.

Indeed, since the Hartle–Thorne metric is axisymmetric, we can define a coordinate transformation that does not involve the azimuthal coordinate as
\begin{equation}\label{eq:HarmonicCoordTransfD4}
 R=r(R, \Theta)\ , \qquad \Theta=\theta(R, \Theta)\ ,\qquad \Phi=\phi \ . 
\end{equation}
The system in Eq.~\eqref{app:HarmonicBoxd3} therefore reduces to two independent partial differential equations, which can be solved by adopting a perturbative ansatz for the coordinate transformation. We expand the unknown functions in powers of both the spin parameter $a$ and the gravitational constant $G_N$, according to
\begin{equation}\label{app:d3Transformation}
\begin{aligned}
    r(R, \Theta)&=R\sum_{i=0}^{n\text{PM}}\left(\frac{G_N m}{R}\right)^i\sum_{j=0}^{\lfloor \ell/2\rfloor}\left(\frac{a}{R}\right)^{2j}\sum_{k=0}^{j}\mathcal{C}_{i, 2j, k}^{(R)}P_{2k}(\cos\Theta)\ ,\\
    \cos\theta(R, \Theta)&=\cos(\Theta)\sum_{i=0}^{n\text{PM}}\left(\frac{G_N m}{R}\right)^i\sum_{j=0}^{\lfloor \ell/2\rfloor}\left(\frac{a}{R}\right)^{2j}\sum_{k=0}^{j}\mathcal{C}_{i, 2j, k}^{(\Theta)}P_{2k}(\cos\Theta)\ ,
\end{aligned}
\end{equation}
where $P_n$ are the Legendre polynomials, $n$PM denotes the PM order, $\ell$ indicates the expansion order in spin, and $\lfloor\cdot\rfloor$ stands for the integer part. The structure of this ansatz is designed to preserve the fundamental symmetries of the rotating configuration while ensuring the correct asymptotic behavior, and can be summarized as follows. First, in the asymptotically flat limit $R \to +\infty$, the two coordinate systems must coincide, implying that $\mathcal{C}^{(R)}_{0,0,0} = 1$. Second, in the non–rotating limit $a \to 0$, the transformation must reproduce the Schwarzschild solution, which requires $\theta = \Theta$ in Eq.~\eqref{app:d3Transformation}. Moreover, at a fixed spin order, the angular dependence appearing in the metric is captured by Legendre polynomials of order not exceeding the corresponding spin power, which motivates the restriction $k \leq j$. Finally, time–reversal invariance, under which $t \to -t$ and $\Phi \to -\Phi$, forbids odd powers of the angular momentum, so only even powers of $a$ appear in the expansion. This construction ensures that the ansatz not only satisfies the harmonic condition order by order but also remains consistent with the symmetries of the Hartle–Thorne spacetime. In practice, substituting Eq.~\eqref{app:d3Transformation} into the harmonic constraint \eqref{app:HarmonicBoxd3} leads to an algebraic hierarchy that determines the coefficients $\mathcal{C}_{i,2j,k}^{(R)}$ and $\mathcal{C}_{i,2j,k}^{(\Theta)}$ recursively in the PM expansion. These coefficients encode the deviation from the Boyer–Lindquist frame and fully characterize the coordinate transformation up to the desired PM and spin order.

Finally, performing the coordinate transformation of Eq.~\eqref{eq:HarmonicCoordTransfD4} and imposing the harmonic condition~\eqref{app:HarmonicBoxd3}, we can determine the metric in harmonic gauge order by order by solving the corresponding algebraic system for the coefficients appearing in Eq.~\eqref{app:d3Transformation}. The resulting equations define the coordinate transformation recursively in both the PM and spin expansions. It is worth noting that this procedure exhibits a certain degree of gauge redundancy, as not all coefficients are uniquely fixed by the harmonic condition. At first and second PM order the residual gauge freedom is parametrized by two coefficients, namely $\mathcal{C}^{(R)}_{1,2,0}$ and $\mathcal{C}^{(R)}_{1,2,2}$. At 3PM order, two additional unfixed parameters appear, $\mathcal{C}^{(R)}_{3,0,0}$ and $\mathcal{C}^{(R)}_{3,2,2}$, corresponding to new degrees of freedom in the spatial redefinition of the harmonic coordinates. While the complete 3PM expression for the metric can be found in~\cite{Gambino:2024uge}, the explicit 1PM result for the Hartle–Thorne metric in harmonic coordinates reads
\begin{equation}\label{eq:HartleThorneMetricHarmonic}
    \begin{aligned}
    g_{tt}&=-1+\frac{2 G_N m}{R}-\frac{a^2 G_N m \zeta (3
   \cos (2 \Theta )+1)}{2 R^3}+\mathcal{O}(G_N^2, a^3)\ ,\\
    g_{t\Phi}&=-\frac{2 a G_N m \sin^2(\Theta)}{R}+\mathcal{O}(G_N^2, a^3)\ ,\\
    g_{RR}&=1+ \frac{2 G_N m}{R}-a^2G_N m\frac{   8\mathcal{C}^{(R)}_{1, 2, 0}+(3 \cos (2 \Theta )+1)(\zeta+2\mathcal{C}^{(R)}_{1, 2, 2}) }{2
   R^3}+\mathcal{O}(G_N^2, a^3)\ ,\\
   g_{\Theta\Theta}&=R^2+2G_N m R+a^2G_N m\frac{\zeta (3\cos(2\Theta)+1)+\mathcal{C}^{(R)}_{1, 2, 2}(3\cos(2\Theta)-1)-4\mathcal{C}^{(R)}_{1, 2, 0}}{2 R}+\mathcal{O}(G_N^2, a^3)\ ,\\
    g_{R\Theta}&=\frac{3G_N m a^2\sin^2(\Theta)\mathcal{C}^{(R)}_{1, 2, 2}}{4R^2}+\mathcal{O}(G_N^2, a^3)\ ,\\
    g_{\Phi\Phi}&=R^2\sin^2(\Theta)+2G_N m R\sin^2(\Theta) \\
    &-a^2G_N m\sin^2(\Theta)\frac{\zeta (3\cos(2\Theta)+1)-4\mathcal{C}^{(R)}_{1, 2, 0}+2\mathcal{C}^{(R)}_{1, 2, 2}}{2 R}+\mathcal{O}(G_N^2, a^3)\ .
    \end{aligned}
\end{equation}
Once expressed in harmonic form, the metric can be easily rewritten in Cartesian coordinates, allowing a direct and transparent comparison with the amplitude–based construction discussed in the thesis. This perturbative strategy for obtaining harmonic coordinates is conceptually similar to the one described in~\cite{Aguirregabiria:2001vk}, although important differences arise in the treatment of residual gauge freedoms. Restricting our analysis to terms up to 3PM and quadratic order in spin, we find a larger number of unfixed parameters than reported in that reference. Specifically, at 1PM order our construction reveals the presence of two independent gauge coefficients, whereas~\cite{Aguirregabiria:2001vk} identifies only one. This discrepancy is not accidental, and as shown in section~\ref{chapter:RotatingMetrics}, the amplitude-based formulation naturally predicts two gauge degrees of freedom in four dimensions, both associated with the structure of the dressed graviton–source vertex. The present result therefore confirms the internal consistency between the harmonic-coordinate approach and the amplitude formalism. We verified that this matching remains valid up to order $O(S^7)$ in the spin expansion, providing a strong cross-check of both frameworks and confirming the robustness of the harmonic representation at high multipolar order.

\section{Myers-Perry metric in $d=4$ in harmonic gauge}\label{App:MPinHarm}

We now turn the attention to the case of the five–dimensional Myers–Perry solution, corresponding to $d=4$ spatial dimensions. Such spacetime geometry is described by
\begin{equation}\label{app:MPD5Metric}
\begin{aligned}
    ds^2=&-dt^2+\frac{\mu}{\Sigma}(dt+\mathfrak{a}_1\sin^2\theta\ d\phi_1+\mathfrak{a}_2\cos^2\theta\ d\phi_2)+\frac{r^2\Sigma}{\Pi-\mu r^2}dr^2\\
    &+\Sigma d\theta^2+(r^2+\mathfrak{a}_1^2)\sin^2\theta\ d\phi_1^2+(r^2+\mathfrak{a}_2^2)\cos^2\theta\ d\phi_2^2\ ,
\end{aligned}
\end{equation}
where 
\begin{equation}
    \Sigma=r^2+\mathfrak{a}_1^2\cos^2\theta+\mathfrak{a}_2^2\sin^2\theta\ , \qquad \Pi=(r^2+\mathfrak{a}_1^2)(r^2+\mathfrak{a}_2^2)\ ,
\end{equation}
and $\mathfrak{a}_1$ and $\mathfrak{a}_2$ are two independent spin parameters with 
\begin{equation}
    \mu=\frac{16 \pi G_N m}{(d-1)\Omega_{d-1}}\ .
\end{equation}
Notice that the spin parameters appearing in Eq.~\eqref{app:MPD5Metric} are related to the physical angular momenta through the relation 
\begin{equation}\label{eq:MPd4SpinRelations}
    \mathfrak{a}_1=\frac{3}{2}a_1\ ,\qquad \mathfrak{a}_2=\frac{3}{2}a_2\ , 
\end{equation}
coherently with Eq.~\eqref{eq:MPvsPhys}. Analogously to what we did in the $d=3$ case for the Hartle–Thorne metric, our goal is to express the Myers-Perry spacetime in harmonic coordinates by introducing an appropriate set of Cartesian variables that respect the two–plane rotational symmetry of the solution. To this end, we define the harmonic Cartesian coordinates as  
\begin{equation}\label{eq:HarmCartesianD5}
    \begin{cases}
    X_1=R\sin\Theta\cos\Phi_1\ ,\\
    Y_1=R\sin\Theta\sin\Phi_1\ ,\\
    X_2=R\cos\Theta\cos\Phi_2\ ,\\
    Y_2=R\cos\Theta\sin\Phi_2\ ,
    \end{cases}
\end{equation}
which are related to the original spheroidal coordinates through the transformation 
\begin{equation}\label{eq:HarmonicCoordinatesD5}
  R=r(R, \Theta)\ , \qquad \Theta=\theta(R, \Theta)\ ,\qquad \Phi_1=\phi_1\ ,\qquad \Phi_2=\phi_2 \ , 
\end{equation}
and satisfy the harmonic condition  
\begin{equation}\label{eq:BoxD5}
    g^{\mu\nu}D_\mu \partial_\nu  (X_1, Y_1, X_2, Y_2)=0\ ,
\end{equation}
where each of the four spatial coordinates $(X_1, Y_1, X_2, Y_2)$ is treated as a scalar field on the background geometry.  

As in the lower–dimensional example, Eq.~\eqref{eq:BoxD5} leads to two independent partial differential equations that can be solved by introducing a perturbative ansatz for the coordinate transformation. In this case, however, the presence of two independent angular momenta $\mathfrak{a}_1$ and $\mathfrak{a}_2$ makes the structure considerably richer. The ansatz is constructed to preserve the symmetries of the Myers-Perry metric while allowing for a systematic expansion in both the spin parameters and the PM order, and reads
\begin{equation}\label{app:d4TransformationMP}
\begin{aligned}
    r(R, \Theta)&=R\sum_{i=0}^{n\text{PM}}(Gm\rho)^i \sum_{\sigma(p,q)}\mathcal{A}_{i}^{(p, q)}(\Theta)\ ,\\
    \cos \theta(R, \Theta)&=\cos(\Theta)\sum_{i=0}^{n\text{PM}}(Gm\rho)^i \sum_{\sigma(p,q)}\mathcal{B}_{i}^{(p, q)}(\Theta)\ .
\end{aligned}
\end{equation}
Here $\rho$ is the harmonic radial function defined in Eq.~\eqref{eq:ScalarHarmonicFunction}, while $\mathcal{A}_i^{(p,q)}$ and $\mathcal{B}_i^{(p,q)}$ denote the expansion coefficients for the radial and polar transformations, respectively. The dependence on the indices $(p,q)$ organizes the spin expansion in terms of the two angular momenta, with $p+q$ indicating the total spin order. Explicitly, each coefficient is expanded as  
\begin{equation}
\begin{aligned}
    \mathcal{A}_{i}^{(p, q)}(\Theta)&=\sum_{k=0}^{n_k}\left(\frac{\mathfrak{a}_1^{p}\mathfrak{a}_2^{q}}{R^{p+q}}\right) \mathcal{C}_{i, p, q, 2k}^{(R)}P_{2k}\Big(f_{\sigma(p, q)}(\Theta)\Big)\ ,\\
     \mathcal{B}_{i}^{(p, q)}(\Theta)&=\sum_{k=0}^{n_k}\left(\frac{\mathfrak{a}_1^{p}\mathfrak{a}_2^{q}}{R^{p+q}}\right) \mathcal{C}_{i, p, q, 2k}^{(\Theta)}P_{2k}\Big(f_{\sigma(p, q)}(\Theta)\Big)\ ,
\end{aligned}
\end{equation}
where the functions $P_{2k}$ are the Legendre polynomials and $f_{\sigma(p,q)}(\Theta)$ encodes the angular dependence of the transformation, defined piecewise as
\begin{equation}
    f_{\sigma(p, q)}\Theta=\begin{cases}
        \cos\Theta & p>q\ , \\
        \cos\Theta\sin\Theta & p=q\ ,\\
        \sin\Theta & p<q\ .
    \end{cases}
\end{equation}
This definition ensures that every term in the expansion respects the intrinsic discrete symmetries of the solution, including reflection and exchange of the two rotation planes. At a given order in spin, all possible combinations of $(p,q)$ such that $p+q$ is even must be considered, since odd powers would violate time–reversal invariance. For example, at quadratic order in the spins, $O(S^2)$, the relevant combinations are $\sigma(p,q)=\{(2,0), (0,2), (1,1)\}$, corresponding respectively to the terms proportional to $\mathfrak{a}_1^2$, $\mathfrak{a}_2^2$, and $\mathfrak{a}_1\mathfrak{a}_2$. At quartic order $O(S^4)$, the set extends to $\sigma(p,q)=\{(4,0),(3,1),(2,2),(1,3),(0,4)\}$, and so on. This hierarchical construction guarantees that, order by order in the spin expansion, all cross–terms between the two rotation parameters are properly accounted for and that the resulting metric preserves the expected $\mathfrak{a}_1 \leftrightarrow \mathfrak{a}_2$ exchange symmetry.

Although the ansatz is formulated to respect the key features of the Myers–Perry geometry, several coefficients remain undetermined after imposing the harmonic condition. These free parameters correspond to gauge redundancies in the definition of harmonic coordinates, similar to those encountered in the four–dimensional case. They can be fixed by choosing a convenient gauge that simplifies the comparison with the amplitude–based results. Following this procedure, the metric in harmonic coordinates can be derived up to the desired order in the PM and spin expansions. In particular, by matching the resulting expression with the general EMT construction introduced in Eq.~\eqref{eq:GenericEMT} for $d=4$, we were able to determine all relevant coefficients unambiguously up to $O(S^7)$. This matching provides a complete fixing of the form factors appearing in Eq.~\eqref{eq:ExplicitMPFF} and confirms the internal consistency between the coordinate–space and amplitude–based descriptions of the black hole. It is important to notice that, differently from the $D = 4$ case, in five dimensions logarithmic contributions appear starting from the second PM order. Such terms, although omitted in Eq.~\eqref{app:d4TransformationMP} for simplicity, must be consistently included when working beyond 1PM. This behavior is perfectly consistent with the amplitude–based description, where one observes the emergence of infrared singularities in $D = 5$ even when the harmonic gauge is adopted. These divergences are renormalized through the insertion of higher–loop counterterm vertices, which in the classical limit manifest precisely as logarithmic corrections in the coordinate transformation. Moreover, the structure of the transformation is constructed to preserve the discrete symmetries of the original Myers-Perry metric, particularly under the combined operations $\mathfrak{a}_1 \leftrightarrow \mathfrak{a}_2$ and $\Theta \to \Theta + \pi/2$, which interchange the two rotation planes. Following the same logic as in the Hartle–Thorne analysis, substituting the ansatz of Eq.~\eqref{app:d4TransformationMP} into the harmonic condition allows one to determine the coefficients of the transformation order by order. In line with what is found from the amplitude perspective, the coordinate transformation exhibits several independent gauge redundancies. At order $\mathcal{O}(G_N S^2)$, two free parameters remain unfixed, while two additional redundancies appear at orders $\mathcal{O}(G_N^2 S^0)$ and $\mathcal{O}(G_N^2 S^2)$. The latter are directly related to the logarithmic terms introduced above and correspond, at the classical level, to the presence of the higher–loop counterterm vertex in Eq.~\eqref{eq:CounterTermVertex}. This close correspondence between the coordinate–space and momentum–space descriptions demonstrates the consistency of the harmonic gauge approach with the amplitude framework and highlights how gauge redundancies in one formalism translate into renormalization freedoms in the other.

Finally, at first PM order, and after using the gauge freedom to set $g_{R\Theta}=0$ for simplicity, we obtain the metric in harmonic coordinates as
\begin{equation}\label{eq:MPmetricHarmonic}
    \begin{aligned}
    g_{tt}&=-1+\frac{8 G_N m}{3\pi R^2}-\frac{8 G_N m(\mathfrak{a}_1^2-\mathfrak{a}_2^2) \cos (2 \Theta )}{3\pi R^4}+\mathcal{O}(G_N^2, S^3)\ ,\\
    g_{t\Phi_1}&=\frac{8 \mathfrak{a}_1 G_N m \sin^2(\Theta)}{3\pi R^2}+\mathcal{O}(G_N^2, S^3)\ ,\\
    g_{t\Phi_2}&=\frac{8 \mathfrak{a}_2 G_N m \cos^2(\Theta)}{3\pi R^2}+\mathcal{O}(G_N^2, S^3)\ ,\\
    g_{RR}&=1+\frac{4G_Nm}{3\pi R^2}-\frac{4G_N m(\mathfrak{a}_1^2-\mathfrak{a}_2^2)\cos(2\Theta)}{3\pi R^4}+\mathcal{O}(G_N^2, S^3)\ ,\\
    g_{\Theta\Theta}&=\frac{4G_Nm}{3\pi}+R^2-\frac{4G_Nm}{9\pi R^2}\Big(\mathfrak{a}_1^2+\mathfrak{a}_2^2+3(\mathfrak{a}_1^2-\mathfrak{a}_2^2)\cos(2\Theta)\Big)+\mathcal{O}(G_N^2, S^3)\ ,\\
    g_{\Phi_1\Phi_1}&=\frac{4G_Nm\sin^2(\Theta)}{3\pi}+R^2\sin^2(\Theta)\\
    &+\frac{2G_Nm\sin^2(\Theta)}{9\pi R^2}\Big(\mathfrak{a}_1^2+\mathfrak{a}_2^2-3(3 \mathfrak{a}_1^2-\mathfrak{a}_2^2)\cos(2\Theta)\Big)+\mathcal{O}(G_N^2, S^3)\ ,\\
    g_{\Phi_2\Phi_2}&=\frac{4G_Nm\cos^2(\Theta)}{3\pi}+R^2\cos^2(\Theta)\\
    &+\frac{2G_Nm\cos^2(\Theta)}{9\pi R^2}\Big(\mathfrak{a}_1^2+\mathfrak{a}_2^2-3(\mathfrak{a}_1^2-3\mathfrak{a}_2^2)\cos(2\Theta)\Big)+\mathcal{O}(G_N^2, S^3)\ ,\\
    g_{\Phi_1\Phi_2}&=\frac{2\mathfrak{a}_1\mathfrak{a}_2G_Nm\sin^2(2\Theta)}{3\pi R^2}+\mathcal{O}(G_N^2, S^3)\ .
    \end{aligned}
\end{equation}
To the best of our knowledge, Eq.~\eqref{eq:MPmetricHarmonic} represents the first explicit derivation of the five–dimensional Myers–Perry metric in harmonic coordinates available in the literature~\cite{Gambino:2024uge}. By rewriting this expression in Cartesian form, one can perform a direct comparison with the metric reconstructed from the amplitude–based approach, finding complete agreement up to third PM order. This match provides a highly non–trivial consistency check of the entire construction, confirming that the coordinate–space harmonic formulation faithfully reproduces the amplitude–derived geometry. Although all the physical information is encoded in the leading (tree–level) contribution, the inclusion of higher–order terms plays a crucial conceptual role,  validating the robustness of the harmonic gauge procedure and demonstrates its capability to capture the correct ultraviolet and infrared behavior of the Myers–Perry spacetime in higher dimensions.

\section{Myers-Perry metric in $d=5$ in ACMC coordinates}\label{App:MPinACMC}

We now move to the case of a bi–axially symmetric spacetime in $d = 5$. In contrast to the previous examples, the higher dimensionality of the background introduces additional angular structure. Although the metric still possesses only two independent angular momenta, it now depends on two polar angles. Consequently, a fully harmonic coordinate transformation analogous to Eq.~\eqref{app:d4TransformationMP} would require the use of higher–dimensional spherical harmonics, rather than the three–dimensional ones that lead to Legendre polynomials in axisymmetric configurations. Carrying out such a general transformation would be algebraically cumbersome and, from a physical standpoint, would not provide new information beyond what is already captured in the lower–dimensional cases. For the purpose of comparing with amplitude–based results and extracting the multipolar structure of the spacetime, it is therefore sufficient to work in ACMC coordinates. This section is devoted to that construction.

Following this logic, our goal is to rewrite the five–dimensional MP metric in generalized ACMC coordinates~\cite{Thorne:1980ru}, which allows for a direct identification of the gravitational multipole moments. We start from a set of spherical coordinates $(t, r, \theta, \psi, \phi_1, \phi_2)$ defined as
\begin{equation}
\begin{cases}
    y_1=r\sin\theta\sin\psi\sin\phi_1\ ,\\
    x_1=r\sin\theta\sin\psi\cos\phi_1\ ,\\
    y_2=r\cos\theta\sin\psi\sin\phi_2\ ,\\
    x_2=r\cos\theta\sin\psi\cos\phi_2\ ,\\
    z=r\cos\psi\ ,
\end{cases}
\end{equation}
where $r^2 = y_1^2 + x_1^2 + y_2^2 + x_2^2 + z^2$, $\phi_k$ are the azimuthal angles associated with each rotation plane, and $(\theta, \psi)$ play the role of polar coordinates, while the explicit metric expression in standard oblate-spheroidal coordinates can be recovered from Eq.~\eqref{eq:MPoblatesphODD} in the $d=5$ case. Expanding the metric in powers of $G_N$ and of the spin parameters $\mathfrak{a}_1$ and $\mathfrak{a}_2$, one can readily verify that the resulting expression is not in ACMC form. For instance, the $rr$–component reads
\begin{equation}
g_{rr}=1-\frac{\Big(\mathfrak{a}_2^2\cos^2\theta+\mathfrak{a}_1^2\sin^2\theta\Big)\sin^2\psi}{r^2}+\cdots
\end{equation}
and lacks the correct power–law dependence $\rho(r)/r^2$ required for a genuine quadrupole term, as defined in Eq.~\eqref{eq:MultipoleExpandedMetric}.  

To bring the metric into ACMC form, we perform a coordinate transformation of the type
$(t, r, \theta, \psi) \to (t, R, \Theta, \Psi)$, with the angular coordinates $(\phi_1, \phi_2)$ left unchanged. This transformation is designed to eliminate the non–multipolar terms in the asymptotic expansion of the metric, following the spirit of the method developed in~\cite{Heynen:2023sin}. Explicitly, the transformation is defined as
\begin{equation}\label{eq:MPACMCmetricExpr}
\begin{aligned}
&r=R + \frac{1}{R} \left( -\frac{1}{4} \left( (\mathfrak{a}_1^2 + \mathfrak{a}_2^2) + (\mathfrak{a}_2^2 - \mathfrak{a}_1^2) \cos 2 \Theta \right) \sin^2\Psi \right)\\ 
&+ \frac{1}{R^3} \left( 
\frac{1}{4} \left( \mathfrak{a}_1^4 + \mathfrak{a}_2^4 + (\mathfrak{a}_2^4 - \mathfrak{a}_1^4) \cos2 \Theta \right) \sin^2\Psi 
- \frac{5}{32} \left( \mathfrak{a}_1^2 + \mathfrak{a}_2^2 + (\mathfrak{a}_2^2 - \mathfrak{a}_1^2) \cos 2 \Theta \right)^2 \sin^4\Psi 
\right)\ ,\\
&\cos\theta=\cos\Theta 
- \frac{(\mathfrak{a}_2^2 - \mathfrak{a}_1^2) \cos\Theta \sin^2\Theta}{2 R^2} \\
&+ \frac{(\mathfrak{a}_2^2 - \mathfrak{a}_1^2) \cos\Theta \sin^2\Theta}{16 R^4} 
\left( 
3 \mathfrak{a}_1^2 + \mathfrak{a}_2^2 
+ 5 (\mathfrak{a}_1^2 - \mathfrak{a}_2^2) \cos2\Theta
+ \left( 4 \mathfrak{a}_2^2 \cos^2\Theta + 4 \mathfrak{a}_1^2 \sin^2\Theta \right) \cos2\Psi 
\right)\ ,\\
&\cos\psi=\cos\Psi
+ \frac{1}{R^2} \left( 
\frac{1}{8} \left( \mathfrak{a}_1^2 + \mathfrak{a}_2^2 + (-\mathfrak{a}_1^2 + \mathfrak{a}_2^2) \cos2\Theta \right) 
\sin\Psi \sin2\Psi
\right) \\
&+ \frac{\sin\Psi}{R^4}
 \bigg( 
-\frac{1}{2} \left( 
\frac{1}{128} \left( 
11 \mathfrak{a}_1^4 - 14 \mathfrak{a}_1^2 \mathfrak{a}_2^2 + 11 \mathfrak{a}_2^4 
- 4 (\mathfrak{a}_1^4 - \mathfrak{a}_2^4) \cos2\Theta
- 7 (\mathfrak{a}_1^2 - \mathfrak{a}_2^2)^2 \cos4\Theta 
\right) \right) \sin2\Psi \\
&-\frac{1}{4} \left( 
\frac{7}{64} \left( \mathfrak{a}_1^2 + \mathfrak{a}_2^2 + (-\mathfrak{a}_1^2 + \mathfrak{a}_2^2) \cos2\Theta \right)^2 
\right) \sin4\Psi
\bigg) \ .
\end{aligned}
\end{equation}
Applying this change of variables, the five–dimensional Myers–Perry metric assumes the ACMC–5 form, which means that all non–physical dipolar terms are removed and the expansion is well–defined up to the fifth multipolar order. In this coordinate system, the gravitational multipoles can be read directly from the asymptotic expansion of the metric components. Extracting the relevant coefficients, we obtain the following set of form factors
\begin{equation}
    \begin{aligned}
      &F_{0, 1}=1\ ,   \qquad &F_{2, 1}=-\frac{3}{5}\ ,\qquad &F_{4, 1}=\frac{4}{35}\ ,\\
      &F_{0, 2}=0\ , \qquad  &F_{2, 2}=-\frac{1}{5}\ ,\qquad &F_{4, 2}=\frac{2}{35}\ ,\\
        &F_{1, 3}=1\ ,\qquad &F_{3, 3}=-\frac{2}{5}\ ,\qquad &F_{5, 3}=\frac{2}{35}\ .
    \end{aligned}
\end{equation}
These coefficients, computed directly from the ACMC–5 version of the $d=5$ Myers–Perry metric, perfectly match those predicted by Eq.~\eqref{eq:dGenericSphericalB} and, in turn, by the general expression in Eq.~\eqref{eq:FFMParbitraryD}.  
The agreement confirms the validity of the multipolar structure derived in the amplitude formalism, thereby providing an independent geometric check of the conjecture connecting classical metrics and source form factors. From these form factors then, the corresponding gravitational multipole moments can be obtained straightforwardly using Eq.~\eqref{eq:GravitationalMultipoles}. Finally, we note that this procedure is not limited to the five–dimensional case. Indeed the same methodology can be systematically extended to higher dimensions, providing a powerful and general tool to extract multipole moments and to test the consistency of our amplitude–based framework in any spacetime dimension.

\section{Black rings with one angular momentum in harmonic gauge}\label{app:BR}

In spacetime dimensions $D>4$, the uniqueness theorem for black holes no longer holds. When non–spherical horizon topologies are allowed, the Myers-Perry family does not exhaust all possible rotating vacuum solutions. A notable example is provided by the black ring solutions of five–dimensional GR~\cite{Emparan:2001wn}, which describe rotating horizons of topology $S^1 \times S^2$. Restricting attention to the case with a single non–vanishing angular momentum, aligned along the $\varphi_1$–direction, the metric written in the standard ring coordinates $(x,y,\varphi_1,\varphi_2)$ takes the form
\begin{equation}\label{eq:BRoriginalMetric}
\begin{aligned}
ds^2=&-\frac{A(y)}{A(x)}\left(dt-C\, \mathcal{R}\frac{1+y}{A(y)}d\varphi_1\right)^2\\
&+\frac{\mathcal{R}^2}{(x-y)^2}A(x)\left(-\frac{B(y)}{A(y)}d\varphi_1^2-\frac{dy^2}{B(y)}+\frac{dx^2}{B(x)}+\frac{B(x)}{A(x)}d\varphi_2^2\right)\ ,
\end{aligned}
\end{equation}
where   
\begin{equation}
    A(z)=1+\lambda z\ ,\qquad B(z)=(1-z^2)(1+\nu z)\ ,\qquad C=\sqrt{\lambda(\lambda-\nu)\frac{1+\lambda}{1-\lambda}}\ ,
\end{equation}
and the dimensionless parameters $0<\nu\le\lambda<1$ control the shape and rotation of the ring, while $\mathcal{R}$ sets its characteristic radius. To remove naked conical singularities from the metric, one must impose the equilibrium condition
\begin{equation}\label{eq:BRequilibrium}
    \lambda=\frac{2\nu}{1+\nu^2}\ ,
\end{equation}
which ensures a perfect balance between centrifugal repulsion and gravitational self-attraction. However, in what follows we will not enforce this relation, so that the black ring solution remains characterized by three independent parameters $(\mathcal{R},\lambda,\nu)$. Our objective is to express this geometry in harmonic coordinates. To achieve this, we first introduce an intermediate transformation to asymptotically flat coordinates, defined by
\begin{equation}
\begin{gathered}
    x=-\left(\frac{1-\lambda}{1-\nu}\right)\frac{r^2-2\left(\frac{1-\lambda}{1-\nu}\right)\mathcal{R}^2\cos^2(\theta)}{r^2}\ , \qquad y=-\left(\frac{1-\lambda}{1-\nu}\right)\frac{r^2+2\left(\frac{1-\lambda}{1-\nu}\right)\mathcal{R}^2\sin^2(\theta)}{r^2}\ ,\\
    (\varphi_1, \varphi_2)=\frac{\sqrt{1-\lambda}}{1-\nu} (\phi_1, \phi_2)\ ,
    \end{gathered}
\end{equation}
so that the metric can be rewritten in the coordinates $(t,r,\theta,\phi_1,\phi_2)$, whose asymptotic structure approaches that of flat five–dimensional spacetime. At this stage, it is convenient to reparametrize the solution in terms of its physical parameters, namely the total mass $m$ and angular momentum $J$. From the large–$r$ behavior of the metric one reads  
\begin{equation}
    m=\frac{3\pi \mathcal{R}^2}{4G_N}\ ,\qquad J=\frac{\pi \mathcal{R}^3}{2G_N}\frac{\sqrt{\lambda(\lambda-\nu)(1+\lambda)}}{(1-\nu)^2}\ ,
\end{equation}
which correspond respectively to the mass monopole and the spin dipole of the configuration. Keeping $\lambda$ as a free dimensionless shape parameter, we can trade $(\mathcal{R},\nu)$ for $(m,J)$, so that the black ring is completely specified by the set $(m,J,\lambda)$.

We can now move to harmonic coordinates by employing the transformation~\eqref{eq:HarmonicCoordinatesD5}, combined with the definition of harmonic Cartesian coordinates in Eq.~\eqref{eq:HarmCartesianD5}, and imposing the harmonic condition~\eqref{eq:BoxD5}. Because of the ring’s more intricate topology, the ansatz for the harmonic transformation differs from the Myers–Perry case. For the reasons discussed in section~\ref{sec:BlackRing}, we restrict the expansion to first PM order and to quadratic order in the spin. Accordingly, the coordinate transformation is written as 
\begin{equation}\label{eq:BRansatz}
\begin{aligned} 
    &r(R, \Theta)=R\sum_{i=0}^{1}\left(G_N m\rho(R)\right)^i\sum_{j=0}^{1}\sum_{k=0}^{2}\mathcal{C}_{i, 2j, 2k}^{(R)}(\lambda)\left(\frac{a}{R}\right)^{2j}P_{2k}(\cos\Theta)+\mathcal{O}(G_N^2, a^3)\ , \\
    &\cos\theta(R, \Theta)=\cos\theta\sum_{i=0}^{1}\left(G_N m\rho(R)\right)^i\sum_{j=0}^{1}\sum_{k=0}^{2}\mathcal{C}_{i, 2j, 2k}^{(\Theta)}(\lambda)\left(\frac{a}{R}\right)^{2j}P_{2k}(\cos\Theta)+\mathcal{O}(G_N^2, a^3)\ ,
\end{aligned}
\end{equation}
where the coefficients $\mathcal{C}_{i,2j,2k}^{(R)}(\lambda)$ and $\mathcal{C}_{i,2j,2k}^{(\Theta)}(\lambda)$ now depend explicitly on the shape parameter $\lambda$. Among the various constraints that must be imposed on this ansatz, a key requirement is that the monopole part of the spatial metric remain independent of $\lambda$, since it encodes only the physical mass. This additional condition, not automatically satisfied by Eq.~\eqref{eq:BRansatz}, reduces the number of gauge–redundant parameters to two at 1PM order, in perfect analogy with the previous cases. After solving for the coefficients and applying the transformation, we obtain the black-ring metric in harmonic coordinates expressed in a convenient gauge, for which neglecting non–spin–induced quadrupoles reads
\begin{equation}\label{app:BRinHarm}
    \begin{aligned}
    g_{tt}&=-1+\frac{8 G_N m}{3\pi R^2}-\frac{12 G_N m a^2\lambda\cos (2 \Theta )}{\pi R^4(1+\lambda)}+\mathcal{O}(G_N^2, a^3)\ ,\\
    g_{t\Phi_1}&=-\frac{4 a G_N m \sin^2(\Theta)}{\pi R^2}+\mathcal{O}(G_N^2, a^3)\ ,\\
    g_{RR}&=1+\frac{4G_N m}{3\pi R^2}-\frac{6G_N ma^2\lambda\cos(2\Theta)}{\pi R^4(1+\lambda)}+\mathcal{O}(G_N^2, a^3)\ ,\\
    g_{\Theta\Theta}&=\frac{4G_N m}{3\pi}+R^2-\frac{2G_N ma^2(1+3\lambda\cos(2\Theta))}{\pi R^2(1+\lambda)}+\mathcal{O}(G_N^2, a^3)\ ,\\
    g_{\Phi_1\Phi_1}&=\frac{4G_N m\sin^2(\Theta)}{3\pi}+R^2\sin^2(\Theta)\\
    &-\frac{G_N ma^2\sin^2(\Theta)}{\pi R^2}\Big(-1+3(1+3\lambda)\cos(2\Theta)\Big)+\mathcal{O}(G_N^2, a^3)\ ,\\
    g_{\Phi_2\Phi_2}&=\frac{4G_N m\cos^2(\Theta)}{3\pi}+R^2\cos^2(\Theta)\\
    &-\frac{G_N ma^2\cos^2(\Theta)}{\pi R^2}\Big(-1+3(-1+3\lambda)\cos(2\Theta)\Big)+\mathcal{O}(G_N^2, a^3)\ .
    \end{aligned}
\end{equation}
The resulting expression can now be compared directly with the amplitude–based metric. We find complete agreement up to the considered order, confirming the consistency between the classical harmonic–coordinate reconstruction and the amplitude formulation. This match provides a stringent test of the amplitude–geometry correspondence for objects with non-trivial horizon topology. Although the essential physical information is already contained in the leading terms, the higher–order structure serves as a powerful validation of the method, demonstrating that the harmonic-coordinate framework correctly captures the long–range multipolar behavior of black rings and can be systematically extended to more general higher-dimensional solutions.

\backmatter

\cleardoublepage

\phantomsection 

\addcontentsline{toc}{chapter}{\bibname}

\bibliographystyle{sapthesis}
\bibliography{biblio} 
\end{document}